UNIVERSITÉ PARIS-SUD

ÉCOLE DOCTORALE 534 :
MODÉLISATION ET INSTRUMENTATION EN PHYSIQUE, ÉNERGIES, GÉOSCIENCES ET ENVIRONNEMENT

Laboratoire :
*Centre de Sciences Nucléaires et de Sciences de la Matière (Orsay, France)*
*Département de la Physique des Leptons - Institut de Recherche Nucléaire NASU (Kiev, Ukraine)*

# THÈSE DE DOCTORAT

PHYSIQUE

par

**Dmitry CHERNYAK (Dmytro CHERNIAK)**

## DÉVELOPPEMENT DE DETECTEURS CRYOGENIQUES A FAIBLE BRUIT DE FOND COMPOSÉS DE CRISTAUX SCINTILLATEURS ENRICHIS EN MOLYBDATE DE ZINC POUR LA RECHERCHE DE LA DOUBLE DESINTEGRATION BETA SANS NEUTRINOS DU $^{100}Mo$

**Date de soutenance : 08/07/2015**

**Composition du jury** :

| | | |
|---|---|---|
| Directeur de thèse : | Andrea GIULIANI | DR1 (CSNSM) |
| Co-directeur de thèse : | Fedor DANEVICH | Professeur (Institute for Nuclear Research) |
| Rapporteurs : | Corinne AUGIER | Professeur (UCBL Lyon 1 et IPNL) |
| | Alessandra TONAZZO | Directeur-adjoint du laboratoire (Laboratoire APC) |
| Examinateurs : | David LUNNEY | DR1 (CSNSM) |
| | Oleg PONKRATENKO | Professeur (Institute for Nuclear Research) |

**Institute for Nuclear Research**
**National Academy of Sciences of Ukraine**
**Joint Ph.D. thesis with the University of Paris Sud 11**

# DEVELOPMENT OF CRYOGENIC LOW BACKGROUND DETECTOR BASED ON ENRICHED ZINC MOLYBDATE CRYSTAL SCINTILLATORS TO SEARCH FOR NEUTRINOLESS DOUBLE BETA DECAY OF $^{100}$Mo

A thesis submitted for the degree of

DOCTOR of PHILOSOPHY in PHYSICS

by

Dmitry CHERNYAK (Dmytro CHERNIAK)

Host laboratories:
Lepton Physics Department – Institute for Nuclear Research NASU (Kiev, Ukraine)
Centre de Sciences Nucléaires et de Sciences de la Matière (Orsay, France)

*Supervisor from France: Prof. A. Giuliani*
*Supervisor from Ukraine: Dr. F. Danevich*

# ACKNOWLEDGEMENTS

First of all, I would like to express my deepest appreciation to my supervisors Dr. Fedor Danevich and Prof. Andrea Giuliani for their constant support, patience, wise guidance and useful suggestions during my PhD study.

I am grateful to all of my colleagues from the Lepton Physics Department of the Institute for Nuclear Research of the NASU, especially to Dr. Vladimir Tretyak, Dr. Vladislav Kobychev, Dr. Denys Poda and Valentyna Mokina for their valuable consultation, cooperation and assistance.

I would like to thank the staff of the Centre de Sciences Nucléaires et de Sciences de la Matière, who kindly supported me for three half-years of my work in the laboratory. Many thanks to Claire Marrache-Kikuchi, Stefanos Marnieros, Emiliano Olivieri, Rejane Bodson, Margherita Tenconi, Michele Mancuso and all of the group of the Physique des solides.

I gratefully acknowledge the cooperation with the Laboratoire Souterrain de Modane, CEA Saclay and Institut de Physique Nucleaire de Lyon. Special gratitude to Claudia Nones and Antoine Cazes for answering my questions.

I want to sincerely thank my dear friends Ruslan Podviyanuk and Eleonora Yerechshenko for the encouragement, support and advices.

My deepest gratitude to my family who make my scientific work possible: mother Mariya Chernyak and older brother Oleksandr.

Finally, the results presented in this PhD thesis were achieved only through the great efforts of many scientists from different countries, institutes and laboratories, so I would like also to thank them all for the hard work, persistence and fruitful cooperation.

# CONTENTS

# LIST OF ABBREVIATIONS

| | | |
|---|---|---|
| **δ** | — | isotopic concentration |
| **0ν2β** | — | neutrinoless double beta decay |
| **2ν2β** | — | two-neutrino double beta decay |
| **AAS** | — | atomic absorption spectroscopy |
| **ADC** | — | analog-to-digital converter |
| **AMoRE** | — | Advanced Mo based Rare process Experiment |
| **BEGe** | — | broad energy germanium detector |
| **BoPET** | — | biaxially-oriented polyethylene terephthalate |
| **BR** | — | branching ratio |
| **C.L.** | — | confidence level |
| **CSNSM** | — | Centre de Sciences Nucléaires et de Sciences de la Matière (Orsay, France) |
| **CUORE** | — | Cryogenic Underground Observatory for Rare Events |
| **EC** | — | electron capture |
| **EDELWEISS** | — | Expérience pour DEtecter Les Wimps En Site Souterrain (experiment to detect WIMPs in an underground site) |
| **ELEGANT** | — | ELEctron GAmma-ray Neutrino Telescope |
| **EXO** | — | Enriched Xenon Observatory |
| **FWHM** | — | full width at half maximum |
| **g.s.** | — | ground state of nucleus |
| **GERDA** | — | GERmanium Detector Array |
| **HM** | — | Heidelberg-Moscow experiment |
| **HPGe** | — | high-purity germanium detectors |
| **HPXe** | — | high-pressure xenon gas |
| **ICP-MS** | — | inductively coupled plasma mass-spectrometry |
| **IGEX** | — | International Germanium EXperiment |
| **IGP** | — | Institute of General Physics (Moscow, Russia) |
| **ISMA** | — | Institute for Scintillation Materials (Kharkhiv, Ukraine) |
| **KamLAND** | — | Kamioka Liquid Scintillator Antineutrino Detector |
| **KamLAND-Zen** | — | KamLAND Zero-Neutrino Double-Beta Decay |
| **LAr** | — | liquid argon |
| **LNGS** | — | Laboratori Nazionali del Gran Sasso (Gran Sasso National Laboratory, Assergi, Italy) |
| **LSM** | — | Laboratoire Souterrain de Modane (Modane Underground Laboratory, Modane, France) |
| **LTG Cz** | — | low-thermal-gradient Czochralski technique for crystal growing |
| **LUCIFER** | — | Low-background Underground Cryogenic Installation For Elusive Rates |
| **LUMINEU** | — | Luminescent Underground Molybdenum Investigation for NEUtrino mass and nature |
| **LXe** | — | liquid xenon |
| **m.w.e.** | — | meters of water equivalent |
| **MMC** | — | metallic magnetic calorimeter |
| **NEMO** | — | Neutrino Ettore Majorana Observatory |
| **NEXT** | — | Neutrino Experiment with a Xenon TPC |
| **NIIC** | — | Nikolaev Institute of Inorganic Chemistry (Novosibirsk, Russia) |
| **NME** | — | nuclear matrix element |
| **NTD** | — | Neutron Transmutation Doped germanium thermistor |
| **PMNS** | — | Pontecorvo-Maki-Nakagawa-Sakata matrix |
| **PMT** | — | photomultiplier tube |
| **ppb** | — | parts-per-billion, $10^{-9}$ |
| **ppm** | — | parts-per-million, $10^{-6}$ |
| **ppt** | — | parts-per-trillion, $10^{-12}$ |
| **PSD** | — | pulse-shape discrimination |
| **PTFE** | — | polytetrafluoroethylene |
| $\boldsymbol{Q_{2\beta}}$ | — | energy of the 2β decay |
| **R&D** | — | research and development |
| **RE** | — | rejection efficiency of randomly coinciding signals in cryogenic bolometers |
| **ROI** | — | region of interest |
| **SM** | — | Standard model |
| **SNO** | — | Sudbury Neutrino Observatory |
| $\boldsymbol{T_{1/2}}$ | — | half-life. For the 2ν2β and 0ν2β decay modes the half-life designated as $T_{1/2}^{2\nu2\beta}$ and $T_{1/2}^{0\nu2\beta}$ respectively |
| **TPC** | — | time projection chamber |

**TSL** — thermo-stimulated luminescence

**wt%** — weight percentage

**Xe-LS** — xenon-loaded liquid scintillator

**$ZnMoO_4$** — zinc molybdate scintillating crystal

# INTRODUCTION

Double beta decay (2β) is a rare nuclear transition which changes the nuclear charge *Z* by two units. There exist 69 naturally-occurring even-even nuclei for which such a process is possible, in particular there are 35 $2\beta^-$ isotopes among them. While the transition involving two electrons and two (anti)neutrinos (2ν2β) is allowed by the Standard Model (SM) and has been already observed for 11 isotopes, this is not the same for the neutrinoless double beta decay (0ν2β), which can occur only if neutrino is a massive Majorana particle. Observations of neutrino oscillations give a clear evidence that neutrino is a massive particle but only the measurement of 0ν2β decay could establish the Majorana nature of the neutrino, help solving neutrino hierarchy problem, measure the effective Majorana mass and test lepton number conservation. Moreover, this process could clarify the presence of right-handed currents admixture in weak interaction, and prove the existence of Majorons. 0ν2β decay can be mediated by many hypothetical processes beyond the Standard Model. Taking into account the uncertainty of the theoretical estimations of 0ν2β decay probability, development of experimental methods for different 2β isotopes is highly requested.

$^{100}$Mo is one of the most promising 2β isotopes because of its large transition energy $Q_{2\beta}$ = 3034.40(17) keV [1] and a considerable natural isotopic abundance δ = 9.8%. From the experimental point of view a large $Q_{2\beta}$ value simplifies the problem of background induced by natural radioactivity and cosmogenic activation.

At present the best sensitivity to 0ν2β decay of $^{100}$Mo was reached by the NEMO-3 experiment that, with ≈ 7 kg of enriched $^{100}$Mo, has obtained a half-life limit $T_{1/2}^{0\nu2\beta}$ > 1.1 × $10^{24}$ yr at 90% confidence level (C.L.). Despite this valuable result, the NEMO technique has two disadvantages that limit its sensitivity: a low detection efficiency (≈ 14% to 0ν2β events) and rather poor energy resolution (≈ 10% at the energy of $Q_{2\beta}$ of $^{100}$Mo). The detection efficiency can be improved up to 80%−90% in so called "source=detector" approach by using detector containing molybdenum. Only cryogenic bolometers and semiconductor high-purity germanium (HPGe) detectors can provide high enough energy resolution (a few keV at $Q_{2\beta}$). However, the energy region above 2615 keV (the highest gamma line in natural radioactivity) is dominated by alpha particles which can cause a significant background. Nevertheless, the simultaneous detection of phonon and scintillation signals in cryogenic scintillating bolometers allows efficient particle discrimination important to reject background caused by α particles. Therefore, the cryogenic scintillating bolometers are highly promising detectors to search for neutrinoless double beta decay in the next-generation experiments.

There are several inorganic scintillators containing molybdenum. The most promising of them are molybdates of Calcium ($CaMoO_4$), Cadmium ($CdMoO_4$), Lead ($PbMoO_4$), Strontium ($SrMoO_4$), Lithium ($Li_2MoO_4$) and Zinc ($ZnMoO_4$). However $CaMoO_4$ contains the 2ν2β active isotope $^{48}$Ca which, even if present in natural calcium with a very small abundance of δ = 0.187%, creates background at $Q_{2\beta}$ energy of $^{100}$Mo. $CdMoO_4$ contains the β active $^{113}$Cd ($T_{1/2}^{0\nu2\beta}$ = 8.04 × $10^{15}$ yr, δ = 12.22%) which, besides being beta active, has a very high cross section to capture thermal neutrons. A potential disadvantage of $PbMoO_4$ is that $^{100}$Mo would be only 27% of the total mass. Current possibilities of $SrMoO_4$ production are far from the required mass and quality of the crystals and possible contamination by anthropogenic $^{90}Sr$. The important advantages of $Li_2MoO_4$ are the highest concentration of molybdenum among the other molybdate crystal scintillators (55.2% in weight), absence of natural long-living radioactive isotopes and comparatively easy crystal growth process [2]. However, the light yield of $Li_2MoO_4$ scintillators is the lowest among all the mentioned molybdate crystals. The important advantage of $ZnMoO_4$ crystals is the absence of heavy elements and high concentration of molybdenum (43% in weight).

Development of large volume zinc molybdate scintillators with a low level of radioactive contamination, high bolometric properties, scintillation efficiency and optical quality from enriched $^{100}Mo$ is required to search for 0ν2β decay of $^{100}Mo$. The production of improved quality $ZnMoO_4$ scintillators calls for development of deep purification techniques of molybdenum and optimization of crystal growth technique. Zinc molybdate crystal scintillators with the mass of a few hundred grams were developed only very recently. $ZnMoO_4$ scintillators with a mass of ~ 0.3 kg, as well as $Zn^{100}MoO_4$ crystals enriched in the isotope $^{100}Mo$ were produced for the first time by using the low-thermal-gradient Czochralski technique. The optical and luminescent properties of the produced crystals were studied to estimate the progress in crystal growth quality. The low-temperature tests with a 313 g $ZnMoO_4$ scintillator and two enriched $Zn^{100}MoO_4$ were performed aboveground in the Centre de Sciences Nucléaires et de Sciences de la Matière (CSNSM, Orsay, France). The low background measurements with three $ZnMoO_4$ and two enriched detectors installed in the EDELWEISS set-up at the Laboratoire Souterrain de Modane were also carried out as a part of this PhD thesis.

To optimize the light collection in $ZnMoO_4$ cryogenic scintillating bolometers, we have simulated the collection of scintillation photons in a detector module for different geometries by Monte Carlo method using the GEANT4 package. Response to the 2ν2β decay of $^{100}Mo$ was simulated for the enriched $Zn^{100}MoO_4$ detectors of different shape and mass to understand the dependence of 2ν2β decay spectra on crystal shape. Finally we simulated 48 $Zn^{100}MoO_4$ crystals with a size of Ø60 × 40 mm installed in the EDELWEISS cryostat. The contribution to background from the internal radioactive contamination of the crystals, cosmogenic activation and radioactive contamination of the set-up were simulated.

Taking into account the poor time resolution of the low temperature bolometers, we also studied contribution to background at the $Q_{2\beta}$ energy of random coincidences of signals, in particular of 2ν2β decay, which is one of the most valuable sources of background in cryogenic bolometers. Methods of the randomly coinciding events rejection were developed and compared. We have also analyzed dependence of the rejection efficiency on a cryogenic detector performance (in particular on the signal to noise ratio and the sampling frequency of the data acquisition).

# CHAPTER 1

# THEORY AND EXPERIMENTAL STATUS OF 2β DECAY

## 1.1. Double beta decay

Neutrino was first postulated by W. Pauli in 1930 to explain beta decay. However, even after more than 80 years, mass and nature of neutrino (Dirac or Majorana particle?) are still unknown.

Several neutrino oscillation experiments finally proved that neutrinos are massive particles. Also these results demonstrated that the Standard Model (SM) of electroweak interactions is incomplete and should be extended to include massive neutrinos.

Neutrinoless double beta decay (0ν2β) is one of the most promising ways to understand nature and properties of neutrino. Experiments to search for this extremely rare decay process can probe lepton number conservation, investigate the Dirac/Majorana nature of the neutrinos and their absolute mass scale with unprecedented sensitivity.

### 1.1.1. Fundamentals of double beta decay theory

Double beta decay (2β) is a rare nuclear process in which an initial nucleus with a mass $A$ decays to a member of the same isobaric multiplet with the change of the nuclear charge $Z$ by two units (see Fig. 1.1). Such transition was first suggested by M. Goeppert-Mayer in 1935 and the half-life of such process was estimated, by using Fermi theory of β decay, to be > $10^{17}$ years [3]. 2β decay is a second-order process in the Standard Model of electroweak interactions, which explains a very low decay probability.

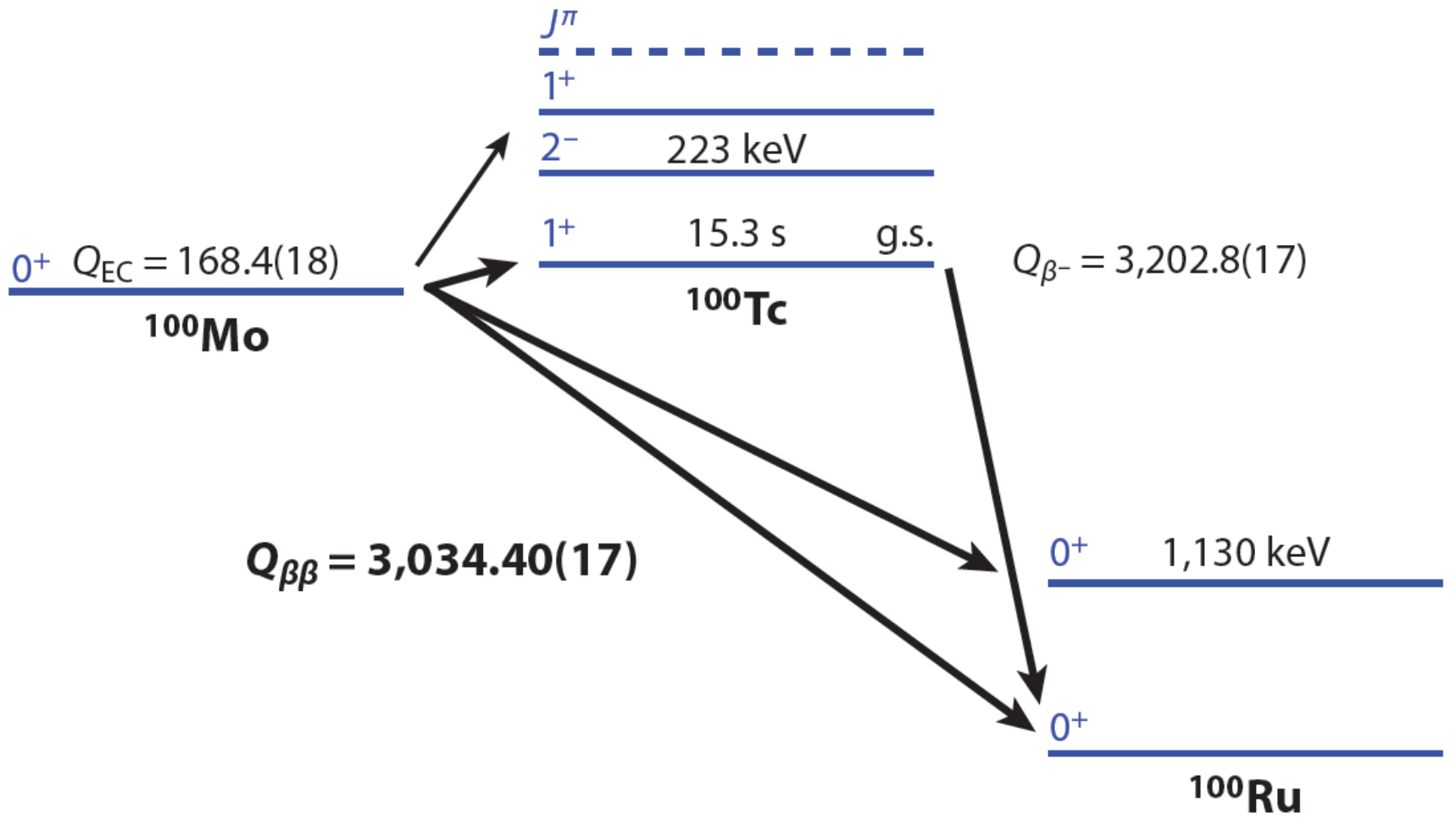

Fig. 1.1. Nuclear-level diagram of the double beta decay of the isotope $^{100}$Mo to the ground state and the first excited $0^+$ level of $^{100}$Ru [4].

The 2β decay is possible when the decay to the intermediate nucleus is energetically forbidden due to the pairing interaction, or if an ordinary β decay is suppressed by a large difference in spin between these nuclei. The decay can proceed between two even-even isobars (see Fig. 1.2) from the initial nucleus to the ground state (g.s.) or to the first excited states of the daughter nucleus. Therefore the 2β decay is possible for 69 naturally occurring even-even nuclei, in particular there are 35 $2\beta^-$ isotopes [5, 6].

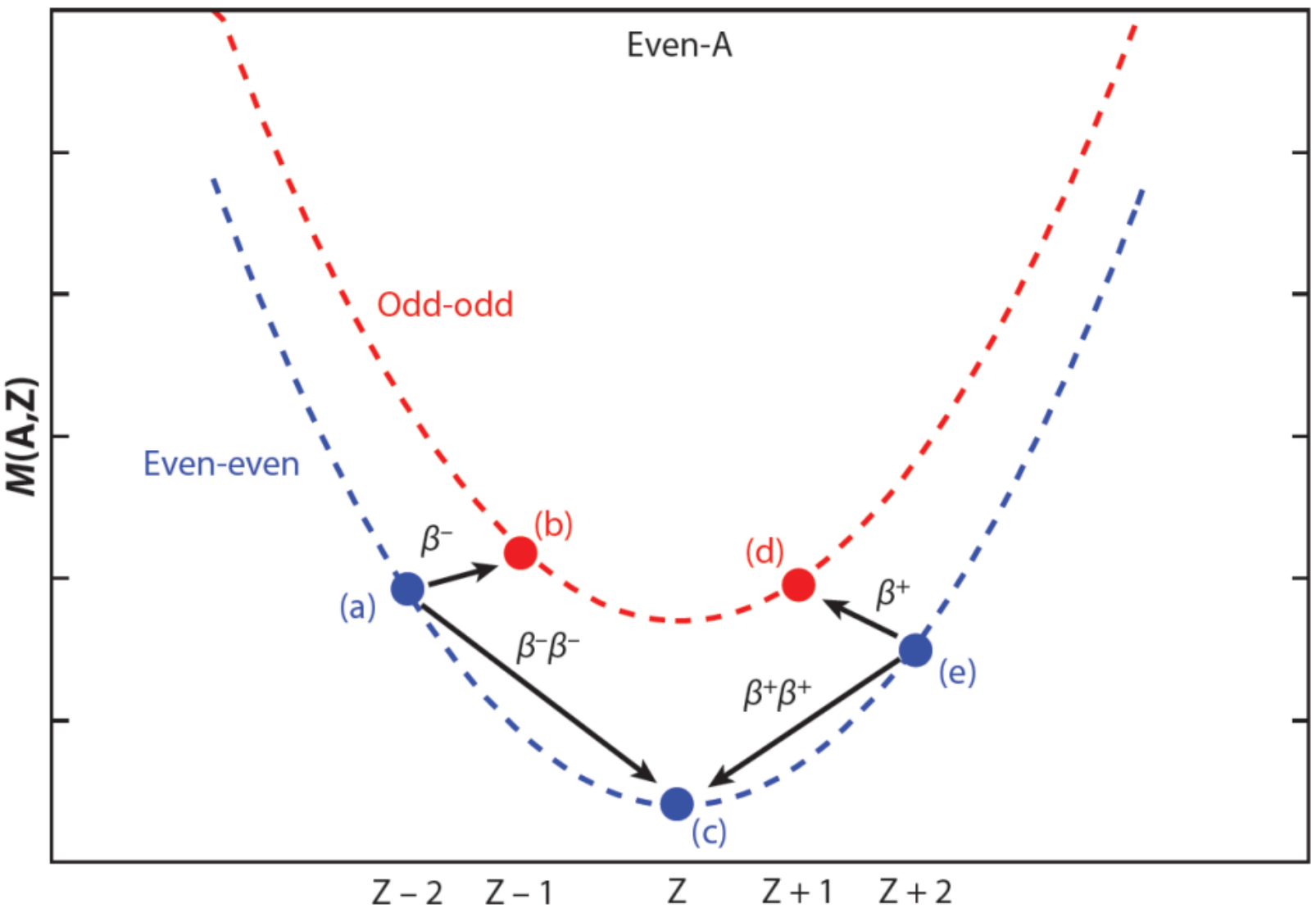

Fig. 1.2. Mass parabolas for nuclear isobars with even *A*. Due to the pairing interaction in the semiempirical mass formula, even-even nuclei have lower masses than odd-odd nuclei. Therefore, $\beta^-$ decay is impossible from point (a) to point (b), though $2\beta^-$ decay is energetically possible as a second-order process from point (a) to point (c). Likewise, $2\beta^+$ decay, double-electron capture or electron capture with emission of one positron processes can take place between point (e) and point (c) [7].

The two-neutrino double beta decay is allowed in the Standard Model of electroweak interactions with the following possible channels:

$$2\beta^-:\quad (A,Z) \to (A,Z+2) + 2e^- + 2\bar{\nu}_e, \tag{1.1}$$

$$2\beta^+:\quad (A,Z) \to (A,Z-2) + 2e^+ + 2\nu_e, \tag{1.2}$$

$$\mathrm{ECEC}: 2e^- + (A,Z) \to (A,Z-2) + 2\nu_e, \tag{1.3}$$

$$\mathrm{EC}\beta^+:\; e^- + (A,Z) \to (A,Z-2) + e^+ + 2\nu_e, \tag{1.4}$$

where the electron capture (EC) is expected mainly on a K-shell. The energy released in these decays is distributed between the leptons and recoil of the nucleus, which is negligible. The $Q_{2\beta}$ energy of the 2ν2β decay for the different modes is as following:

$$2\beta^-:\quad M(A,Z) - M(A,Z+2), \tag{1.5}$$

$$2\beta^+:\quad M(A,Z) - M(A,Z-2) - 4m_e c^2, \tag{1.6}$$

$$\mathrm{ECEC}: M(A,Z) - M(A,Z-2) - 2\varepsilon, \tag{1.7}$$

$$\mathrm{EC}\beta^+:\; M(A,Z) - M(A,Z-2) - 2m_e c^2 - 2\varepsilon, \tag{1.8}$$

where *M(A,Z)* is the atomic mass of the 2β decay isotope, and *ε* is the excitation energy of the atomic shell of the daughter nucleus.

Moreover, there are also considered neutrinoless $2\beta^-$ decay modes (we will not mention here $2\beta^+$, ECEC and $\mathrm{EC}\beta^+$ possible processes):

- neutrinoless double beta decay (0ν2β)

$$^{A}_{Z}X \to {}_{Z+2}^{\;\;A}X + 2e^-, \tag{1.9}$$

- 2β decay with emission of one or more neutral bosons $\chi$ (Majorons)

$$^{A}_{Z}X \to {}_{Z+2}^{\;\;A}X + 2e^- + \chi. \tag{1.10}$$

The neutrinoless double beta decay (1.9) is the most interesting process since no neutrinos are emitted (see Fig. 1.3). This hypothetical process was first considered by Racah [8] and Furry [9] at the end of 1930s to understand if neutrino is Majorana (particle ≡ antiparticle) or Dirac (particle ≠ antiparticle) type. The 0ν2β decay violates the lepton number by two units and is forbidden in the Standard Model. Furthermore, such process is possible only when neutrinos are massive Majorana particles. Indeed, the antineutrino $\bar{\nu}_R$ is right-handed in the emission process $n_1 \to p_1 + e_1^- + \bar{\nu}_R$, while in the absorption process $n_2 + \nu_L \to p_2 + e_2^-$ the neutrino $\nu_L$ is left-handed (see Fig. 1.3). Therefore, when neutrino has a non-zero mass, it would not have a definite helicity in the two processes and a match of their helicities becomes possible.

There considered also other mechanisms of neutrinoless double beta process, for example, mediated by right-handed currents with the exchange of both light and heavy neutrinos, or the exchange of other exotic particles.

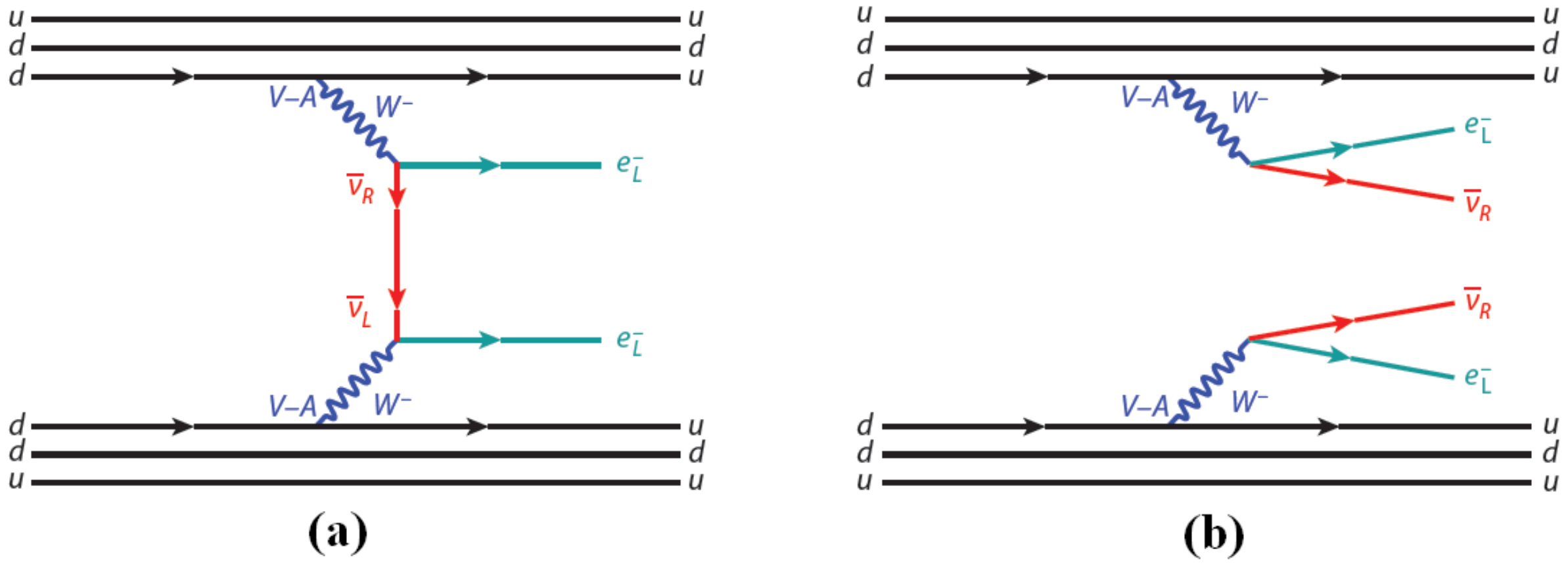

Fig. 1.3. Feynman diagrams of (a) 0ν2β decay with the exchange of light Majorana neutrino, and (b) 2ν2β decay.

The transition (1.10) is interesting from the point of existence of the so-called Majorons, the Nambu-Goldstone bosons that arise upon a global breakdown of B-L symmetry [10]. The Majorons can be described as a hypothetical neutral pseudoscalar particle with zero (or very small) mass, which can be emitted in 0ν2β decay [11, 12, 13]. The energy spectra of the process (1.10) are continuous, since Majorons do not interact with an ordinary matter and cannot be detected. However, the energy distributions have different maxima energies compare to the two-neutrino 2β decay mode, so the different channels of 2β processes can be separated by analysis of their energy spectra (see Fig. 1.4).

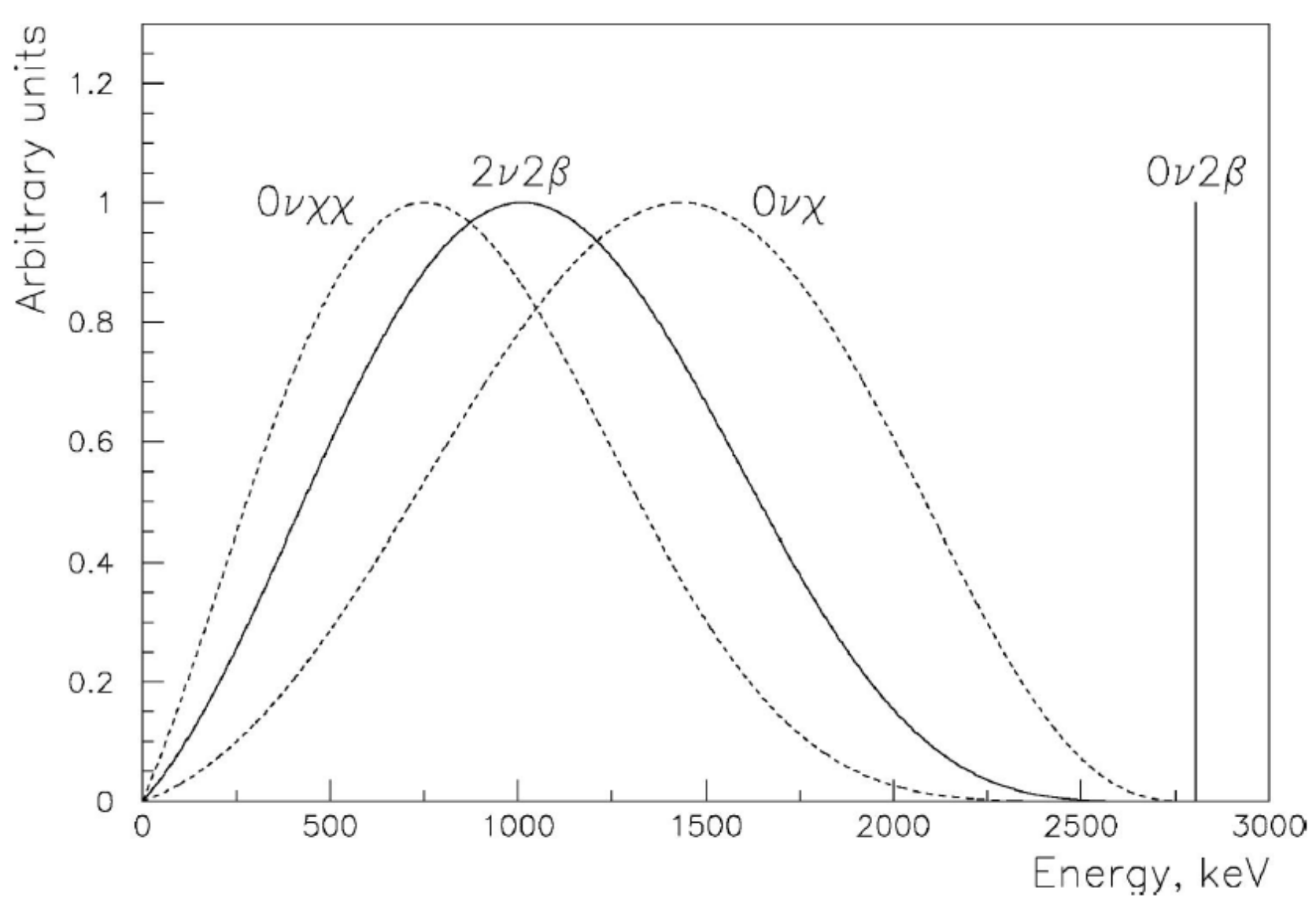

Fig. 1.4. The electron sum energy spectra calculated for the different double beta decay modes of the isotope $^{116}Cd$ [14].

### 1.1.2. Properties of neutrino and weak interactions

The first observation of the solar neutrinos in the 1960s of the last century by Ray Davis [15] showed a clear disagreement with the Standard Solar Model [16]. The chlorine experiment was sensitive only to the electron neutrinos and the measured neutrino flux was lower than expected. To explain the solar neutrino problem the hypothesis of neutrino oscillations was proposed [17]. In the following past decades many experiments (Homestake [18], GALLEX/GNO [19], SAGE [20], Super-Kamiokande [21], SNO [22]) were performed to investigate the neutrino disappearance problem, and they confirmed that the electron neutrino flux decreases, which can be explained by oscillation of the electron neutrinos into muon and tau neutrinos. The following studies of the neutrinos demonstrated the oscillation from muon neutrino into tau and electron neutrinos (K2K [23] and T2K [24] accelerator experiments, OPERA experiment [25]), as well as the oscillation of the electron antineutrino (KamLAND [26] reactor experiment). Thanks to these experiments, today we know that neutrinos are massive particles and the flavors (electron, muon and tau) are mixing among mass eigenstates.

The mixing of neutrinos can be described with a Pontecorvo-Maki-Nakagawa-Sakata matrix (PMNS) [17, 27, 28] with the following equation:

$$\nu_l = \sum_i U_{li}\, \nu_i, \qquad (1.11)$$

where $l = e, \mu, \tau$; $i$ starts from 1 to the number of mass eigenstates and $U_{li}$ is a PMNS matrix element. In the most general case there are three mass eigenstates, and PMNS is a 3 × 3 unitary matrix characterized by the three mixing angles ($\theta_{12}$, $\theta_{23}$ and $\theta_{13}$) and three CP-violating phases ($\delta$, $\alpha_1$ and $\alpha_2$):

$$U_{li} = \begin{pmatrix} c_{12}c_{13} & s_{12}c_{13} & s_{13}e^{-i\delta} \\ -s_{12}c_{23} - c_{12}s_{23}s_{13}e^{i\delta} & c_{12}c_{23} - s_{12}s_{23}s_{13}e^{i\delta} & s_{23}c_{13} \\ s_{12}s_{23} - c_{12}c_{23}s_{13}e^{i\delta} & -c_{12}s_{23} - s_{12}c_{23}s_{13}e^{i\delta} & c_{23}c_{13} \end{pmatrix} \times \mathrm{diag}(1, e^{i\alpha_1}, e^{i\alpha_2}) \qquad (1.12)$$

where $c_{ij} \equiv \cos\theta_{ij}$ and $s_{ij} \equiv \sin\theta_{ij}$. It should be stressed, that in case of the Dirac neutrino particles there are only 4 free parameters in PMNS matrix, due to the $\alpha_1$ and $\alpha_2$ Majorana phases can be reabsorbed by a rephasing of the neutrino fields.

The probabilities of the neutrino oscillation are described with mixing angles and square mass differences of the eigenstates: $\Delta m_{ij}^2 = m_j^2 - m_i^2$. The difference $\Delta m_{12}^2 > 0$ and mixing angle $\theta_{12}$ describes the solar neutrino mixing ($\nu_e \to \nu_{\mu,\tau}$), while $\Delta m_{23}^2$ and $\theta_{23}$ describes the mixing observed with atmospheric neutrinos ($\nu_\mu \to \nu_\tau$). From the results of the oscillation experiments we obtained that $\Delta m_{12}^2 \ll \Delta m_{23}^2$, which allows three different mass patterns (see Fig. 1.5 and Fig. 1.6): direct hierarchy ($m_1 < m_2 \ll m_3, \Delta m_{23}^2 > 0$), inverted hierarchy ($m_3 \ll m_1 < m_2, \Delta m_{23}^2 < 0$), and degenerate hierarchy ($m_1 \simeq m_2 \simeq m_3$) [29, 30, 31, 32].

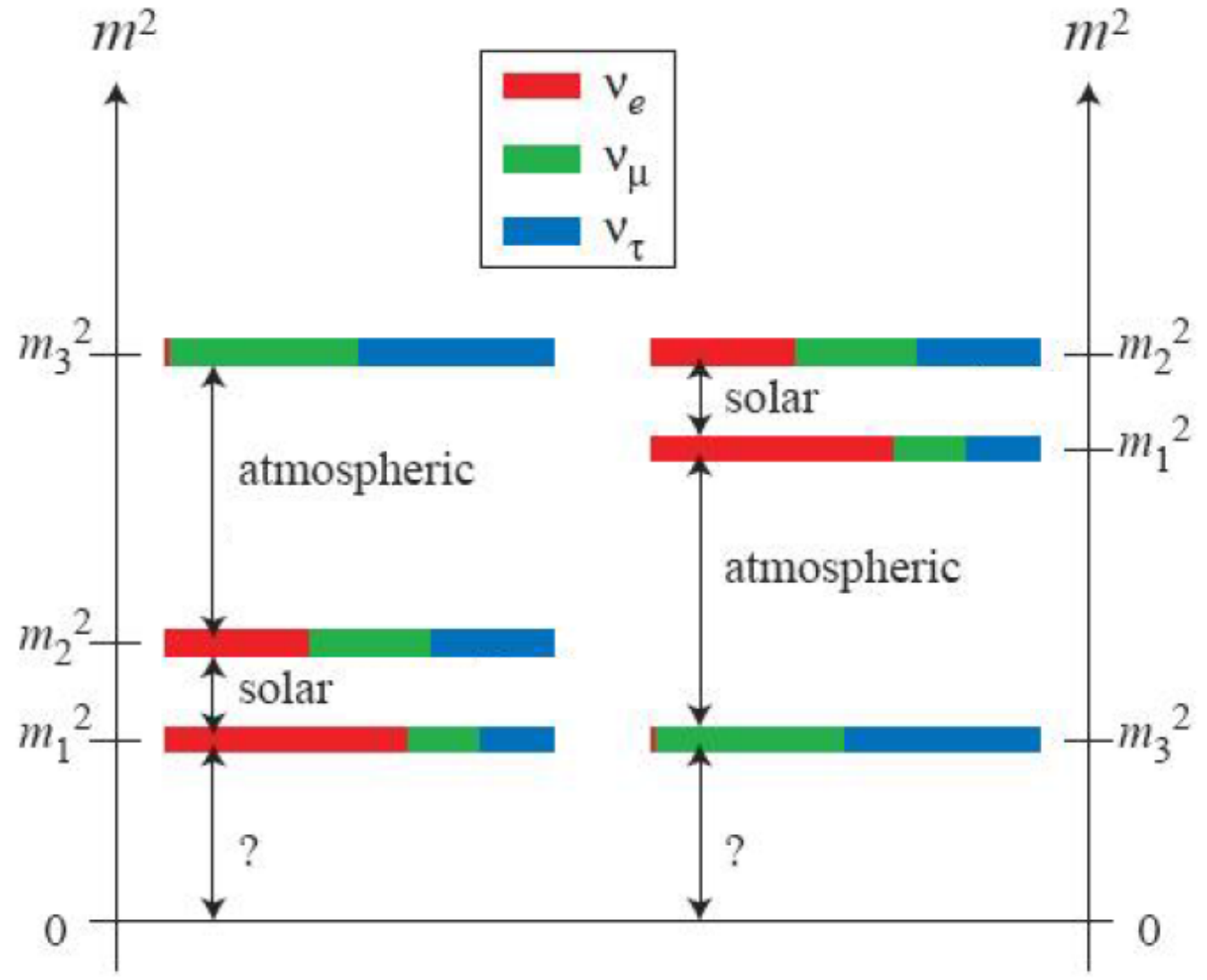

Fig. 1.5. Normal and inverted hierarchies of the mass patterns and neutrino flavor composition of the mass eigenstates.

The neutrino mass hierarchy problem can be solved in the future oscillation experiments [33]. However they will not provide information on the absolute scale of neutrino mass which is presently provided by experimental measurements of the effective sum of neutrino masses (Σ), electron neutrino mass ($m_\beta$) and effective Majorana mass ($|\langle m_\nu\rangle|$).

The upper limits on the sum of neutrino masses Σ comes from the cosmology by fitting the experimental data with different complex models. Therefore the limits are strongly dependent on the model, and most recent results of the Σ upper limit ranges from ~ 1 eV to 0.23 eV depending on the set of data and models used in calculations [34]. The lower limit of the Σ on the level of ~ 0.04 eV was obtained by oscillation experiments.

The electron neutrino mass $m_\beta$ can be measured in beta decay by studying the end point of Kurie plot. The tritium experiments give the upper limit on $m_\beta$ at the level of 2 eV at 95% confidence level (C.L.) [35, 36]. In near future another spectrometer experiment KATRIN should improve the sensitivity to ~ 0.2 eV [37]. There are also calorimetric experiments to measure the neutrino mass $m_\beta$ [38, 39].

Finally, the value of the effective Majorana mass $|\langle m_\nu\rangle| = \left|\sum_i U_{li}^2 m_i\right|$ can be measured if neutrino is Majorana particle in neutrinoless double beta decay. Even if 0ν2β

decay will not be experimentally observed with an upper limit on the effective Majorana mass lower than ~ 0.01 eV (see Fig. 1.6), then the normal hierarchy of neutrino mass pattern will be confirmed.

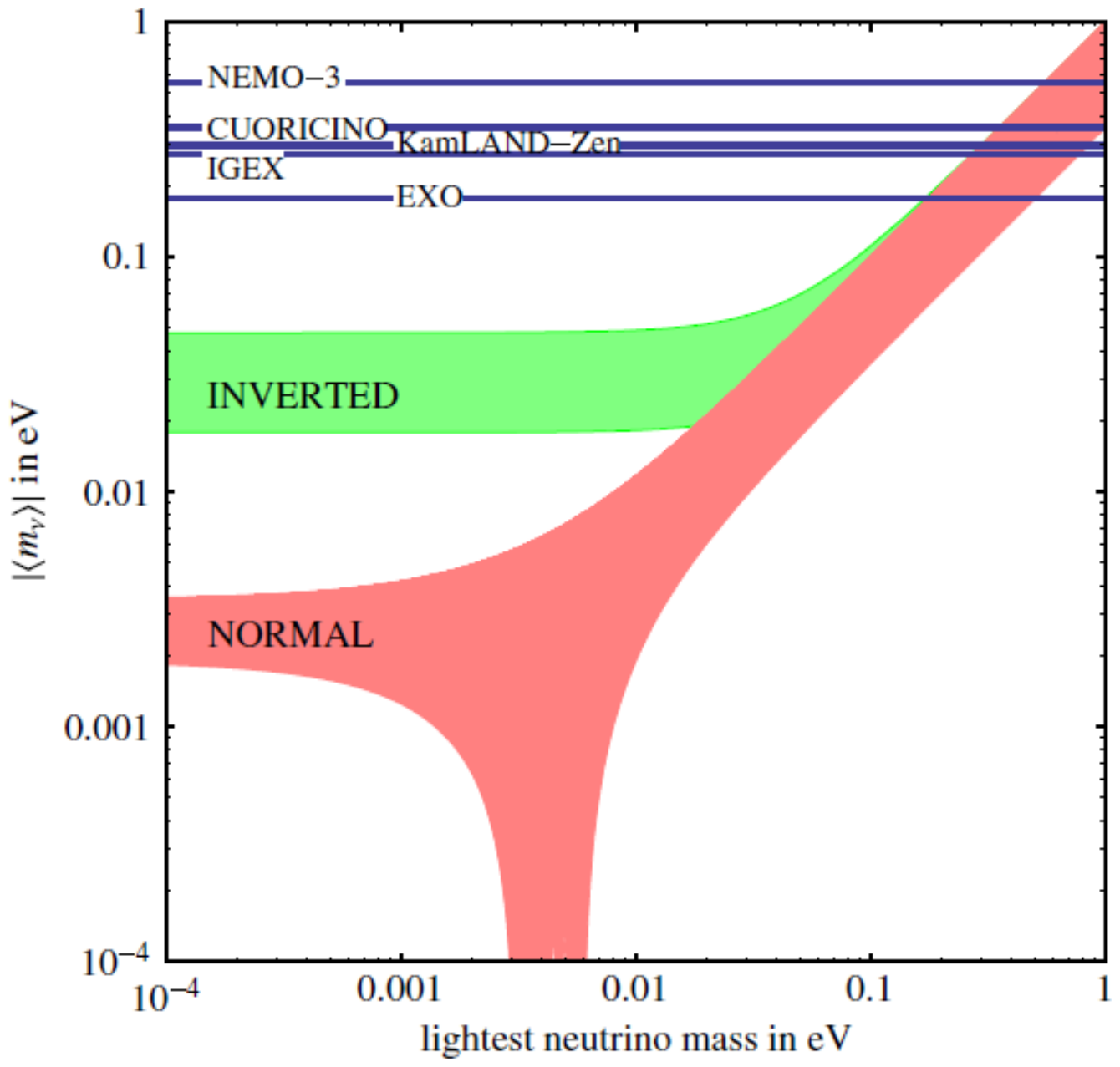

Fig. 1.6. The effective Majorana mass as a function of the smallest of the three mass eigenvalues. The upper limits on $\langle m_\nu \rangle$ are taken from CUORICINO [40], IGEX [41], NEMO-3 [42], KamLAND-Zen [43], EXO [44] experiments, and IBM-2 nuclear matrix elements [45].

### 1.1.3. Probability of 2β decay

The probability of two-neutrino 2β decay can be expressed through the half-life with the following formula [12, 13, 46]:

$$\left(T_{1/2}^{2\nu}\right)^{-1} = G^{2\nu} \cdot |M^{2\nu}|^2, \tag{1.13}$$

where $G^{2\nu}$ is the phase space factor, and $M^{2\nu}$ is the nuclear matrix element (NME) of the 2ν2β decay. The phase space factor can be precisely calculated [11, 13], while the NME can be obtained either from calculations, either directly by measuring the rate of 2ν2β decay [12, 13]. This gives a possibility to test the details of nuclear structure by comparing the calculated and the measured NME values. It should be also noted, that the probability of $2\beta^-$ decay (1.1) is higher than for the other 2ν2β decay modes (1.2–1.4), which have smaller phase space $G^{2\nu}$ and, as a result, are more difficult to observe experimentally.

The neutrinoless 2β decay rate can be written as

$$\left(T_{1/2}^{0\nu}\right)^{-1} = G^{0\nu} \cdot |M^{0\nu}|^2 \cdot |\langle m_\nu \rangle|^2 / {m_e}^2, \tag{1.14}$$

where $G^{0\nu}$ is the phase space integral of the 0ν2β decay, $M^{0\nu}$ is the nuclear matrix element, $m_e$ is the electron mass, and $\langle m_\nu \rangle$ is the effective Majorana mass of neutrino, which can be expressed through the elements of PMNS matrix as

$$\langle m_\nu \rangle = c_{12}^2 c_{13}^2 m_1 + s_{12}^2 c_{13}^2 e^{i\alpha_1} m_2 + s_{13}^2 e^{i\alpha_2} m_3. \tag{1.15}$$

As we can see from the formula (1.14), the effective Majorana mass of neutrino can be obtained from experimentally measured half-life, theoretically evaluated phase space factor and nuclear matrix element. However, while the phase space integral $G^{0\nu}$ can be precisely calculated [47, 48, 49], only approximate evaluations of the NME were obtained. Moreover, these estimations are often in disagreement among each other, which leads to the significant uncertainties of $\langle m_\nu \rangle$.

The difficulties in the calculations of NME are due to the nuclear many-body problem. There were several methods to calculate nuclear matrix elements for 0ν2β decay: the quasi-particle random phase approximation (QRPA, Renormalized QRPA – RQRPA, proton-neutron pairing – pnQRPA, etc.) [50, 51, 52], the nuclear shell model (NSM) [53, 54], the interacting boson model (IBM) [55], the generating coordinate method (GCM) [56], projected-Hartree-Fock-Bogoliubov model (PHFB) [57]. The results of 2ν2β decay can provide an important information used to calibrate NME calculations [58], because two-neutrino 2β decay is the closest nuclear process to 0ν2β decay. However, as we can see from the theoretical estimations of 0ν2β half-lifes presented in Fig. 1.7 (for axial-vector coupling constant $g_A$ = 1) and Fig. 1.8 (for $g_A$ = 1.25), the results of NME calculations are still into significant disagreement. Moreover, it is not possible to understand which method is closer to the reality at this moment, thus the experimental studies of 2β decay processes in variety of nuclei are necessary.

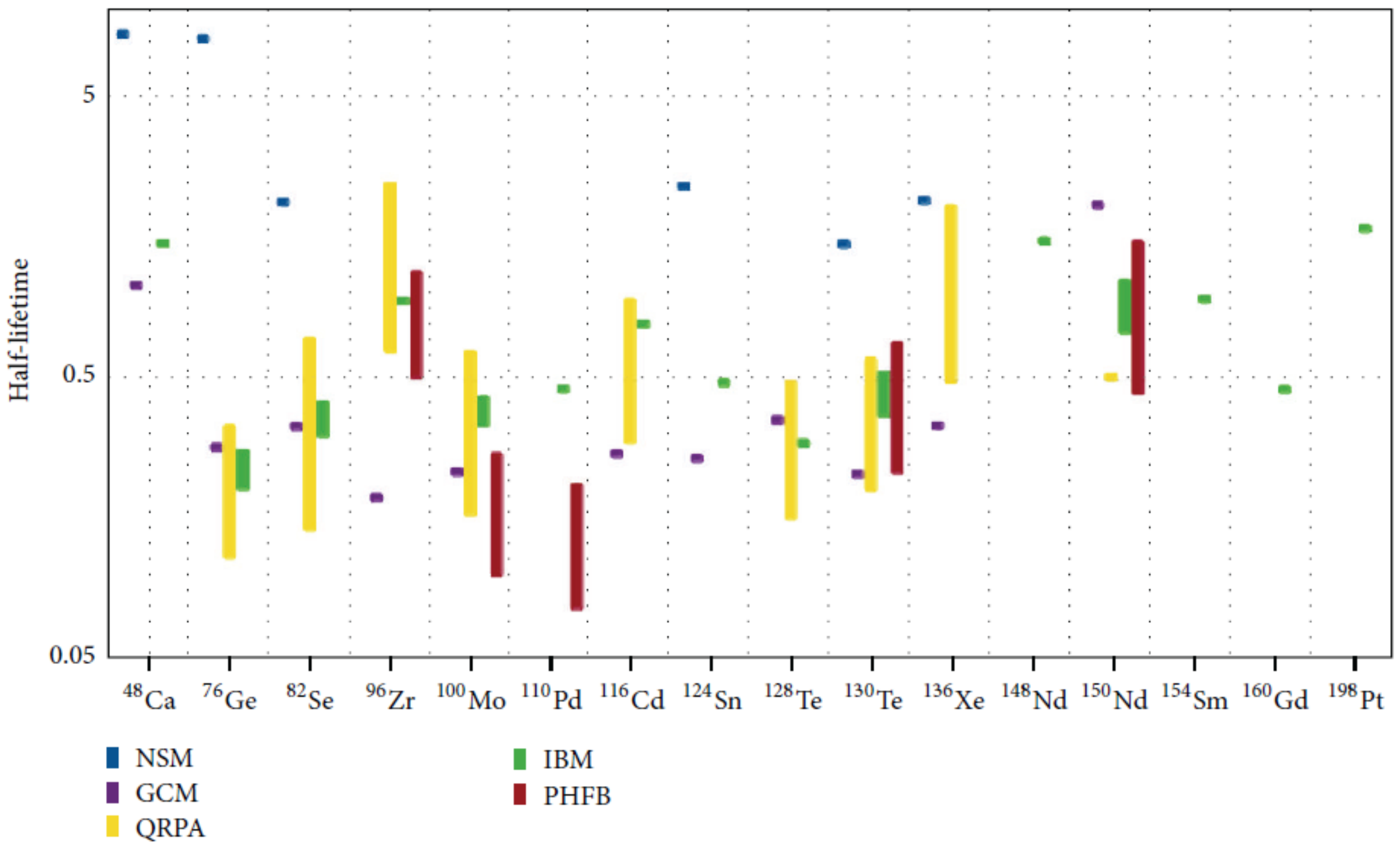

Fig. 1.7. The half-lifes of 0ν2β decay (in units of $10^{26}$ yr), evaluated for the different NME calculation methods [59] for the effective Majorana mass $|\langle m_\nu \rangle| = 50$ meV and the axial-vector coupling constant $g_A$ = 1. Bars refer to ranges of calculated values for the same model with different parametrizations. Single calculations are marked with dots. Discrepancies among different calculations are of the order of a factor 2-3.

It should be stressed that nuclear matrix elements are required not only to extract the value of effective Majorana mass of neutrino, but also to compare sensitivities and results of the experiments with different nuclei. Therefore we can conclude that NME are one of the crucial problems of 0ν2β decay.

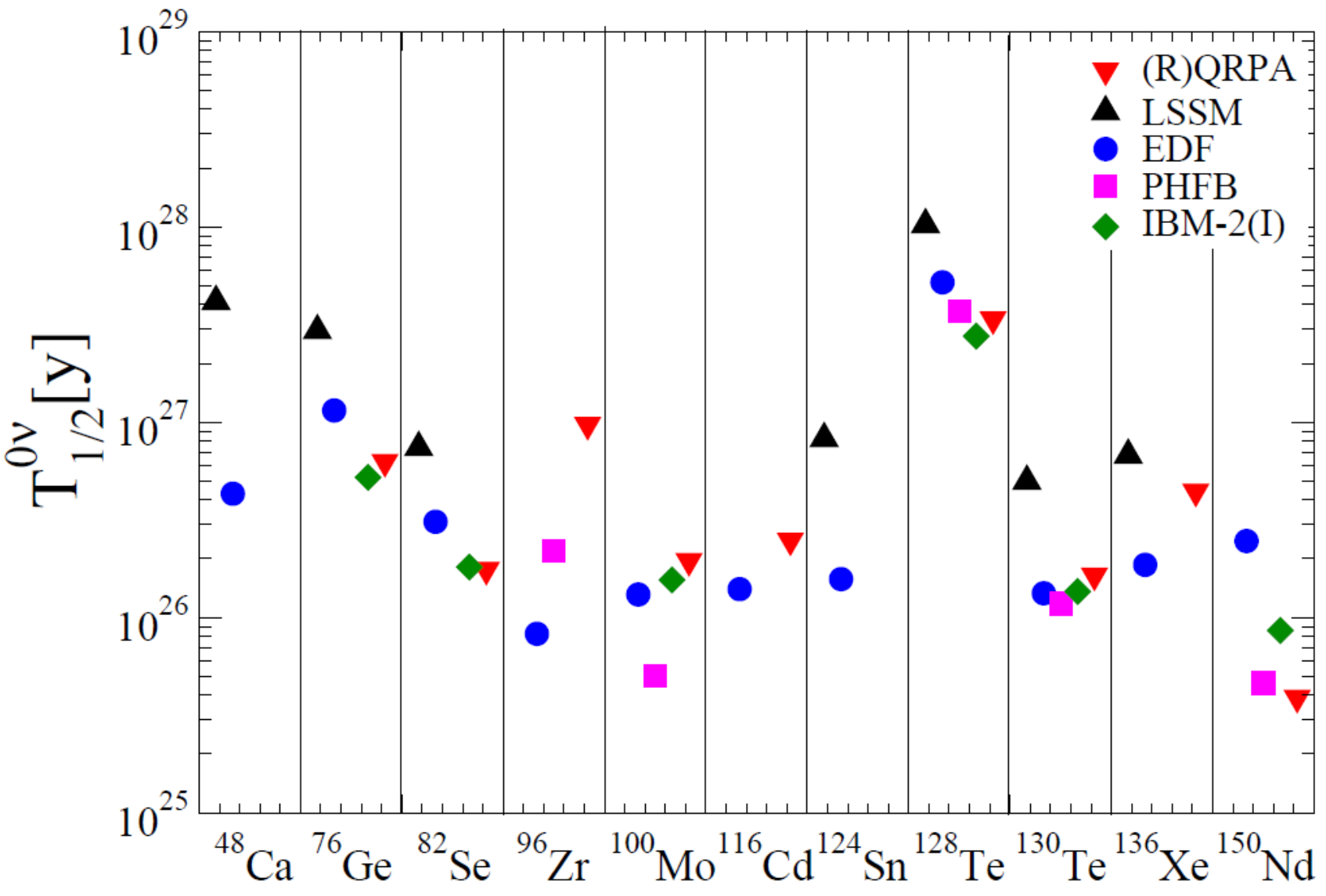

Fig. 1.8. The 0ν2β decay half-lives estimated for the different approaches of NME calculation [60] for the effective Majorana mass $|\langle m_\nu \rangle| = 50$ meV and the axial-vector coupling constant $g_A = 1.25$.

### 1.2. Current status of 2β experiments

The first experiment to search for double beta decay was performed by R. Fireman in 1948 [61]. He used Geiger counters to measure 2β decay of $^{124}$Sn and set the lower limit on such process as $3 \times 10^{15}$ yr. Afterwards, there were many experiments to study 2β decay and nowadays 2ν2β decay was measured for 11 nuclei and the half-life varies from $10^{18}$ to $10^{22}$ years [62].

In general, the experimental approaches to search for 2β decay can be divided into two categories: direct (counting) and indirect experiments. Direct experiments are focused on the direct detection of the two electrons emitted during 2β decay by a particle detector. From the other side, indirect approach is based on measurement of an excess of daughter isotopes in a material containing the parent 2β decay isotopes.

The first 2β decay observation with indirect method was achieved in 1950 [63], while the direct experimental registration of 2ν2β decay was measured only in 1987 [64]. However, indirect experiments are not available to distinguish between the two-neutrino and neutrinoless modes of 2β process. Therefore, the 2β decay studies are currently focused on direct experimental methods.

#### 1.2.1. Geochemical and radiochemical experiments

In geochemical method an old mineral containing 2β isotope is analyzed by counting the number of daughter atoms accumulated over the long time by 2β decay process. In such minerals an amount of daughter nuclides exceeds the natural abundance, and can be measured with a help of mass spectrometry. Taking into account that the age of the samples ranges from $10^6$ yr to more than $10^9$ yr, such experiments are very sensitive. The probability of the decay $\lambda_{\beta\beta}$ depends on the age of the mineral $T$ and abundances measured for parent $N(Z, A)$ and daughter $N(Z \pm 2, A)$ nuclei:

$$\lambda_{\beta\beta} \cong \frac{N(Z\pm2,A)}{N(Z,A)} \cdot \frac{1}{T}. \qquad (1.16)$$

The main disadvantage of such method is the impossibility to distinguish between the two-neutrino and neutrinoless 2β decay modes, therefore $\lambda_{\beta\beta}$ is the total probability of 2ν2β and 0ν2β decay modes: $\lambda_{\beta\beta} = \lambda_{2\nu} + \lambda_{0\nu}$. Thus the sensitivity of such experiments to 0ν2β process is limited by half-life of the two-neutrino 2β decay.

Moreover, only restricted amount of 2β isotopes can be studied with a geochemical method since the minerals used in experiments should satisfy several geological and chemical requests. First of all, concentration of the studied element in the mineral should be high. Next, the decay product should be generated by 2β decay only, and not by other processes like evaporation of the light decay products. Besides, the age of the mineral should be correctly determined from geological surroundings.

As a result of these limitations the geochemical methods for a long time were used only for selenium and tellurium ores. In both cases the daughter products are chemically inert gases krypton and xenon, whose initial abundance in minerals was very small. High sensitivity in the study of these gases can be achieved by using mass spectrometer, which allows to register even scanty excess of the daughter nucleus accumulated over a long geological period. 2β decay was observed for the first time in the geochemical experiment with a tellurium ore of the age of $1.5 \pm 0.5 \times 10^{6}$ years. The half-life of the isotope $^{130}$Te measured using mass spectrometer was $1.4 \times 10^{21}$ years [63]. The results of the geochemical experiments for all studied isotopes are presented in Table 1.1.

Table 1.1
The half-lives relatively to 2β decay of different isotopes measured in radiochemical and geochemical experiments.

| Isotope | $Q_{2\beta}$, keV | Decay mode | $T_{1/2}$, years |
|---|---|---|---|
| $^{82}$Se | 2995 | $2\beta^-$ | $= (1.2 \pm 0.1) \times 10^{20}$ [65] |
| $^{96}$Zr | 3350 | $2\beta^-$ | $= (3.9 \pm 0.9) \times 10^{19}$ [66]<br>$= (9.4 \pm 3.2) \times 10^{18}$ [67] |
| $^{100}$Mo | 3034 | $2\beta^-$ | $= (2.1 \pm 0.3) \times 10^{18}$ [68] |
| $^{128}$Te | 868 | $2\beta^-$ | $= (1.8 \pm 0.7) \times 10^{24}$ [65]<br>$= (7.7 \pm 0.4) \times 10^{24}$ [69]<br>$= (2.2 \pm 0.4) \times 10^{24}$ [70] |
| $^{130}$Te | 2533 | $2\beta^-$ | $\sim 0.8 \times 10^{21}$ [70]<br>$= (2.7 \pm 0.1) \times 10^{21}$ [71] |
| $^{130}$Ba | 2611 | $2\beta^+$, $\varepsilon\beta^+$, $2\varepsilon$<br>$2\varepsilon$ | $> 4.0 \times 10^{21}$ [72]<br>$= (2.1^{+3.0}_{-0.8}) \times 10^{21}$ [72]<br>$= (2.2 \pm 0.5) \times 10^{21}$ [73] |
| $^{132}$Ba | 840 | $2\varepsilon$ | $> 3.0 \times 10^{20}$ [72]<br>$> 2.2 \times 10^{21}$ [73] |
| $^{238}$U[1] | 1150 | $2\beta^-$ | $= (2.0 \pm 0.6) \times 10^{21}$ [74] |

[1] The radiochemical method was applied to measure 2β decay half-life of the isotope $^{238}$U.

Although geochemical experiments are unable to provide direct information about the decay mode, the half-life of $^{130}$Te relatively to the 2ν2β decay mode can be estimated from the ratio of decay probabilities of the two tellurium isotopes ($^{128}$Te and $^{130}$Te) [75, 76]. This result is based mainly on the difference in the energy dependence of the phase space factors ($G^{2\nu} \sim Q_{2\beta}^{11}$, $G^{0\nu} \sim Q_{2\beta}^{5}$). The geochemical experiments can give only model dependent information on the probability of 0ν2β decay; however they cannot provide sufficiently reliable evidence of the neutrinoless decay.

Radiochemical experiment to search for 2β decay of $^{238}$U was also based on measurements of daughter nuclei in the sample of $^{238}$U [74]. The decay product is radioactive, thus the much higher sensitivity in comparison to mass spectrometry methods to detect small amounts of atoms can be achieved. Moreover, radiochemical method do not require long-term (geochemical) accumulation of decay product, the studied material is well known, therefore, this method have not uncertainties associated with the age of the mineral, an initial concentration of daughter nuclides and a possible effects of dissipation or accumulation of 2β decay product by long geological period. After the deep purification of the uranium salt, the decay product $^{238}$Pu was accumulated for 33 years, and then extracted by chemical methods from the sample. Afterwards, it was placed into low-background α particle counters to measure the amount of α particles with the energy of 5.51 MeV from $^{238}$Pu. As a result, the radiochemical measurement with $^{238}$U gave the half-life $T_{1/2}$ = (2.0 ± 0.6) × $10^{21}$ years [74].

### 1.2.2. Counting experiments

Counting experiments are based on the direct detection of 2β decay process. Therefore they can distinguish the two-neutrino and neutrinoless double beta decay modes. Most of the experimental results on 2β decay were achieved in counting experiments. These experiments can be divided into two groups: detectors with "active" and "passive" sources.

In the experiments with active source (also called calorimeter experiments) a sensitive detector contains the nuclei of 2β isotope, i.e. detector is equal to source. The decisive advantage of this technique is close to 100% detection efficiency to the 2β process, since all the energy of β-particles emitted in the decay is absorbed in the detector. Disadvantage of this method is an inability to prove that the observed event is related to two electrons.

The isotope under study and detector are not the same in the experiments with a passive source. An important advantage of such experiments is ability to obtain information about the kinematical characteristics of the emitted β particles (γ quanta) such as the energy of the particles, their coordinates, tracks and charge. Another advantage of the method is possibility to study a wide range of isotopes, since the only requirement is to produce quite thin foils (tens of mg/cm$^2$). However, strong disadvantage of the passive sources is the low efficiency of registration (typically ~ 10%), and deterioration of the energy resolution due to the energy losses in the sample.

Among the counting experiments with active and passive sources there is also an intermediate type of detectors which combine the main properties of these two groups. In such experiments the source and detector are the same, and electron tracks can be reconstructed. Moreover, registered electrons can be discriminated from α and γ particles, as well as positrons in some cases. However, the measured angular information is very limited, and individual electron energies cannot be obtained. The most famous examples of such detector type are the time projection chambers (TPC) filled with a xenon gas and pixelated CdZnTe detectors.

The first direct detection of 2ν2β decay was achieved with a passive source of isotope $^{82}$Se [64]. The experiment was based on the TPC where the central electrode was the 14 g source of $^{82}$Se enriched up to 97% and deposited on a thin Mylar film. The TPC was located in the 700 G magnetic field, and electrons emitted in 2ν2β decay were measured. As a result, only 36 candidates for two-neutrino double beta events were detected, and the measured half-life was $1.1^{+0.8}_{-0.3} \times 10^{20}$ years [64]. This experiment had a huge value for the study of 2β decay, and started the era of 2β decay observations in counting experiments.

### 1.2.3. Most sensitive 2β decay experiments

Among the 35 naturally-occurring $2\beta^-$ candidates the experiments are concentrated on the most promising isotopes with a high $Q_{2\beta}$ value (see Table 1.2). The $Q_{2\beta}$ energy is one of the most important characteristics due to the phase space factor depends on 2β decay energy as $G^{2\nu} \sim Q_{2\beta}^{11}$ and $G^{0\nu} \sim Q_{2\beta}^{5}$ for two-neutrino and neutrinoless double beta decay, respectively [13]. Therefore, taking into account Eq. (1.13) and Eq. (1.14), the $Q_{2\beta}$ value directly influences the probability of double beta decay processes. Moreover, suppression of background becomes easier when the energy of 2β decay exceeds maximal natural gamma quanta energy 2615 keV (gamma quanta of $^{208}$Tl).

Table 1.2
The most promising double beta candidates for next-generation experiments [77].

| Double beta isotope | $Q_{2\beta}$ energy (MeV) | Isotopic abundance (%) | Enrichable by centrifugation |
|---|---|---|---|
| $^{48}$Ca | 4.27226 (404) | 0.187 | No |
| $^{76}$Ge | 2.03904 (16) | 7.8 | Yes |
| $^{82}$Se | 2.99512 (201) | 9.2 | Yes |
| $^{96}$Zr | 3.35037 (289) | 2.8 | No |
| $^{100}$Mo | 3.03440 (17) | 9.6 | Yes |
| $^{116}$Cd | 2.81350 (13) | 7.5 | Yes |
| $^{130}$Te | 2.52697 (23) | 33.8 | Yes |
| $^{136}$Xe | 2.45783 (37) | 8.9 | Yes |
| $^{150}$Nd | 3.37138 (20) | 5.6 | No |

The sensitivity of the detector can be expressed in terms of a lower half-life limit as following [14]:

$$T_{1/2}^{0\nu} \sim \delta \cdot \varepsilon \cdot \sqrt{\frac{m \cdot t}{R \cdot B}}, \tag{1.17}$$

where $\delta$ is the abundance of 2β isotope in the detector; $\varepsilon$ is the detection efficiency; $m$ is the mass of detector; $t$ is the time of measurements; $R$ is the energy resolution (FWHM) of the detector at $Q_{2\beta}$; and $B$ is the background rate in the energy region of 0ν2β decay.

As we can see from Eq. (1.17) the isotopic concentration and the detection efficiency are the most important parameters, because any other characteristics are under the square root.

The natural abundances for the most promising double beta candidates are presented in Table 1.2. The typical abundances are on the level of few percent, except the case of $^{130}$Te and $^{160}$Gd with 33.8% and 21.86% isotopic abundances, respectively. So, detectors can be manufactured without application of isotopically enriched materials. However, enriched detectors are necessary to explore the inverted hierarchy region of neutrino mass pattern; meanwhile the detector volume should be as low as possible to minimize background. This requirement immediately limits the list of candidate nuclei to $^{76}$Ge, $^{82}$Se, $^{100}$Mo, $^{116}$Cd, $^{130}$Te and $^{136}$Xe, which can be enriched by gas-centrifuge technique.

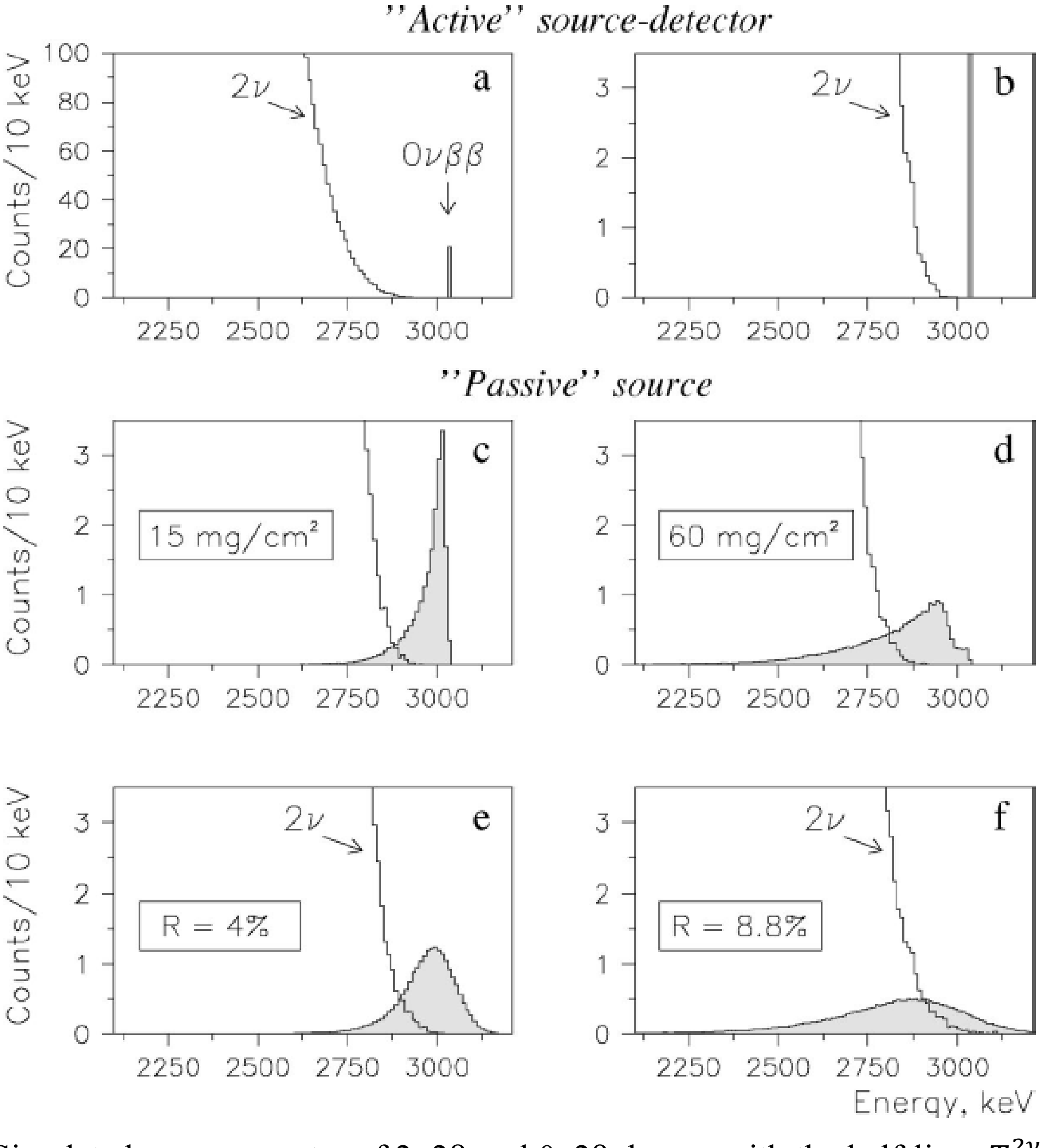

Fig. 1.9. Simulated energy spectra of 2ν2β and 0ν2β decays with the half-lives $T_{1/2}^{2\nu}$ = $10^{19}$ yr and $T_{1/2}^{0\nu}$ = $10^{24}$ yr in a zero background experiment utilizing 1 kg of $^{100}$Mo over 5 years of measurements [14]. (a) "Active" source technique: $^{100}$Mo contained in the detector with 100% of the detection efficiency and the energy resolution of 10 keV. (b) The same as (a) with a scaled-up vertical axis. (c) "Passive" source technique: $^{100}$Mo foil source with thicknesses 15 mg/cm$^2$ in the same detector. (d) The same as (c) with the foil thickness of 60 mg/cm$^2$. (e) The same as (c) but the energy resolution of the detector at 3 MeV is 120 keV. (f) The same as (d) but the energy resolution is 264 keV.

The detection efficiency should be close to the hundred percent in order to reach a highest possible sensitivity. As it was already mentioned in subsection 1.2.2, this is feasible only for the active source-detector technique. In the "passive source" experiments the sensitivity is restricted by the reverse relation between source mass and detection efficiency. The mass of 2β isotope can be increased by using thicker passive source; however this leads to lower detection efficiency and energy resolution due to absorption of electrons inside the source (see Fig. 1.9 (c,d)).

Another very important characteristic of the detector is the energy resolution, since the high-energy tail of 2ν2β spectrum can be an irremovable background in the energy window of 0ν2β decay peak. The higher energy resolution, the smaller part of two-neutrino tail will fall in the neutrinoless 2β decay peak (see Fig. 1.9), and thus the background level will be lower [78]. This feature is especially important for the isotopes with a fast 2ν2β decay process like $^{100}$Mo ($T_{1/2}^{2\nu}$ = 7.16 × $10^{18}$ yr [79]), for which an energy resolution less than 1% is required (< 30 keV at 3 MeV) [77].

The lowest possible background level is requested for the next generation double beta decay experiments. Experimental set-up should be placed deep underground to reduce cosmic ray flux, and shielded by passive and active shields against environmental radioactivity. Highly radiopure materials should be used to produce the detectors and shielding materials. Taking into account the typical time of measurements of 5–10 years, the detector should be stable enough during the experiment.

The experimental studies of double beta decay started in 1940s and continue nowadays with a step by step improvement of experimental technologies. For the last 20 years a significant progress on this subject was achieved. The best results on half-lives of the most interesting 2β isotopes are presented in Table 1.3. Here we will briefly describe the most sensitive 2β decay experiments.

Table 1.3
The best reported results on 2β decay processes. The limits on 0ν2β decay are at 90% C.L.

| Isotope | Experiment | $T_{1/2}^{2\nu}$, years | $T_{1/2}^{0\nu}$, years |
|---|---|---|---|
| $^{48}$Ca | NEMO-3 | $(4.4^{+0.5}_{-0.4}(\text{stat}) \pm 0.4(\text{syst})) \times 10^{19}$ [79] | $> 1.3 \times 10^{22}$ [80] |
| | ELEGANT VI | | $> 5.8 \times 10^{22}$ [81] |
| $^{76}$Ge | IGEX | $(1.45 \pm 0.15) \times 10^{21}$ [82] | $\geq 1.57 \times 10^{25}$ [85] |
| | HM | $(1.74^{+0.18}_{-0.16}) \times 10^{21}$ [83] | $\geq 1.9 \times 10^{25}$ [86] |
| | | | $= (2.23^{+0.44}_{-0.31}) \times 10^{25}$ [87] |
| | GERDA-I | $(1.84^{+0.14}_{-0.10}) \times 10^{21}$ [84] | $> 2.1 \times 10^{25}$ [88] |
| $^{82}$Se | NEMO-3 | $(9.6 \pm 0.1(\text{stat}) \pm 1.0(\text{syst})) \times 10^{19}$ [79] | $> 3.6 \times 10^{23}$ [80] |
| $^{96}$Zr | NEMO-3 | $(2.35 \pm 0.14\ (\text{stat}) \pm 0.16\ (\text{syst})) \times 10^{19}$ [89] | $> 9.2 \times 10^{21}$ [80] |
| $^{100}$Mo | NEMO-3 | $(7.16 \pm 0.01\ (\text{stat}) \pm 0.54\ (\text{syst})) \times 10^{18}$ [79] | $> 1.1 \times 10^{24}$ [90] |
| $^{116}$Cd | Solotvina | $(2.9^{+0.4}_{-0.3}) \times 10^{19}$ [91] | $> 1.7 \times 10^{23}$ [91] |
| | NEMO-3 | $(2.88 \pm 0.04\ (\text{stat}) \pm 0.16\ (\text{syst})) \times 10^{19}$ [79] | $> 1.6 \times 10^{22}$ [80] |
| $^{130}$Te | NEMO-3 | $(7.0 \pm 0.9\ (\text{stat}) \pm 1.1\ (\text{syst})) \times 10^{20}$ [92] | $> 1.0 \times 10^{23}$ [80] |
| | CUORICINO | | $\geq 2.8 \times 10^{24}$ [93] |
| $^{136}$Xe | EXO-200 | $(2.165 \pm 0.016\ (\text{stat}) \pm 0.059\ (\text{syst})) \times 10^{21}$ [94] | $> 1.1 \times 10^{25}$ [95] |
| | KamLAND-Zen | $(2.38 \pm 0.02\ (\text{stat}) \pm 0.14\ (\text{syst})) \times 10^{21}$ [43] | $> 1.9 \times 10^{25}$ [96] |
| $^{150}$Nd | NEMO-3 | $(9.11^{+0.25}_{-0.22}(\text{stat}) \pm 0.63(\text{syst})) \times 10^{18}$ [97] | $> 1.8 \times 10^{22}$ [80] |

The detectors with high-pure germanium (HPGe) were one of the most sensitive 2β decay experiments at the end of last century. The intrinsic purity of germanium crystals, availability of enriched isotope $^{76}$Ge, low level of radioactive contamination and high energy resolution were the key features of such experiments. The first observation of 2ν2β decay in $^{76}$Ge with HPGe detectors gave the half-life of (0.9 ± 0.1) × $10^{21}$ years [98]. Afterwards, the International Germanium Experiment (IGEX) and the Heidelberg-Moscow (HM) experiments performed more accurate measurements.

The IGEX experiment was installed in the Laboratorio Subterráneo de Canfranc (Canfranc Underground Laboratory, Canfranc, Spain) at the depth of 2450 meters of water equivalent (m.w.e.). Three HPGe detectors with a mass of 2 kg each enriched in $^{76}$Ge up to 86% were used in the experiment. The shielding of the set-up consisted of 2.5 tons of archeological lead, ~ 10 tons of the 70 years old low-activity lead, active muon veto from plastic scintillators, and ~ 1.5 tons of an external polyethylene neutron moderator. Pulse-shape discrimination (PSD) technique was applied for the data analyzing. The achieved background level in the energy range of 2.0 MeV–2.5 MeV was ≈ 0.06 counts/(keV·kg·yr) [99]. The energy resolution at the energy of expected 0ν2β decay peak of $^{76}$Ge ($Q_{2\beta}$ = 2039.04(16) keV) was ~ 4 keV. The data were collected with a total exposure of 8.87 kg × year. The half-life relatively to the 2ν2β decay of $^{76}$Ge was measured as (1.45 ± 0.15) × $10^{21}$ yr [82]. A limit was set on the 0ν2β decay of $^{76}$Ge as $T_{1/2}^{0\nu} \geq 1.57 \times 10^{25}$ yr (see Fig. 1.10) [85].

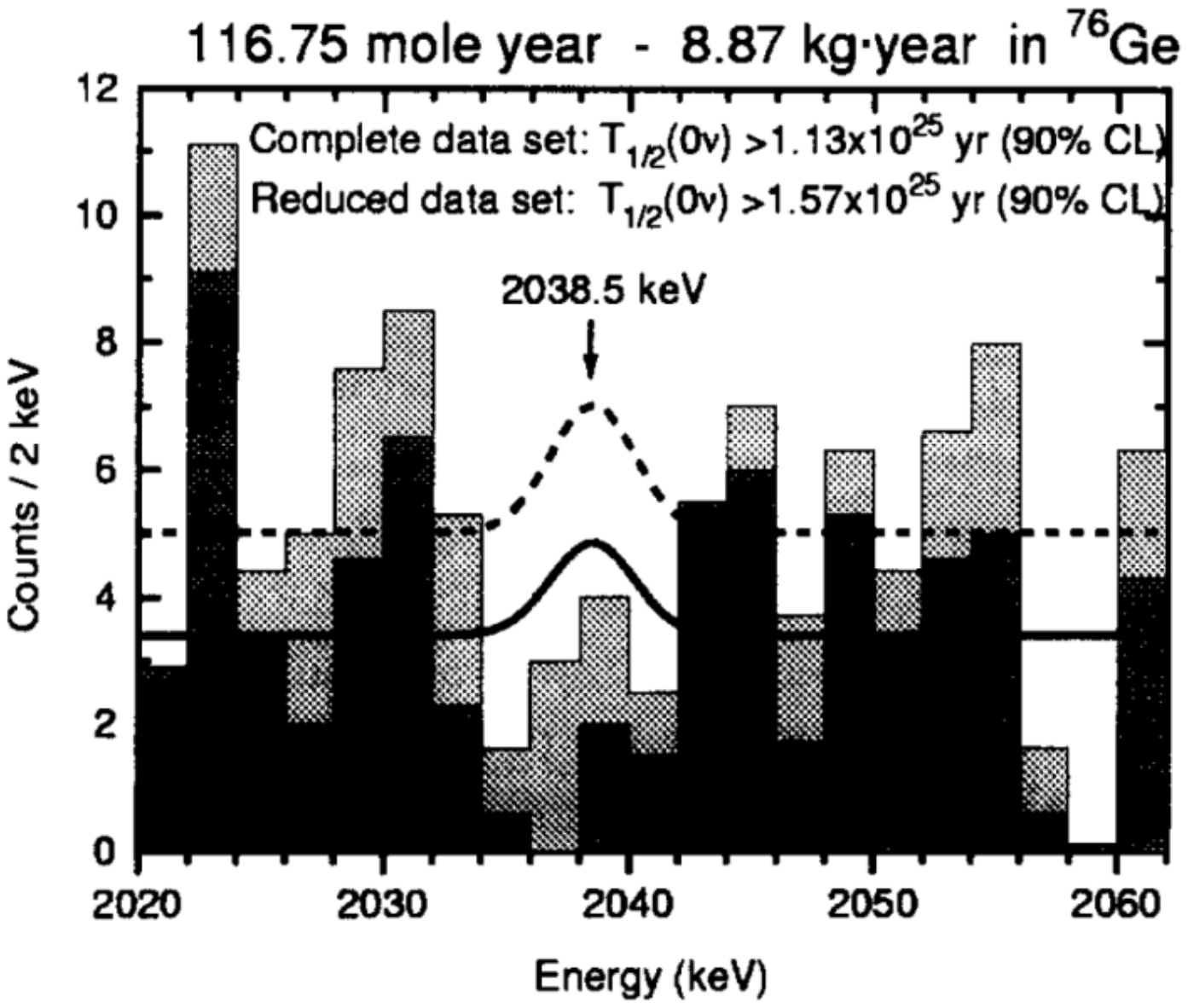

Fig. 1.10. The energy spectrum accumulated in the IGEX experiment in the vicinity of neutrinoless double beta decay of $^{76}$Ge. Fits of the data and excluded peaks are shown for the data sets with and without application of PSD.

The Heidelberg-Moscow (HM) experiment was based on 5 HPGe detectors enriched in $^{76}$Ge at the level of 86%. The detectors with a total mass of 10.96 kg were operated at 3600 m.w.e. at the Laboratori Nazionali del Gran Sasso (LNGS, Assergi, Italy). The background level at $Q_{2\beta}$ was ≈ 0.06 counts/(keV·kg·yr), thanks to a shielding and pulse shape analysis. The achieved energy resolution was 3.9 keV at the energy 2038.5 keV. After analysis of the data from 35.5 kg × year exposure, the half-life limit on 0ν2β decay of $^{76}$Ge was set as $T_{1/2}^{0\nu} \geq 1.9 \times 10^{25}$ yr (90% C.L.) [86]. However, in 2001 H.V. Klapdor-Kleingrothaus with a few co-authors has claimed the discovery of 0ν2β decay. The announced half-life time of 0ν2β decay was $T_{1/2}^{0\nu} = (0.8 - 18.3) \times 10^{25}$ yr (95% C.L.), with a best value of $1.5 \times 10^{25}$ yr [100]. The derived value of the effective neutrino mass was $\langle m_\nu \rangle$ = (0.11–0.56) eV at 95% C.L., with a best value of 0.39 eV. Taking into account the uncertainties in the nuclear matrix elements the claimed range for the effective neutrino mass can be wider by at most a factor 2. The small peak at the energy $Q_{2\beta}$ of $^{76}$Ge was identified as due to the 0ν2β decay of $^{76}$Ge. However, the scientific community (including the most part of the HM collaboration) harshly criticized the claim [101, 102, 103, 104]. The skepticism raised due to some peaks couldn't be identified, the claimed 0ν2β decay peak was strongly dependent on the spectral window chosen for the analysis, and its statistical significance seems to be lower than the claimed 2.2σ.

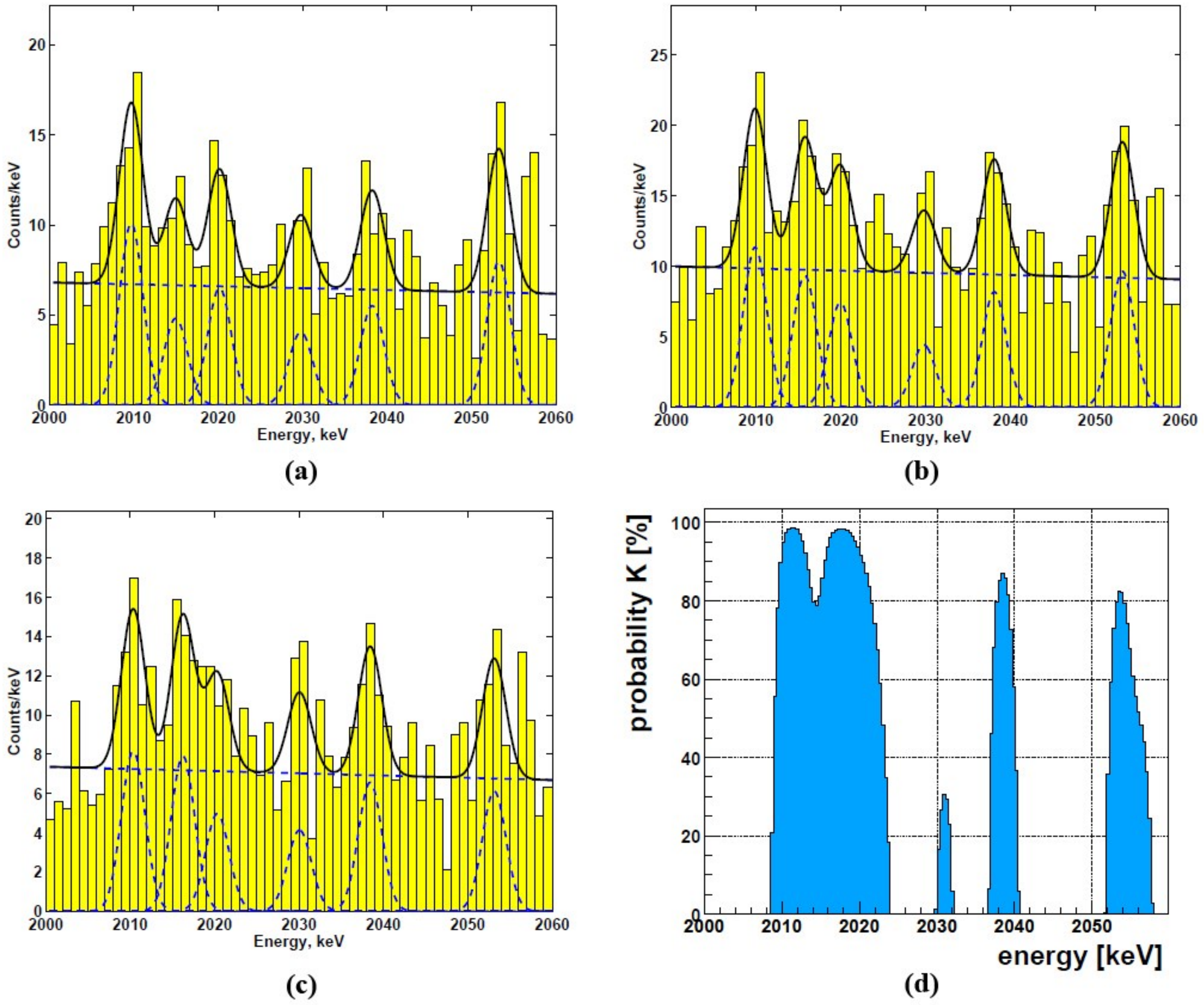

Fig. 1.11. The total sum spectrum of 5 HPGe detectors enriched in $^{76}$Ge (total detectors mass 10.96 kg) measured in the Heidelberg-Moscow experiment. The energy spectra are in the range 2000 keV–2060 keV with its fits for the periods: (a) August 1990 to May 2000 (50.57 kg × yr); (b) August 1990 to May 2003 (71.7 kg × yr); (c) November 1995 to May 2003 (56.66 kg × yr). (d) Search for the lines in the energy spectrum (c) by using the maximum likelihood method. The Bi lines at 2010.7, 2016.7, 2021.8 and 2052.9 keV can be seen, with the signal at ~ 2039 keV [105].

After reanalysis of the data, resulting in increase of the total exposure to 71.7 kg × yr, H.V. Klapdor-Kleingrothaus and co-authors have published a new result trying to prove the observation of 0ν2β decay (see Fig. 1.11 and Fig. 1.12). The pulse-shape analysis applied to the data provides a value of the claimed half-life $T_{1/2}^{0\nu} = (1.19^{+0.37}_{-0.23}) \times 10^{25}$ at 4.2σ (99.9973% C.L.) [105], which later was updated as $T_{1/2}^{0\nu} = (2.23^{+0.44}_{-0.31}) \times 10^{25}$ (6.4σ level) [87]. The claimed neutrinoless double beta half-life of $^{76}$Ge corresponds to the effective neutrino mass $\langle m_\nu \rangle = (0.32 \pm 0.03)$ eV.

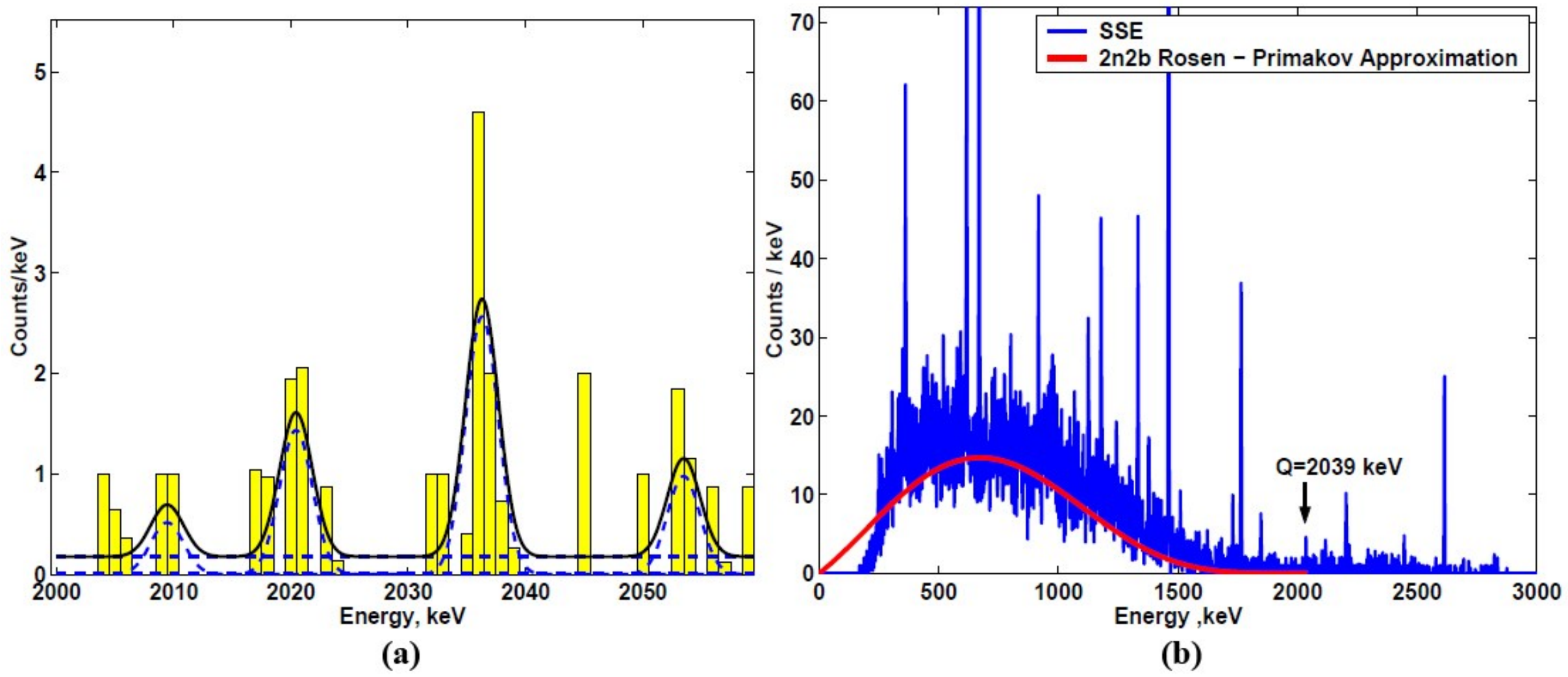

Fig. 1.12. The energy spectrum of the HPGe detectors (56.66 kg × yr exposure) obtained in HM experiment after applying the pulse-shape analysis: (a) in the range 2000 keV–2060 keV; (b) in the energy range of 100 keV–3000 keV. The solid curve shows the approximation of 2ν2β spectrum shape [105].

The GERmanium Detector Array (GERDA) is an experiment to investigate 2β decay of $^{76}$Ge. It is designed for two phases and located at the LNGS underground laboratory. The goal of the first phase GERDA-I was to verify the claim of H.V. Klapdor-Kleingrothaus and co-authors. The operation of the GERDA-I was finished in 2013. The detector consisted of the coaxial HPGe crystals taken from the HM and IGEX experiments (total mass ~ 18 kg, concentration of $^{76}$Ge is 86%) and broad energy germanium (BEGe) diodes (~ 3.6 kg of ~ 88% enriched Ge). The germanium detectors were placed in a stainless-steel cryostat filled with ~ 100 tons of liquid argon (LAr) used as a passive shield and cooling medium [106]. The inner wall of the cryostat was covered by ultra-pure copper to reduce gamma radiation from the steel container. The cryostat was surrounded by a water Cherenkov detector acted as a shield for the inner part of the set-up and as a muon veto. The schematic view of the detector is presented in Fig. 1.13.

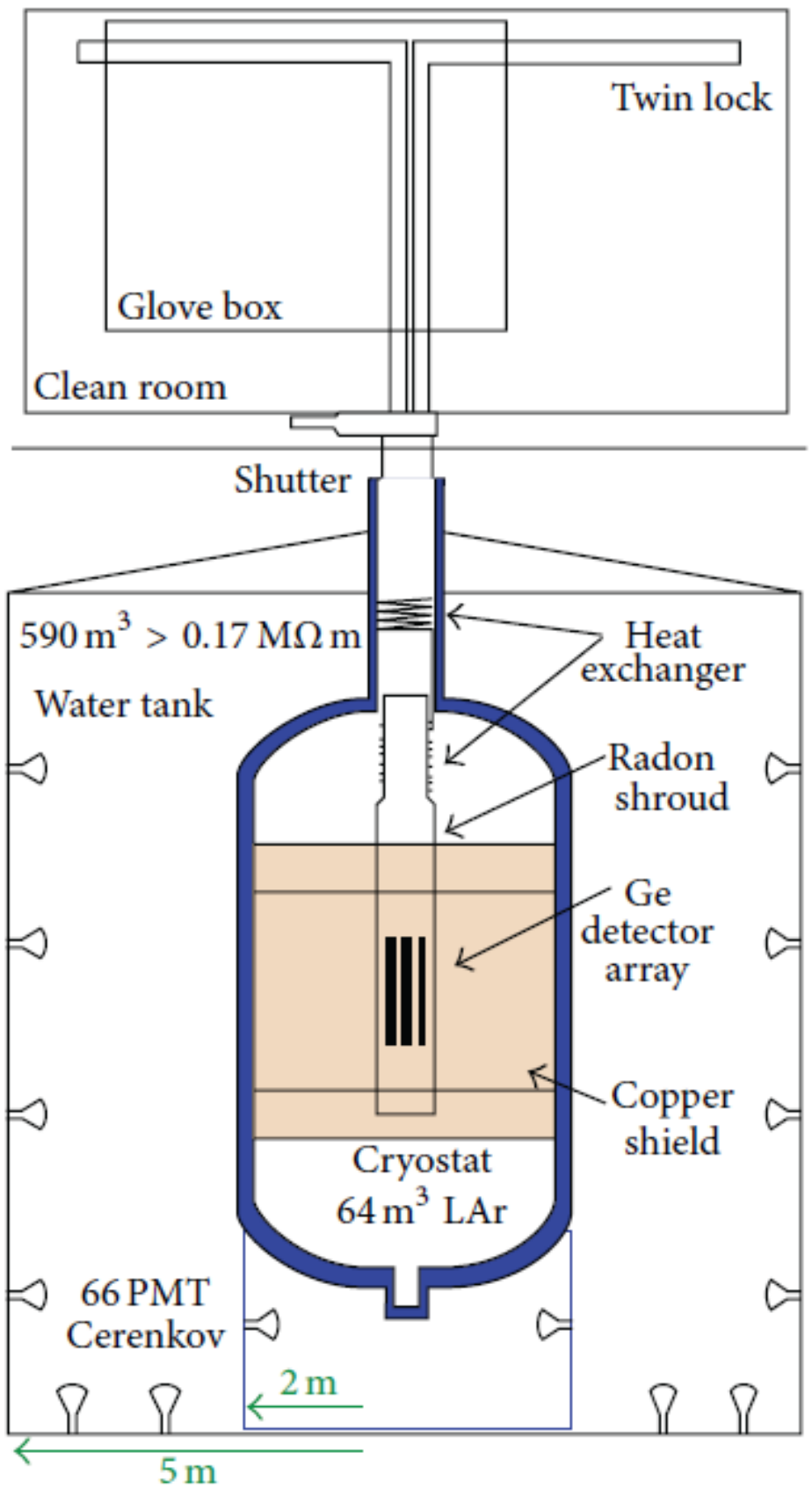

Fig. 1.13. Schematic view of the main components of the GERDA-I experiment to investigate 2β decay of the isotope $^{76}$Ge.

The average energy resolution at $Q_{2\beta}$ of $^{76}$Ge was 4.8 keV and 3.2 keV for the HPGe and the BEGe detectors respectively. The half-life of the two-neutrino double beta decay of $^{76}$Ge was measured with exposure of 5.04 kg × yr as $1.84^{+0.14}_{-0.10} \times 10^{21}$ [84]. The obtained value is slightly higher than in the IGEX and the HM experiments, however the results are consistent.

The total exposure of the first phase of GERDA experiment was 21.6 kg × yr. The background level after the pulse-shape analysis was about 0.01 counts/(keV·kg·yr). As a result no 0ν2β decay signal was detected (see Fig. 1.14), and the lower limit for the half-life of neutrinoless double beta decay of $^{76}$Ge was set as $T^{0\nu}_{1/2} > 2.1 \times 10^{25}$ (90% C.L.) [88].

The GERDA result was combined with the data from HM and IGEX experiments and improved the half-life limit of neutrinoless double beta decay of $^{76}$Ge to $T^{0\nu}_{1/2} > 3.0 \times 10^{25}$ (90% C.L.) [88]. Therefore, the claim for the observation of 0ν2β decay in $^{76}$Ge by a part of the Heidelberg-Moscow collaboration is strongly disfavored.

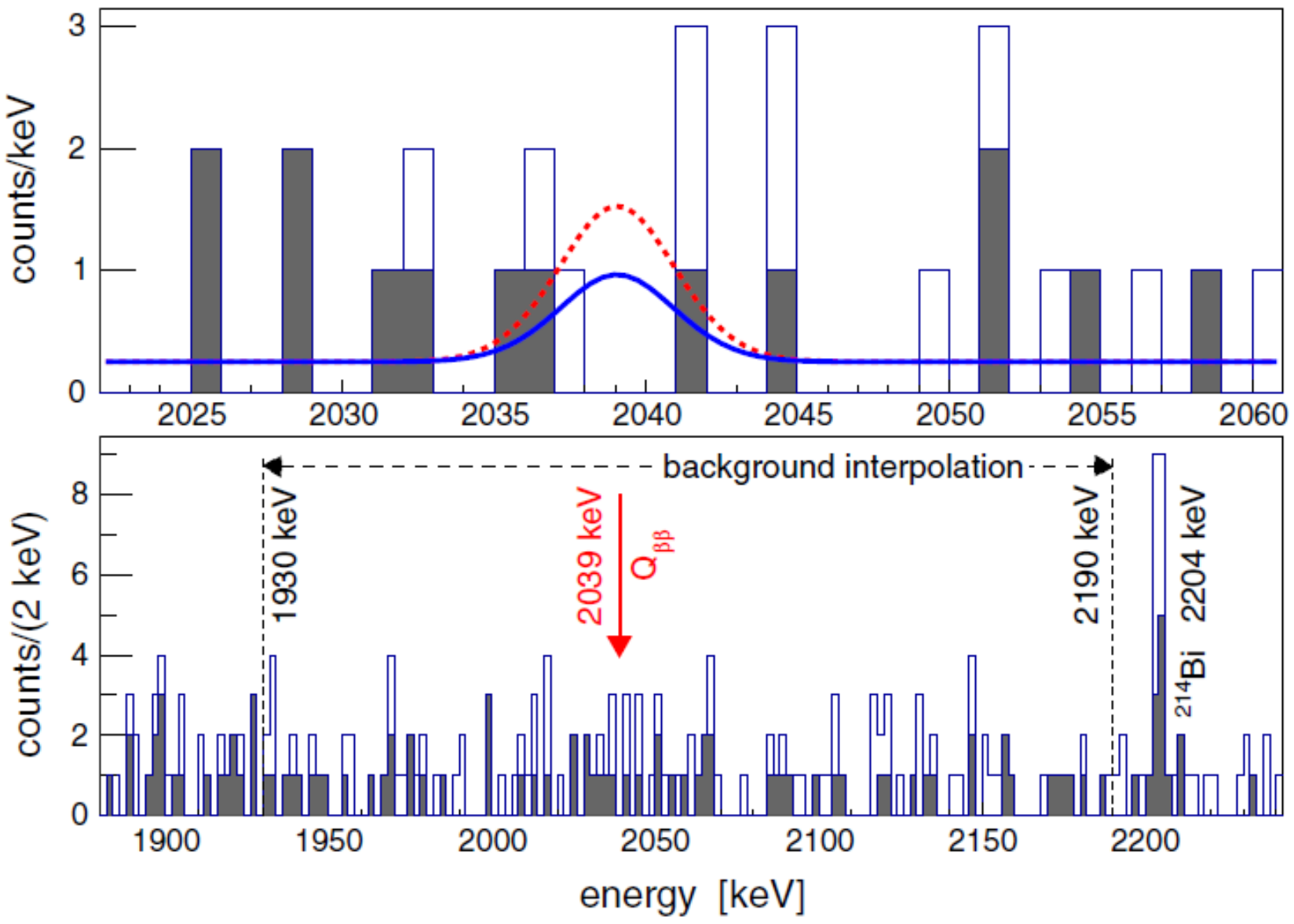

Fig. 1.14. The combined energy spectrum from all enriched Ge detectors obtained in the GERDA-I experiment for the total exposure of 21.6 kg × yr. The spectrum without (with) PSD is shown by the open (filled) histogram. The lower panel demonstrates the region used for the background interpolation. The upper panel shows the spectrum in $Q_{2\beta}$ region with the expectations (with PSD selection) based on the HM half-life claim $T_{1/2}^{0\nu} = (1.19^{+0.37}_{-0.23}) \times 10^{25}$ yr from the Ref. [105] (red dashed), and with the GERDA-I limit $T_{1/2}^{0\nu} > 2.1 \times 10^{25}$ yr (blue solid) [88].

One of the most remarkable results of 2β decay investigations were achieved in the Neutrino Ettore Majorana Observatory (NEMO) experiments. The NEMO-3 experiment was based on the tracking detector with passive sources and operated at the underground laboratory "Laboratoire Souterrain de Modane" (LSM, Modane, France) from February 2003 to January 2011.

The set-up has cylindrical design and was divided into 20 equal sectors filled with ≈ 10 kg of enriched isotopes in form of thin foils (see Fig. 1.15). The NEMO-3 experiment simultaneously studied the following seven 2β isotopes: $^{100}$Mo (enrichment level 95.14%–98.95%, total mass of isotope $^{100}$Mo was 6.9 kg), $^{82}$Se (96.82%–97.02%, 0.93 kg), $^{130}$Te (33.8%–89.4%, 0.62 kg), $^{116}$Cd (93.2%, 405 g), $^{150}$Nd (91%, 36.5 g), $^{96}$Zr (57.3%, 9.43 g), and $^{48}$Ca (73.2%, 6.99 g) [107]. The source foils had the thickness in the range of 30–60 mg/cm$^2$ and were surrounded by a tracker and a calorimeter. The tracking volume was composed of 6180 open octagonal drift cells operated in Geiger mode. Energy and time-of-flight measurements were acquired with the 1940 large blocks of plastic scintillators coupled to a low radioactivity photomultiplier tubes (PMTs) through the light guides. The detector was also surrounded by a solenoid which produced a 25 Gauss magnetic field parallel to the foils with 2β isotopes, in order to identify the particle charge. The set-up was covered with a 20 cm of low radioactivity iron shielding to protect the detector from γ-rays and thermal neutrons. To thermalize fast neutrons and capture thermal neutrons the external shield of the set-up was also covered by borated water tanks and wood.

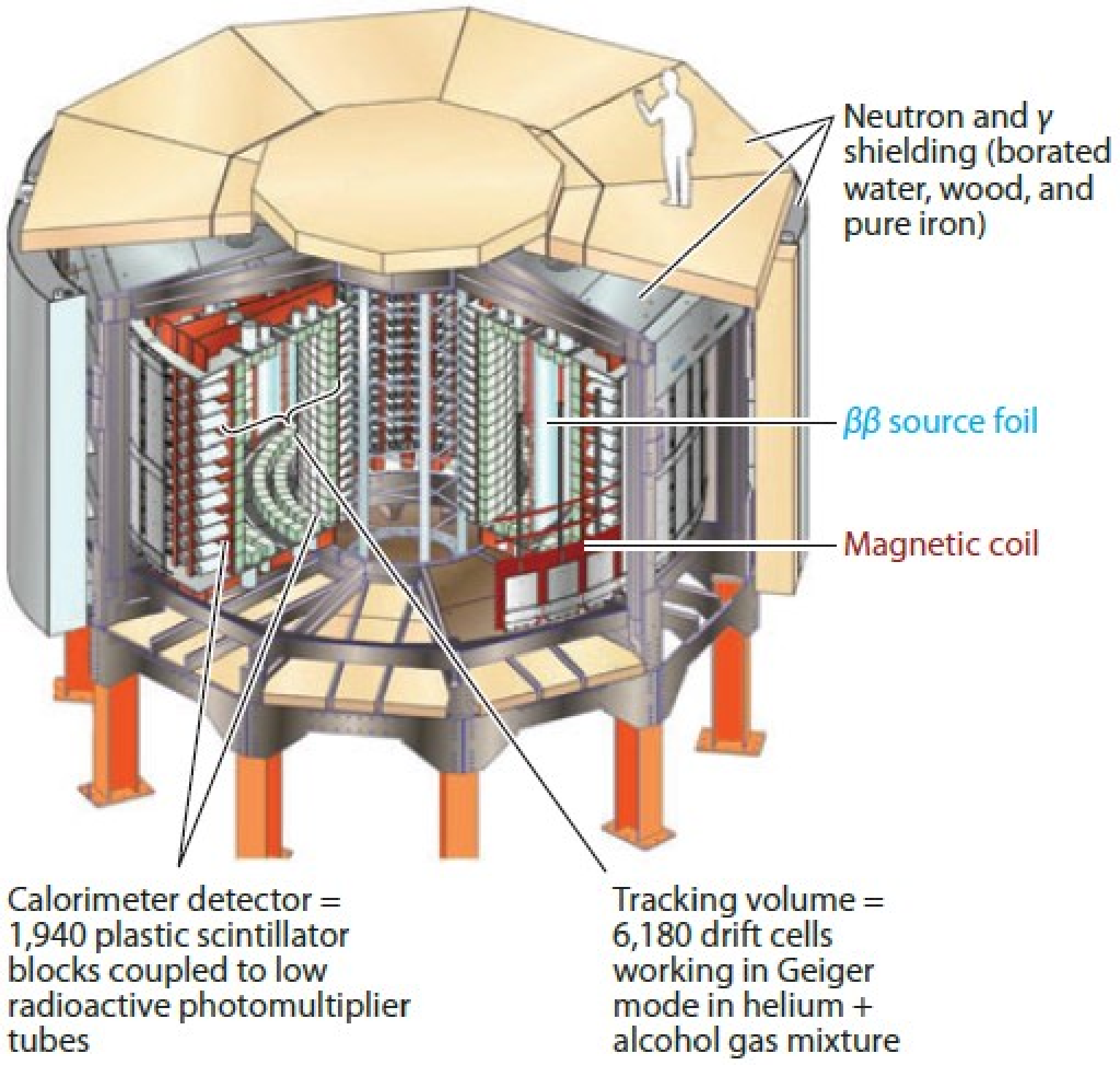

Fig. 1.15. The main components of the NEMO-3 detector with its shielding. The paraffin shield under the central tower is not presented on the figure.

The NEMO-3 detector was capable to identify different particle types with high discrimination efficiency between the signal and background events by using the tracking, charge, energy, and timing information. An example of a double beta decay candidate event is shown in Fig. 1.16. Events were selected under the requirement that two reconstructed electron tracks were emitted from the same vertex in the source foil.

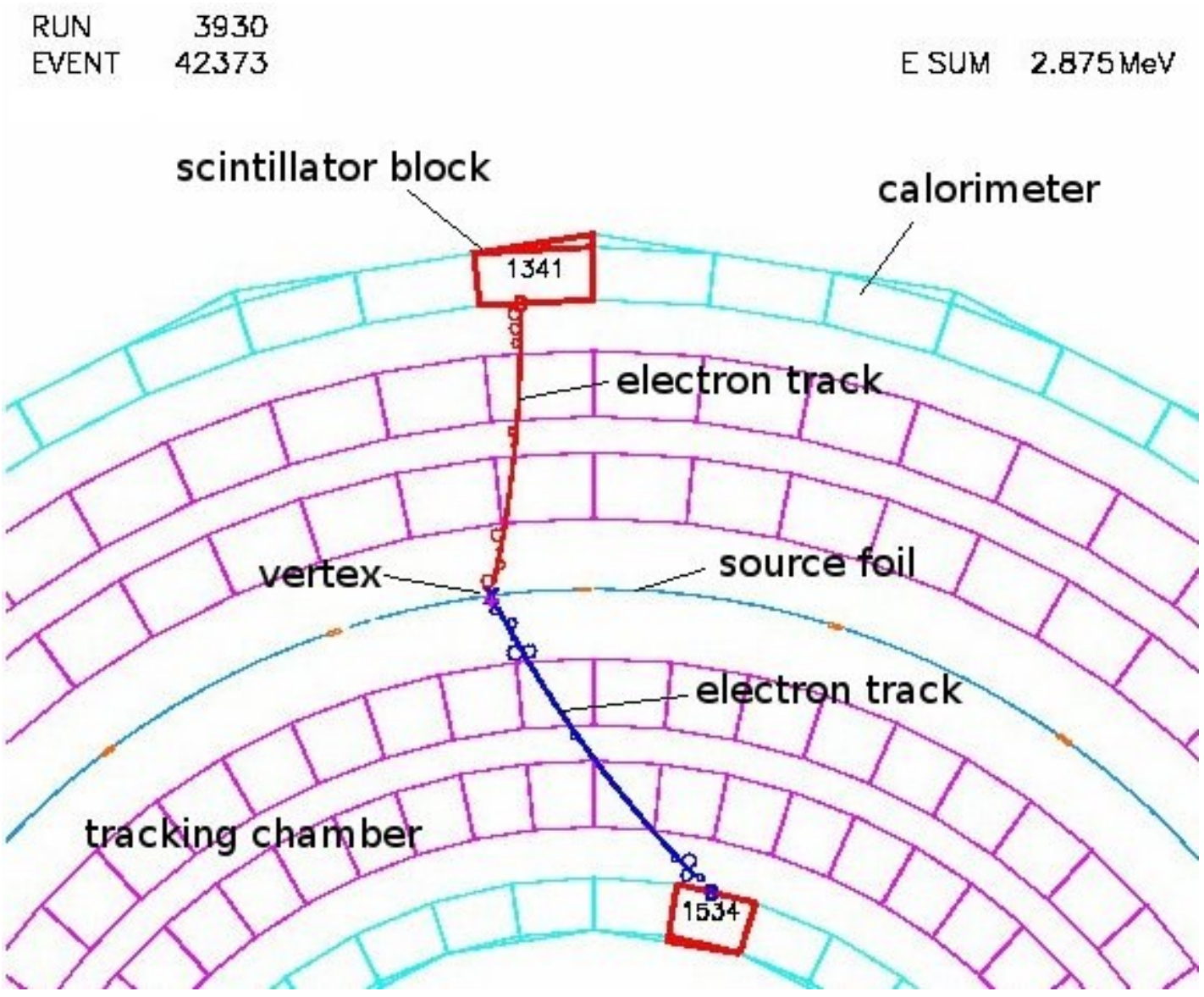

Fig. 1.16. A typical candidate for the event from two-neutrino double beta decay of $^{100}Mo$ obtained with the NEMO-3 detector.

The disadvantages of the NEMO-3 experiment were the same as for all passive source experiments: low energy resolution and poor detection efficiency. The achieved energy resolution of the calorimeter ranged from 14.1% to 17.6% for electrons with the energy 1 MeV [80]. The Monte-Carlo simulated detection efficiency to the 0ν2β decay of $^{100}$Mo was only 13% [79]. Despite these disadvantages the most precise measurements of two-neutrino 2β decay half-lives were obtained. The NEMO-3 results are presented in Table 1.3.

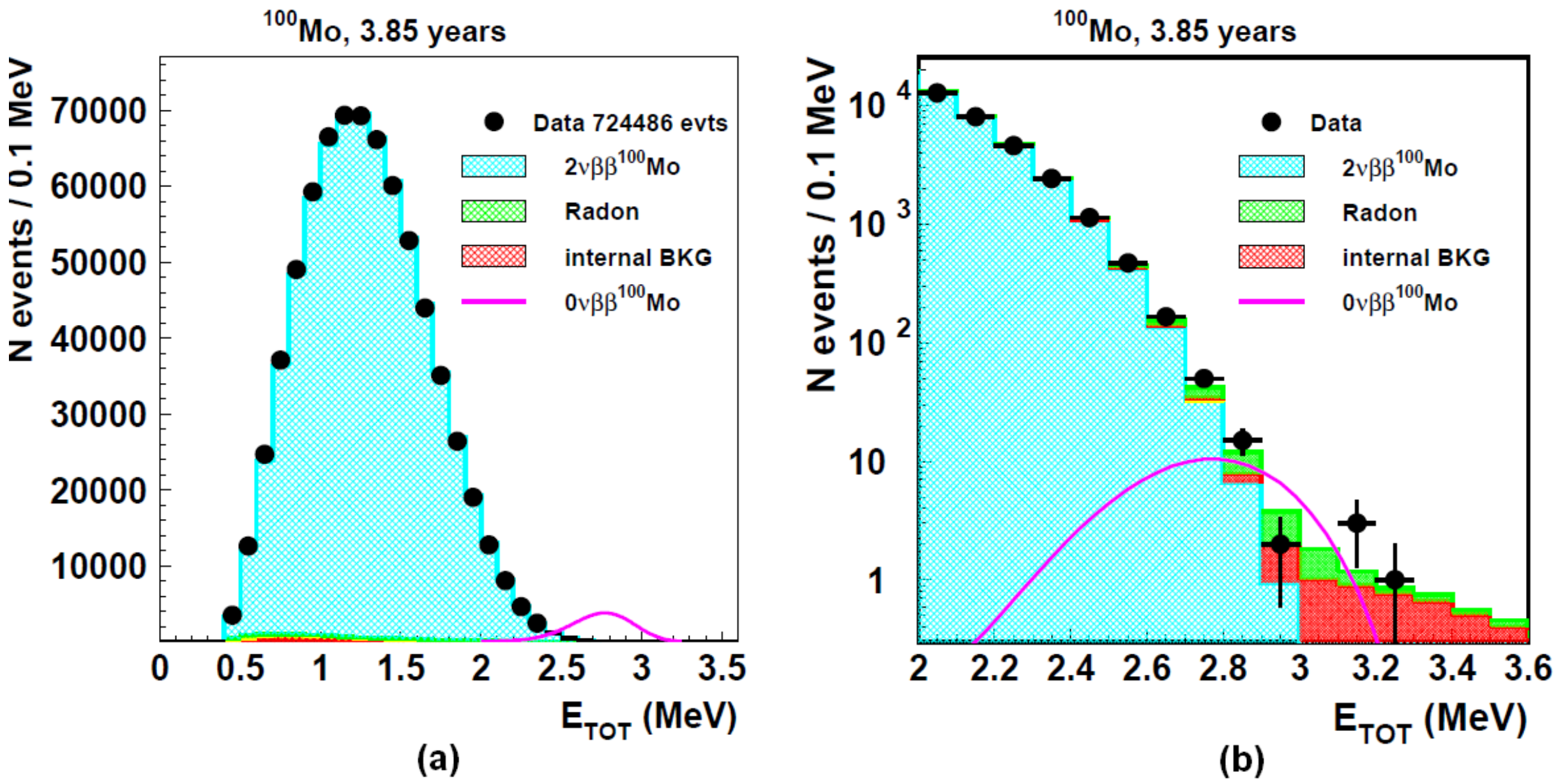

Fig. 1.17. Distribution of the energy sum of two electrons for $^{100}$Mo: (a) in the energy interval 0–3.5 MeV, and (b) in the region around the $Q_{2\beta}$ value of $^{100}$Mo. The data were accumulated with the NEMO-3 detector during 1409 days of measurements. An expected 0ν2β decay peak is shown by the curve in arbitrary units. The different components of background are displayed as the histogram.

It should be also stressed, that the most remarkable results were obtained with $^{100}$Mo ($Q_{2\beta}$ = 3034.40(17) keV), which was the main isotope of the NEMO-3 experiment. The two-electron energy distribution measured during 1409 days is shown in Fig. 1.17. The half-life time of 2ν2β decay of $^{100}$Mo was determined as (7.16 ± 0.01 (stat.) ± 0.54 (syst.)) × $10^{18}$ years [79]. There was no evidence of 0ν2β decay in the data accumulated with exposure of 34.7 kg × yr. The lower limit on neutrinoless 2β decay half-life time of $^{100}$Mo was set as $T^{0\nu}_{1/2} > 1.1 \times 10^{24}$ years (90% C.L.) [90]. This value corresponded to the limit on the effective neutrino mass $\langle m_\nu \rangle$ < (0.3–0.9) eV.

Double beta experiment to search for double beta decay of $^{116}$Cd was carried out in the Solotvina Underground Laboratory (Ukraine) at the depth of 1000 m.w.e. [108]. The Solotvina detector was aimed to study 2β decay of $^{116}$Cd with the help of cadmium tungstate ($CdWO_4$) crystal scintillators [109]. The set-up was based on four $^{116}CdWO_4$ scintillating crystals with a total mass of 330 g, enriched in the isotope $^{116}$Cd up to 83%. The scintillators were viewed by a low background PMT through the Ø10 × 55 cm light guide, glued from 25 cm of high purity quartz and 30 cm of plastic scintillator. The $^{116}CdWO_4$ crystals were surrounded by 15 $CdWO_4$ scintillators of a large volume (with total mass 20.6 kg) viewed by a low background PMT through an active plastic light guide Ø17 × 49 cm. The detectors were located in the additional active shield made of plastic scintillator with dimensions 40 × 40 × 95 cm. Therefore, the so-called 4$\pi$ active shield was provided for the $^{116}$CdWO$_4$ scintillating crystals. The outer passive shield consisted of high purity copper (3–6 cm), lead (22.5–30 cm), and polyethylene (16 cm). Two plastic scintillators (220 × 130 × 3 cm) were installed above the passive shield to provide cosmic muon veto. The schematic view of the Solotvina set-up is shown in Fig. 1.18.

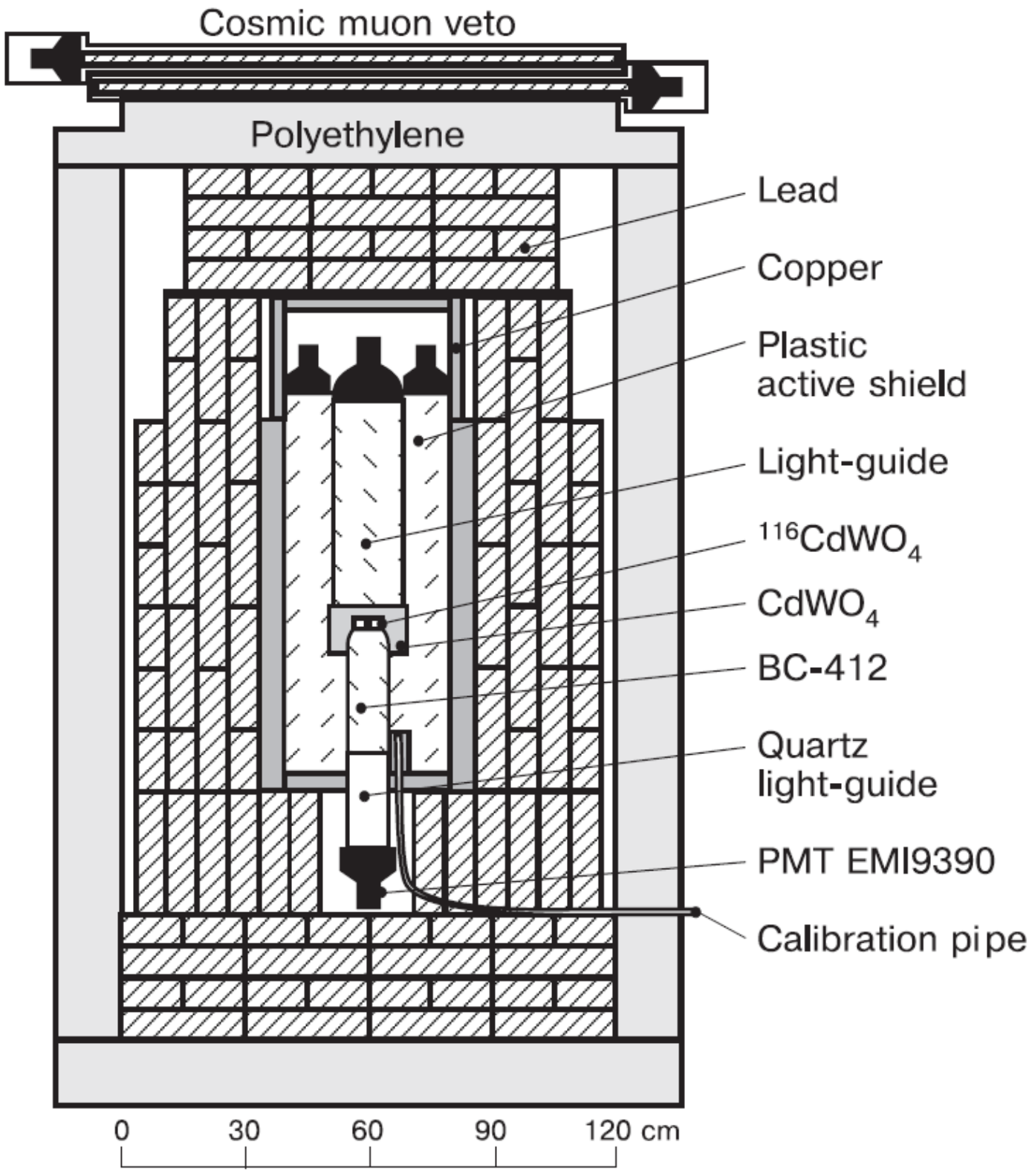

Fig. 1.18. The schematic view of the Solotvina detector aimed to study 2β decay of $^{116}$Cd by using four enriched $^{116}$CdWO$_4$ crystal scintillators with a total mass of 330 g.

The background level achieved for the $^{116}$CdWO$_4$ detector with the described above active and passive shields, as well as the applied time-amplitude and pulse-shape analysis of the data, was ~ 0.04 counts/(keV·kg·yr) in the energy region 2.5–3.2 MeV ($Q_{2\beta}$ of $^{116}$Cd is 2813.50(13) keV). The energy resolution of the enriched cadmium tungstate crystal scintillators was 8.8% at 2.8 MeV. No peculiarity, which could be ascribed to the 0ν2β decay of $^{116}$Cd, was observed in the data accumulated over 14183 hours. The half-life limit on the neutrinoless 2β decay was established as $T_{1/2}^{0\nu} > 1.7 \times 10^{23}$ yr at 90% C.L., which corresponds to the restriction on the effective Majorana neutrino mass $\langle m_\nu \rangle$ < 1.7 eV (90% C.L.) calculated on the basis of Ref. [110]. The half-life value of the two-neutrino double beta decay of $^{116}$Cd was measured as $T_{1/2}^{2\nu} = 2.9^{+0.4}_{-0.3} \times 10^{19}$ years [91]. In addition, new half-life limits were set for the different 2β decay processes in $^{106}$Cd, $^{108}$Cd, $^{114}$Cd, $^{180}$W and $^{186}$W nuclei [91]. The measured spectrum for the $^{116}$CdWO$_4$ crystals in the energy region 1–4 MeV is shown in Fig. 1.19.

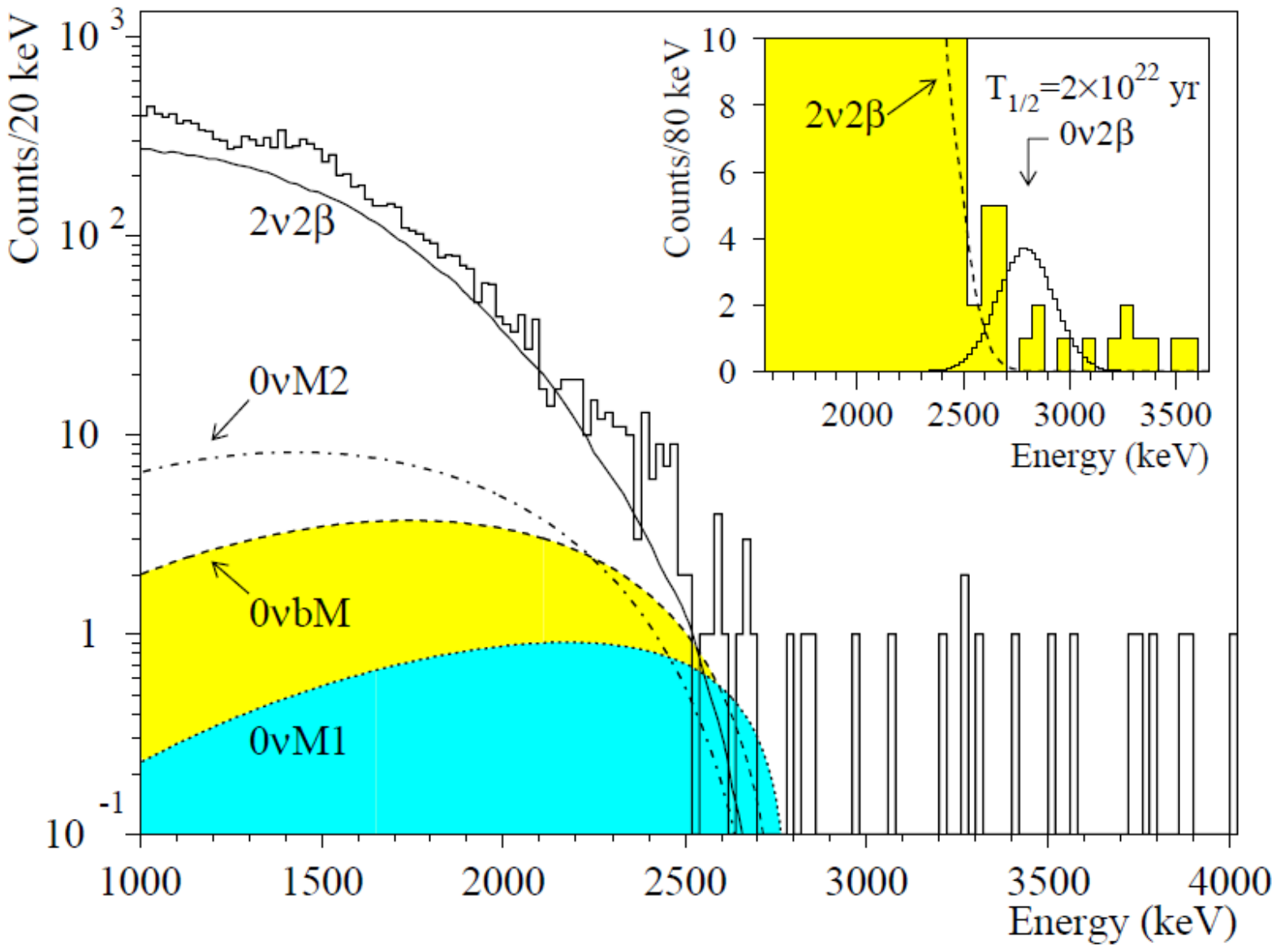

Fig. 1.19. The energy spectrum of the $^{116}CdWO_4$ detector in the interval 1–4 MeV measured during 14183 hours in the Solotvina experiment (histogram). The fit from 2ν2β decay of $^{116}$Cd is displayed by a solid curve. The smooth curves 0νM1, 0νM2, and 0νbM are excluded with 90% C.L. distributions of 0ν2β decay of $^{116}$Cd with emission of one, two, and bulk Majorons, respectively. The expected peak from 0ν2β decay with $T_{1/2}^{0\nu} > 2.0 \times 10^{22}$ yr is shown in the inset.

The ELEGANT VI (ELEctron GAmma-ray Neutrino Telescope) experiment searched for neutrinoless double beta decay of $^{48}$Ca with calcium fluoride crystal scintillators. The detector was installed in the Oto Cosmo Observatory (Japan) at the depth of 1400 m.w.e. The ELEGANT VI set-up consisted of 23 $CaF_2$(Eu) scintillating crystals (45 × 45 × 45 mm$^3$) with a total mass of 6.66 kg, which contained 7.6 g of the studied 2β isotope due to natural abundance of $^{48}$Ca which is only δ = 0.187% [111]. 46 of pure $CaF_2$ crystals (45 × 45 × 75 mm$^3$), connected from the two faces to $CaF_2$(Eu) crystals, served as light guides. Photomultiplier tubes with quartz window were attached to these light guides. For further background rejection the calcium fluoride crystals were covered by 38 CsI(Tl) scintillators (65 × 65 × 250 mm$^3$). Thus, the 4π active shield was achieved for the $CaF_2$(Eu) detectors. The detectors were placed in the air-tight box flushed by pure $N_2$ gas to remove radon from the air. The passive shield consisted of oxygen-free high-conductivity copper (5 cm), lead (10 cm), LiH-loaded paraffin (15 mm), Cd sheet (0.6 mm) and $H_3BO_3$-loaded water tank moderator. A schematic drawing of the ELEGANT VI set-up is presented in Fig. 1.20.

The extrapolated energy resolution of the ELEGANT VI detector at $Q_{2\beta}(^{48}Ca)$ = (4272.26 ± 4.04) keV was 4–6% depending on the $CaF_2$(Eu) scintillator. The background rate at the $Q_{2\beta}$ energy was ≈ 0.3 counts/(keV·kg·yr). The effect corresponding to 0ν2β decay was not observed during the exposure of 4.23 kg × yr. The lower limit on the half-life of $^{48}$Ca was set as $T_{1/2}^{0\nu} > 1.4 \times 10^{22}$ yr (90% C.L.) [112].

Thereafter, the measurement continued with the same apparatus except for the new flash ADC (analog-to-digital converter), which allowed to decrease the background level at the energy region of 0ν2β decay to ~ 0.08 counts/(keV·kg·yr). The exposure of the updated measurement was 9.3 kg × yr, and the derived half-life limit of 0ν2β decay was $T_{1/2}^{0\nu} > 2.7 \times 10^{22}$ yr at 90% C.L. [81]. The measured energy spectrum is presented in Fig. 1.21.

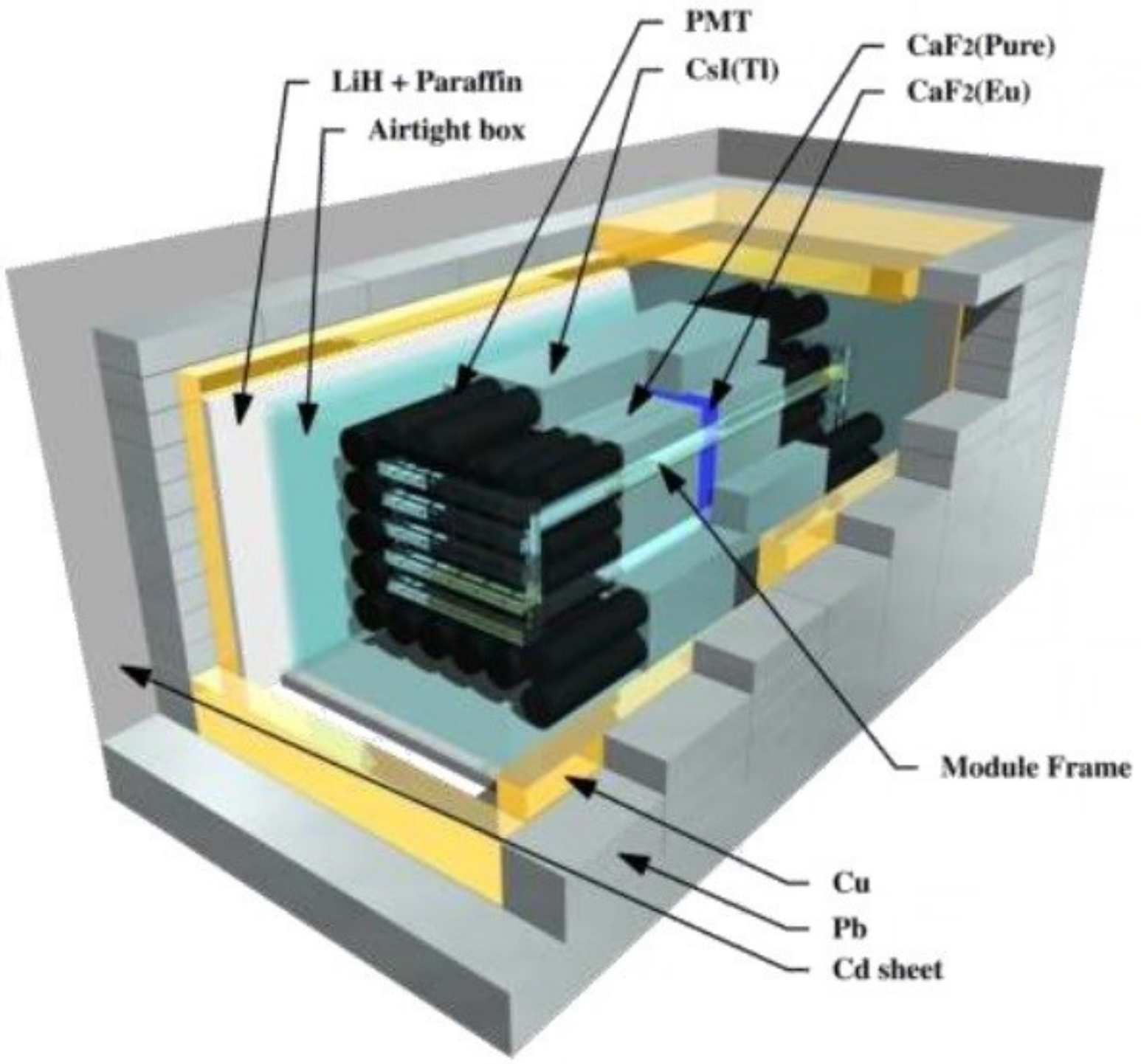

Fig. 1.20. Main elements of the ELEGANT VI set-up. The $H_3BO_3$-loaded water tanks which surrounded the set-up are not shown in the picture.

Combination of all the data accumulated in the ELEGANT experiments allowed to obtain the half-life limit on 0ν2β decay of $^{48}$Ca $T_{1/2}^{0\nu} > 5.8 \times 10^{22}$ years at 90% C.L. [81]. This corresponds to the upper limit on the effective Majorana neutrino mass $\langle m_\nu \rangle$ < (3.5–22) eV (90% C.L.) by using the nuclear matrix elements given in the Refs. [13, 113].

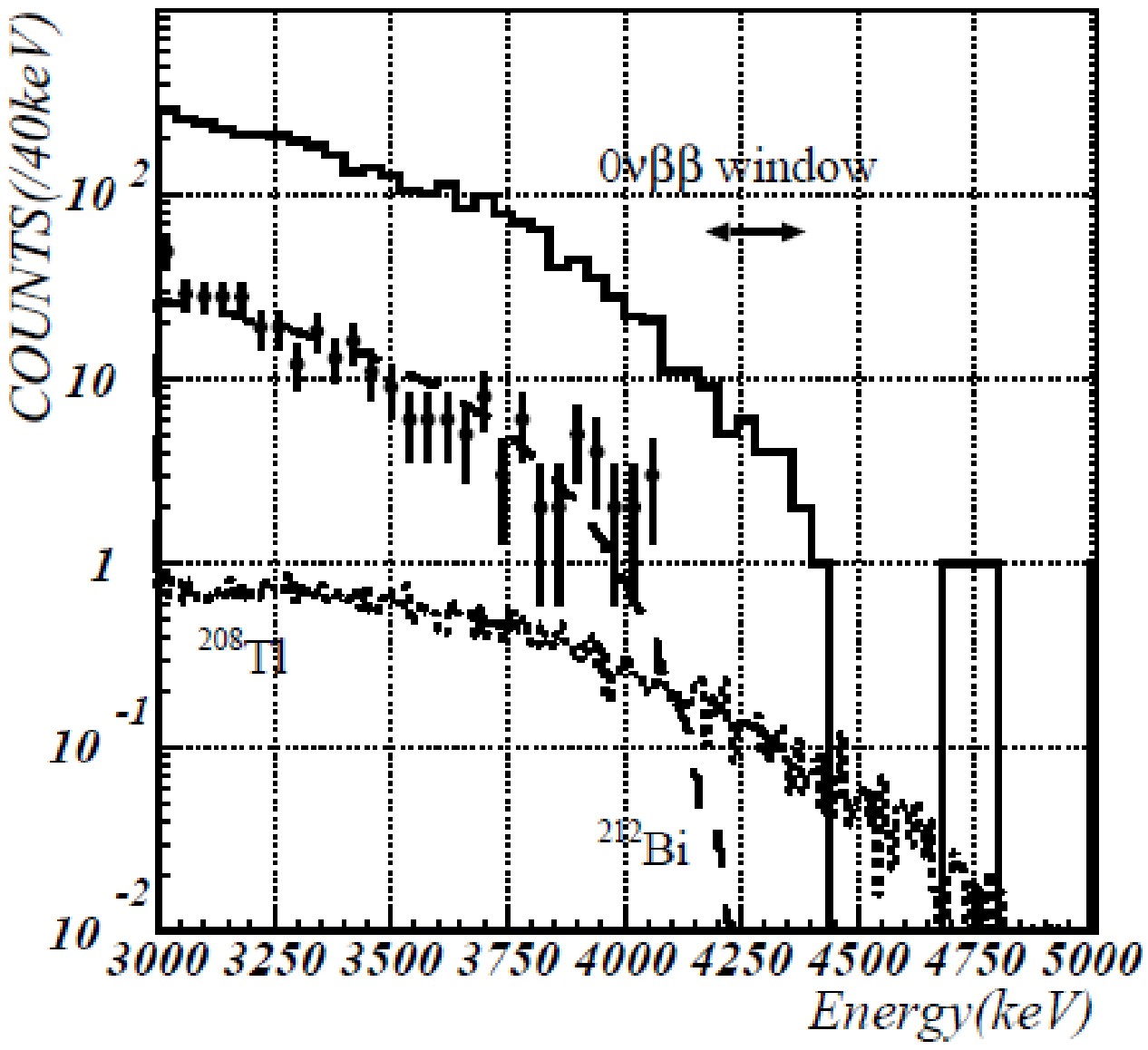

Fig. 1.21. The energy spectra obtained in the ELEGANT VI experiment with the $CaF_2$(Eu) crystal scintillators. The solid histogram represents raw energy spectrum. The experimental data after the pulse-shape analysis is presented by filled circles. The dashed and dotted lines correspond to the expected background after pulse-shape analysis from $^{212}$Bi and $^{208}$Tl, respectively.

The bolometric technique was used in the CUORICINO experiment to search for neutrinoless double beta decay of $^{130}$Te. The CUORICINO was a precursor of the larger experiment called CUORE (Cryogenic Underground Observatory for Rare Events), which will be discussed in the subsection 1.3.5.2. The CUORICINO experiment was located at the LNGS underground laboratory and operated from April 2003 till June 2008.

The detector was a tower array of 62 $TeO_2$ crystals which worked as cryogenic bolometers in the dilution refrigerator at the temperature of 10 mK. The tower consisted of 13 planes: the upper ten planes and the lowest one were filled with a four 5 × 5 × 5 cm$^3$ $TeO_2$ crystals on each plane, while the 11th and 12th planes have nine crystals 3 × 3 × 6 cm$^3$. Properties of tellurium dioxide crystals used in the set-up are presented in Table 1.4.

Table 1.4
The main properties of $TeO_2$ crystals used in the CUORICINO experiment. The isotopic abundance and enrichment of the crystals in $^{130}$Te is denoted as δ.

| Crystal type | Amount of crystals | Size of crystal, cm$^3$ | Crystal mass, g | δ, % | Mass of $^{130}$Te, g | Exposure (kg × yr) |
|---|---|---|---|---|---|---|
| Natural | 44 | 5 × 5 × 5 | 790 | 33.8 | 217 | 16.74 |
| Natural | 14 | 3 × 3 × 6 | 330 | 33.8 | 91 | 2.114 |
| $^{130}$Te-enriched | 2 | 3 × 3 × 6 | 330 | 75 | 199 | 0.895 |
| $^{128}$Te-enriched[2] | 2 | 3 × 3 × 6 | 330 | — | — | — |

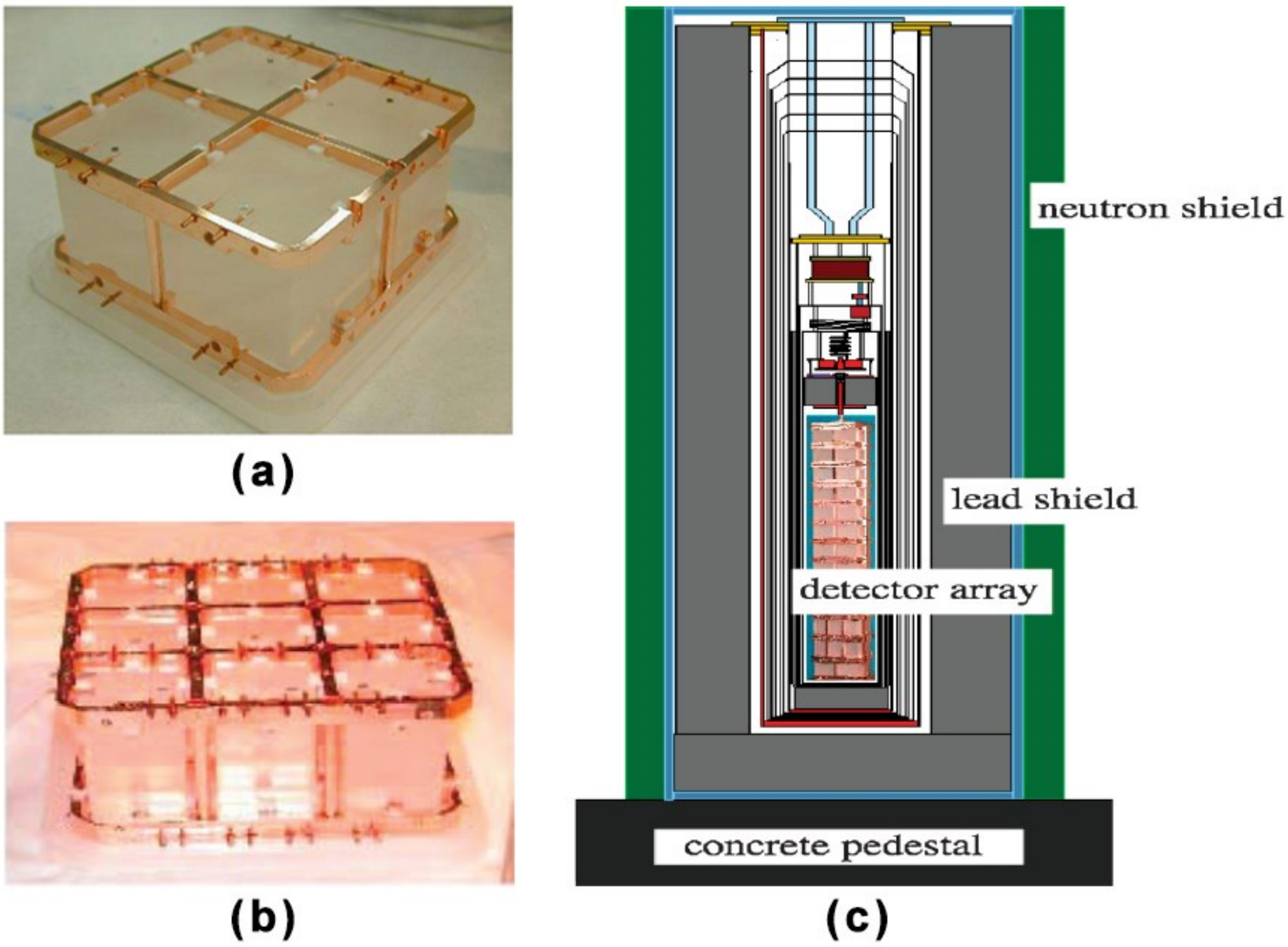

Fig. 1.22. The CUORICINO detector modules with (a) four 5 × 5 × 5 cm$^3$ and (b) nine 3 × 3 × 6 cm$^3$ $TeO_2$ crystals. (c) A schematic view of the CUORICINO set-up demonstrating the tower hanging from the mixing chamber and the different heat shields with the external shielding.

[2] The content of the isotope $^{130}$Te in the $^{128}TeO_2$ crystals was so low that they were not considered in the 0ν2β analysis of $^{130}$Te.

The crystals were fixed with polytetrafluoroethylene (PTFE) supports in a copper holder which acted also as a thermal bath (see Fig. 1.22 (a,b)). Neutron transmutation doped (NTD) germanium thermistors were glued to each $TeO_2$ crystal to measure the thermal pulses. A silicon heater was also attached to the bolometers to periodically heat the crystals with a fixed amount of energy. The obtained heat pulses were used for offline gain correction, and were periodically supplied by the calibrated ultra-stable pulser [114].

To protect the bolometers from an internal radioactive contamination of the dilution unit the detector tower was shielded by low-activity Roman lead (activity in $^{210}$Pb is < 4 mBq/kg). The inner layer of 10 cm thick Roman lead was placed inside the cryostat from above the tower. The intrinsic thermal shield was covered by Roman lead of 1.2 cm thickness. Moreover, the thermal shields made from the electrolytic copper acted as an additional shield with a minimum thickness of 2 cm. The cryostat was also surrounded by two 10 cm thick layers of lead. The inner layer was made from low radioactive lead with activity 16 ± 4 Bq/kg of $^{210}$Pb, while the outer one was made from ordinary low-activity lead. To reduce the background from neutrons, an external 10 cm shield of borated polyethylene was installed. The set-up was placed in a Plexiglas box flushed with pure $N_2$ gas to remove radon. Finally, a Faraday cage surrounded the detector to reduce electromagnetic interference. A sketch of the CUORICINO set-up is presented in Fig. 1.22 (c). More details on the set-up and detector features can be found in Ref. [40] and references therein.

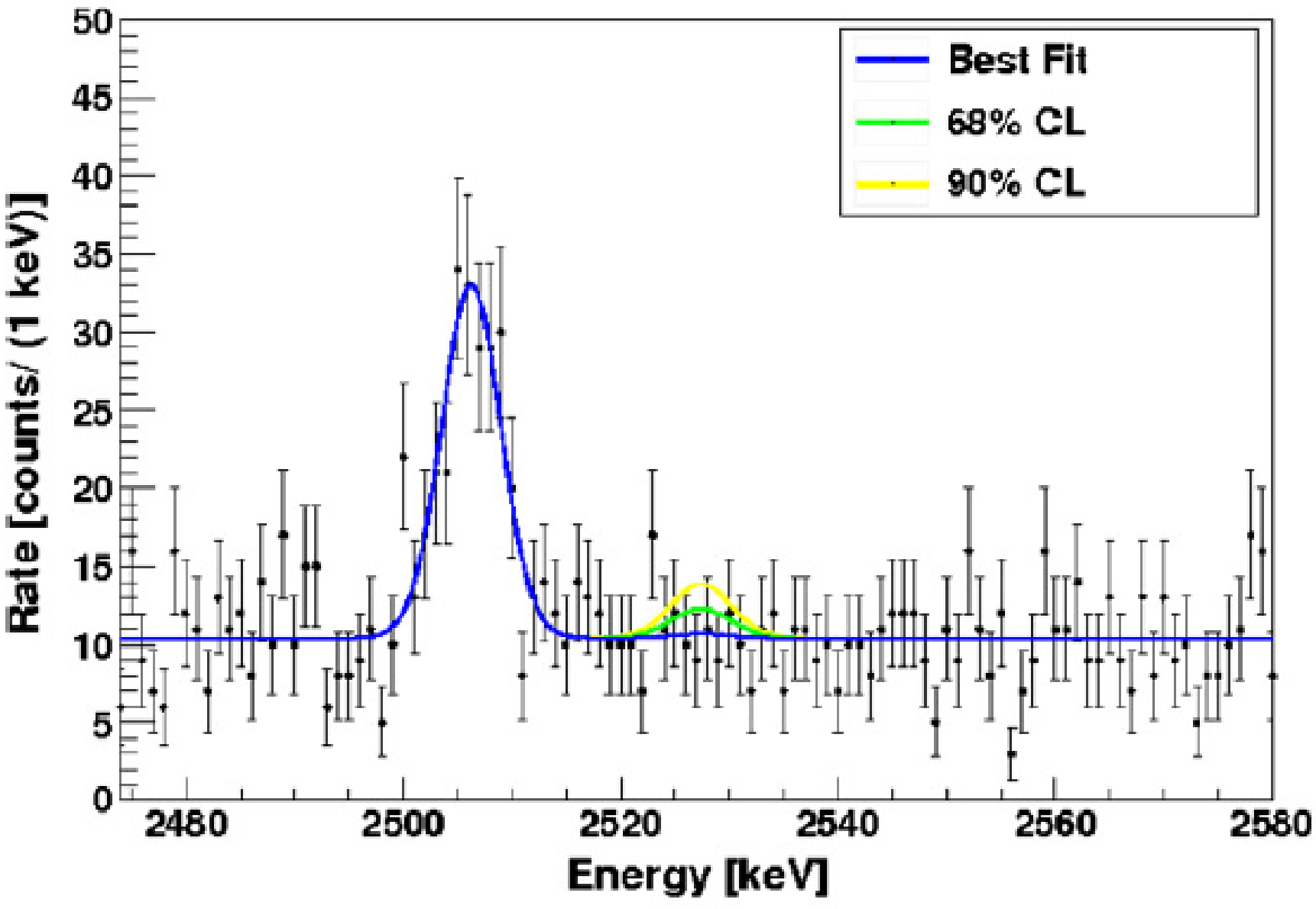

Fig. 1.23. The CUORICINO sum spectrum in the energy region of 0ν2β decay of $^{130}$Te. The experimental data is marked with dots; blue line represents the best fit curve; green and red lines denote to 0ν2β decay peak with the half-life lower limits at 68% and 90% C.L., respectively.

The average background level in the energy region of 0ν2β decay ($Q_{2\beta}$ of $^{130}$Te is (2526.97 ± 0.23) keV) was (0.18 ± 0.01) counts/(keV·kg·yr) for the 5 × 5 × 5 cm$^3$ bolometers, and (0.20 ± 0.04) counts/(keV·kg·yr) for the 3 × 3 × 6 cm$^3$ crystals [115]. The energy resolution was evaluated using 2615 keV $^{208}$Tl peak. The average energy resolution was 6.3 keV for 5 × 5 × 5 cm$^3$ crystals, 9.9 keV for 3 × 3 × 6 cm$^3$ natural crystals, and 13.9 keV for 3 × 3 × 6 cm$^3$ enriched crystals. No evidence of 0ν2β decay was observed in the CUORICINO experiment (see Fig. 1.23). A half-life limit on 0ν2β decay of $^{130}$Te $T_{1/2}^{0\nu} \geq 2.8 \times 10^{24}$ yr at 90% C.L. was obtained by analysis of the 19.75 kg × yr exposure. The corresponding upper limit for the effective neutrino mass was set between 0.30 to 0.71 eV depending on the theoretically calculated nuclear matrix elements [93].

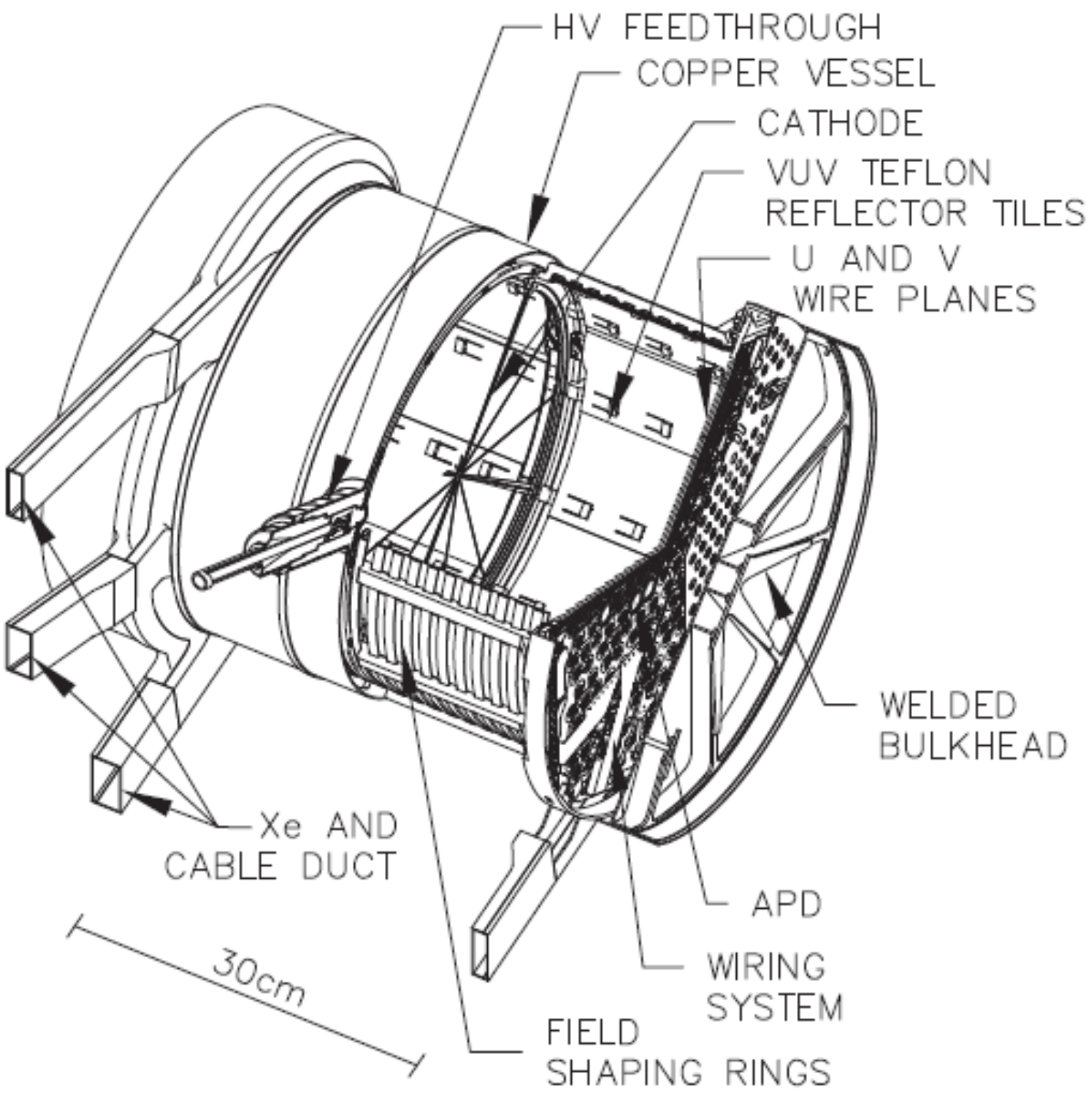

Fig. 1.24. Schematic view of the EXO-200 time projection chamber used to study 2β decay of $^{136}$Xe.

The Enriched Xenon Observatory (EXO) aims to search for neutrinoless double beta decay of $^{136}$Xe with the help of time projection chamber filled with the liquid xenon (LXe). The EXO-200 is a pilot stage experiment running underground at the depth of $1585^{+11}_{-6}$ m.w.e. in the Waste Isolation Pilot Plant (Carlsbad, USA) [116].

The EXO-200 uses 200 kg of xenon enriched in the isotope $^{136}$Xe up to 80.6%. 175 kg of this xenon are in liquid phase. Meanwhile, only 110 kg of the LXe are located in the active volume of the detector. The TPC has cylindrical shape with sizes ≈ 40 cm in diameter and ≈ 44 cm in length. The cathode grid placed in the middle of the cylinder divides it into two identical drift volumes. The ionization signal is recorded at each end of the chamber by wire planes. There exist two parallel wire grids oriented at the angle of 60 degrees between each other and spaced by the distance of 6 mm. The drifting ionization comes through the first (induction, V) wire grid, and is collected by the second (anode, U) wire grid, which enables the two-dimensional localization of the ionization cloud. The large area avalanche photodiodes (APD, see Ref. [117] for details) are located 6 mm behind the charge collection anode planes, and used to read-out the scintillation signal. The simultaneous collection of the charge and light provides the three-dimensional position sensitivity by calculation of the electron drift time. The EXO-200 time projection chamber is operating at the temperature of 167 K and the pressure of 147 kPa. A schematic view of the TPC is presented in Fig. 1.24.

The time projection chamber is mounted inside the double-walled vacuum insulated copper cryostat. The cryostat was made from 5901 kg of specially selected low-background copper and provides the shielding of 5 cm in copper, and 50 cm in the liquid refrigerator (ultra-clean fluid HFE-7000 [118]). The outer shielding layer of 25 cm of lead is covering the cryostat in all directions. The detector with its shielding is housed inside the clean room, which is surrounded on four sides by a cosmic ray veto system. The veto system is based on 29 panels of 5 cm thick Bicron BC-412 plastic scintillators, viewed by 8 PMTs and supported by 4 cm of borated polyethylene. The main components of the set-up are shown in Fig. 1.25. The details of the EXO-200 detector can be found in Ref. [119].

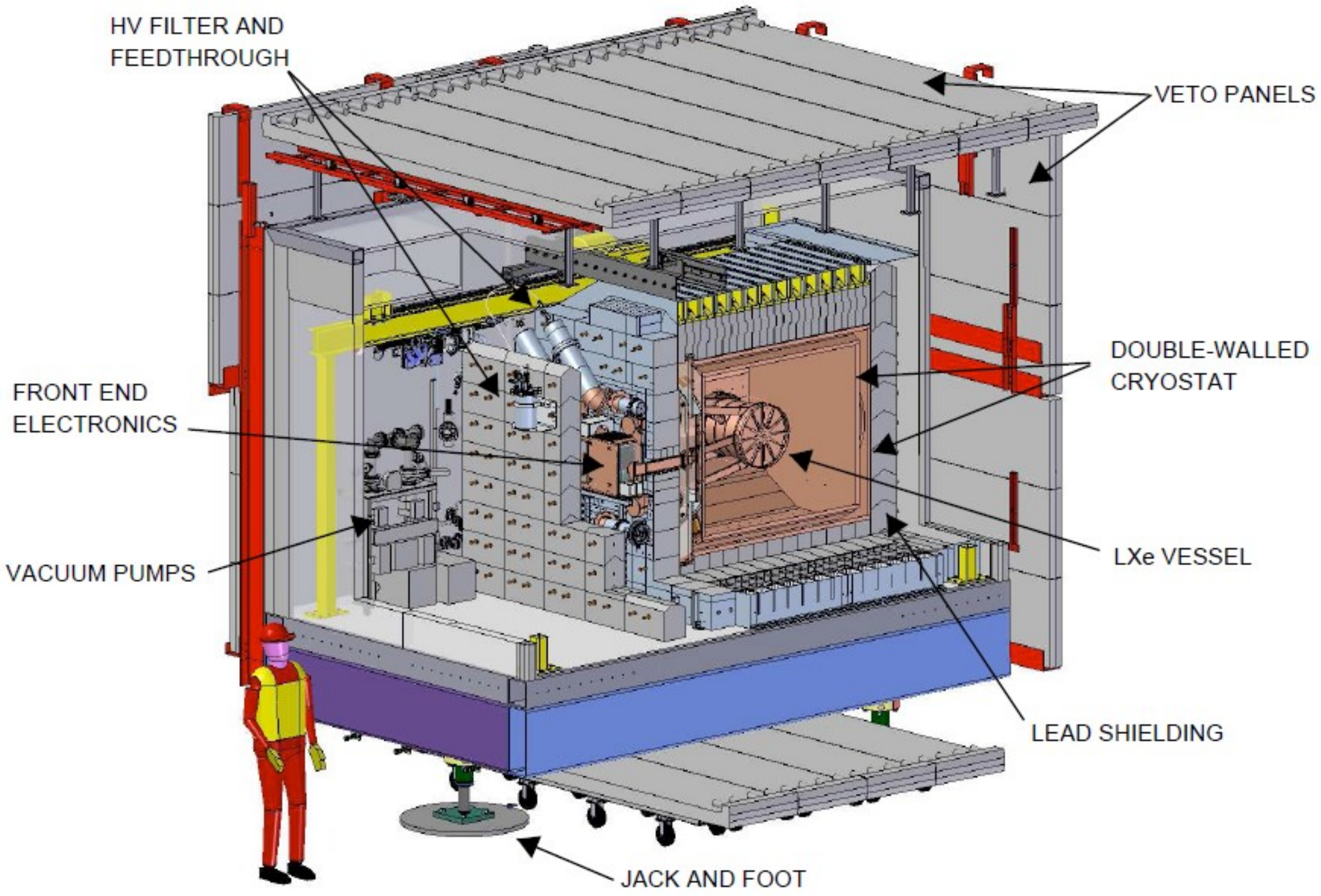

Fig. 1.25. Schematic view of the primary elements of the EXO-200 set-up.

The events in the EXO-200 detector are divided into two groups depending on the number of detected charge deposits: single-site (SS) or multi-site (MS) events. The achieved energy resolution at the 2β decay energy of $^{136}$Xe ($Q_{2\beta}$ = 2457.83 ± 0.37 keV) is (1.53 ± 0.06)% for SS events, and (1.65 ± 0.05)% for MS. The background counting rate in the region of interest (ROI) ±2σ was evaluated as (1.7 ± 0.2) × $10^{-3}$ counts/(keV·kg·yr). The half-life value of 2ν2β decay of $^{136}$Xe was measured during 127.6 days of measurements as (2.165 ± 0.016(stat) ± 0.059(syst)) × $10^{21}$ yr [94]. The latest results of EXO-200 experiments based on 100 kg × yr exposure of $^{136}$Xe indicated no statistically significant evidence of 0ν2β decay (see Fig. 1.26). The lower limit on neutrinoless double beta half-life was set as 1.1 × $10^{25}$ yr at 90% C.L. [95]. This value corresponds to the upper limit on the effective Majorana neutrino mass as $\langle m_\nu \rangle \le$ (0.19–0.45) eV, using the nuclear matrix elements from references [53, 56, 120, 121] and phase space factor from Ref. [49].

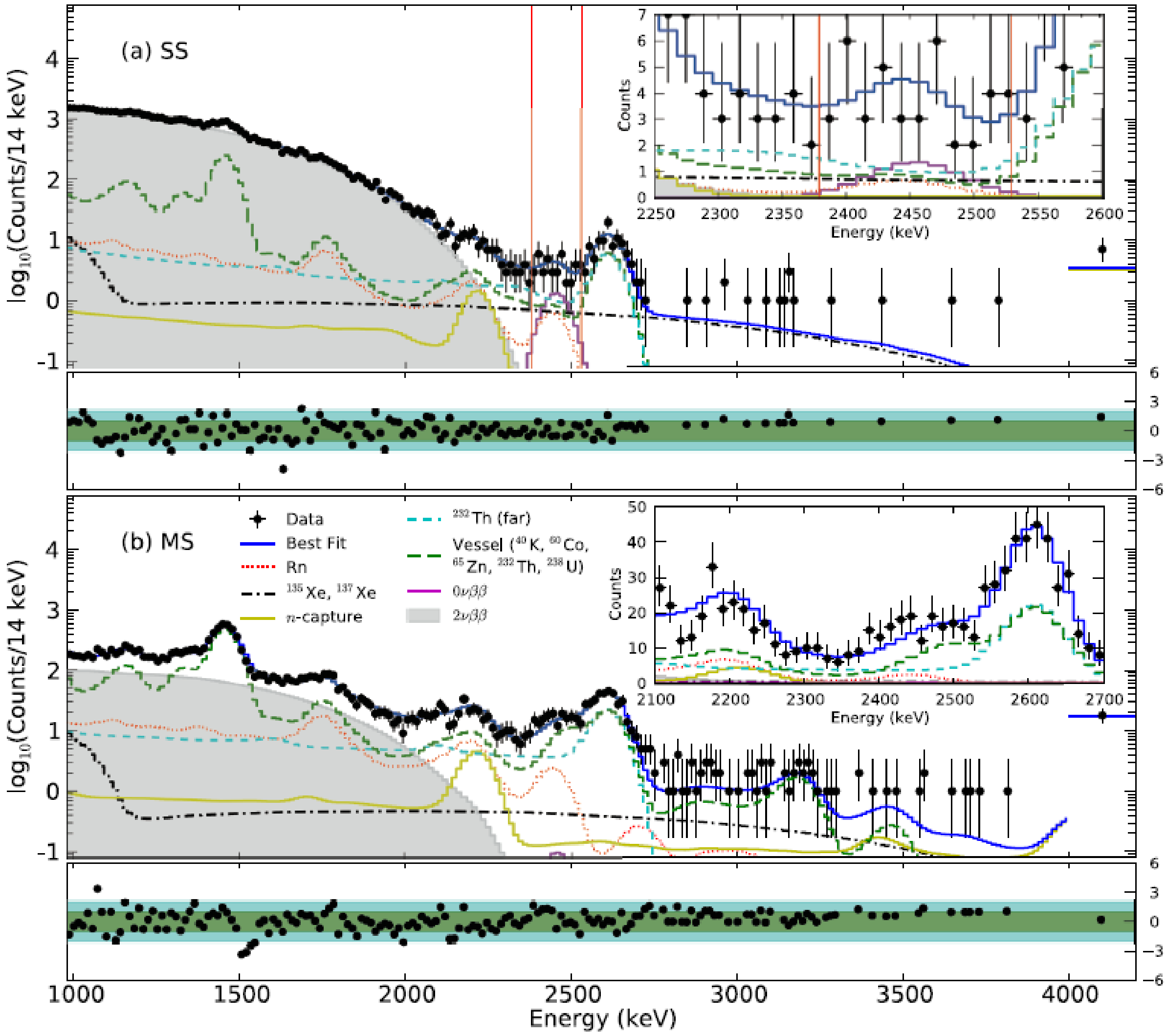

Fig. 1.26. The energy spectra measured during 100 kg × yr exposure of $^{136}$Xe in the EXO-200 experiment [95]. The upper panel (a) is for single-site events, while the lower one (b) is for multi-site events. The spectra around the $Q_{2\beta}$ = 2.45783(37) keV region are displayed in the insets. The size of one bin is 14 keV. The experimental data is marked with black points, as well as the residuals between data and the best fit normalized to the Poisson error. The data error bars indicate ±1 standard deviation intervals. There are 7 (18) events in the energy interval 4.0–9.8 MeV in the SS (MS) spectrum collected into the one overflow bin. Two red vertical lines in the SS spectra represent the ±2σ region of interest.

The KamLAND-Zen (KamLAND Zero-Neutrino Double-Beta Decay) is another currently running experiment to investigate double beta decay of $^{136}$Xe. The KamLAND-Zen detector is based on the KamLAND (Kamioka Liquid Scintillator Antineutrino Detector) which was modified in the middle of 2011. The set-up is located in the Kamioka underground laboratory (Hida, Japan) at the depth of approximately 2700 m.w.e.

13 tons of xenon-loaded liquid scintillator (Xe-LS) are contained in the spherical inner balloon of 3.08 m in diameter, and serves both as source and detector (see Fig. 1.27). The amount of enriched xenon gas in the Xe-LS, measured by gas chromatography, is (2.52 ± 0.07)% by weight. The isotopic enrichment of xenon gas in the isotope $^{136}$Xe is (90.93 ± 0.05)%. This corresponds to the total mass 296 kg of $^{136}$Xe in the KamLAND-Zen detector. The inner balloon is produced from the transparent nylon film of 25 μm in thickness, and is hanged in the center of the detector with the help of 12 film straps of the same material. The outer balloon of 13 m in diameter, constructed from the 135 μm thick nylon/EVOH (ethylene vinyl alcohol copolymer) composite film, surrounds the inner balloon. The outer balloon is filled with 1 kton of liquid scintillator which has 3% lower light yield and 0.10% lesser density than the Xe-LS. The liquid scintillator acts as an active shield against external gamma quanta and neutrons, and is also used to detect an internal contamination of the Xe-LS and the inner balloon material. The outer balloon is surrounded by the spherical stainless-steel containment tank of 18 m in diameter. The volume between them is filled with the buffer oil to protect the liquid scintillator from external radiation. Scintillation light is detected by 1325 17-inch and 554 20-inch photomultiplier tubes mounted on the stainless-steel containment tank, which provides 34% of solid-angle coverage. The stainless-steel containment tank is surrounded by a 3.2 kton water-Cherenkov detector for cosmic-ray muon identification. The detailed description of the KamLAND detector is given in Ref. [122].

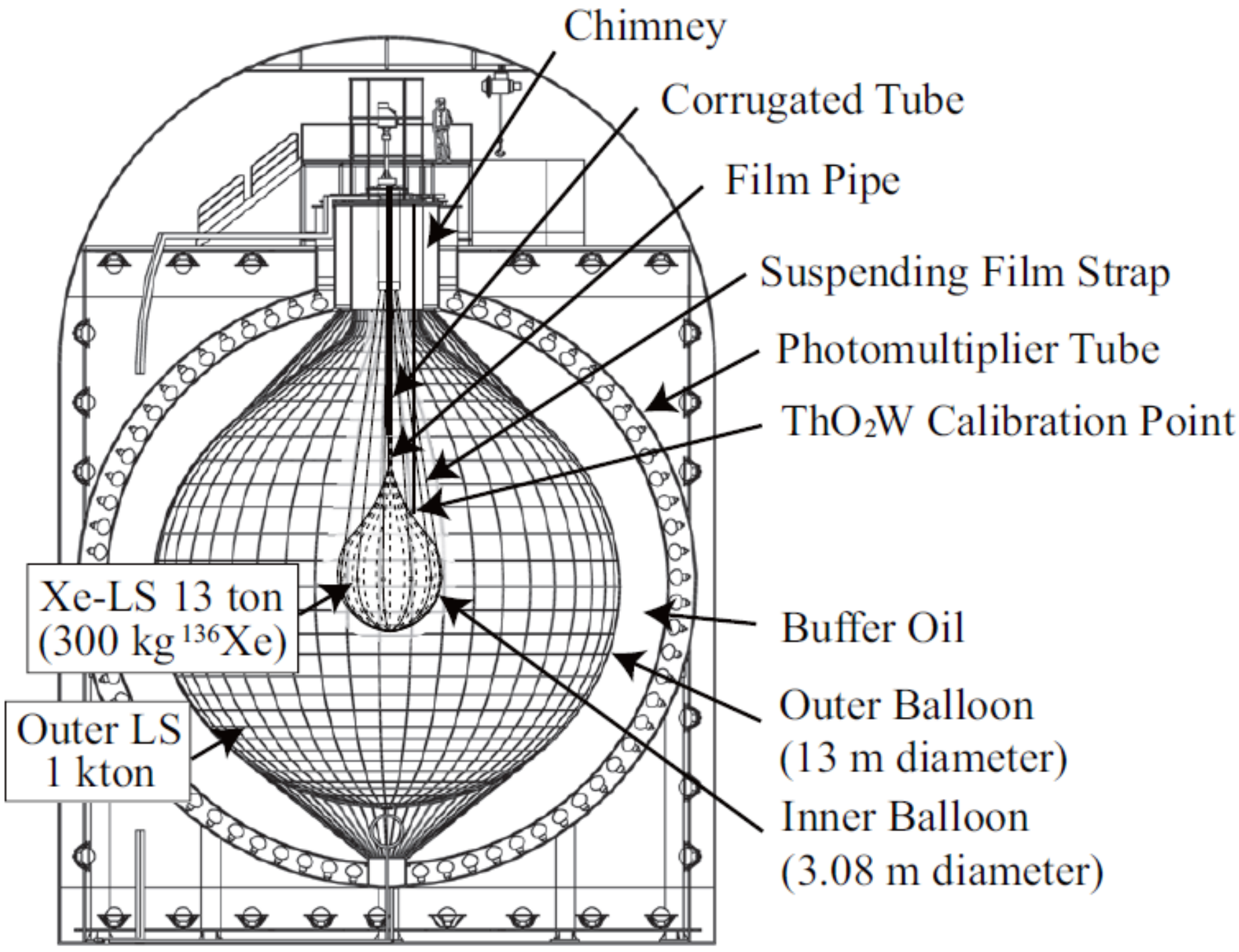

Fig. 1.27. Schematic diagram of the KamLAND-Zen detector used to search for 0ν2β decay of $^{136}$Xe.

The first KamLAND-Zen measurements demonstrated insufficient background level in the energy region of 0ν2β decay due to the strong contribution by $^{110m}$Ag. Assuming that impurities may occur due to suspended dust or fine particulate in the Xe-LS, the KamLAND-Zen collaboration passed 37 m$^3$ of the xenon-loaded liquid scintillator (corresponding to 2.3 full volume exchanges) through a 50 nm PTFE-based filter. Therefore the data is divided into two sets: one taken before the filtration (DS-1) and another one taken after it (DS-2). The details on each data set are presented in Table 1.5.

Table 1.5
The main characteristics of the two data sets of the KamLAND-Zen detector.

| | DS-1 | DS-2 | Total |
|---|---|---|---|
| Live time (days) | 112.3 | 101.1 | 213.4 |
| Fiducial Xe-LS mass (ton) | 8.04 | 5.55 | — |
| Xe concentration (wt%) | 2.44 | 2.48 | — |
| $^{136}$Xe mass (kg) | 179 | 125 | — |
| $^{136}$Xe exposure (kg × yr) | 54.9 | 34.6 | 89.5 |

The performed Xe-LS filtration was not enough to remove $^{110m}$Ag impurities, which are still dominant in 0ν2β decay region. The best-fit on $^{110m}$Ag rates in the Xe-LS are (0.19 ± 0.02) (ton × day)$^{-1}$ for the DS-1, and (0.14 ± 0.03) (ton × day)$^{-1}$ for the DS-2. The detector background measured with a total exposure of 89.5 kg × yr is presented in Fig. 1.28.

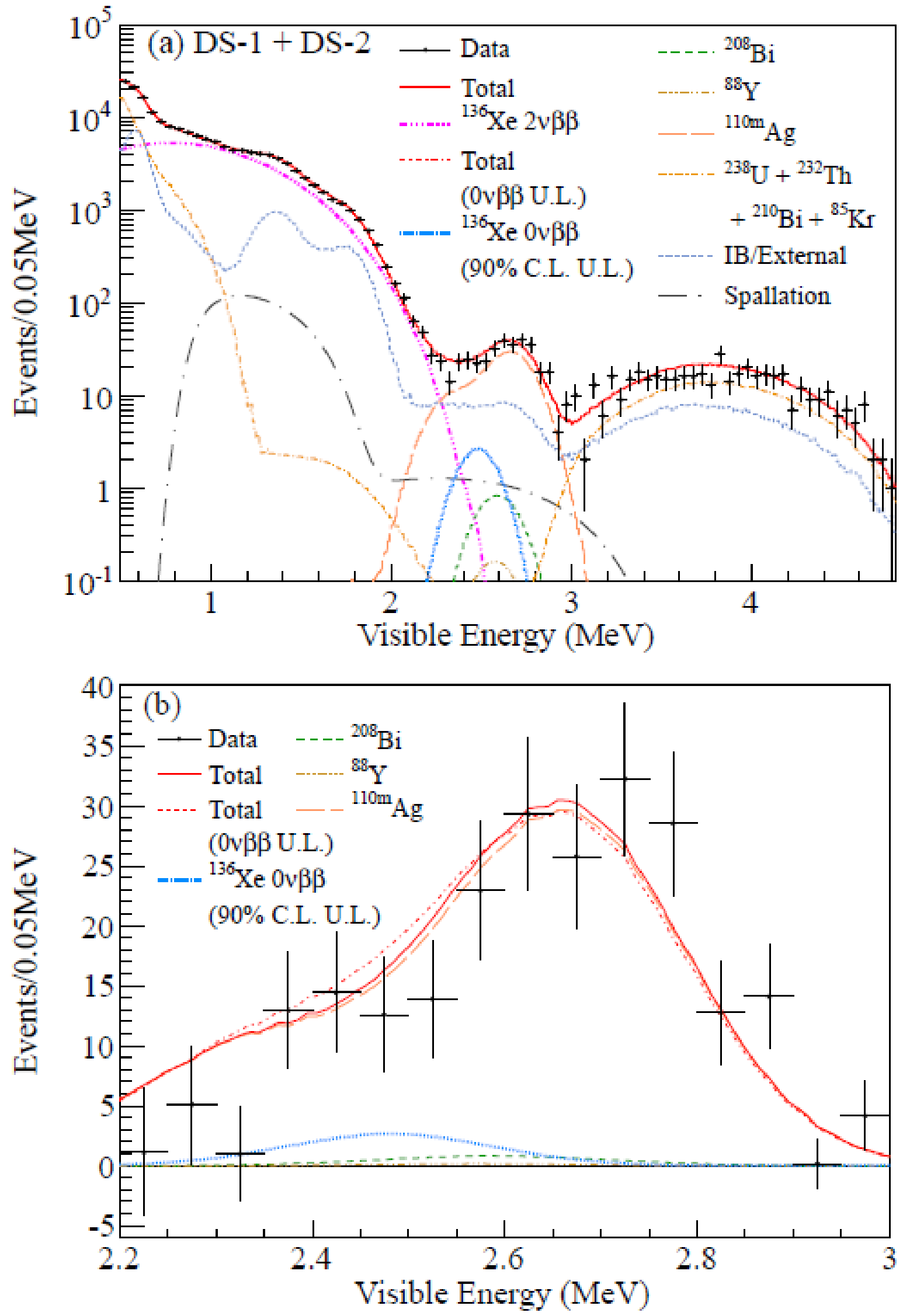

Fig. 1.28. (a) The KamLAND-Zen energy spectrum measured with a total $^{136}$Xe exposure of 89.5 kg × yr. (b) The same spectra in the energy region 2.2 < E < 3.0 MeV after the subtraction of known background contributions.

The KamLAND-Zen vertex resolution is $\sigma \sim 15\ \text{cm}/\sqrt{\text{E(MeV)}}$, and the energy resolution is $\sigma = (6.6 \pm 0.3)\%/\sqrt{\text{E(MeV)}}$. The two-neutrino double beta decay of $^{136}$Xe was studied with the exposure of 27.4 kg × yr. The obtained half-life value is (2.38 ± 0.02(stat) ± 0.14(syst)) × $10^{21}$ yr [43], which is consistent with the EXO-200 result. Taking into account the total exposure of 89.5 kg × yr, the lower limit on neutrinoless double beta decay half-life was set as $T_{1/2}^{0\nu} > 1.9 \times 10^{25}$ yr at 90% C.L. [96]. However, the KamLAND-Zen result is still suppressed by the background from $^{110m}$Ag. There exist two possible explanations of $^{110m}$Ag presence in the Xe-LS: either the inner balloon was contaminated during manufacturing by the radioactive contamination resulted from the Fukushima-I power plant fallout, or the $^{110m}$Ag is of cosmogenic origin and was produced in the xenon by spallation of the hadronic component of cosmic rays.

Finally, the combined results from the KamLAND-Zen and the EXO-200 experiments give the lower limit on 0ν2β decay half-life as $T_{1/2}^{0\nu} > 3.4 \times 10^{25}$ yr at 90% C.L. This corresponds to the upper limit on effective Majorana neutrino mass of $\langle m_\nu \rangle$ < (0.12−0.25) eV, depending on the nuclear matrix element calculations [96].

### 1.3. Next-generation 2β decay experiments

Nowadays experiments to search for neutrinoless double beta decay approaching a new phase of the development to advance sensitivity high enough to reach the inverted hierarchy region. Large-scale detectors operating a few hundred kilograms of 2β isotopes, with an extremely low background level are requested for this purpose.

#### 1.3.1. Experiments with $^{136}$Xe

The nEXO project ("next EXO") is an upgrade of the existing EXO-200 experiment (see subsection 1.2.3). The collaboration has performed an extensive research and development (R&D) to design a large xenon detector and to study the possibility of tagging the barium single ion (Ba-tagging) produced by the 2β decay in order to reduce background further [123]. The nEXO plans to use ~ 5000 kg of LXe, isotopically enriched in $^{136}$Xe up to 90%. The TPC will have single drift region in order to maximize the advantage of LXe self-shielding. The energy resolution of the detector should be improved to ~ 1% at $Q_{2\beta}$. Moreover, massive improvements of electronics, charge read-out and light detection are also foreseen. The nEXO background is evaluated as 3.7 $\text{ROI}^{-1}\cdot\text{ton}^{-1}\cdot\text{yr}^{-1}$ in the region of interest (ROI = $Q_{2\beta} \pm 0.5\cdot$FWHM). The estimated sensitivity of the detector is $T_{1/2}^{0\nu} > 6.6 \times 10^{27}$ yr at 90% C.L. for 5 years of measurements. This corresponds to the upper limit on effective Majorana neutrino mass as 7–18 meV depending on nuclear matrix elements calculations (see Fig. 1.29). In case if the Ba-tagging technique will be applied to the nEXO detector, the experimental sensitivity $\langle m_\nu \rangle \leq$ 3–8 meV will be high enough to completely cover the inverted hierarchy region [124].

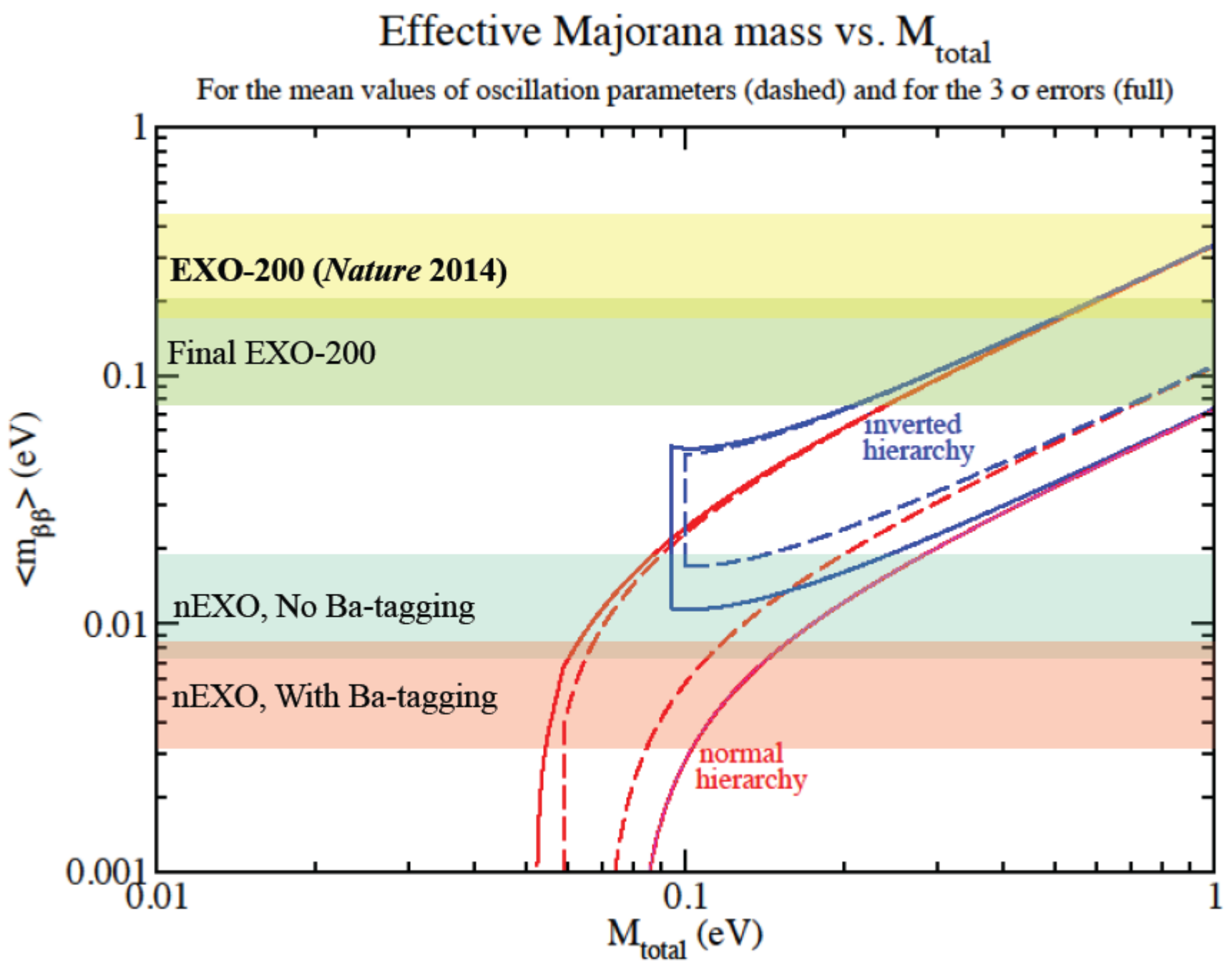

Fig. 1.29. The expected sensitivity of the nEXO experiment to search for neutrinoless double beta decay of $^{136}Xe$ with 5000 kg of the enriched liquid xenon.

The KamLAND-Zen is a currently running experiment, the Phase-1 of which has been already discussed in subsection 1.2.3. The Phase-2 was started in December 2013 after several significant improvements. The background (especially from $^{110m}Ag$) was decreased by using a new Xe-LS purification system which was working in the circulation mode and consists of distillation system, $N_2$ purging system and density controlling. The concentration of xenon was increased up to (2.96 ± 0.01) wt%, which corresponds to ~ 347 kg of $^{136}Xe$. The preliminary result on 2ν2β half-life is (2.32 ± 0.05(stat) ± 0.08(syst)) × $10^{21}$ yr from 114.8 days of measurements [125]. The lower limit on neutrinoless double beta decay half-life in Phase-2 is $T^{0\nu}_{1/2} > 1.9 \times 10^{25}$ yr at 90% C.L. The combined result of the two phases is currently $T^{0\nu}_{1/2} > 2.6 \times 10^{25}$ yr at 90% C.L. [125].

The next phase of the KamLAND-Zen experiment is an expansion of the detector up to ~ 600–700 kg of xenon. There are also several ongoing R&D for the KamLAND2-Zen experiment. In particular, the energy resolution is planned to be improved by increasing the light collection efficiency. Moreover, the light yield should be ~ 40% higher by using liquid scintillator with higher scintillation efficiency. New high quantum efficiency PMTs are expected to be utilized to improve the quantum efficiency up to 31%. There are also studies on dead-layer free scintillation balloon, and other innovative activities on the KamLAND2-Zen detector [125]. The actual and expected sensitivities to 0ν2β decay are presented in Fig. 1.30.

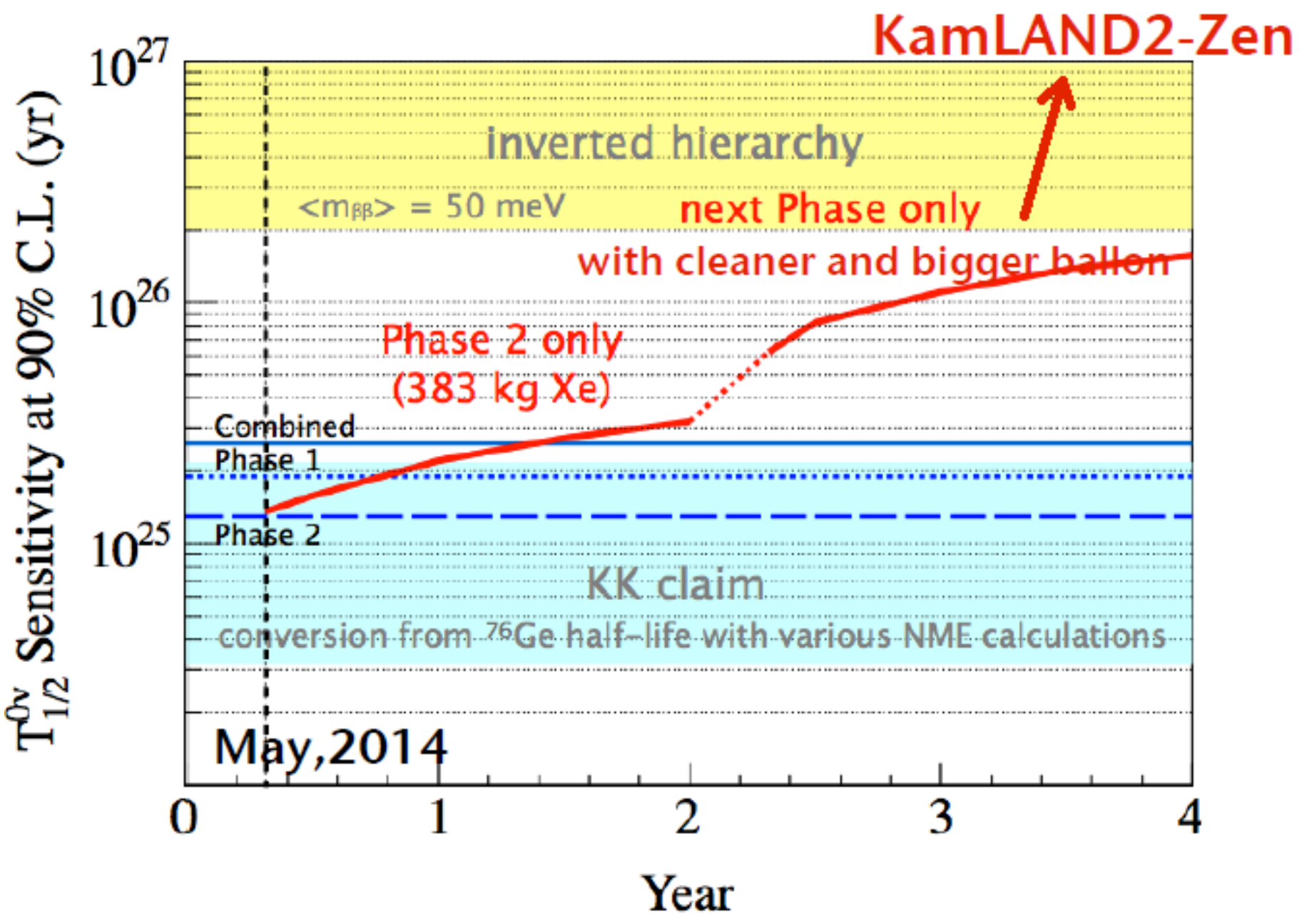

Fig. 1.30. The current and expected KamLAND-Zen detector sensitivities to the half-life of neutrinoless double beta decay of $^{136}$Xe.

The NEXT project (Neutrino Experiment with a Xenon TPC) has a similar concept to the EXO experiment, except that the high-pressure xenon gas (HPXe) will be used to search for $0\nu2\beta$ decay of $^{136}$Xe. The first phase of the experiment, called NEXT-100, will use 100 kg of HPXe enriched in $^{136}$Xe up to 90%. The set-up will be located in the Canfranc Underground Laboratory (Spain) at the depth of 2450 m.w.e. The NEXT-100 operation is planned to start before the end of this year.

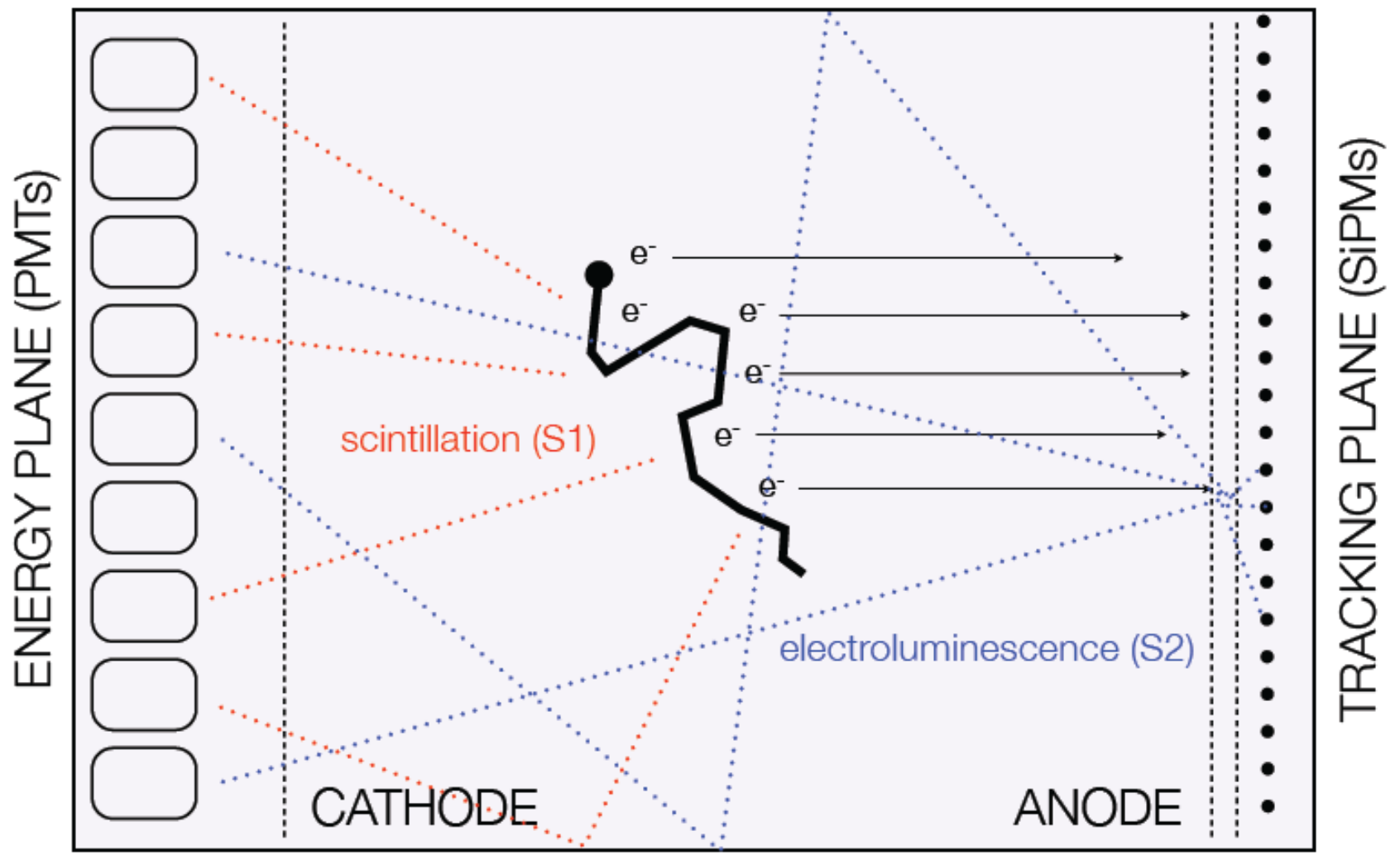

Fig. 1.31. The main idea of the NEXT detector: the anode generates the electroluminescence light which is detected by the silicon photomultipliers (SiPMs) plane used for tracking; and is also read-out by the PMTs plane behind the transparent cathode for precise energy measurement [126].

The NEXT-100 detector reads out scintillation and ionization in the TPC as light signals, which allows to obtain high energy and space resolutions. The expected energy resolution of the detector is 1% at 662 keV, which corresponds to FWHM = 0.5% at $Q_{2\beta}$. The particle tracks will be reconstructed with an uncertainty on the level of 5–10 mm. The basic concept of the detector is presented in Fig. 1.31. The main features of the NEXT-100 detector are described in Ref. [127].

The predicted background rate in the energy region of 0ν2β decay of $^{136}Xe$ is 5 × $10^{-4}$ counts/(keV·kg·yr). The expected sensitivity of the NEXT-100 detector is $T_{1/2}^{0\nu} > 5.9 \times 10^{25}$ yr at 90% C.L. after the 5 years of data taking [127], which corresponds to the upper limit on the effective Majorana neutrino mass as $\langle m_\nu \rangle \leq$ 102–129 meV (see Fig. 1.32).

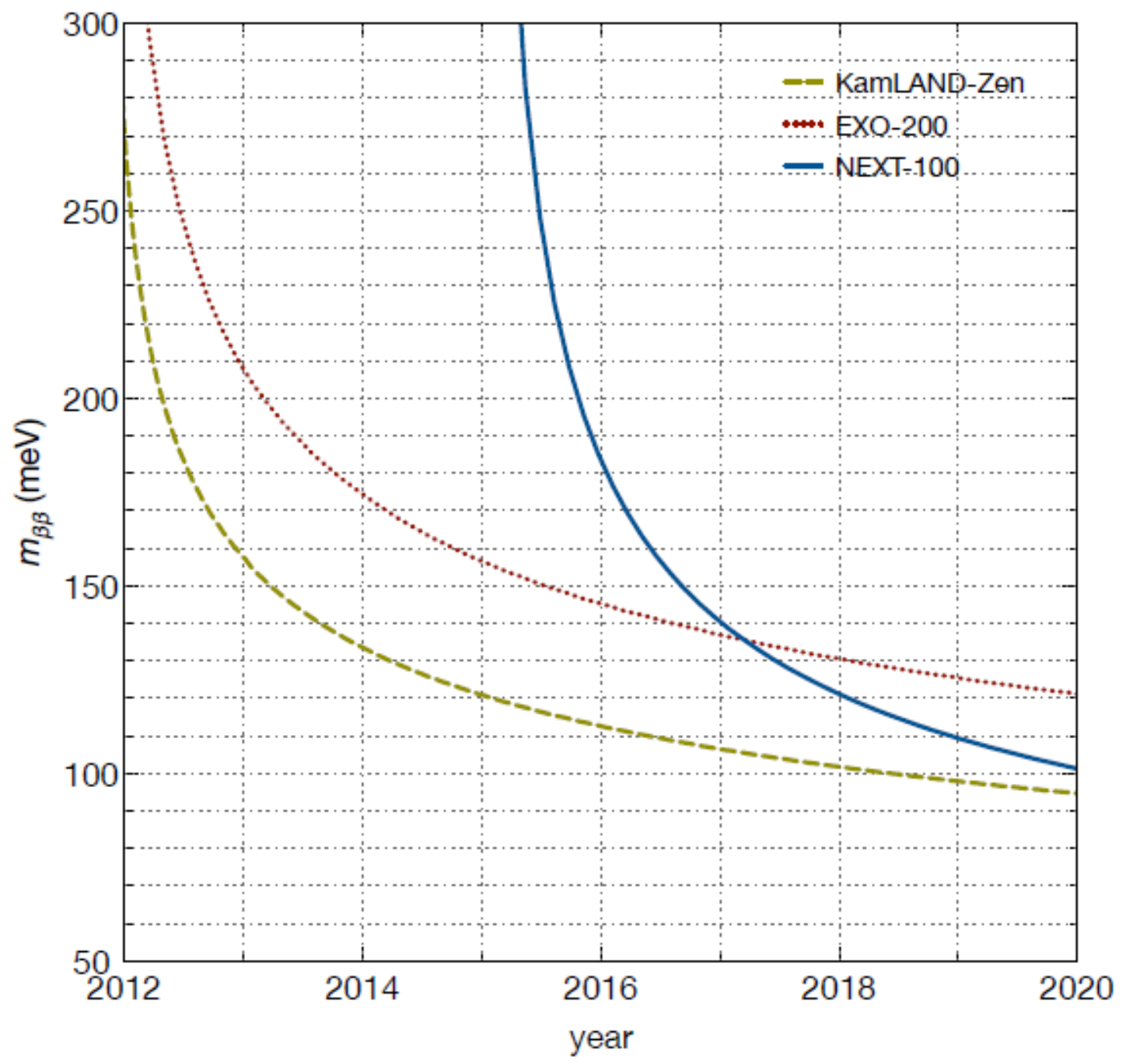

Fig. 1.32. The expected sensitivities of the EXO-200, the KamLAND-Zen and the NEXT-100 experiments to search for 0ν2β decay of $^{136}Xe$ as a function of the running time [126].

### 1.3.2. SuperNEMO tracking detector

The SuperNEMO collaboration is going to use the NEMO-3 tracking and calorimetry technique (see subsection 1.2.3). The detector should consist of 20 modules containing in total ≈ 100 kg of enriched 2β isotope in the form of a thin foil. The single modules are of a planar design (see Fig. 1.33) unlike the cylindrical shape of the NEMO-3 detector. Otherwise, the design is very similar to the NEMO-3 experiment: the thin (40 mg/$cm^2$) foil (3 × 4.5 m) source will be placed in the middle plane of a gas tracking volume and will be surrounded by ~ 500–700 plastic scintillator blocks used as a calorimeter. The tracking detector will consist of ~ 2000 wire drift cells operated in Geiger mode and located in the magnetic field of 25 Gauss to perform particle identification and vertex reconstruction, as well as to measure angular correlations of electrons. The possible double beta candidates for the SuperNEMO experiment are $^{82}Se$, $^{150}Nd$ and $^{48}Ca$ [128].

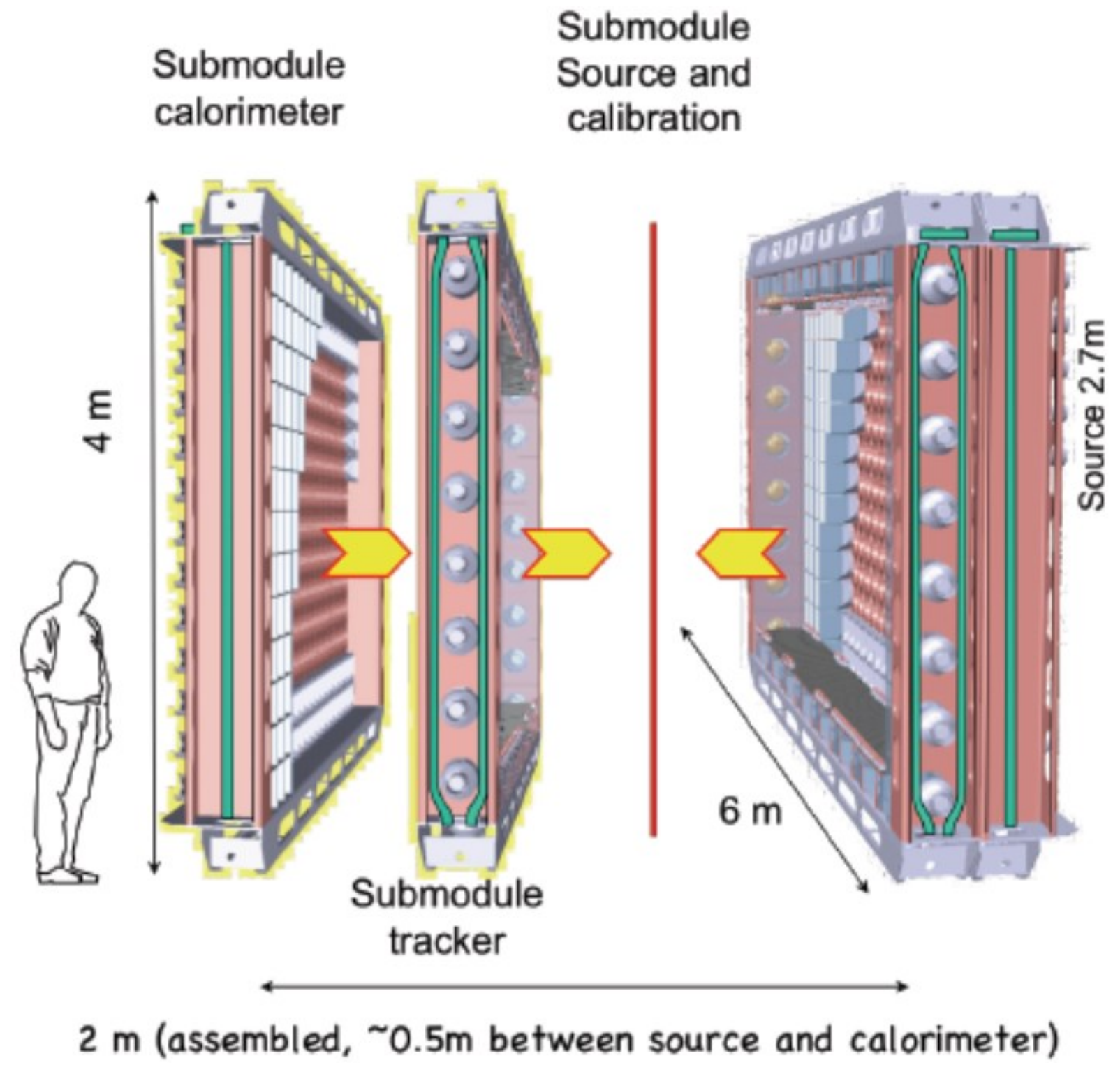

Fig. 1.33. The single module of the SuperNEMO tracking detector.

The expected energy resolution is ~ 4% at 3 MeV, while the predicted spatial resolution is ~ 0.7 mm in the direction perpendicular to the 2β source foils and ~ 10 mm in the parallel one. The detection efficiency should be increased up to ~ 30%. Initially, the SuperNEMO experiment will operate only with a single module containing ~ 7 kg of $^{82}$Se. This demonstrator is now under construction in the LSM and should start data taking before the end of this year. The expected half-life sensitivity of the SuperNEMO demonstrator is $T^{0\nu}_{1/2} > 6.6 \times 10^{24}$ yr at 90% C.L., taking into account the predicted zero background and two years of measurements [129]. The 5-year sensitivity of the complete SuperNEMO detector with ~ 100 kg of $^{82}$Se is evaluated as ~ $10^{26}$ yr for 0ν2β decay half-life or $\langle m_\nu \rangle \leq$ 40–140 meV for effective Majorana neutrino mass [129].

### 1.3.3. High-purity germanium detectors enriched in $^{76}$Ge

The GERDA project is one of the most promising experiments to search for 0ν2β decay of $^{76}$Ge with germanium detectors. The GERDA-II is an upgrade of the completed phase I which had been already discussed in subsection 1.2.3. The phase II will perform different modifications to detector assembly, improve electronics and apply liquid argon scintillation veto. The amount of the BEGe detectors will be increased up to 30 BEGe with 20.8 kg of germanium enriched in $^{76}$Ge up to 88%. Taking into account 18 kg of the old coaxial HPGe detectors (enriched in $^{76}$Ge to 86%), the total mass of germanium for the GERDA-II is expected to be 34 kg of isotope $^{76}$Ge.

The background is expected to be on the level of $10^{-3}$ counts/(keV·kg·yr), which is ten times lower than the counting rate obtained in phase I. The energy resolution of BEGe detectors is less than 3 keV at $Q_{2\beta}$. The predicted GERDA-II half-life sensitivity to the 0ν2β decay of $^{76}$Ge after the total exposure of 100 kg × yr is $T_{1/2}^{0\nu} > 1.4 \times 10^{26}$ yr at 90% C.L. [130]. The dependence of the expected GERDA-II sensitivity from the exposure time is presented in Fig. 1.34.

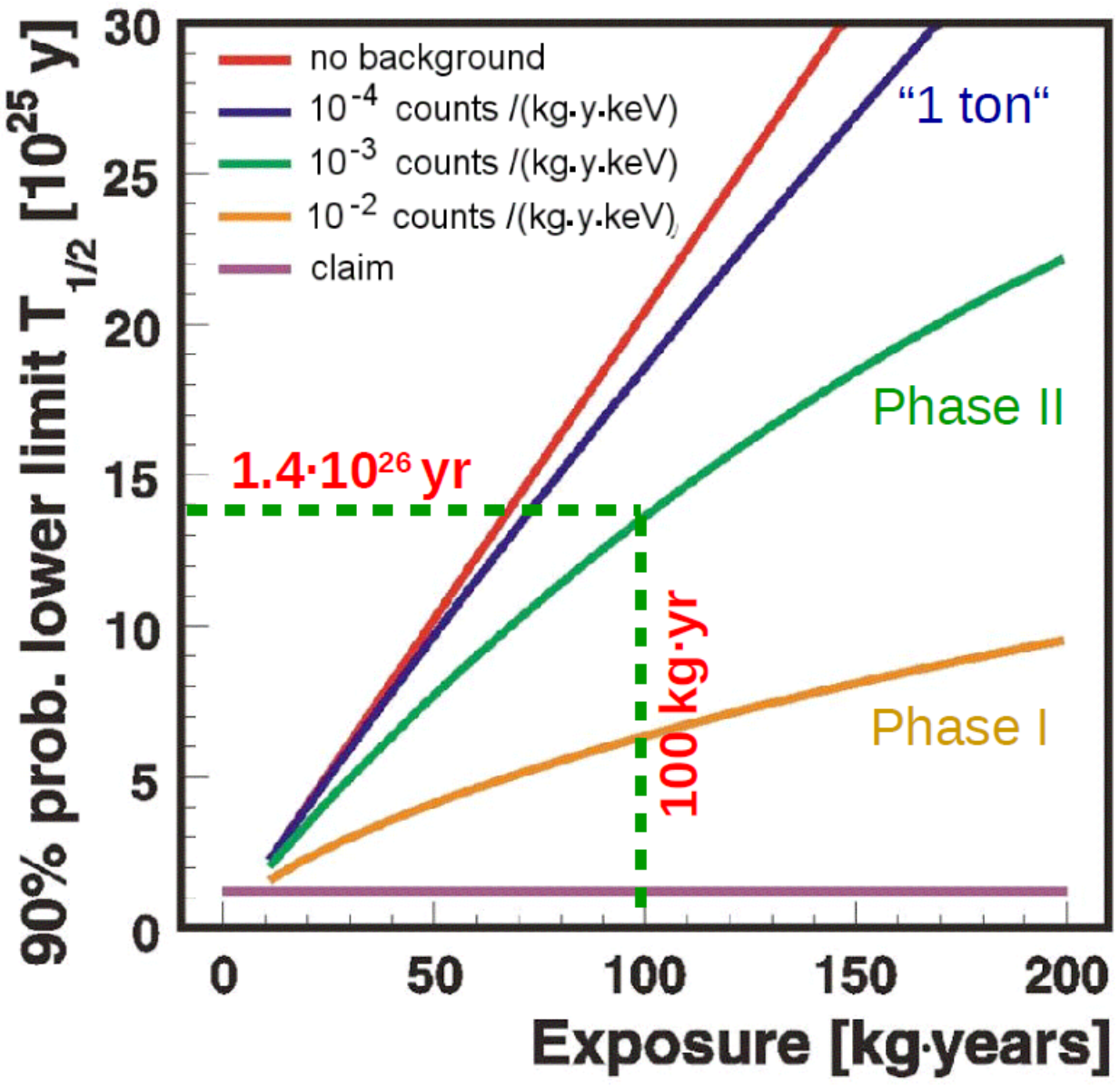

Fig. 1.34. The dependence of the expected GERDA-II sensitivity to 0ν2β decay of $^{76}$Ge on exposure.

The Majorana experiment is another project to search for neutrinoless double beta decay of $^{76}$Ge with germanium detectors. The basic idea is to use close-packed arrays of HPGe as source and detector in ultraclean cryostat.

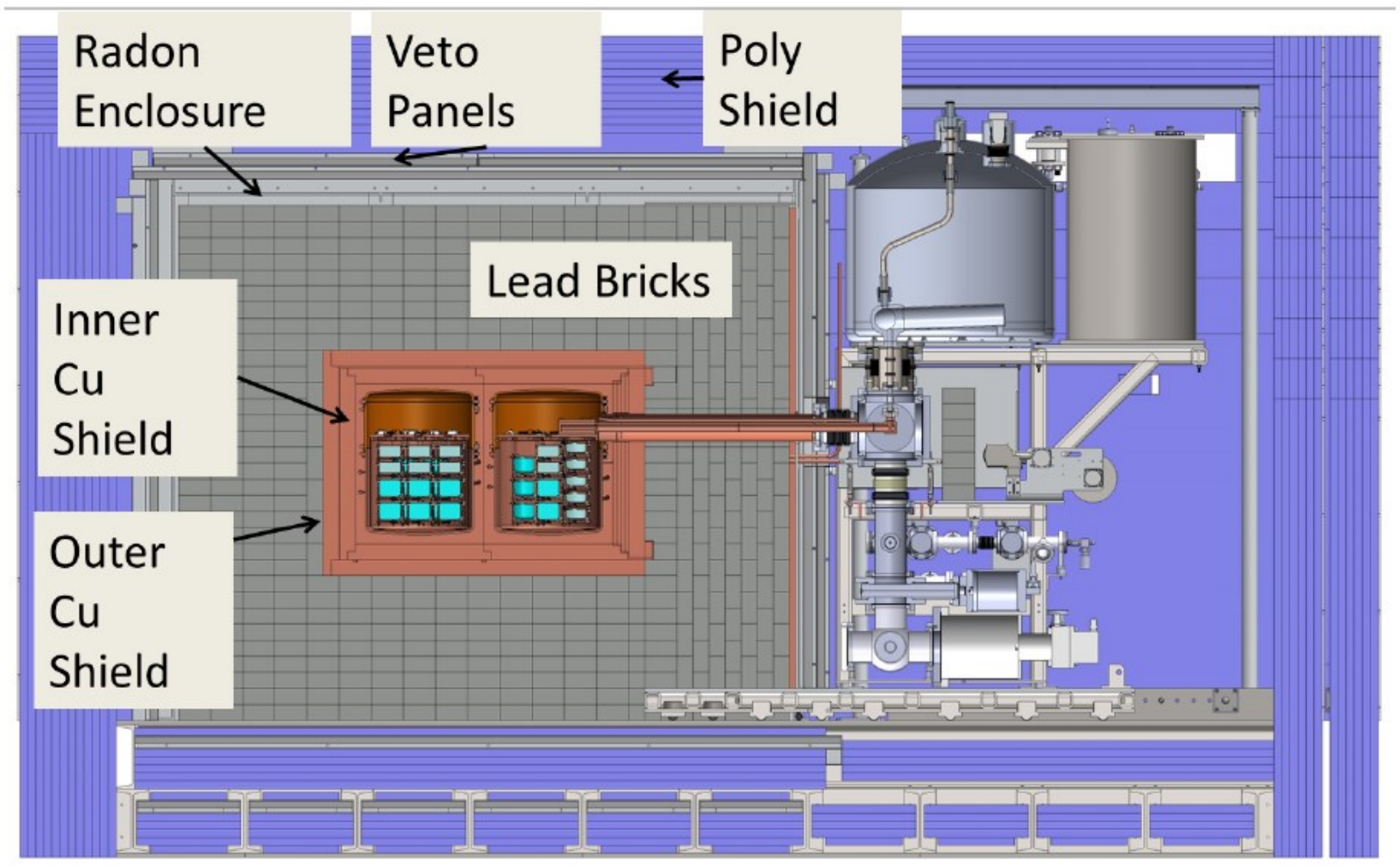

Fig. 1.35. The schematic view of the Majorana Demonstrator detector with its shielding.

The so-called Majorana Demonstrator is the initial phase of the experiment that will use 40 kg of germanium, 30 kg of the detectors will be produced from germanium enriched in $^{76}$Ge to 86%. The collaboration will use P-type, Point-Contact Ge detectors with a mass of 0.6–1.0 kg. The Majorana Demonstrator will consist of two cryostats built from the ultrapure electroformed copper, with each cryostat capable to house 20 kg of germanium detectors. The cryostats will be surrounded by 5 cm of electroformed copper as an inner layer, another 5 cm of oxygen-free high thermal conductivity copper as an outer layer, and 45 cm of high-purity lead. The cryostats, copper, and lead shielding will be contained in the radon exclusion box. The shielding is also supplemented by an active muon veto and 30 cm of polyethylene. The main components of the Majorana Demonstrator are presented in Fig. 1.35. The experiment will be located at the depth of 4260 m.w.e. in the Sanford Underground Research Facility in Lead, South Dakota, USA.

The expected background for the Majorana Demonstrator is 0.75 × $10^{-3}$ counts/(keV·kg·yr) in the 4 keV ROI around 2039 keV. The predicted half-life sensitivity after one year of measurements (26.6 kg × yr of exposure) is $T^{0\nu}_{1/2} > 3 \times 10^{25}$ yr at 90% C.L. (see Fig. 1.36). The experiment is expected to be sensitive enough to probe the claim of 0ν2β detection in $^{76}$Ge [87]. After 5 years of running the experimental sensitivity will reach a value of $T^{0\nu}_{1/2} > 1.2 \times 10^{26}$ years at 90% C.L. [131].

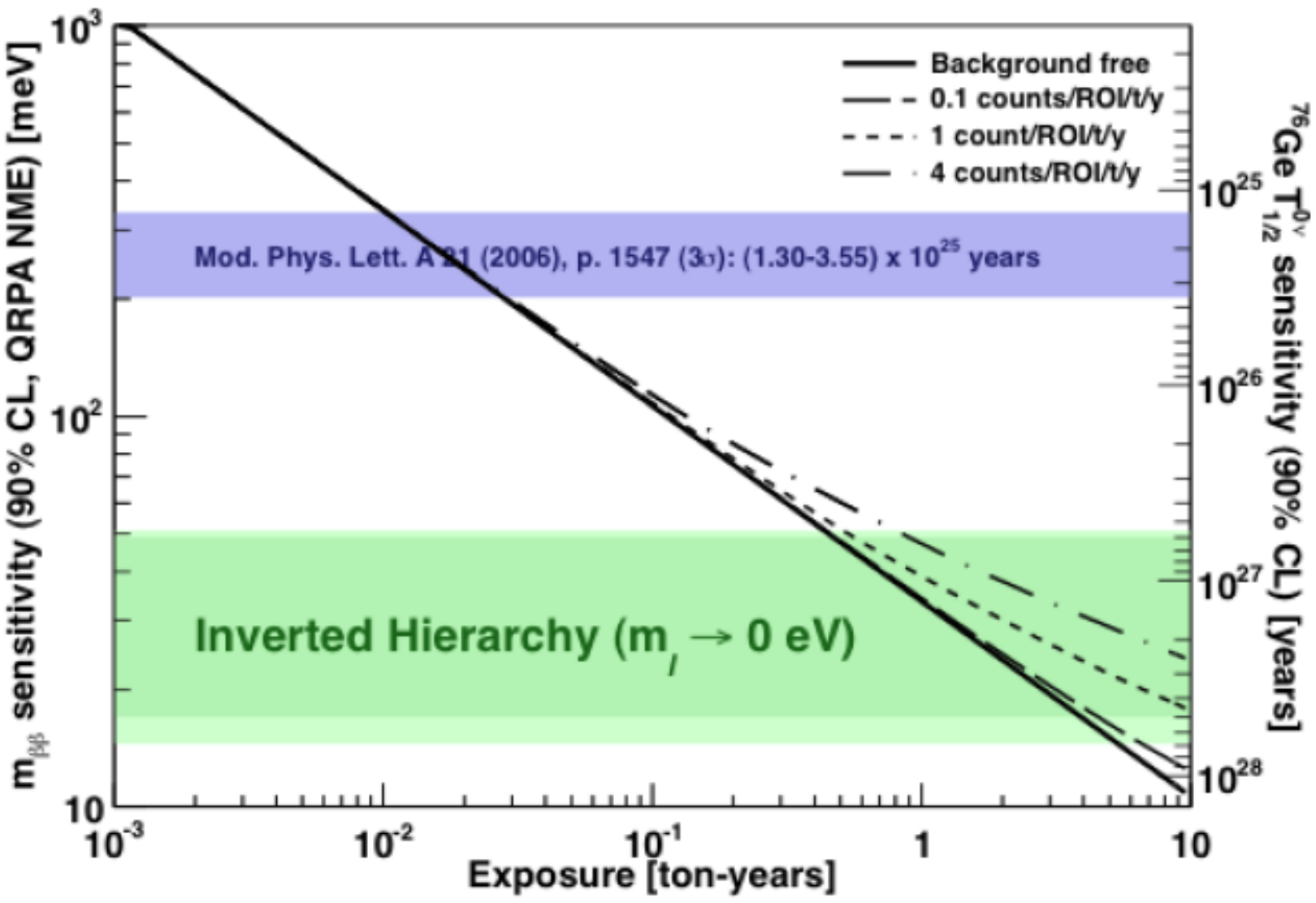

Fig. 1.36. Dependence of the expected Majorana Demonstrator sensitivity to 0ν2β decay of $^{76}$Ge on exposure for different background counting rate. To convert the half-life to neutrino mass the nuclear matrix elements from Ref. [132] were used. The blue band displays the region of the claim [87].

### 1.3.4. SNO+ project

The SNO+ experiment is a modification of the Sudbury Neutrino Observatory (SNO) to search for 0ν2β decay of $^{130}$Te in the SNOLAB underground laboratory (6070 m.w.e.) in the Creighton Mine near Sudbury, Ontario, Canada.

The detector consists of 12 m diameter acrylic sphere filled with ~ 780 tons of tellurium loaded liquid scintillator. The liquid scintillator will be a linear alkylbenzene with 2 g/L of PPO and 0.3% of natural tellurium. The mass of $^{130}$Te is ~ 800 kg. The detector will be surrounded by internal ~ 1700 tons ultra-pure water shield viewed by ~ 9500 PMTs, and external one of ~ 5700 tons which will act as veto for cosmic rays and external background radiation. The schematic view of the SNO+ detector is presented in Fig. 1.37.

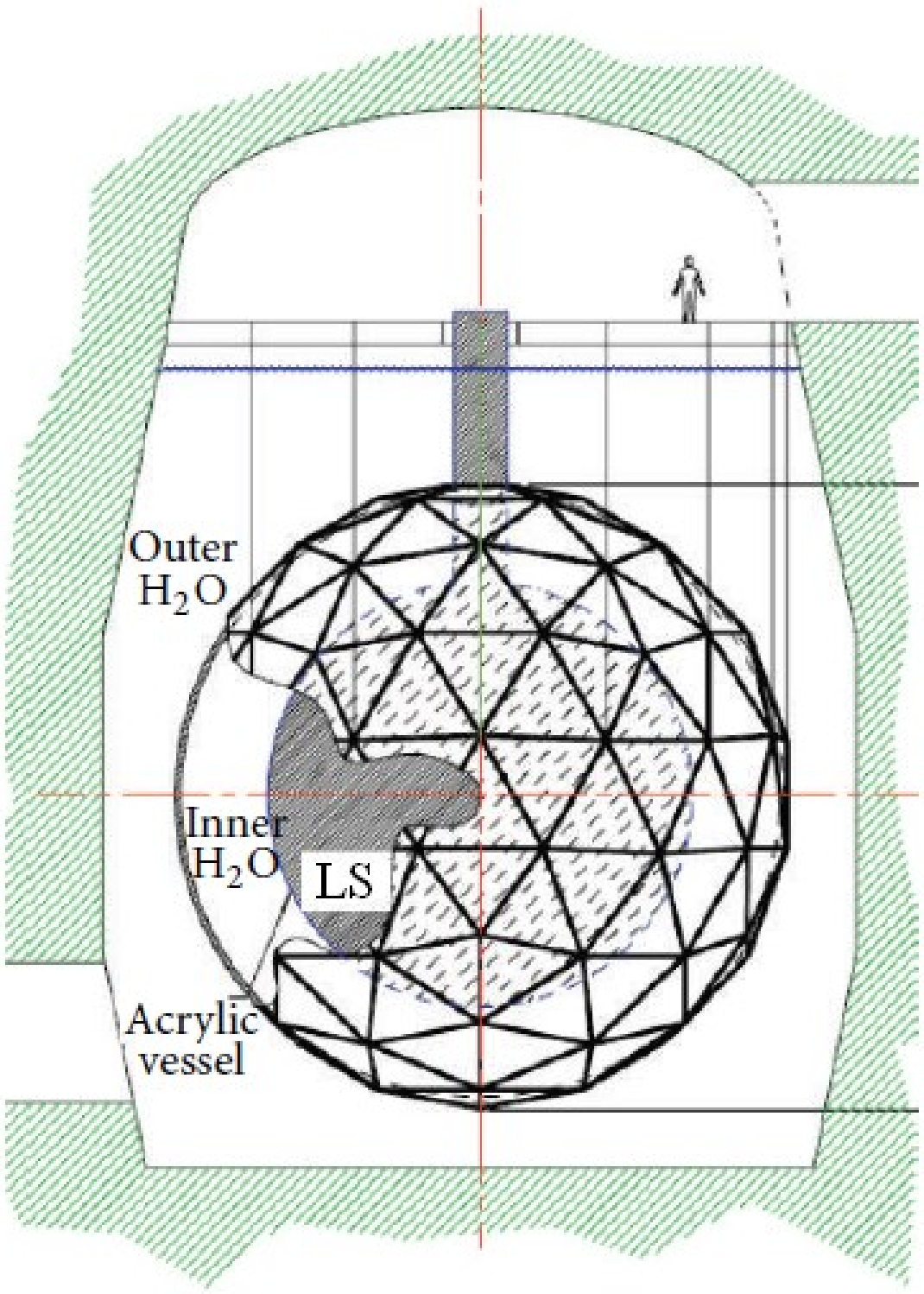

Fig. 1.37. The schematic diagram of the SNO+ detector. The LS refers to Te-loaded liquid scintillator.

The expected background rate in the ROI is ~ 3 × $10^{-4}$ counts/(keV·kg·yr). The first step of the SNO+ experiment is to reach a sensitivity that enters the inverted hierarchy region. The further one will be an increasing of the tellurium loading up to 3% (~ 8 tons of $^{130}$Te) to cover completely the inverted hierarchy region. The expected sensitivity of the SNO+ experiment after 1 and 5 years of measurements is on the level of $T_{1/2}$ ~ 4.27 × $10^{25}$ yr and 9.84 × $10^{25}$ years at 90% C.L. [133], respectively (see Fig. 1.38).

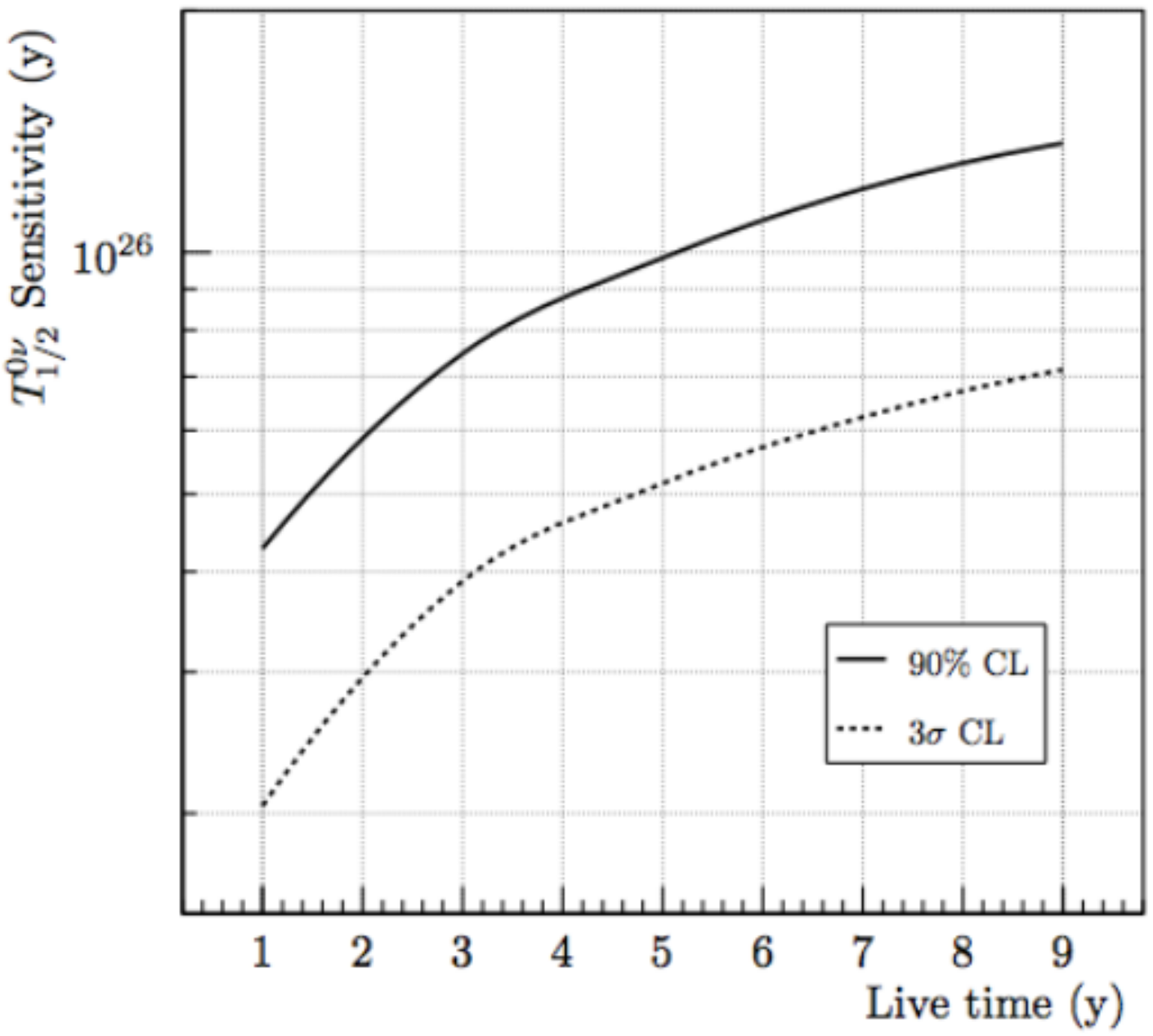

Fig. 1.38. The expected sensitivity of the SNO+ experiment to search for 0ν2β decay of $^{130}$Te as a function of the running time.

### 1.3.5. Bolometric technique for 2β decay

Cryogenic bolometers are one of the most promising approaches to search for 0ν2β decay. Bolometers can contain 2β candidate in its sensitive volume, i.e. able to realize the "source=detector" technique, which allows to achieve a high detection efficiency. The scintillating bolometer is a cryogenic detector with simultaneous and independent read-out of the heat and scintillation signals, which allows to discriminate the α/γ(β) events, and therefore to suppress background effectively. Moreover, bolometers provide an excellent energy resolution of a few keV required to investigate the normal hierarchy of the neutrino mass pattern [40].

#### 1.3.5.1. Basic principles of bolometers

Bolometer is a low temperature detector which consists of an energy absorber thermally coupled to a temperature sensor, sensitive to excitations produced by an absorber when the particle energy is released [134, 135]. In case of 0ν2β applications the absorber is a single diamagnetic dielectric crystal, for which the heat capacity below the Debye temperature decreases inversely as the cube of the absolute temperature. Thus the heat capacity becomes so small, that the crystal's increase in temperature for a given heat input may be relatively large. A nuclear event in the absorber causes the crystal lattice vibrations, which increase its temperature measured by the temperature sensor. The bolometers are operating in a cryostat at milli-Kelvin temperatures (usually < 20 mK) to prevent a significant noise from thermal phonons due to a very small intensity of the heat signals (<< 10 meV).

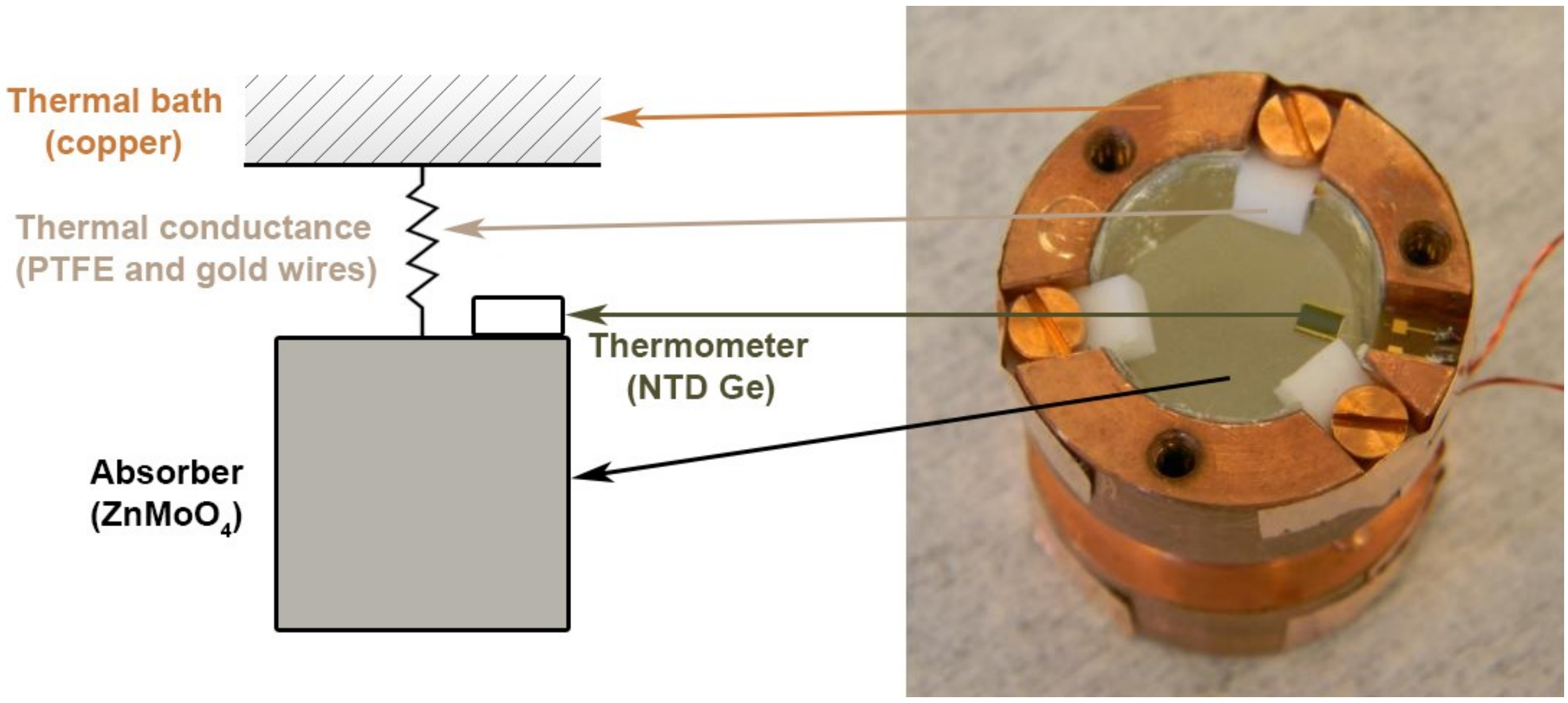

Fig. 1.39. The schematic diagram of the typical bolometer (left) and the photograph of the detector based on 23.8 g $ZnMoO_4$ crystal (right).

A typical bolometer is shown in Fig. 1.39. A crystal is assembled inside a copper holder, serving as a thermal bath, with the help of PTFE elements, used also for thermal conductance. The temperature sensor is thermally connected to the crystal to read-out the temperature rise. In Fig. 1.39 a NTD Ge thermistor attached to the absorber surface by using epoxy glue spots is shown. The high-sensitivity thermistors are either properly doped semiconductor thermistors, or superconductive films kept at the transition edge. The most commonly used thermometers in 2β decay bolometers are Ge or Si semiconductor thermistors with a dopant concentration slightly below the metal-insulator transition [136, 137]. Such type of thermistors is also used in the form of thin amorphous films, like NbSi thermal sensors [138].

The metallic magnetic calorimeter (MMC) is another possibility to measure small changes of temperature. It operates at temperatures lower than 100 mK and uses a paramagnetic temperature sensor in a weak magnetic field [139]. A rise of temperature causes change in the magnetization of the sensor. The measurement of this change with a help of low-noise SQUID (superconducting quantum interference device) magnetometer, provides an accurate value of the deposited energy.

Scintillating bolometers are particularly interesting detectors. In this case the absorber is made of a scintillating crystal containing 2β isotope. The scintillation light from the crystal is read-out by another bolometer, since commonly used photomultipliers and photodiodes are inappropriate to use in vacuum at the very low temperature. A thin silicon or germanium wafer with attached temperature sensor is used for this purpose. The optical photons from the crystal can be detected thanks to small volume of the wafer leading to an extremely low heat capacity. In order to improve the light collection, the crystal scintillator is usually surrounded by reflecting foil. The schematic view of the scintillating bolometer is presented in Fig. 1.40.

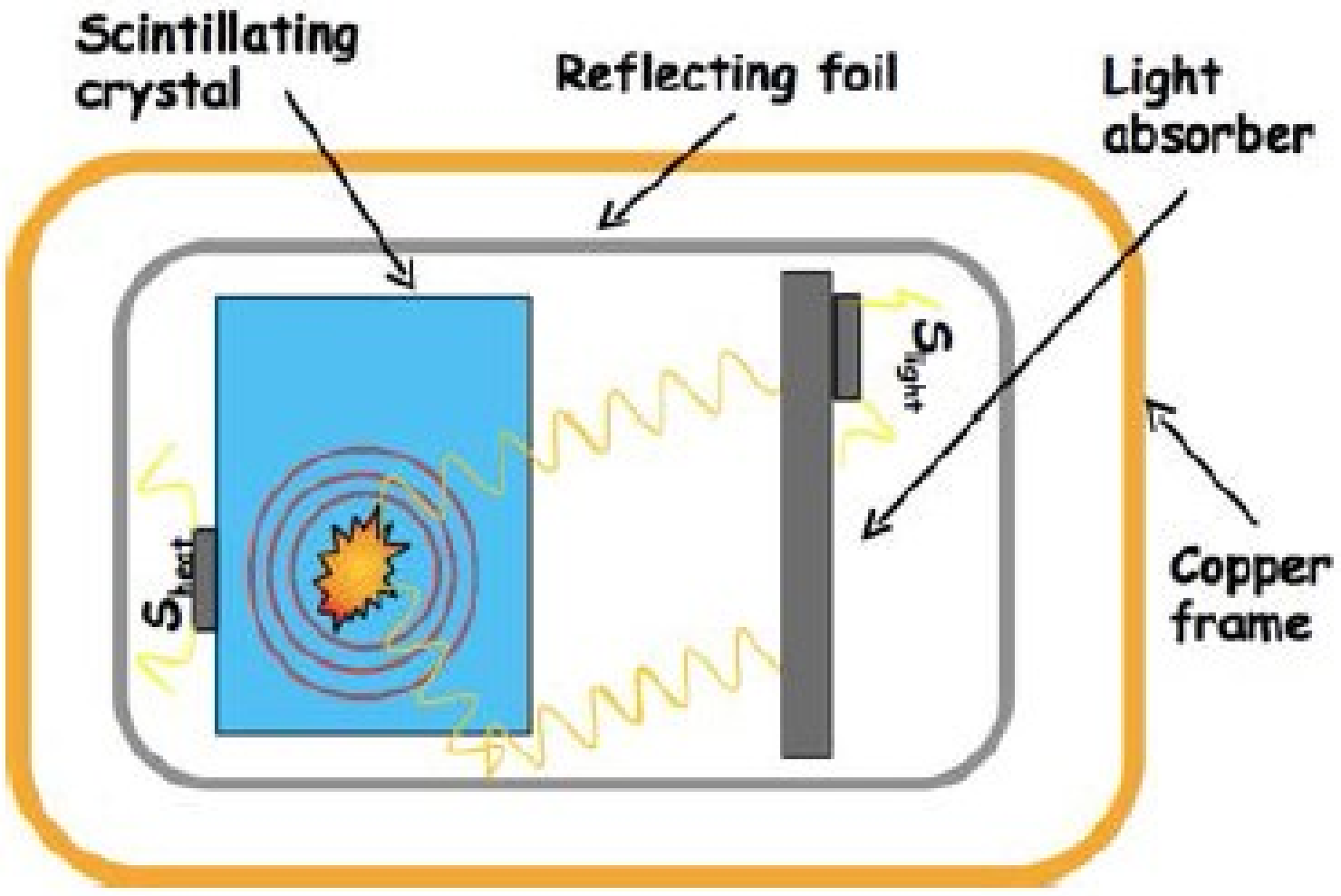

Fig. 1.40. Schematic view of cryogenic scintillating bolometer.

An important advantage of the scintillating bolometers is simultaneous detection of the heat and scintillation signals. This feature allows to identify and reject α particles with a very high efficiency (99.9% and more) due to quenching of scintillation efficiency for α particles. Moreover, the pulse shape analysis in the heat channel can also provide effective α/γ(β) discrimination [140]. The capability to significantly suppress α background is important for neutrinoless double beta decay searches by using 2β candidates with a $Q_{2\beta}$ larger than 2.6 MeV, since α decays in the crystal and surrounding materials surface layer provide significant background counting rate. Thanks to particle discrimination in scintillating bolometers a rather low level of background $10^{-4}$ counts/(keV·kg·yr) could be achieved [141].

### 1.3.5.2. Bolometric 2β decay experiments

The CUORE (Cryogenic Underground Observatory for Rare Events) is a next-generation experiment to search for 0ν2β decay of $^{130}$Te by using $TeO_2$ bolometers [142]. The project is based on the successfully finished CUORICINO pilot experiment (see subsection 1.2.3).

The CUORE detector will consist of 19 towers each one with 13 planes of 4 natural $TeO_2$ crystals, leading to the total array of 988 bolometers. Each crystal with a size of 5 × 5 × 5 cm$^3$ and a mass of 750 g will contain ~ 0.2 kg of $^{130}$Te. The total mass of $^{130}$Te in the CUORE detector is expected to be ~ 206 kg. NTD Ge thermistors will be thermally coupled to $TeO_2$ crystals and operated in a $^3$He/$^4$He dilution refrigerator at the temperature of about 10 mK. The detector array will be surrounded by 6 cm thick internal shield of the low activity Roman lead, and by external shield of a 30 cm thick low activity lead. The cryostat will be also protected from the environmental background by a borated polyethylene shield and an air-tight box. The main components of the set-up are shown in Fig. 1.41. The CUORE experiment is planning to start the data-taking in 2015 at the LNGS underground laboratory.

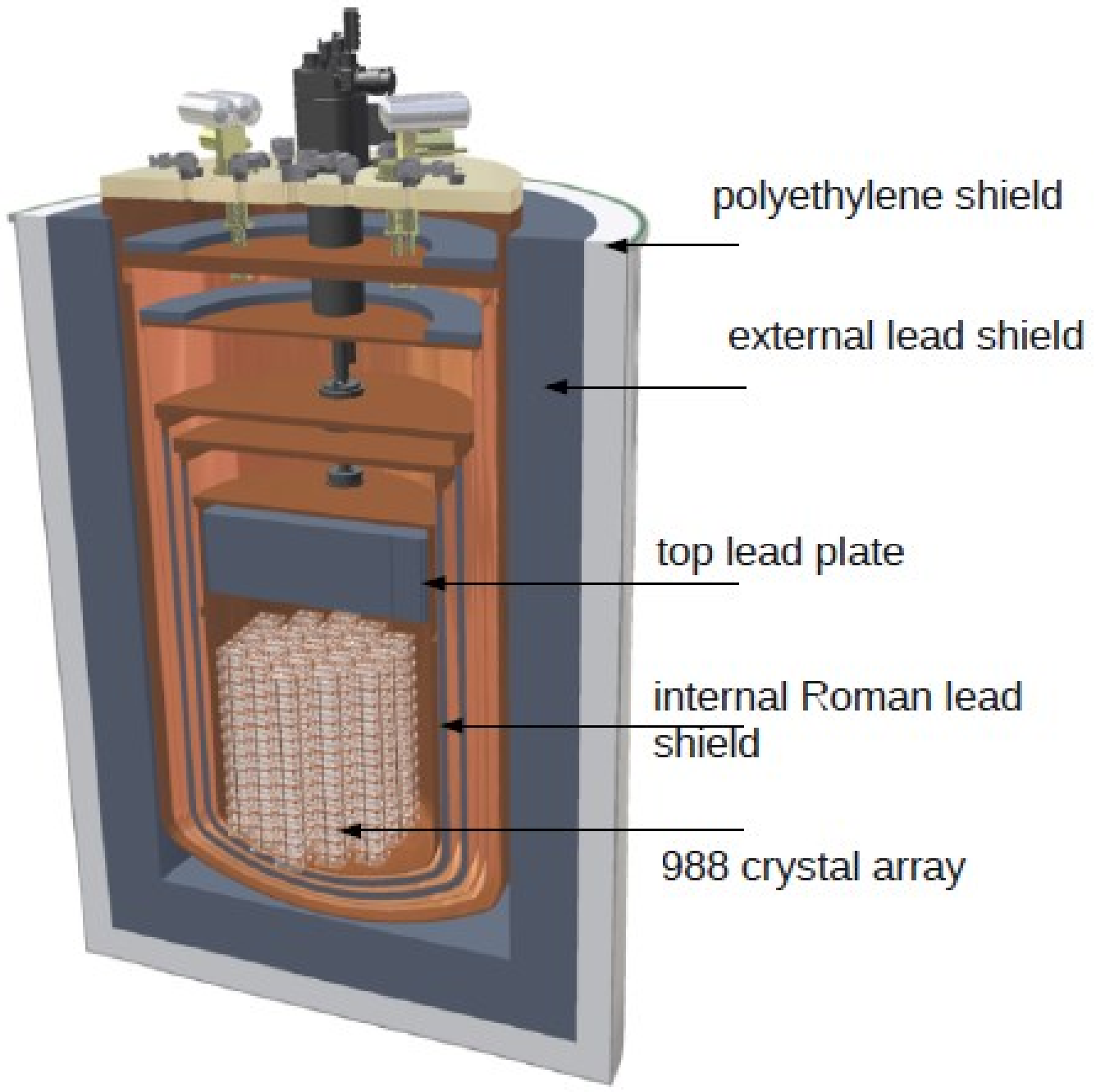

Fig. 1.41. The schematic view of CUORE set-up to search for 0ν2β decay of $^{130}$Te by using $TeO_2$ bolometers.

The expected energy resolution of the CUORE detector is 5 keV at the ROI, while the background level is assumed to be on the level of 0.01 counts/(keV·kg·yr). The predicted half-life sensitivity to 0ν2β decay of the isotope $^{130}$Te is $T_{1/2}^{0\nu} > 9.5 \times 10^{25}$ yr at 90% C.L. after 5 years of measurements [143]. The corresponding sensitivity to effective Majorana neutrino mass is $\langle m_\nu \rangle \leq$ 51–133 meV.

The first step towards the CUORE detector is a demonstrator experiment called CUORE-0 [144]. It was constructed in the CUORICINO cryostat (see subsection 1.2.3) at LNGS to test the detector performances expected for the CUORE. The CUORE-0 detector is a single tower of 52 natural $TeO_2$ crystals (5 × 5 × 5 cm$^3$) operated as bolometers at the base temperature of 13–15 mK. The total detector mass is 39 kg, while the mass of the isotope $^{130}$Te is about 11 kg. The operation of the detector started in spring 2013.

Energy resolution 4.8 keV in the ROI, and background counting rate (0.063 ± 0.006) counts/(keV·kg·yr) were achieved by the CUORE-0 [145]. The CUORE-0 is expected to surpass the CUORICINO sensitivity after about one year of live time measurements. The experiment expected to be finished in 2015 after two years of data-taking. The dependence of the expected half-life sensitivities of the CUORE-0 and CUORE experiments on live time is presented in Fig. 1.42.

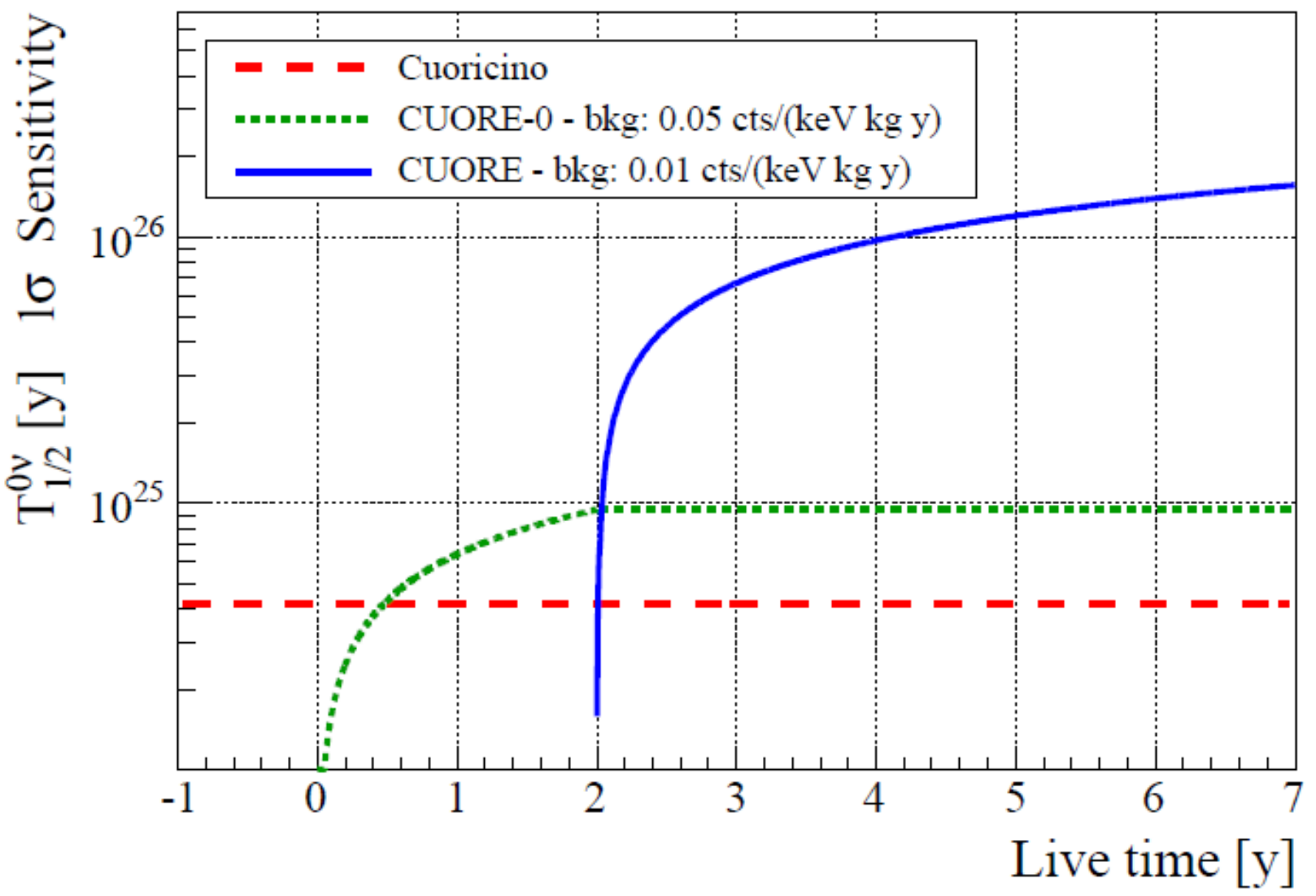

Fig. 1.42. Expected sensitivities of CUORE-0 (dotted green line) and CUORE (solid blue line) experiments to the 0ν2β decay of $^{130}$Te. The CUORICINO sensitivity is shown by a dashed red line.

A very promising technique to search for 0ν2β decay is the scintillating bolometers, extensively developing now within the LUCIFER [146, 147], the AMoRE [148, 149], and the LUMINEU [150] projects.

The LUCIFER R&D is mainly concentrated on Zn$^{82}$Se scintillator; a possibility to use Zn$^{100}$MoO4 and $^{116}$CdWO4 crystals as scintillating bolometers is also discussed. Isotope $^{82}$Se is a promising 2β candidate thanks to the high energy release $Q_{2\beta}$, favorable theoretical estimations of 0ν2β half-life [45, 60], considerable natural isotopic abundance (δ = 8.73(22)% [151]) and the well-established enrichment/purification technology of $^{82}$Se. ZnSe is an interesting material thanks to its good bolometric and scintillating properties, and high mass fraction of selenium (56% in weight).

LUCIFER detector should consist of an array of 36 Zn$^{82}$Se scintillating bolometers (see Fig. 1.43) operated in the CUORICINO cryostat at the LNGS. The total detector mass is expected to be 17 kg of Ø45 × 55 mm Zn$^{82}$Se crystals enriched in the isotope $^{82}$Se up to 95%. The predicted background level in the region of interest is $10^{-3}$ counts/(keV·kg·yr), while the energy resolution is assumed to be about 10 keV. The expected half-life sensitivity to 0ν2β decay of $^{82}$Se after 5 years of measurements is $T^{0\nu}_{1/2} > 6 \times 10^{25}$ yr at 90% C.L. [147], which corresponds to the limit on the effective Majorana neutrino mass as $\langle m_\nu \rangle \leq$ 65–194 meV depending on the NME calculations.

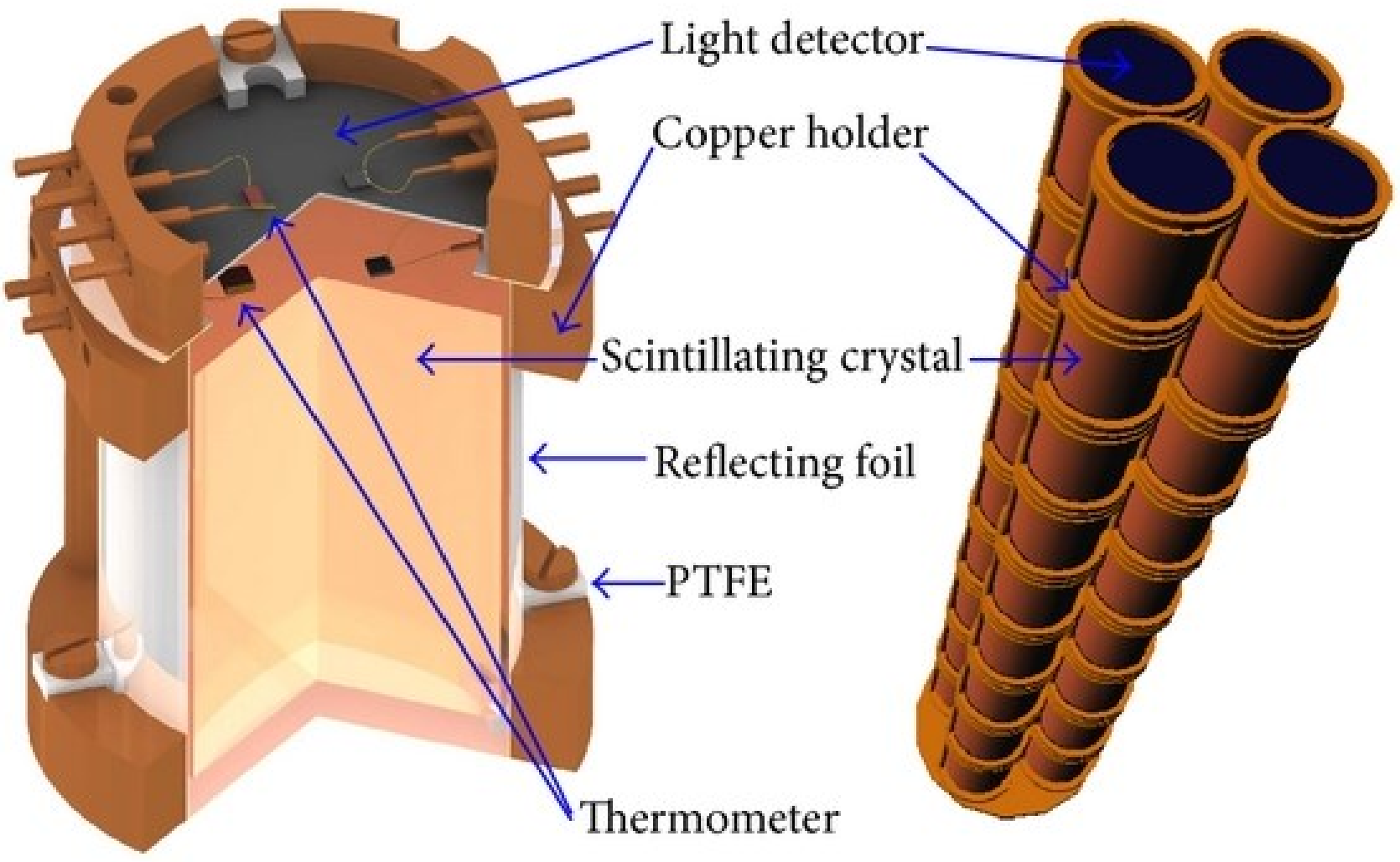

Fig. 1.43. The schematic view of the LUCIFER single module detector (left) and the 9-floor tower containing 36 scintillating bolometers (right).

The AMoRE (Advanced Mo based Rare process Experiment) project is going to use $^{40}Ca^{100}MoO_4$ scintillators to search for 0ν2β decay of $^{100}$Mo. Isotope $^{100}$Mo is a another promising 2β candidate with the high $Q_{2\beta}$ energy, favorable theoretical estimations of $T_{1/2}^{0\nu}$ [45, 60], considerable natural isotopic abundance (δ = 9.82(31)% [151]) and possibility to enrich $^{100}$Mo by centrifugation method. The advantage of the $CaMoO_4$ scintillators is the highest light output at room and low temperatures in comparison to other crystals containing molybdenum [152]. However, an irremovable background source at the $Q_{2\beta}$ of $^{100}$Mo can occur due to 2ν2β decay of $^{48}$Ca with the energy of 4.27 MeV. Therefore, the AMoRE collaboration developed $^{40}Ca^{100}MoO_4$ crystals enriched in $^{100}$Mo to 96% and depleted in $^{48}$Ca (≤ 0.001%). The detector will be scintillating bolometers based on $^{40}Ca^{100}MoO_4$ crystals as the absorber, and MMCs as the sensor. An assembled scintillating bolometer with 216 g $CaMoO_4$ crystal is shown in Fig. 1.44.

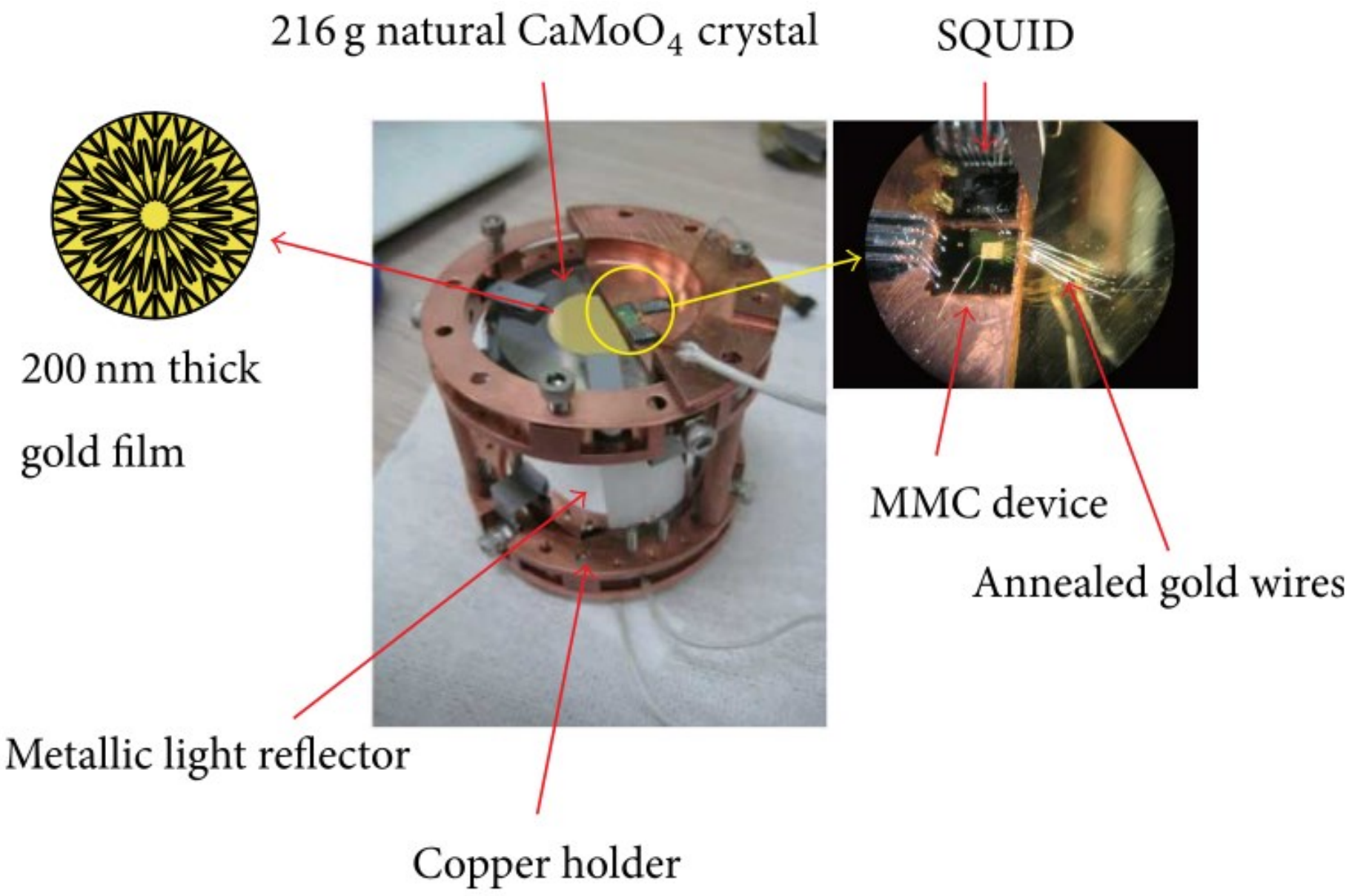

Fig. 1.44. The AMoRE R&D scintillating bolometer based on natural $CaMoO_4$ crystal with a mass of 216 g and MMC detector.

The AMoRE collaboration is going to construct a prototype detector with 10 kg of $^{40}Ca^{100}MoO_4$ scintillating crystals and start experiment in 2016 at the Yangyang underground laboratory located at the depth of ~ 2000 m.w.e at Pumped Storage Power Plant, Yangyang, South Korea. Afterwards, a large scale experiment AMoRE-200 with 200 kg $^{40}Ca^{100}MoO_4$ crystals should be performed in the next 5-6 years. The energy resolution of a few keV and nearly zero background in the ROI are expected. The sensitivity of the AMoRE-200 experiment to the effective Majorana neutrino mass is estimated to be in the range of 20–50 meV, which corresponds to the inverted scheme of the neutrino mass [149].

The LUMINEU (Luminescent Underground Molybdenum Investigation for NEUtrino mass and nature) project is another R&D program to search for 0ν2β decay of $^{100}Mo$ by using scintillating bolometers technique. The LUMINEU project aims to develop high-quality radiopure zinc molybdate ($ZnMoO_4$) crystal scintillators of large mass (300–500 g) and test them as low-temperature detectors. $ZnMoO_4$ crystal was chosen thanks to the absence of heavy elements, high concentration of molybdenum (43% in weight), and promising scintillation and bolometric properties.

The main purpose of the LUMINEU project is to demonstrate feasibility of $ZnMoO_4$ based cryogenic scintillating bolometers for the next-generation 0ν2β decay experiment capable to explore the inverted hierarchy region of the neutrino mass pattern. A post-LUMINEU experiment based on the current R&D project will be a low-background underground array of ~ 48 enriched $Zn^{100}MoO_4$ scintillators with a size of Ø60 × 40 mm, containing ~ 10 kg of the isotope $^{100}Mo$. A sensitivity of such experiment to the 0ν2β decay is expected to be on the level of the best current and planned double beta decay experiments [153]. By scaling up to the 100–1000 kg experiment could completely cover the inverted hierarchy region. The main part of this PhD thesis is devoted to the LUMINEU project.

### 1.4. Conclusions and perspectives

The neutrinoless double beta decay is the most promising way to test lepton number violation, to investigate the nature of neutrinos, their absolute mass scale and hierarchy, to test a range of effects beyond the Standard Model.

The discovery of neutrino oscillations gives a strong motivation to 0ν2β investigations. However, in order to convert observed 0ν2β rates (limits) into the neutrino mass constraints, a precise knowledge of nuclear processes is required. There are several approaches to calculate the nuclear matrix elements for the 0ν2β decay, but there is still a significant disagreement between the calculations.

The most sensitive 2β experiments detected 2ν2β decay in 11 nuclei and set limits on 0ν2β decay with the sensitivity on the level of $\lim T_{1/2} \sim 10^{23}–10^{25}$ years. The next-generation experiments should investigate the inverted hierarchy of neutrino mass region. To achieve this goal with theoretically most favorable 2β candidates one should choose a proper experimental technique and increase the mass scale of the detector to hundreds kg of isotope of interest. At the same time, the detector should have a very high energy resolution and extremely low background. No experiment (project) is able to fulfill all the requirements completely. For example, a large-scale liquid scintillation experiments can achieve a very low background counting rate, however with a rather poor energy resolution. Therefore, the irremovable 2ν2β decay background becomes to be a crucial problem. Cryogenic scintillating bolometers can satisfy almost all of the requirements, however only tracking detectors can guaranty the registration of double beta decay process by the detection of two electrons.

It should be also stressed that if the neutrinoless double beta decay will be discovered in one nucleus, a few other isotopes should be also studied in order to confirm the observation and to improve the theoretical models used for NME calculations. Therefore, the variety of $0\nu2\beta$ experiments with different candidates and techniques should be developed. Moreover, if no $0\nu2\beta$ decay is observed in the inverted neutrino hierarchy region, this would be an important achievement for neutrino physics indicating in favor of the Dirac nature of the neutrino.

Finally, there is no obvious answer on how to approach the direct hierarchy region except the increasing of detector mass. It seems that this task requires some massive R&D activities with revolutionary approaches, and one of them is cryogenic scintillating bolometers.

# CHAPTER 2

# DEVELOPMENT OF ZINC MOLYBDATE CRYOGENIC SCINTILLATING BOLOMETERS

## 2.1. Production of zinc molybdate scintillating crystals

The zinc molybdate crystals are known for more than 50 years [154, 155]. In year 2006 the $ZnMoO_4$ compound in form of poly-crystals was studied and proposed to be used as an appropriate detector material to search for neutrinoless double beta decay of $^{100}Mo$ [156]. Two years later relatively large $ZnMoO_4$ crystals were grown for the first time by using the Czochralski and Kyropoulos methods in Institute of General Physics (IGP, Moscow, Russia) [157]. However the optical quality of the produced crystals was rather poor (see Fig. 2.1).

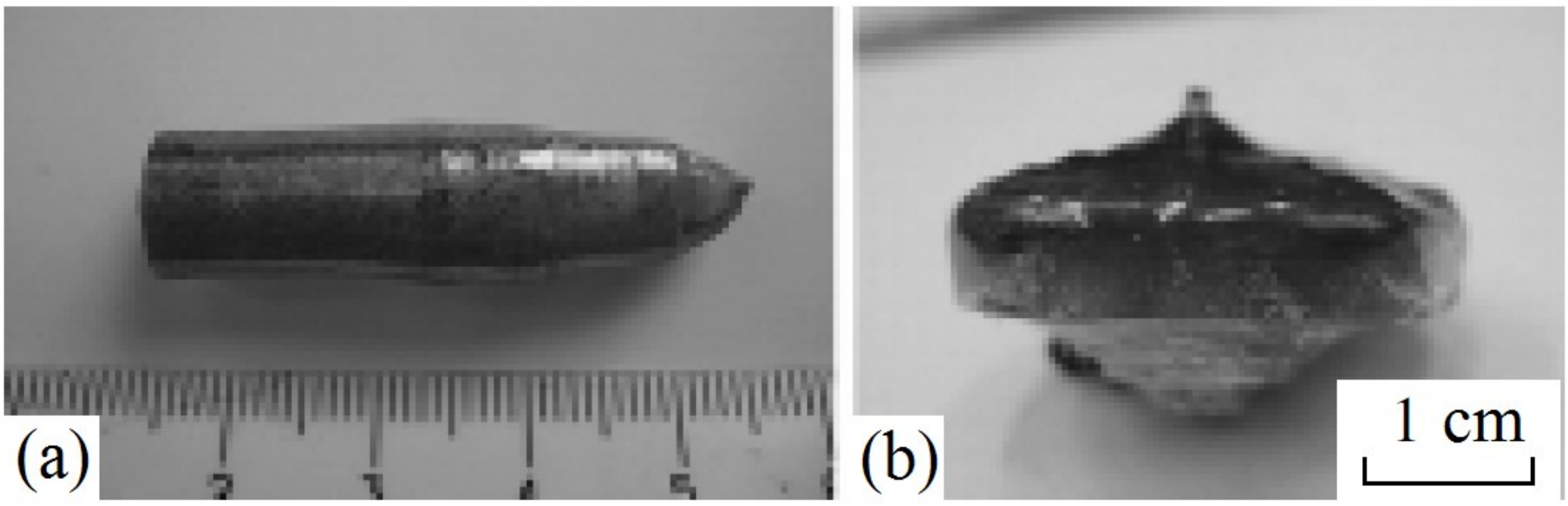

Fig. 2.1. Photos of the $ZnMoO_4$ crystals grown in 2008 year by the (a) Czochralski (40 mm in length and 15 mm in diameter) and (b) Kyropoulos methods (15 mm in length and 30 mm in diameter) [157].

Single crystal samples of improved quality (see Fig. 2.2) were grown in the Institute for Scintillation Materials (ISMA, Kharkhiv, Ukraine) by using the Czochralski technique in 2009 [158]. Thereafter one of these crystals was tested for the first time as a low temperature scintillating bolometer [159]. The obtained results demonstrated applicability of the $ZnMoO_4$ crystals as cryogenic phonon-scintillating detectors. However all of the produced zinc molybdate crystals were of insufficient quality with an intense yellow colour which led to the absorption of scintillation light. As a result scintillating crystals had a lower light output with a nonuniformity of light collection in the light channel of the cryogenic scintillating bolometers.

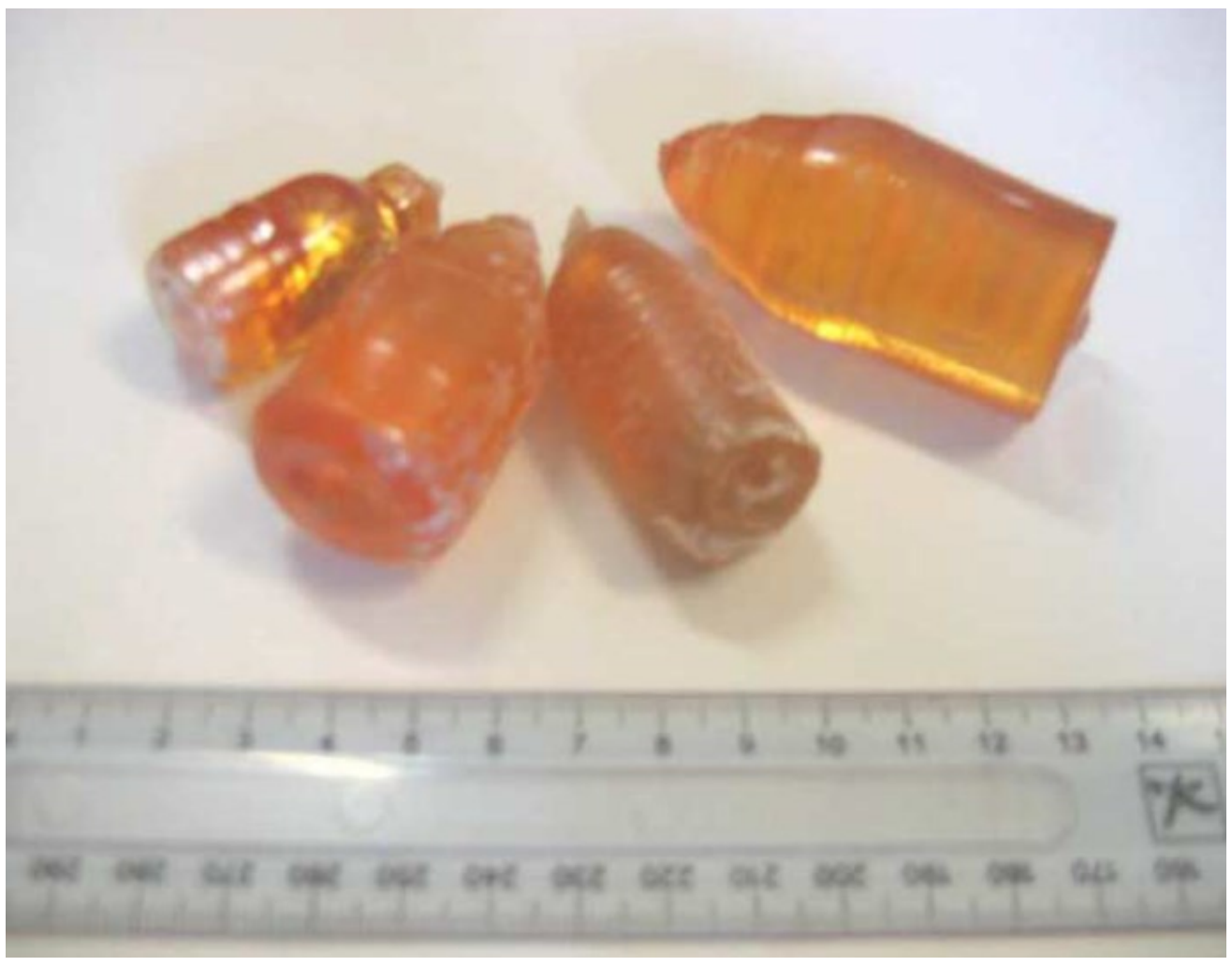

Fig. 2.2. Photo of the $ZnMoO_4$ crystals 20 mm–30 mm in diameter and 30 mm–40 mm in height grown by the Czochralski technique in 2009 year [158]. The scale is in centimeters.

In 2010 high-quality $ZnMoO_4$ crystals were produced for the first time in the Nikolaev Institute of Inorganic Chemistry (NIIC, Novosibirsk, Russia). The crystal boules with a maximum size of Ø25 × 60 mm were grown by using the low-thermal-gradient Czochralski technique (LTG Cz) [160, 161, 162]. The significant improvement of the crystal quality was achieved thanks to developing of the purification technique based on the recrystallization of molybdenum oxide from aqueous solutions by co-precipitation of impurities on zinc molybdate sediment. The developed $ZnMoO_4$ scintillators are presented in Fig. 2.3.

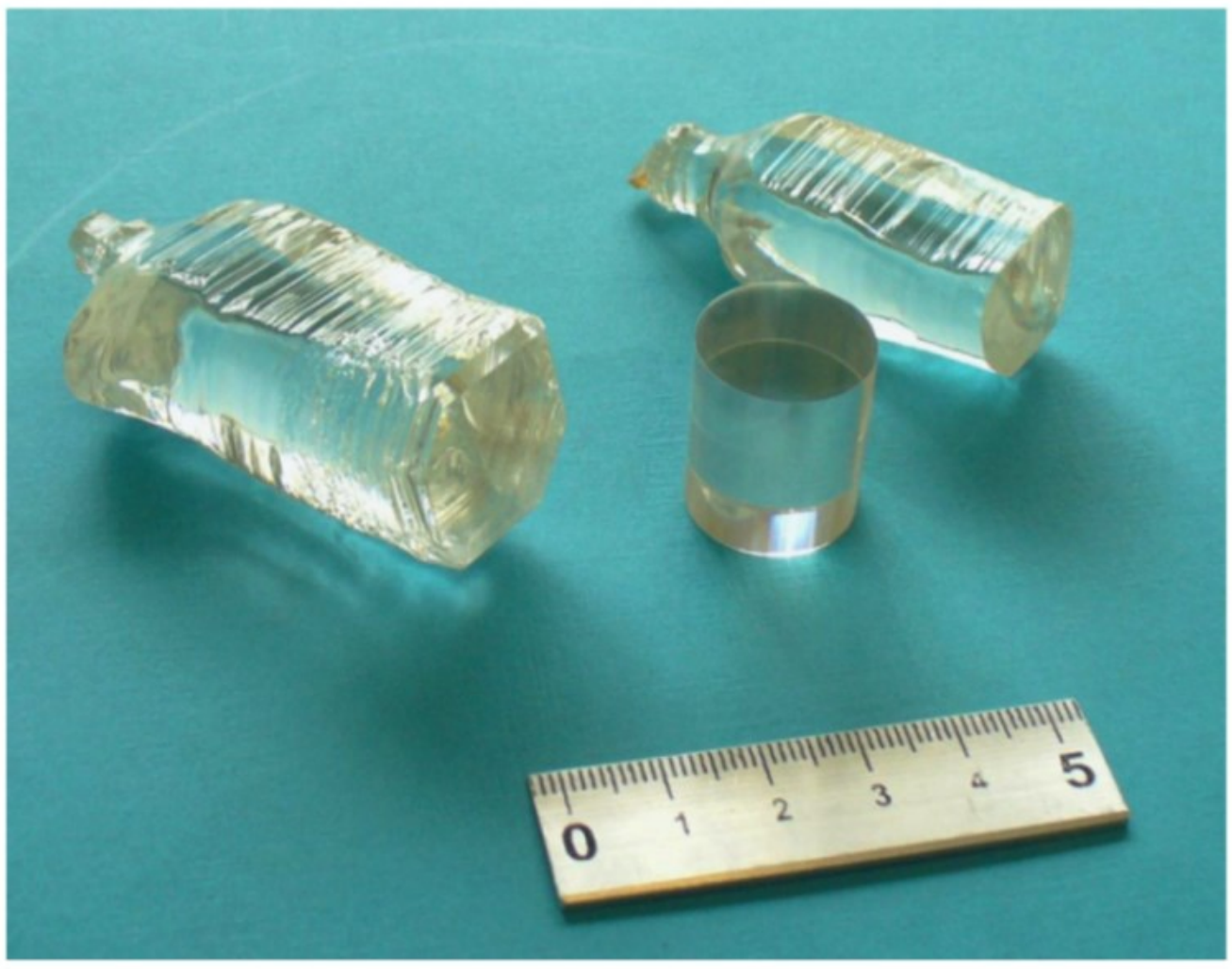

Fig. 2.3. Photo of the $ZnMoO_4$ scintillator and crystal boules produced in Nikolaev Institute of Inorganic Chemistry by the low-thermal-gradient Czochralski technique. The scale is in centimeters.

The further step in $ZnMoO_4$ production was the increase of crystal volume. In 2011 large zinc molybdate crystal boules with a mass of hundreds grams were grown in NIIC (see Fig. 2.4 where a scintillation element produced from one of the crystal boules is shown). The crystal growing and purification techniques were used the same as before, however the overall quality deteriorated after increasing of the crystal volume.

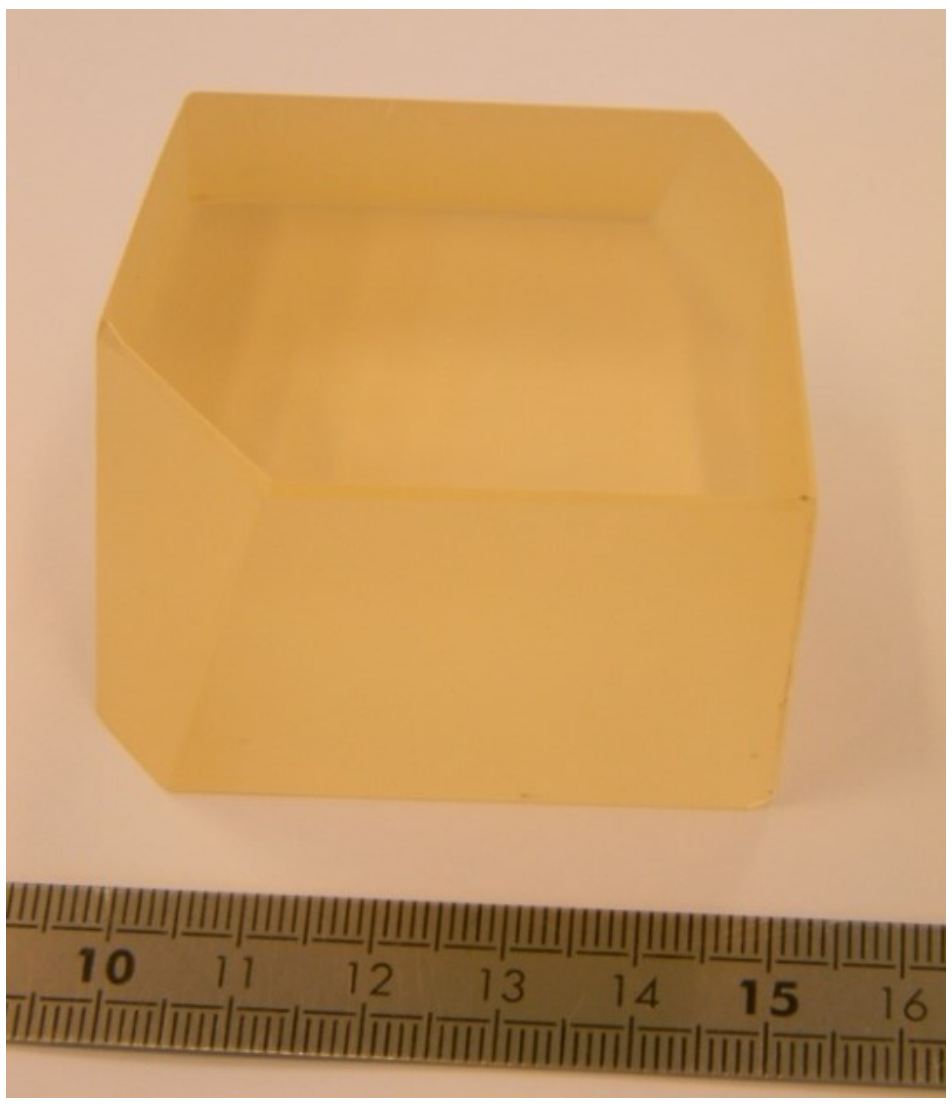

Fig. 2.4. Photo of the 313 g $ZnMoO_4$ scintillator produced in Nikolaev Institute of Inorganic Chemistry by the low-thermal-gradient Czochralski technique. The scale is in centimeters.

In order to apply $ZnMoO_4$ crystals as cryogenic scintillating bolometers in future double beta decay experiments, production of large high quality radiopure zinc molybdate crystals is required. These crystals should have low level of radioactive contamination, high bolometric properties, scintillation and optical quality, large mass and finally should be produced from molybdenum enriched in $^{100}$Mo. Technology to produce advanced quality $ZnMoO_4$ crystal scintillators, including enriched in $^{100}$Mo, has been developed in the Nikolaev Institute of Inorganic Chemistry (NIIC, Novosibirsk, Russia).

### 2.1.1. Purification of molybdenum

To grow high quality radiopure $ZnMoO_4$ crystal scintillators one should use high purity zinc and molybdenum. In case of zinc there is a commercially available high purity zinc oxide, meanwhile there are no commercial molybdenum compounds that complies with the required levels of purity, not to say for radioactive contamination.

The most dangerous contaminators of the zinc molybdate scintillating crystals are thorium and radium: activities of $^{228}$Th and $^{226}$Ra should not exceed the level of 0.01 mBq/kg [153]. The total alpha activity of the nuclides from the U/Th chains (see Annex for decay schemes) should not exceed the mBq/kg level, while the activity of $^{40}$K is requested to be low enough (should not exceed a few mBq/kg level) to avoid random coincidence of events due to the poor time resolution of the cryogenic bolometers. These activities correspond to the concentrations on the level of ~ ppt for thorium and uranium and ~0.01 ppm for potassium. High scintillation and optical properties of the crystals can be achieved with trace impurity level lower than a few ppm of total contamination for transition elements. To improve crystal transparency it is important to decrease the contamination of transition metals such as Fe, V, Cr, Co, etc. To produce enriched $Zn^{100}MoO_4$ scintillating crystals it is extremely important to develop efficient purification methods of the enriched molybdenum with minimal losses, since the contamination of enriched materials is on the level of 10 ppm–100 ppm, while the radioactive contamination (e.g., measured in Ref. [163]) is at the level of 2 mBq/kg of $^{226}$Ra, 0.6 mBq/kg of $^{228}$Th and 36 mBq/kg of $^{40}$K , which is essentially exceeds the acceptable level.

Two-stage purification technique of molybdenum was developed in NIIC. The method consists of the sublimation of molybdenum oxide in vacuum and double recrystallization from aqueous solutions by co-precipitation of impurities on zinc molybdate sediment [164].

Commercially available molybdenum is usually purified by the sublimation of molybdenum oxide under the atmospheric pressure with a subsequent leaching in aqueous solutions with ammonia. However such purification procedure cannot reach the zinc molybdate crystal growth requirements, due to high concentration of impurities, in particular of tungsten, which are still at the level of up to 0.5 wt% (weight percentage) even in the high purity grade materials. Furthermore, additional vacuum sublimation of $MoO_3$ gives no substantial improvement. The problem of tungsten contamination in molybdenum is well-known [165]. Moreover, it is impossible to decrease traces of Na, Ca and Si to concentrations levels less than 20 ppm–70 ppm with the sublimation of molybdenum oxide.

To reduce the concentration of tungsten in molybdenum oxide, and thus separate molybdenum from tungsten, the following exchange reaction during the sublimation at high temperature was assumed:

$$\mathrm{ZnMoO_4} + \mathrm{WO_3} = \mathrm{ZnWO_4} + \mathrm{MoO_3}\uparrow . \qquad (2.1)$$

To prove this assumption, sample of molybdenum oxide powder with 10 wt% of tungsten oxide was prepared in NIIC. Afterwards zinc molybdate was added to the powder before the sublimation. After the sublimation the concentration of tungsten in $MoO_3$ decreased by one order of magnitude to 0.1 wt%.

Another proof of the proposed technique efficiency was obtained by chemical and X-ray diffraction analysis of the rests after the sublimation processes. The typical amount of the rests is about 1 wt%–3 wt% of the initial amount of the purified molybdenum oxide. These residues were collected from a few sublimations and mixed together. Then they were annealed in air atmosphere to oxidize rests of metals. To reduce the concentration of the formed molybdenum oxide the sample was sublimated in vacuum. The elemental composition of the bottoms was studied using atomic emission analysis and gave the following results: Fe — 0.064 wt%; Cu — 0.011 wt%; Mg — 0.026 wt%; Na — 0.13 wt%; Ca — 0.14 wt%; K — 1 wt%; Si — 2.6 wt%; Zn — 14 wt%; W — 18 wt%; Mo — 22 wt%. The X-ray diffraction analysis showed oxides of molybdenum and silicon, tungstate (in form of tungstate-molybdate), zinc molybdate, also $K_2Mo_7O_{22}$ and $K_2MgSi_5O_{12}$ in the rests of $MoO_3$ after the sublimation process. The data obtained in NIIC confirmed efficiency of the proposed method of molybdenum purification from tungsten by the sublimation of molybdenum oxide in vacuum with presence of zinc molybdate.

Table 2.1
Results of molybdenum oxide purification by sublimation processes. Purity level was measured by atomic emission spectrometry.

| Material | Concentration of impurities (ppm) | | | |
|---|---|---|---|---|
| | Fe | K | Si | W |
| Initial $MoO_3$ | 6 | 100–500 | 600 | 200–500 |
| After 1$^{st}$ sublimation | 2–6 | 10–50 | 100–500 | 100–200 |
| After 2$^{nd}$ sublimation | < 1 | 1–8 | 70 | 30–40 |

Following the proposed purification technique (see formula (2.1)), high purity $ZnMoO_4$ in amount of 1 wt% was added to the $MoO_3$. After sublimation process the obtained sublimates consisted of a mixture of molybdenum oxides of different composition and color, which prevents their use for zinc molybdate synthesis. Therefore the sublimates were annealed in air atmosphere to receive a stoichiometric molybdenum oxide of yellow color. Finally the obtained sublimates were studied by atomic emission spectrometry. The results of the analysis are presented in Table 2.1. The double sublimation improves the purity of molybdenum oxide by one-two orders of magnitude depending on the element. The sublimation of molybdenum oxide should also remove metal oxides which have a high vapor pressure at temperatures up to a thousand degrees.

The final stage of molybdenum purification was a double recrystallization of ammonium molybdate in aqueous solutions with the deposition of impurities on zinc molybdate sediment. It is based on a typical technology of purification in aqueous solutions with a gradual release of impurities during the changing of solution composition. The molybdenum oxide was dissolved in solution of ammonia at room temperature. Depending on the mixing ratio of the components mono-molybdates and various poly-compounds and hetero-poly compounds were formed. The composition of the compounds depends on the components concentration and acidity of the solution. In aqueous solutions molybdates form normal molybdate:

$$MoO_3 + 2NH_4OH = (NH_4)_2MoO_4 + H_2O \qquad (2.2)$$

or ammonium hepta-molybdates, $(NH_4)_xH_{(6-x)}[Mo_7O_{24}]$. During long time exposure in presence of impurities the poly-molybdates can form ammonium salts of molybdosilicic and molybdophosphoric acids, for instance $H_8[Si(Mo_2O_7)_6]$ and $H_7[P(Mo_2O_7)_6]$. As the central atom can serve not only Si(IV) and P(V), but also V(V), Ge(IV), Cr(III), etc., while inner sphere ligands can be ions $VO^{3-}$, $WO_4^{2-}$, $CrO_4^{2-}$ and $TeO_4^{2-}$. The poly-molybdates solutions at pH < 6 could dissolve oxides and hydroxides of many metals like ZnO, $Fe(OH)_3$, $Ni(OH)_2$, $Cu(OH)_2$, so during the crystallization of salts a partial co-crystallization of impurities can occur. As a result, the recrystallization of ammonium para-molybdate is not effective enough for molybdenum purification. Moreover, the typical concentration of impurities in high purity molybdenum oxide is on the level of 1 ppm–100 ppm. Thus it is hard to remove the fine microcrystals of the impurity sediments by filtration.

To prepare a high purity molybdenum oxide for $ZnMoO_4$ crystal growth the additional stage of purification in aqueous solutions was applied. Taking into account that a small amount of zinc oxide does not affect the crystals quality, a small amount of ZnO dissolved at pH > 6 in the solution, was used to initiate precipitation. Zinc oxide in amount of 1 g/L–2 g/L was added to the solution of ammonium para-molybdate, and then ammonia was added to increase the pH level to 7–8. Several hours after exposure zinc molybdate started to precipitate absorbing impurities from the solution. By subsequent rising of pH contaminants precipitated in the form of hydroxides. Besides, at the level of pH = 8–9 the best conditions appeared for precipitation of thorium and uranium. After separation of the sediment, the solution was evaporated to 70% of the solution volume at near the boiling point temperature. Finally, to bind the residues of iron impurities the ammonium oxalate was added to the solution. As we can see from the analysis results presented in Table 2.2 the improved technique of molybdenum oxide purification provides deeper purification from contaminants that deteriorate the properties of $ZnMoO_4$ crystals.

The additional (after sublimation) purification by recrystallization in aqueous solution improves quality of molybdenum oxide destroying crystal granules of $MoO_3$ formed during the sublimation process. The granule structure of molybdenum oxide complicates the process of $ZnMoO_4$ synthesis, taking into account that grinding of the granules can contaminate the material.

Table 2.2
Purity level of $MoO_3$ before and after purification by recrystallization and sublimation. Data for commercial 5N5 grade product and enriched $^{100}$Mo material are given for comparison.

| Material | Concentration of impurities (ppm) | | | | | | | |
|---|---|---|---|---|---|---|---|---|
| | Na | Mg | Si | K | Ca | Fe | Zn | W |
| Initial $MoO_3$ | 60 | 1 | 60 | 50 | 60 | 8 | 10 | 200 |
| After recrystallization from aqueous solutions | 30 | < 1 | 30 | 20 | 40 | 6 | 1000 | 220 |
| After sublimation and recrystallization from aqueous solutions | — | < 1 | 30 | 10 | 12 | 5 | 500 | 130 |
| After double sublimation and recrystallization from aqueous solutions | — | < 1 | — | < 10 | < 10 | < 5 | 70 | < 50 |
| 5N5 grade $MoO_3$ used to produce $ZnMoO_4$ crystal studied in [158, 159] | 24 | — | 9 | 67 | 15 | < 18 | — | 96 |
| Samples of enriched isotope $^{100}$Mo used in [163] (before purification, data of producer) | 10 | < 10 | 50–360 | < 30 | 40–50 | 10–80 | — | 200 |
| $ZnMoO_4$ crystal | 0.36 | 0.72 | 0.09 | 0.14 | 0.13 | 0.38 | — | 190 |

Molybdenum oxide obtained by application of the developed in NIIC purification techniques and high purity grade zinc oxide produced by the company UMICORE were used to synthesize zinc molybdate powder for crystal growth.

### 2.1.2. $ZnMoO_4$ crystal growth

The $ZnMoO_4$ scintillating crystals were produced in Nikolaev Institute of Inorganic Chemistry by the low-thermal-gradient Czochralski technique. The crystal boules were grown in the platinum crucibles 40 mm and 80 mm in diameter in air atmosphere. According to the certificates of platinum crucibles, the concentration of iron in platinum was less than 40 ppm. The temperature gradient did not exceed 1 K/cm during the crystal growth; the rotation speed was in the range of 20 rotations per minute at the beginning of the process, decreasing to 5 rotations per minute at the end. The crystals were grown at the speed of 0.6 mm–1.2 mm per hour. The yield of the produced crystal boules was at the level of 80%. It should be noted that such high efficiency cannot be achieved by the ordinary Czochralski method, where the maximum yield of the produced boule is on the level of 30%–45%. The set-up to grow a zinc molybdate crystal boules and schematic view of the puller are shown in Fig. 2.5.

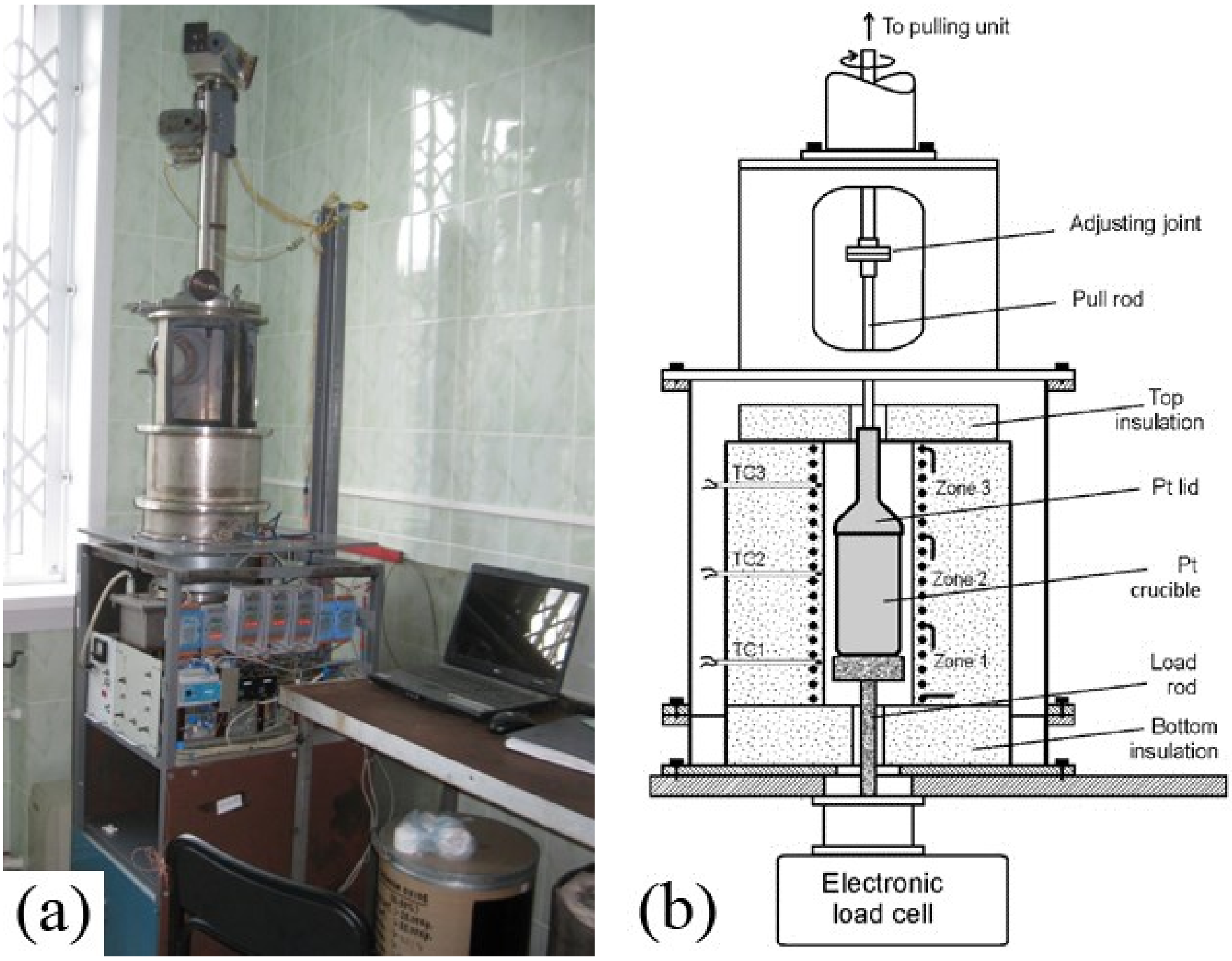

Fig. 2.5. (a) Photo of the set-up developed in Nikolaev Institute of Inorganic Chemistry used to grow zinc molybdate crystals by the low-thermal-gradient Czochralski technique. (b) Schematic view of the puller to grow crystals by the low-thermal-gradient Czochralski technique [162].

Several small samples were cut from the produced crystal boules for optical and luminescent tests. Contamination of a sample of $ZnMoO_4$ crystal was measured by using glow discharge mass spectrometry. The results of the analysis are shown in Table 2.2. Two large volume $ZnMoO_4$ scintillation elements of advanced quality are shown in Fig. 2.6.

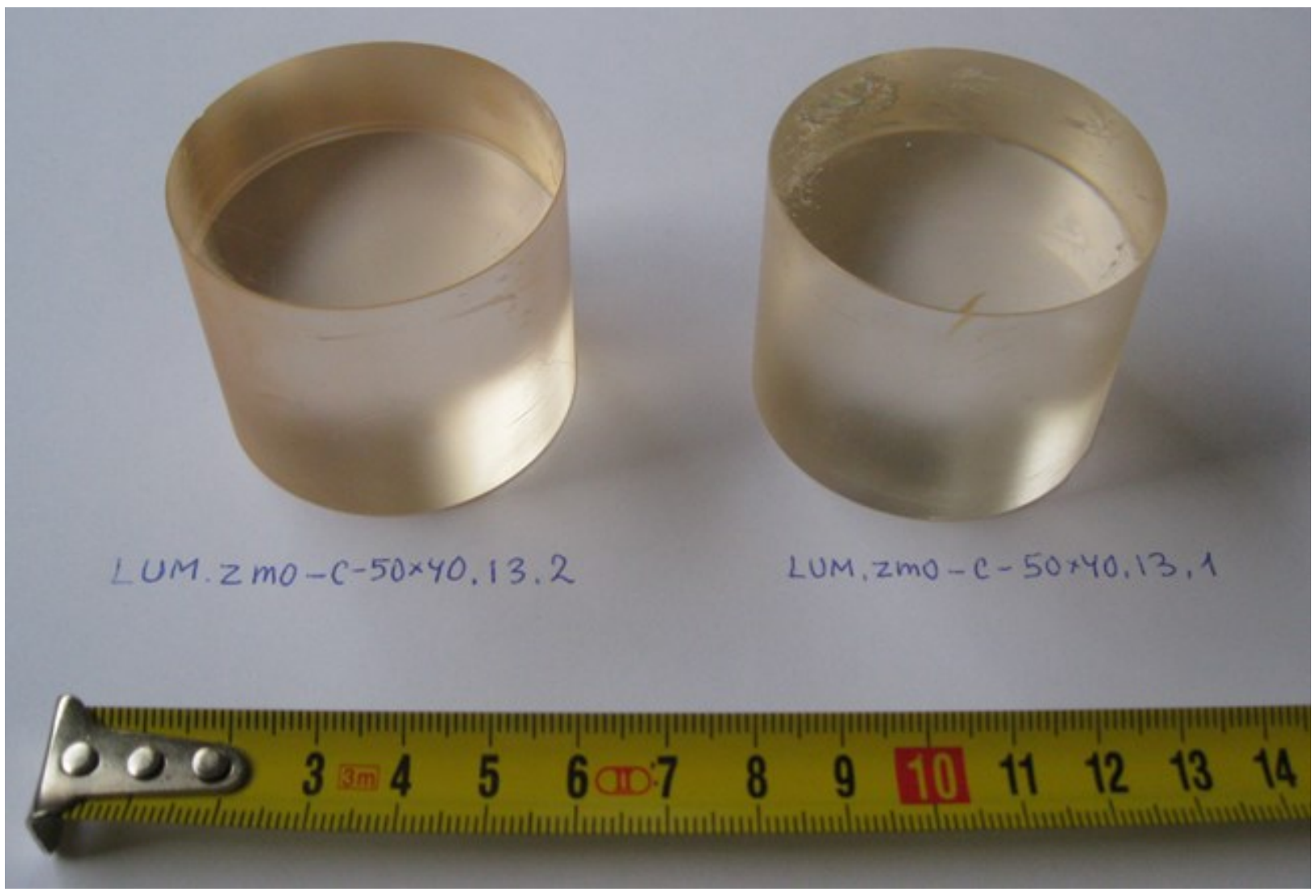

Fig. 2.6. Photo of the $ZnMoO_4$ crystals Ø50 × 40 mm of size with mass 336 g (left) and 334 g (right) produced in the Nikolaev Institute of Inorganic Chemistry by the low-thermal-gradient Czochralski technique in 2013. The scale is in centimeters.

### 2.1.3. Production of enriched $Zn^{100}MoO_4$ crystal scintillator

Enriched in isotope $^{100}$Mo zinc molybdate ($Zn^{100}MoO_4$) crystal scintillators should be developed for high sensitivity experiment to search for 0ν2β decay of $^{100}$Mo. We have used a sample of enriched molybdenum with isotopic concentration of $^{100}$Mo about 99.5% produced in 1980s at the Kurchatov Institute (Moscow, former Soviet Union) for development of $Zn^{100}MoO_4$ crystal scintillator. The material in form of metallic powder was used in the experiment to search for double beta decay of $^{100}$Mo to excited states of $^{100}$Ru at the Modane underground laboratory (France) [166]. The effect was not observed in this experiment.

Thereafter, the enriched molybdenum was again utilized in the ARMONIA (meAsuReMent of twO NeutrIno 2β decAy of $^{100}$Mo to the first excited $0_1^+$ level of $^{100}$Ru) experiment at the Gran Sasso laboratory with the same aim. Additional purification of the material was performed by recrystallization in aqueous solution. The molybdenum was dissolved in 20% ultrapure nitric acid and transformed into molybdenum acid: $^{100}MoO_3 \cdot nH_2O$. Then the compound was rinsed by solution of nitric acid and annealed. As a result a sample of purified molybdenum oxide ($^{100}MoO_3$) with a mass of 1199 g was obtained. Radioactive contamination of the sample was tested in the experiment [163]. Concentration of impurities in the $^{100}MoO_3$ was measured by inductively coupled plasma mass-spectrometry (ICP-MS) and atomic absorption spectroscopy (AAS) methods. As we can see from the results of the analysis presented in Table 2.3, the material should be additionally purified to achieve the crystal growth requirements.

The purification procedure described in detail in subsection 2.1.1 was applied to the $^{100}MoO_3$. Zinc oxide was used for the sublimation process to reduce the concentration of tungsten. Then the rests were analyzed with an atomic emission spectrometry at the NIIC. The following composition of the bottoms after sublimation was obtained: Fe (0.05 wt%), Si (0.6 wt%–0.8 wt%), Mo (11 wt%–12 wt%), Zn (12 wt%–16 wt%) and W (22 wt%–33 wt%). The performed analysis proved the contamination of the initial molybdenum oxide by iron. It should be stressed also that losses of the enriched molybdenum after the sublimation process did not exceed 1.4%.

Table 2.3
Purity level of $^{100}MoO_3$ measured by inductively coupled plasma mass-spectrometry (ICP-MS) and atomic absorption spectroscopy (AAS) methods.

| Element | Concentration of element in $^{100}MoO_3$ (ppm) | |
|---|---|---|
| | ICP-MS | AAS |
| Na | — | < 60 |
| Mg | < 0.5 | < 4 |
| Al | 2.4 | — |
| Si | — | < 500 |
| K | < 15 | < 10 |
| Ca | — | < 10 |
| V | 0.05 | — |
| Cr | 0.2 | < 5 |
| Mn | 0.1 | — |
| Fe | 8 | < 5 |
| Ni | 0.01 | — |
| Cu | 0.1 | — |
| Zn | 0.1 | < 4 |
| Ag | 0.3 | — |
| W | 1700 | 550 |
| Pb | 0.008 | — |
| Th | < 0.0005 | — |
| U | 0.001 | — |

The next stage of molybdenum purification was recrystallization of the obtained molybdenum oxide in aqueous solution. Some losses of the enriched molybdenum during this procedure are expected to be recovered in the course of further $Zn^{100}MoO_4$ mass production. Taking into account the previous purifications by recrystallization from the aqueous solutions of about 20 kg of natural molybdenum performed by the NIIC group, the estimated losses of enriched molybdenum at this stage should not exceed 2%.

Finally, 131.75 g of the purified enriched molybdenum oxide were combined by a solid-phase synthesis with 72.23 g of high-purity zinc oxide produced by the company UMICORE. The obtained 203.98 g of $Zn^{100}MoO_4$ powder were kept for 12 hours at the temperature of ≈ 680 °C in a platinum cup. Afterwards the zinc molybdate crystal boule was produced in Nikolaev Institute of Inorganic Chemistry by the low-thermal-gradient Czochralski technique. The crystal boule was grown in the platinum crucible with a size of Ø40 × 100 mm, the temperature gradient did not exceed 1 K/cm, and the rotation speed was in the range of 4–20 rotations per minute.

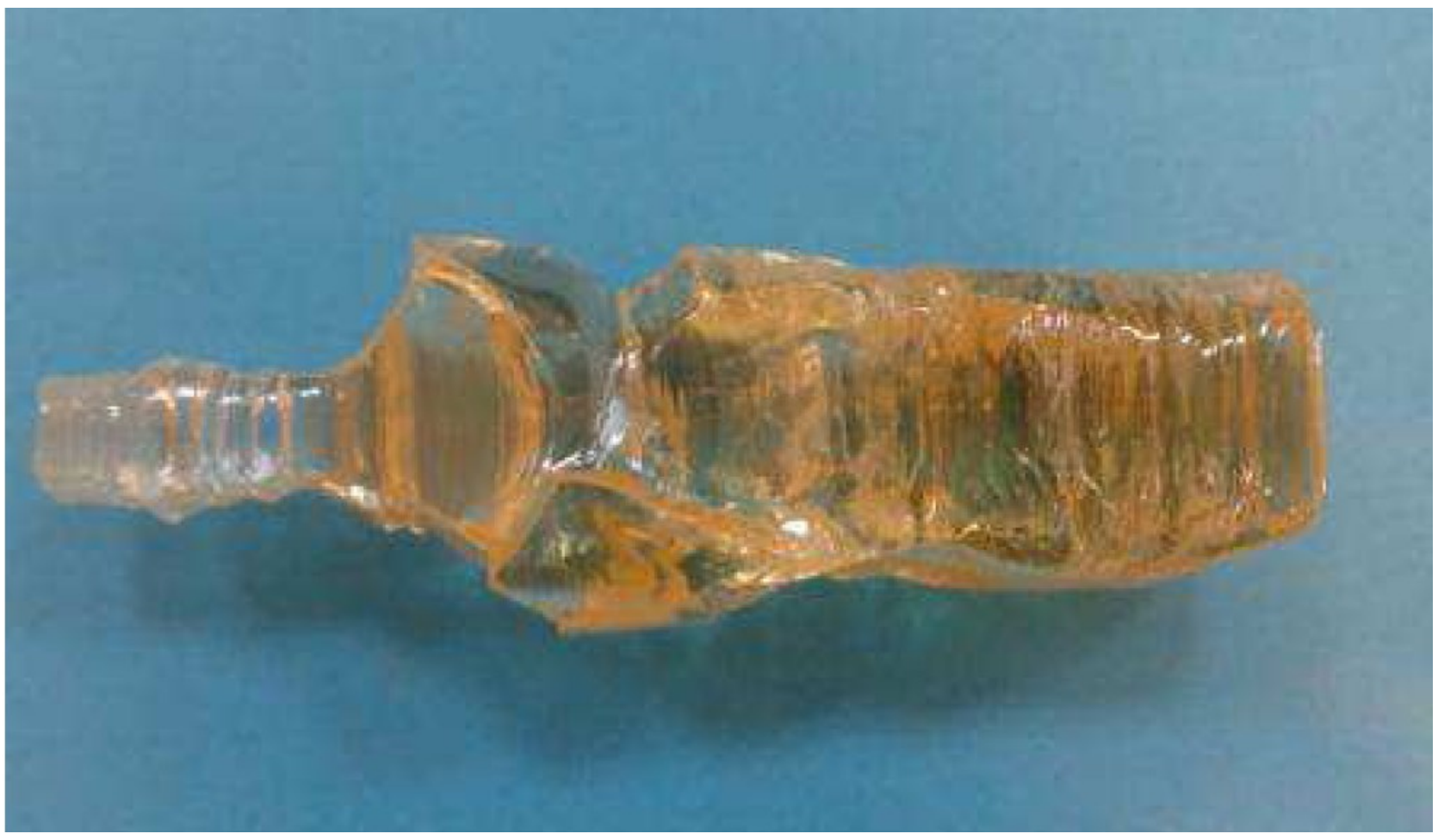

Fig. 2.7. Photo of the $Zn^{100}MoO_4$ crystal boule enriched in $^{100}Mo$ to 99.5%. The boule with the mass of 170.7 g and length of 95 mm was grown for the first time by using low-thermal-gradient Czochralski technique in the Nikolaev Institute of Inorganic Chemistry.

A $Zn^{100}MoO_4$ crystal boule with a mass of 170.7 g was grown [167]. As we can see from the Fig. 2.7 some orange coloration of the sample is present. The coloration is probably due to the traces of iron remained in the enriched zinc molybdate powder used for the crystal growth. We would like to remind that the initial molybdenum oxide was contaminated by iron at the level of 8 ppm. Development of the enriched $Zn^{100}MoO_4$ crystals with an improved optical quality is now in progress in Nikolaev Institute of Inorganic Chemistry.

The yield of the enriched crystal boule was 83.7%. Taking into account that ≈ 1.3 g of the $Zn^{100}MoO_4$ crystal remained in the seed and 30.83 g of the residual melt, the irrecoverable losses of the enriched molybdenum during the crystal growth process can be estimated as ≈ 0.6%. The overall losses of the $^{100}Mo$ during the different stages of crystal production are presented in Table 2.4. The total amount of losses is expected to decrease by a factor 1.5–2, to the level already achieved by using LTG Cz technique for the production of enriched cadmium tungstate crystals from the isotopes of $^{116}Cd$ (≈ 2%) [168] and $^{106}Cd$ (2.3%) [169].

Table 2.4
Irrecoverable losses of the enriched molybdenum in all the stages of the $Zn^{100}MoO_4$ crystal production.

| Stage | Losses of $^{100}Mo$ |
|---|---|
| Sublimation of $^{100}MoO_3$ | 1.4 % |
| Recrystallization from aqueous solutions | 2.0 % |
| Crystal growth | 0.6 % |
| Total | 4.0 % |

## 2.2. Characterization of $ZnMoO_4$ crystals

The main properties of the $ZnMoO_4$ crystals are presented in Table 2.5.

Table 2.5
Properties of $ZnMoO_4$ scintillating crystals.

| Property | Value | Reference |
|---|---|---|
| Density (g/cm$^3$) | 4.3 | [157] |
| Melting point (°C) | 1003 ± 5 | [170] |
| Structural type | Triclinic, *P*1 | [157, 170] |
| Cleavage plane | Weak (001) | [157] |
| Hardness on the Mohs scale | 3.5 | [171] |
| Index of refraction (in the wavelength interval 406–532 nm) | 1.87–2.01 | [171] |
| Wavelength of emission maximum (nm) | 585–625 | [157, 158, 171] |
| Light yield (photon/MeV) at 10 K | $5 \times 10^3$ | [172] |
| Quenching factor for $\alpha$ particles of ~ 5 MeV | 0.15–0.19 | [164, 171, 173] |
| Debye temperature (K) | 625 | [164] |
| Scintillation decay time (μs) | ≈ 1.3, 16, 150 | [156] |

### 2.2.1. Light transmission

Optical properties of three samples of $ZnMoO_4$ crystals with size of 15 × 15 × 5 mm produced by using Czochralski technique in IGP ([157], sample 1) and ISMA ([158], sample 2), and by using LTG Cz technique in NIIC ([171], sample 3) were measured. The transmittance was studied with a help of the commercial fiber optic spectrophotometer Ocean Optics HR2000CG-UV-NIR and the deuterium/tungsten–halogen lamp Ocean Optics DH 2000.

To collect the light transmitted by the sample we used two multimode optical fibers with a 200 μm core coupled to two quartz lenses as collimators. The optical fiber collimator operating as a spatial filter was necessary to use in order to disallow the stray light scattered in the forward direction by crystal surfaces to re-enter the collection optics and affect the measurement results.

The results of the transmission measurements are presented in Fig. 2.8. An essential improvement of the optical quality was achieved thanks to the advanced purification methods and the low-thermal-gradient Czochralski crystal growing technique. Moreover, the $ZnMoO_4$ sample 3 shown much better optical homogeneity in comparison to the samples 1 and 2, which have significantly different light transmission when the light beam was scanned over the crystals surface. It should be also noted that the absorption band at 440 nm–460 nm observed in the sample 1 can be explained as a result of iron contamination [172]. The same problem of crystal contamination by iron can be supposed for the sample 2. However, the advanced purification technique applied to produce the sample 3, allows to get rid of the iron contamination problem and improve optical quality of the $ZnMoO_4$ crystals.

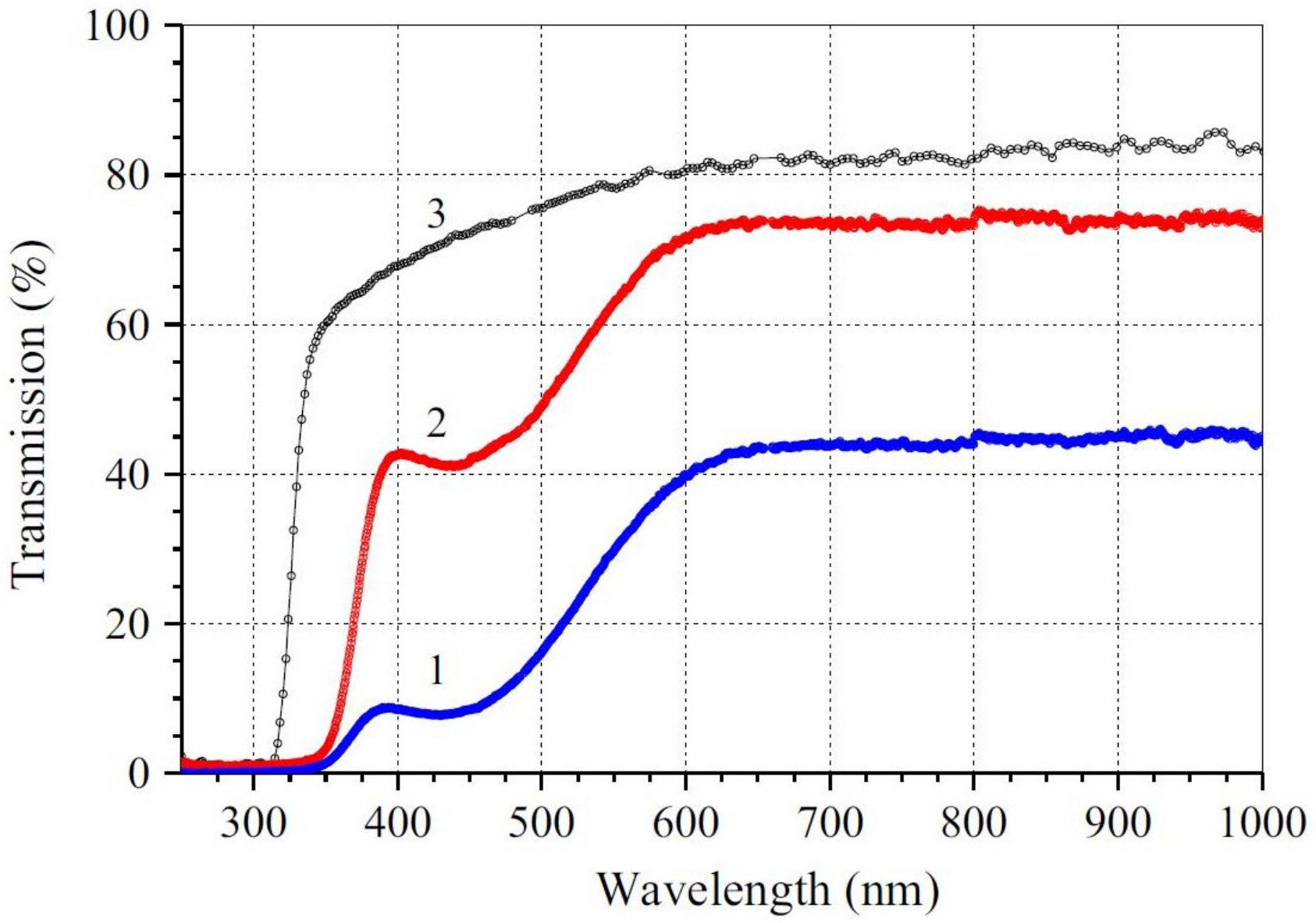

Fig. 2.8. Optical transmission curves of the $ZnMoO_4$ crystal samples with a size of 15 × 15 × 5 mm developed in IGP [157] (1), ISMA [158] (2), and NIIC [171] (3).

The absorption spectra of the 2 mm thick $ZnMoO_4$ single crystal produced by NIIC were measured with a help of the spectrometer Varian Cary 5000. The absorption coefficient was calculated from the formula: $\alpha = -\log T \times \ln 10/t$, where $T$ is a transmission coefficient, and $t$ is a thickness of the crystal sample. The obtained value of the transmission coefficient was higher than 0.5 in the range from 327 nm to 4.96 mm. The measurement results are presented in Fig. 2.9.

In the wavelength region from 400 nm to 2 μm the absorption coefficient reduced from 1.47 to 0.89 $cm^{-1}$. However, the broad absorption band around 440 nm that could correspond to $Fe^{2+}/Fe^{3+}$ impurities as described in Ref. [172] and [174] was not observed. This can be explained by a very low concentration of iron in the zinc molybdate crystal (≈ $1.71 \times 10^{16}$ $cm^{-3}$); while for the safe detection around 440 nm by such transmission experiments the concentration should be on the level of ~ $10^{18}$ atoms of Fe per $cm^3$. The obtained absorption coefficient is much lower than that for the $ZnMoO_4$ crystals produced by IGP in Ref. [157] and ISMA in Ref. [158], where the $\alpha_{abs}(< 550 \text{ nm}) \geq 2.5$ $cm^{-1}$.

To estimate the attenuation length we took the refractive index $n$ = 1.91 at the wavelength 589 nm determined with a Na lamp in Ref. [155]. Afterwards, solving self-consistently $R = (1 + e^{-\alpha t}) \times ((n-1)/(n+1))^2$ and $\alpha = -\log(T + R) \times \ln 10/t$, we obtained the absorption coefficient $\alpha$ (589 nm) = 0.023 $cm^{-1}$, which leads to an attenuation length of 43 cm at this wavelength.

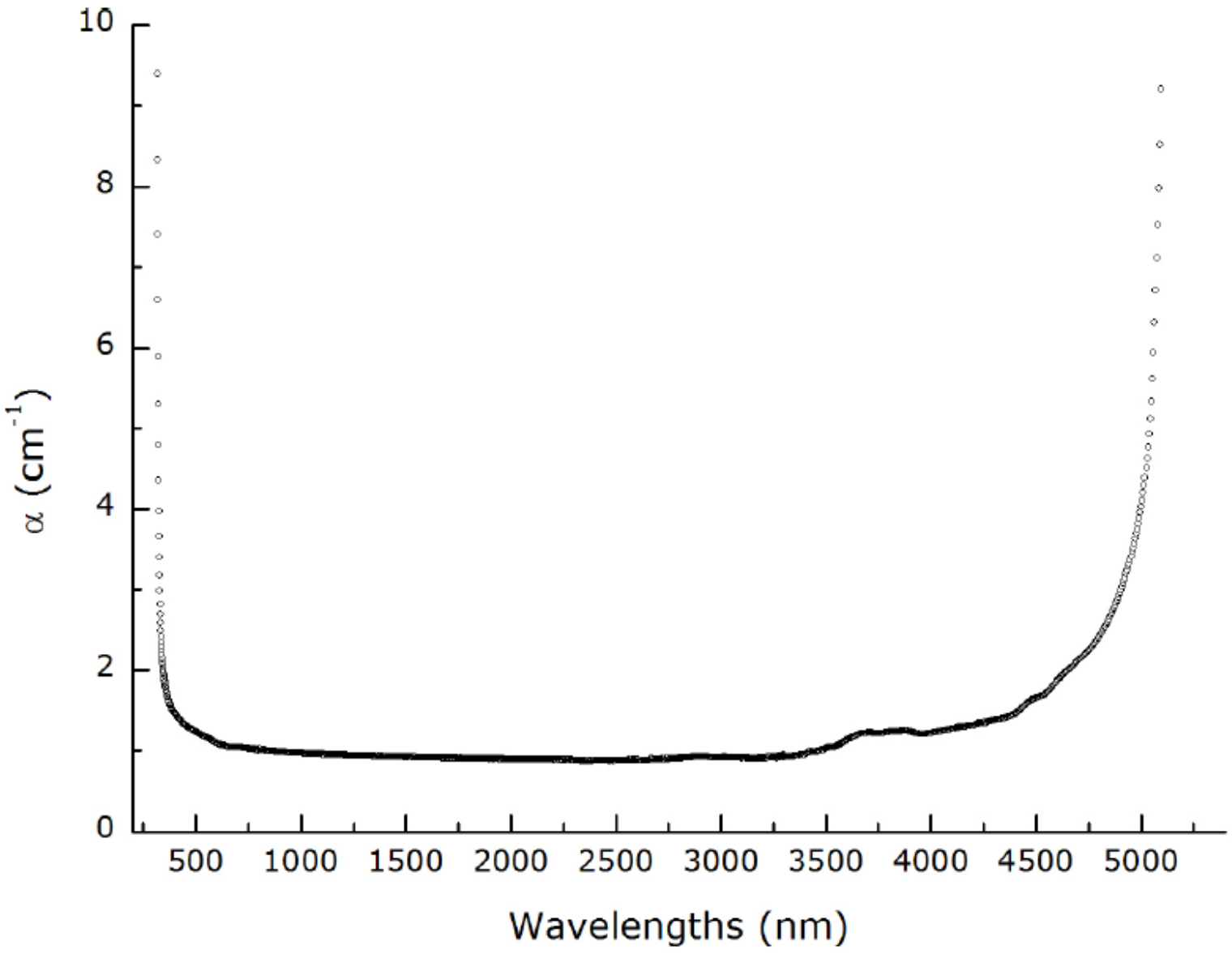

Fig. 2.9. Dependence of the absorption coefficient $\alpha$ from the wavelength measured for the 2 mm thick $ZnMoO_4$ single crystal.

### 2.2.2. Index of refraction

The refractive index was studied for two $ZnMoO_4$ crystals produced by ISMA ([158], sample 2) and NIIC ([171], sample 3) with a size of 15 × 15 × 5 mm and triangular prism shape. The GS-5 goniometer was used in the measurements. The zinc molybdate sample produced by ISMA demonstrated a multiple refraction of light, which can be explained by imperfections of the crystal structure. The polycrystalline structure of the sample (as well as improved quality of the $ZnMoO_4$ crystal produced by NIIC) was proved by the X-ray crystallography measurements.

The measurements of refractive indices were performed for two normal waves with different polarizations. The results obtained for sample 3 are presented in Table 2.6. The values of the refractive indices have difference in the range of 0.02–0.03 for different directions. This variation indicates that the $ZnMoO_4$ crystal is biaxial, which is consistent with a triclinic structure of zinc molybdate. The refractive indices were measured for three wavelengths that correspond to the red, green and blue parts of visible spectrum. The previous data on the index of refraction 1.90–1.92 obtained with a Na light in Ref. [155] are not in contradiction with present results. The difference in results can be explained with a different wavelengths used in measurements and variety of directions in the present study.

Table 2.6
Refractive indices $n_1$ and $n_2$ of a $ZnMoO_4$ crystal produced by NIIC for two light polarizations (see text for details). The error of the values is ±0.01.

| Wavelength (nm) | Refractive indexes | |
|---|---|---|
| | $n_1$ | $n_2$ |
| 406 | 1.94–1.97 | 1.98–2.01 |
| 532 | 1.89–1.91 | 1.94–1.96 |
| 655 | 1.87–1.90 | 1.91–1.93 |

### 2.2.3. Luminescence under X-ray excitation

The luminescence of the $ZnMoO_4$ crystal was studied in the temperature region 8 K–295 K under the X-ray excitation at the Faculty of Physics of Taras Shevchenko National University of Kyiv (Kyiv, Ukraine). A crystal sample with a size of 10 × 10 × 2 mm was produced by the NIIC. The sample was irradiated by X-rays from a BHV7 tube with a rhenium anode (20 kV, 20 mA). Light from the crystal was detected in the visible region by a FEU-106 photomultiplier, which is sensitive in the wide wavelength region of 350 nm–820 nm, and in the near infrared region by a FEU-83 photomultiplier with enhanced sensitivity up to ≈ 1 μm. Spectral measurements were carried out using a high-aperture MDR-2 monochromator. Emission spectra measured at 8 K, 85 K and 295 K are shown in the upper part of the Fig. 2.10. There are at least three emission bands in the spectra with the maximum at 490 nm, 610 nm and a near infrared emission at 700 nm–800 nm. The most intensive luminescence was observed at the liquid nitrogen temperature with a maximum at the wavelength ≈ 610 nm.

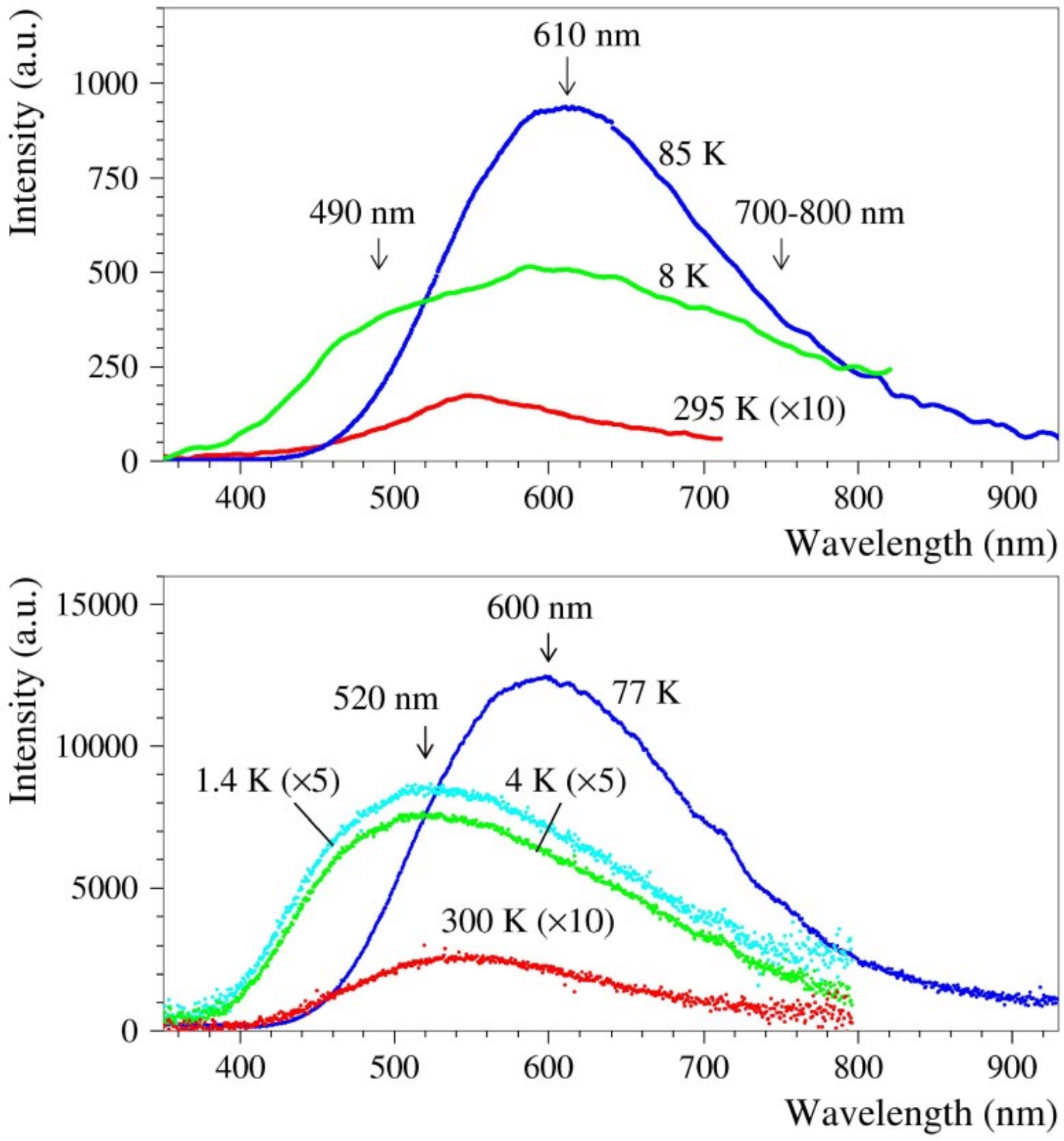

Fig. 2.10. Emission spectra of the $ZnMoO_4$ crystal sample with size of 10 × 10 × 2 mm developed by NIIC under the X-ray excitation measured in Taras Shevchenko National University of Kyiv (upper part) and Institut d'Astrophysique Spatiale (lower part).

The similar luminescence measurements were performed with a Ø20 × 6 mm $ZnMoO_4$ crystal at the Institut d'Astrophysique Spatiale (Orsay, France). The sample was irradiated through a beryllium window and an aluminum thin foil using an X-ray micro-tube (Bullet type by Moxtek; HV = 40 kV; I = 100 μA). The crystal was placed into a reflecting cavity and cooled down using a homemade cryostat. The light was transmitted outside by an optical fiber, directly to the AVANTES 2048 spectrometer. The emission spectra were accumulated over 20 s, after switching the X-ray tube off and stabilization of the light emission level. The results of the measurements are presented in the lower part of the Fig. 2.10. The obtained data are qualitatively similar, however the shape of the spectra and the relative intensities of the bands are a bit different. Possibly this is related to different samples and radiation doses used in the measurements.

The dependence of $ZnMoO_4$ luminescence intensity on temperature was studied in Taras Shevchenko National University of Kyiv. The measurements were performed in the temperature region 8 K–410 K. The obtained data, presented in Fig. 2.11, are in agreement with the previous results [171]. The dependence of the luminescence on temperature in the region above 120 K can be described by the Mott formula:

$$J(T)/J_0 = 1/\left(1 + A \cdot e^{-E_T/kT}\right), \tag{2.3}$$

where $J(T)$ is the luminescence intensity at temperature $T$, $J_0$ is the luminescence intensity at temperature $T \to 0$, $k$ is the Boltzmann constant, $A$ is the coefficient, $E_T$ is the damping energy. A fit of the measured dependence in the interval 120 K–410 K gives rather low damping energy $E_T = 0.120(2)$ eV, the coefficient $A$ is 2020(190), and $\chi^2 = 0.15$. As we can see from the lower part of Fig. 2.11, the dependence of emission on the temperature is different for the short (492 nm) and long (647 nm) wavelengths, which confirms further our assumption about different nature of the emission bands.

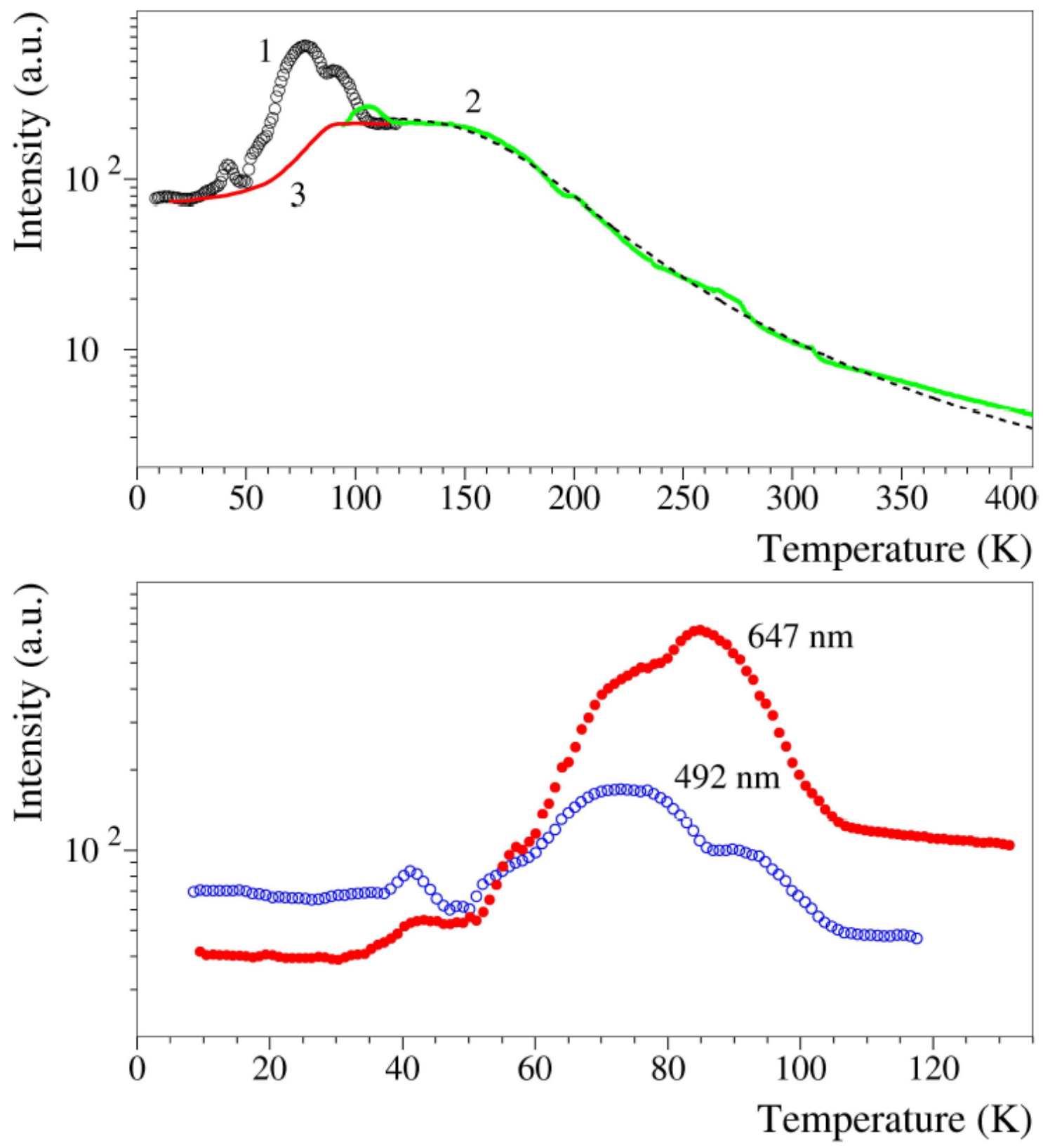

Fig. 2.11. (Upper part) Dependence of $ZnMoO_4$ luminescence intensity under X-ray excitation in a wide interval of the emission spectrum on temperature measured in the temperature interval 8 K–120 K (1) and 85 K–410 K (2). Dependence of luminescence intensity on temperature after subtraction of thermally stimulated luminescence (curve 3, see text and Fig. 2.12). Fit of the temperature dependence in the interval 120 K–410 K obtained by using Mott formula is shown with a dashed line. (Lower part) Dependence of short (492 nm) and long (647 nm) wavelength luminescence on the temperature.

Thermo-stimulated luminescence (TSL) of $ZnMoO_4$ crystal sample was studied after X-ray excitation over 20 min at the temperatures 8 K and 85 K (see Fig. 2.12). The highly intensive TSL was observed at the temperature 78 K after irradiation at 8 K. A strong TSL peak was detected at the temperature 114 K, as well as a few smaller peaks were observed at higher temperatures after irradiation of the crystal at 85 K. It should be stressed that the observed intense TSL with a numerous peaks in a wide temperature interval indicates the presence of point defects in the sample.

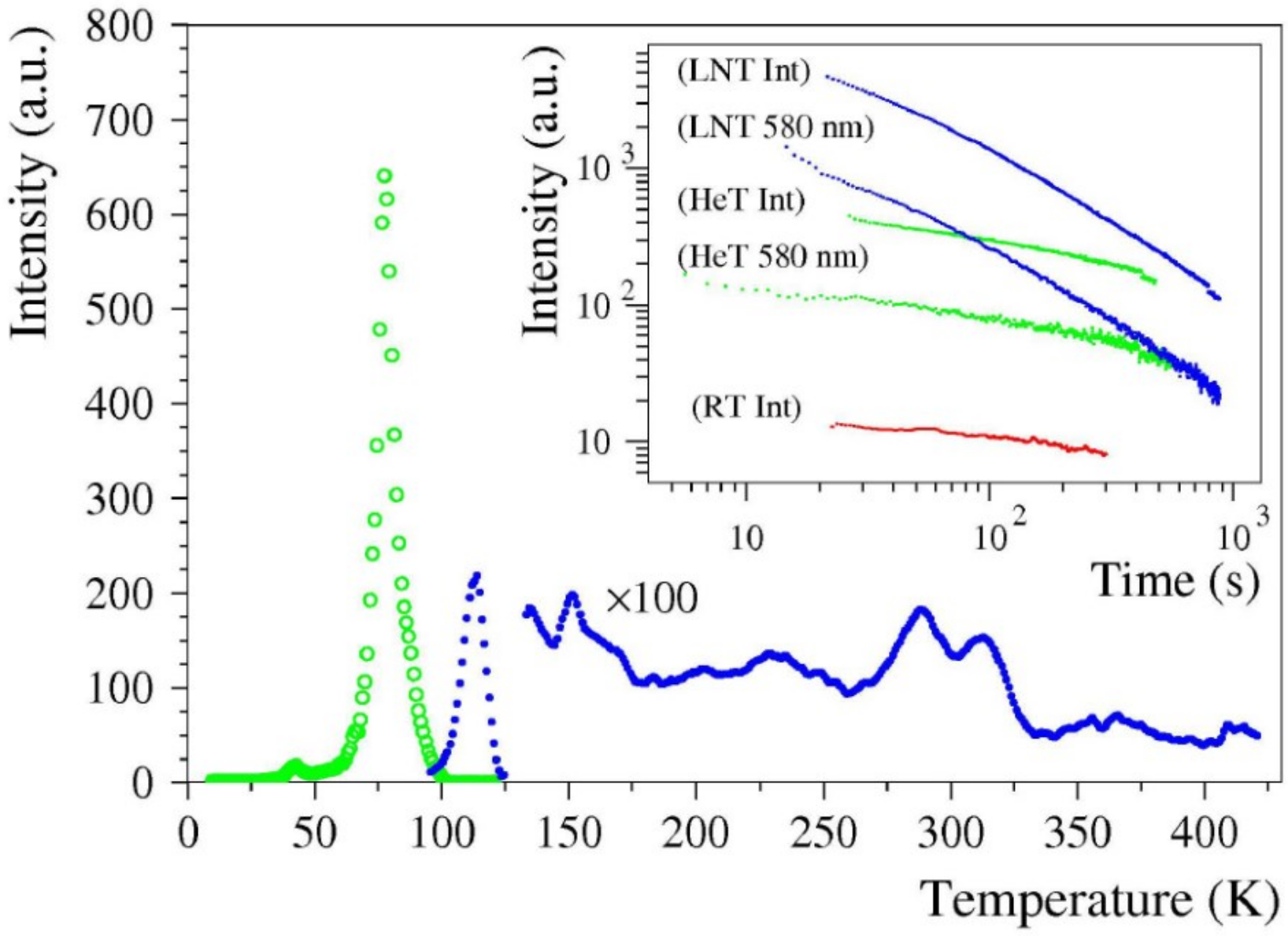

Fig. 2.12. Thermally stimulated luminescence of $ZnMoO_4$ crystal sample after the X-ray excitation during 20 minutes at 8 K (green color, open circles) and at 85 K (blue color, filled circles). (Inset) Phosphorescence after the X-ray irradiation during 20 minutes at 8 K (HeT), 85 K (LNT) and at room temperature 295 K (RT) in a wide interval of emission spectrum (Int) and at 580 nm.

We have also observed rather slow phosphorescence after the irradiation of the $ZnMoO_4$ sample at the temperatures 8 K, 85K and 295K (see inset in Fig. 2.12). This result also proves the presence of shallow traps due to imperfections and defects in the crystal. Moreover, the difference in the phosphorescence decay time confirms a different origin of the optical bands. The traps cause the low scintillation efficiency of the sample. Therefore, in order to improve scintillation properties of the $ZnMoO_4$ crystals further improvement of the material quality is necessary.

Similar thermo-stimulated luminescence and long-time phosphorescence measurements between 1.4 K and 300 K were also performed using the set-up in Orsay, where one intense TSL emission peak was clearly detected around the temperature 50 K.

Finally, it should be stressed that the results of our studies of $ZnMoO_4$ luminescence are in agreement with the data previously published in Refs. [157, 158, 171, 172].

### 2.3. Cryogenic scintillating bolometers based on zinc molybdate scintillators

Zinc molybdate is a promising material to search for neutrinoless double beta decay of $^{100}Mo$ with the bolometric technique [141, 153, 159, 173, 175]. The development of the cryogenic scintillating bolometers based on $ZnMoO_4$ crystals was performed in this work. Large volume $ZnMoO_4$ crystal of improved quality with mass 313 g was tested as scintillating bolometer. Moreover, the natural $ZnMoO_4$ and enriched $Zn^{100}MoO_4$ crystal scintillators produced from the deeply purified compounds (see section 2.1) were studied as a part of the LUMINEU program (see subsection 1.3.5.2).

### 2.3.1. Low-temperature measurements with 313 g $ZnMoO_4$ crystal

The natural $ZnMoO_4$ crystal with mass 313 g (see Fig. 2.4) is a precursor of the LUMINEU program which was produced from the first large volume crystal boule grown by the LTG Cz technique in the Nikolaev Institute of Inorganic Chemistry in 2011. A second $ZnMoO_4$ sample with mass 329 g was obtained from the same boule and tested as a cryogenic scintillating bolometer at the LNGS [173].

The 313 g crystal scintillator was placed inside a copper holder and fixed by using PTFE clamps. In order to improve light collection of the detector a reflector foil 3M VM2000/2002 surrounded the crystal. The scintillation light was read-out by the two ultrapure germanium wafers (Ø50 × 0.25 mm) placed from the opposite face planes of the crystal. The neutron transmutation doped (NTD) germanium thermistor working as a temperature sensor was attached to the crystal surface with the help of six epoxy glue spots and a 25 μm thick Mylar spacer (which was removed after the gluing procedure). The heater based on heavily-doped silicon meander was used to control and stabilize the thermal bolometric response by periodically injecting a certain amount of thermal energy to the crystal. The detector based on the 313 g $ZnMoO_4$ crystal is shown in Fig. 2.13.

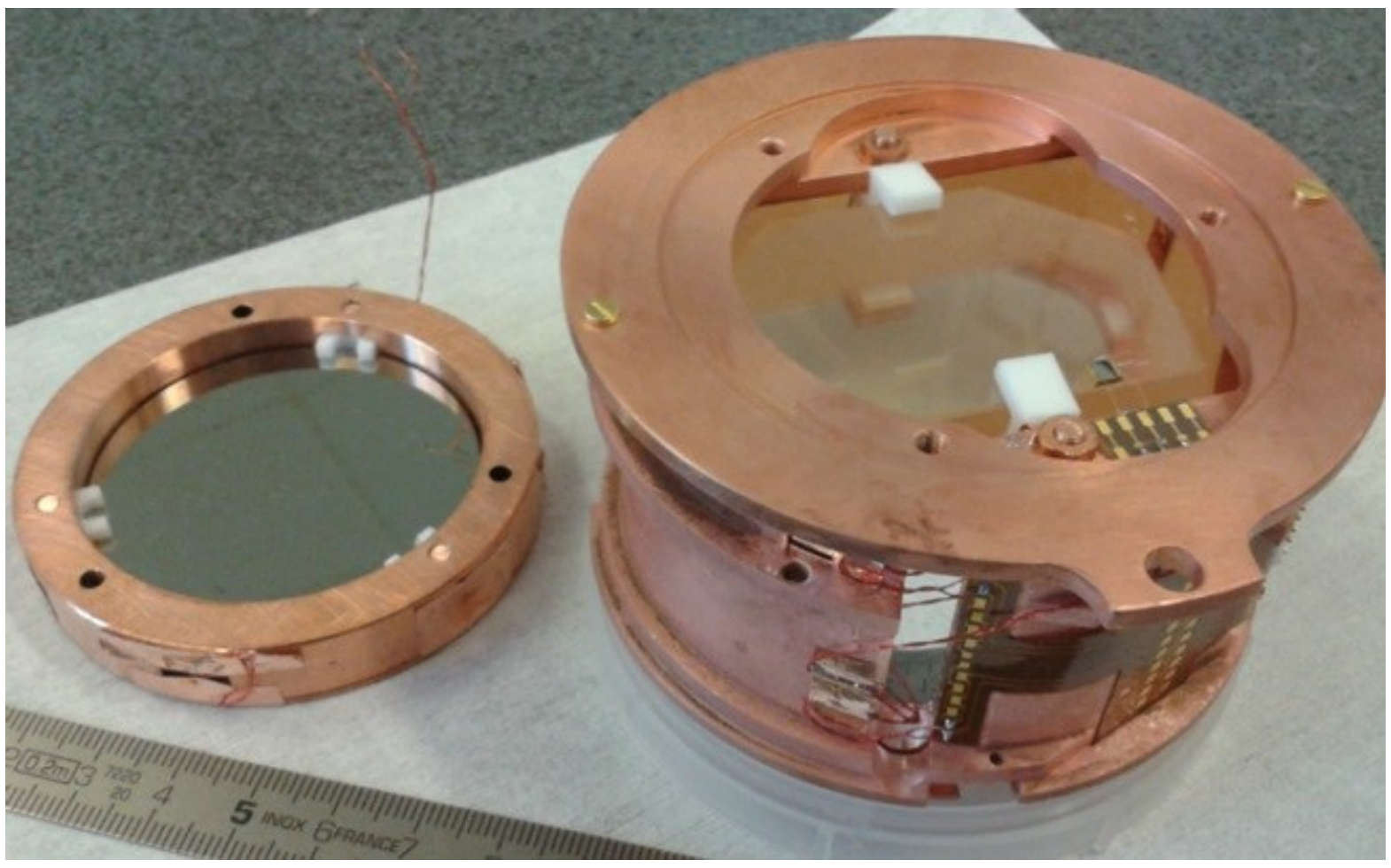

Fig. 2.13. Photo of the scintillating bolometer based on the 313 g $ZnMoO_4$ precursor crystal and of one ultrapure Ge photodetectors (two were used).

The precursor crystal was tested in the aboveground “wet” $^{3}He/^{4}He$ dilution refrigerator at the Centre de Sciences Nucléaires et de Sciences de la Matière (CSNSM, Orsay, France). The cryostat was surrounded by a low activity lead shield used to decrease the rate of pile-up events caused by environmental gamma rays due to the low time resolution of the bolometers. The data were accumulated by using a 16 bit ADC with a sampling frequency 30 kHz. The time of measurements was 38 hours with an operation temperature of 17 mK.

The $ZnMoO_4$ detector was irradiated by gamma quanta from a weak $^{232}$Th source, while the photodetectors were calibrated by using a $^{55}$Fe source placed near the germanium slabs.

The bolometric data (here and in the following subsections) were analyzed by using an optimum filter method [176]. The spectrometric performances of the 313 g $ZnMoO_4$ bolometer were deteriorated by the pile-up effects due to the high counting rate ≈ 2.5 Hz. Nevertheless, the measurements demonstrated normal operability of the detector which gives a possibility to estimate the scintillation light yield (the amount of detected light energy per particle energy) for γ quanta (β particles) and muons, as well as to evaluate quenching of scintillation yield for α and γ(β) particles. The light-to-heat scatter plot obtained with the scintillating bolometer based on the 313 g $ZnMoO_4$ precursor crystal is shown in Fig. 2.14. The results of the aboveground measurements are reported in Table 2.7.

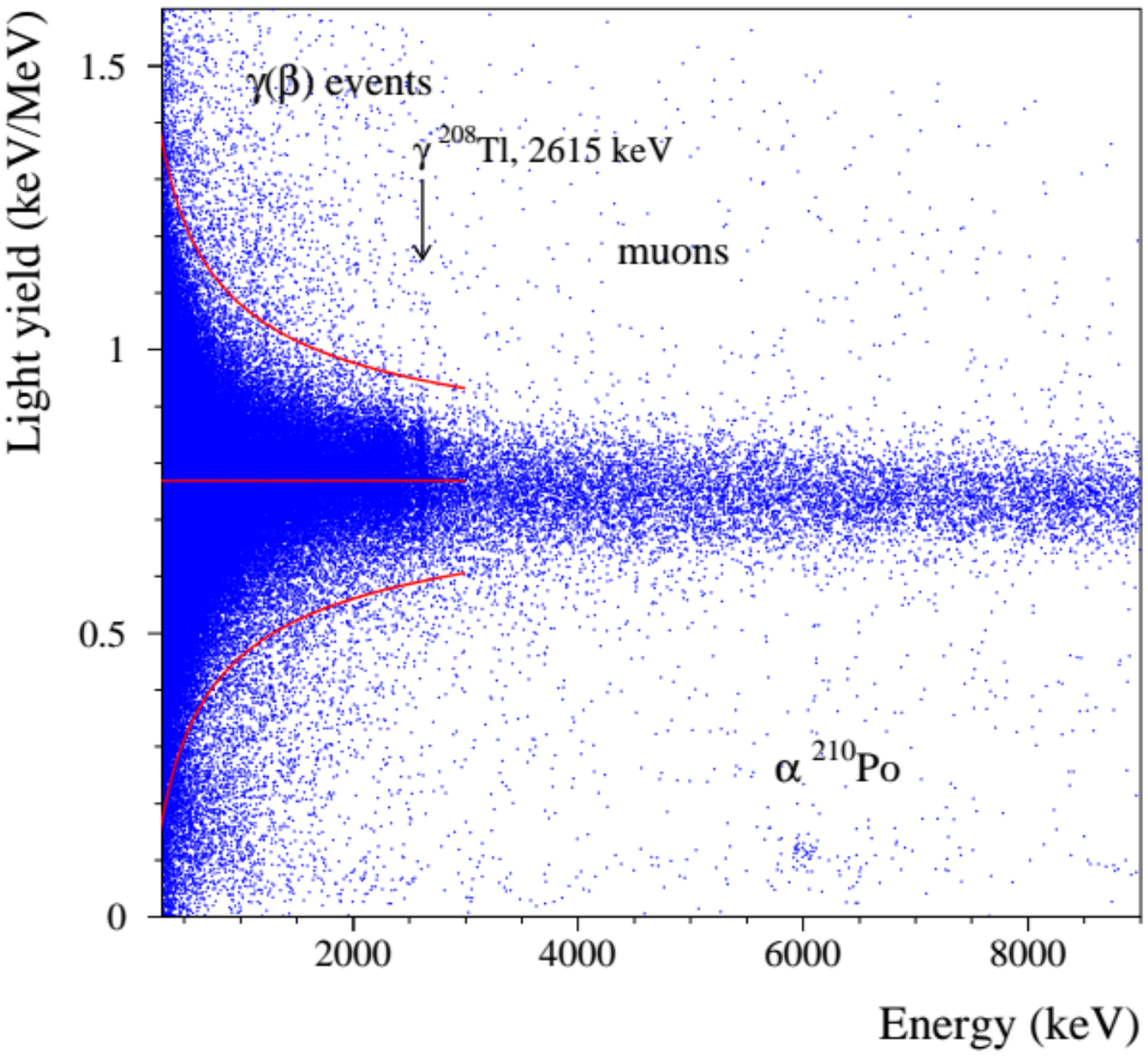

Fig. 2.14. Scatter plot of the light-to-heat signal amplitude ratio as a function of the heat signal amplitude accumulated at 17 mK with the aboveground scintillating bolometer based on 313 g $ZnMoO_4$ precursor crystal operated during 38 h with $^{232}$Th source. The visible band is related to γ(β) events (below 2.6 MeV) and cosmic muons. Three sigma interval of the light yield for the γ(β) band is shown by solid red curves together with the median value.

The 313 g detector was installed in the EDELWEISS-III cryostat located at the depth of 4800 m.w.e. in the Modane Underground Laboratory (LSM, France). The $ZnMoO_4$ bolometer was tested during the EDELWEISS-III commissioning runs together with 15 ultra-pure Ge detectors (≈ 0.8 kg each one) fully covered with interleaved electrodes (FID). The detectors were operated inside the $^3$He/$^4$He inverted dilution refrigerator with a ~ 50 liters of the experimental volume [177]. The EDELWEISS set-up is placed inside a clean room (ISO Class 4) and supplied by deradonized (≈ 30 mBq/m$^3$) air flow. The passive shield consists of low background lead (20 cm thick) and polyethylene (50 cm). The set-up is surrounded by a muon veto (98% coverage [178]) made of 5 cm thick plastic scintillators, and is completed by neutron and radon monitors. The schematic diagram of the EDELWEISS set-up is presented in Fig. 2.15.

Table 2.7
The main results of the aboveground and underground measurements with $ZnMoO_4$ and $Zn^{100}MoO_4$ crystals working as cryogenic scintillating bolometers. The energy resolution FWHM reported for the heat channels is estimated as filtered baseline and measured for γ quanta and α particles of internal $^{210}$Po. The light yield for γ(β) events ($LY_{γ(β)}$) and quenching factor for α particles ($QF_α$) are also presented.

| Detector | | FWHM (keV) | | | | | $LY_{γ(β)}$ (keV/MeV) | $QF_α$ |
|---|---|---|---|---|---|---|---|---|
| Crystal | Mass (g) | Baseline | $^{133}$Ba 356 keV | $^{214}$Bi 609 keV | $^{208}$Tl 2615 keV | $^{210}$Po 5407 keV | | |
| $ZnMoO_4$ | 313 | 1.4(1) | 6.4(1) | 6(1)* | 24(2)*/9(2) | 19(1) | 0.77(11)* | 0.15(2)*/0.14(1) |
| | 336 | 1.5(2) | 6(1) | — | — | 29(4) | — | — |
| | 334 | 1.06(3) | 3.8(4) | — | — | 15(1) | — | 0.19(2) |
| $Zn^{100}MoO_4$ | 59.2 | 1.4(1) | — | 5.0(5)* | 11(3)* | — | 1.01(11)* | ≈ 0.15* |
| | 62.9 | 1.8(1) | — | 10(1)* | 15(3)* | — | 0.93(11)* | ≈ 0.15* |

* — results of the aboveground measurements

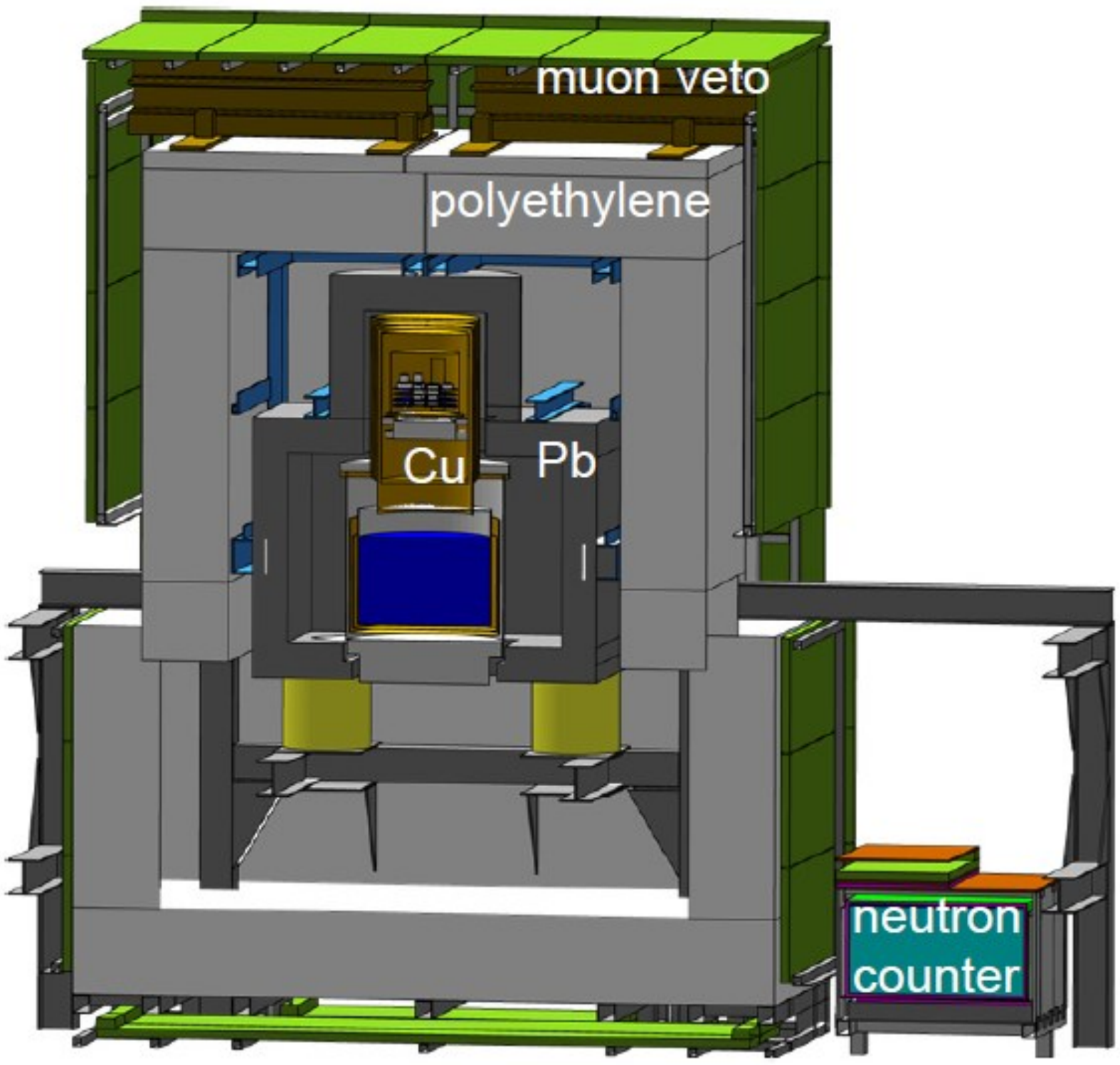

Fig. 2.15. Schematic view of the EDELWEISS set-up used for the underground test of the 313 g $ZnMoO_4$ scintillating bolometer.

The base temperature during the 313 g detector measurements was stabilized around 19 mK. The triggered signals were recorded in 2 seconds window by a 14 bit ADC with 1.9841 kHz sampling frequency. Half of this time window contains the baseline data. It should be stressed that one of the light detectors was not used in the measurements due to a high level of microphonic noise. Energy calibration of the $ZnMoO_4$ detector was performed by using $^{133}$Ba and $^{232}$Th gamma sources over 546 h and 51 h, respectively. The detector's background was accumulated over 305 hours of data acquisition.

An excellent discrimination capability of the 313 g $ZnMoO_4$ scintillating bolometer was achieved. The full separation of γ(β)-induced events from populations of α particles caused by trace (surface) contamination by radionuclides from the U/Th chains is presented in Fig. 2.16(a). Moreover, the energy spectrum obtained using $^{232}$Th gamma source (see Fig. 2.16(b)) demonstrates high spectrometric properties of the detector. The main results of the underground measurements with 313 g $ZnMoO_4$ detector are given in Table 2.7.

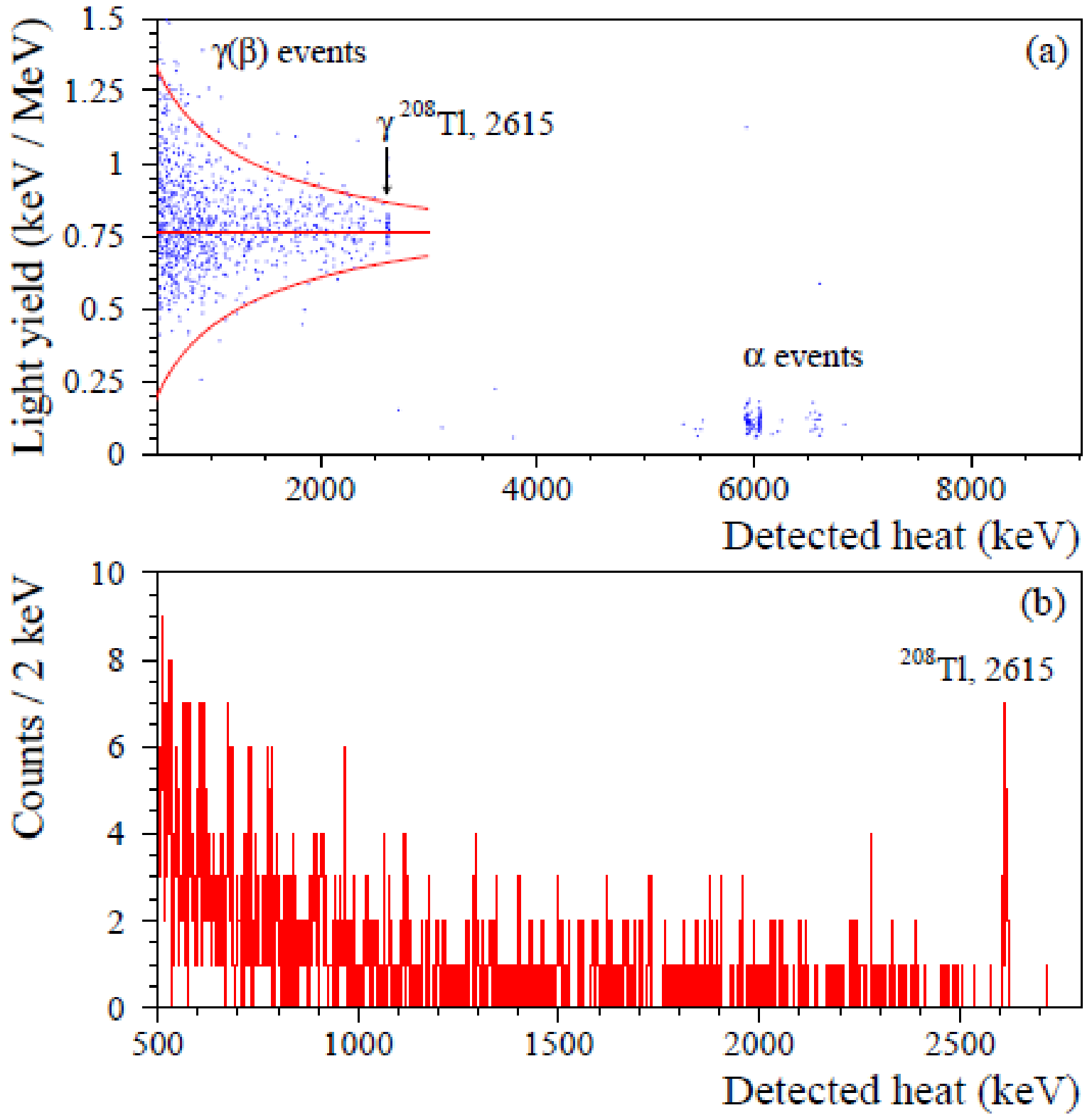

Fig. 2.16. (a) Scatter plot of the light-to-heat signal amplitude ratio as a function of the heat signal amplitude for the 313 g $ZnMoO_4$ scintillating bolometer operated at 19 mK in the EDELWEISS set-up at the LSM underground laboratory. The detector was irradiated by γ quanta from the $^{232}$Th source for over 51 h. Two visible bands correspond to γ(β) events and α particles. The positions of the α events are shifted from the nominal values due to thermal energy overestimation for α particles in case of using calibration data for γ particles. Three sigma interval of the light yield for the γ(β) band and its median value are drawn by solid red lines. (b) Energy spectrum of the $^{232}$Th source obtained with the 313 g $ZnMoO_4$ scintillating bolometer during 51 hours of underground measurements.

### 2.3.2. Scintillating bolometers based on two Ø50 × 40 mm $ZnMoO_4$ crystals

Advanced quality $ZnMoO_4$ crystal boules with mass about 1 kg were produced in 2013 at the Nikolaev Institute of Inorganic Chemistry by using the LTG Cz growing technique and molybdenum purification method containing sublimation in vacuum and double recrystallization from aqueous solutions (see section 2.1 for details). The crystals were recrystallized in order to improve the quality of material. In result, two almost colorless $ZnMoO_4$ cylindrical samples with a size of Ø50 × 40 mm and mass of 336 g and 334 g (see Fig. 2.6) were produced as a part of LUMINEU program.

Two scintillating bolometers based on the advanced quality $ZnMoO_4$ crystals with masses of 336 g and 334 g were constructed as for the 313 g crystal (see subsection 2.3.1), except that only one ultrapure Ge photodetector was used for each detector. The photograph of the two assembled $ZnMoO_4$ bolometers is presented in Fig. 2.17.

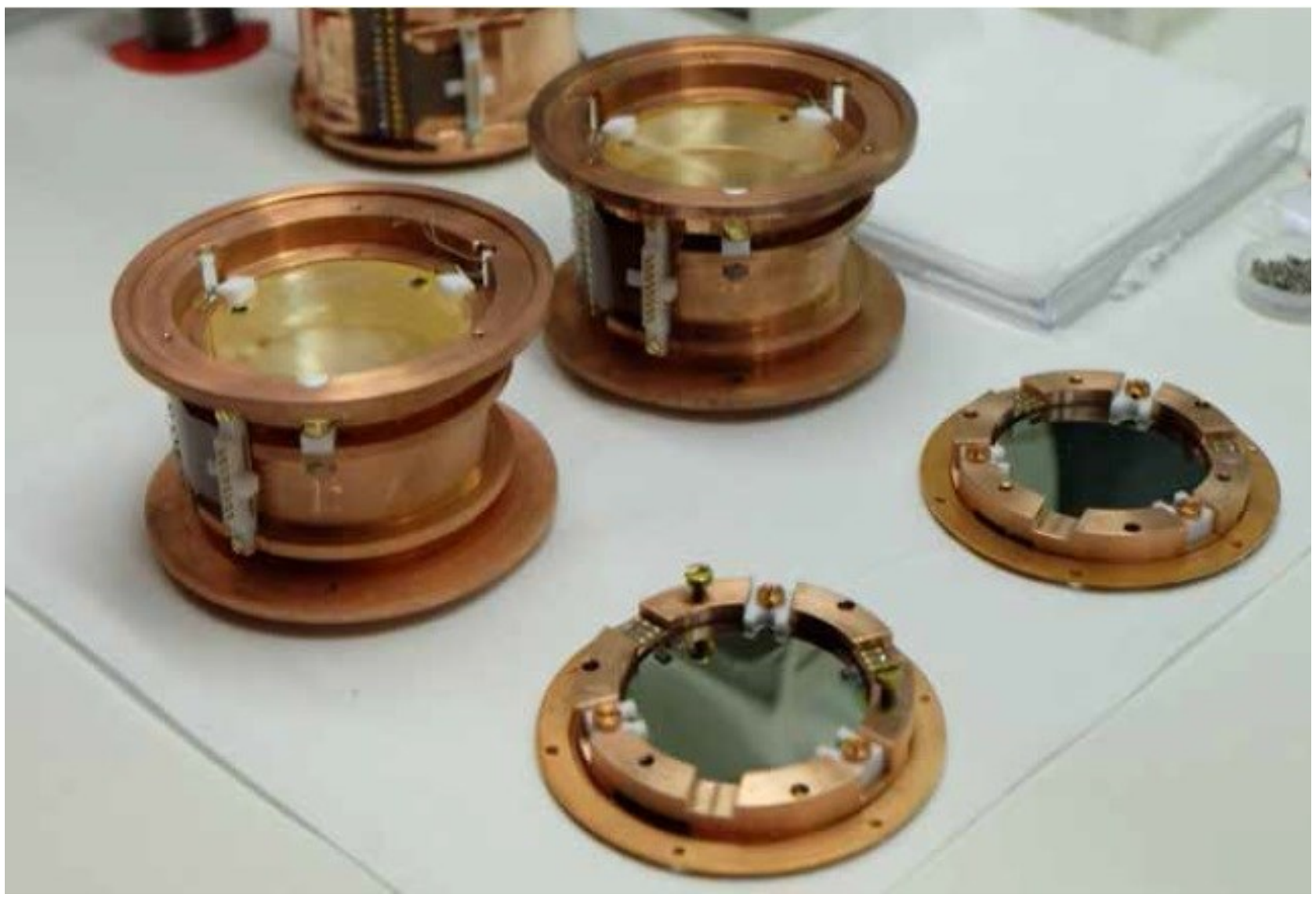

Fig. 2.17. Photo of the scintillating bolometers based on 336 g and 334 g $ZnMoO_4$ crystals of advanced quality with their ultrapure Ge photodetectors.

The two advanced $ZnMoO_4$ (Ø50 × 40 mm) and two enriched $Zn^{100}MoO_4$ crystal scintillators (described in the next subsection 2.3.3) were installed together with 36 FID Ge detectors into the EDELWEISS-III cryostat after completing the commissioning runs. It should be noted that the EDELWEISS set-up was upgraded by installing ultra radiopure NOSV Copper [179] screens and adding polyethylene shield at the 1 K plate. Moreover, all the detectors were provided with an individual low background Copper-Kapton cables. In addition, a pulser system driving the stabilization heaters attached to each element of the scintillating bolometers was implemented recently.

After upgrade of the EDELWEISS set-up the data were accumulated by a 16 bit ADC with 1 kHz sampling rate and time window of 2 seconds. The baseline data were recorded over 1 second. The detectors were operated at a temperature of 18 mK. The energy scale of the bolometers was measured with a $^{133}$Ba gamma source, while the measurements with a $^{232}$Th source are foreseen in a future.

The $ZnMoO_4$ detectors in the EDELWEISS set-up are still under optimization, until at least a significant vibration-induced noise problem will be solved. However, preliminary results obtained with the 334 g natural $ZnMoO_4$ scintillating bolometer are very promising: the detector allows an efficient α/γ(β) separation (see Fig. 2.18(a)) and demonstrates excellent spectrometric properties (see Fig. 2.18(b)). The complete information on the performances of Ø50 × 40 mm $ZnMoO_4$ detectors are reported in Table 2.7.

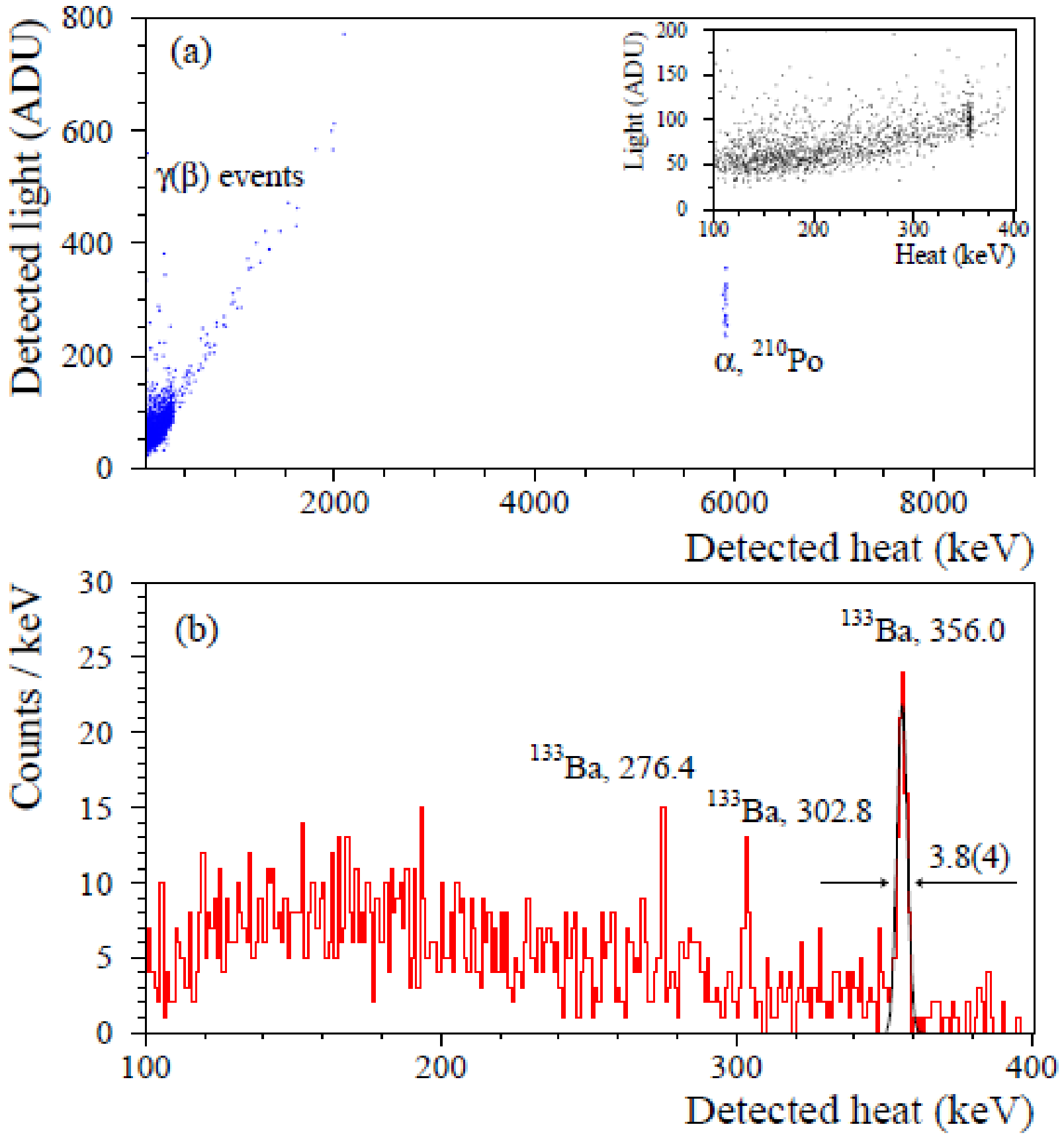

Fig. 2.18. (a) Scatter plot of light versus heat signals measured during 15 h in the calibration run with $^{133}Ba$ source by using 334 g $ZnMoO_4$ scintillating bolometer in the EDELWEISS set-up. ADU denotes to Analogue-to-Digital Units. (Inset) Part of the scatter plot in the energy range of $^{133}Ba$ source. (b) The energy spectrum of $^{133}Ba$ source accumulated over 15 h in the same calibration run by using the scintillating bolometer based on 334 g $ZnMoO_4$ crystal in the EDELWEISS set-up.

#### 2.3.3. First bolometric test of two enriched $Zn^{100}MoO_4$ crystal scintillators

Two enriched $Zn^{100}MoO_4$ crystals with mass of 59.2 g ($Zn^{100}MoO_4$-top) and 62.9 g ($Zn^{100}MoO_4$-bottom) were produced from the crystal boule described in subsection 2.1.3.

The sample $Zn^{100}MoO_4$-top was cut from the top part of the boule, close to its ingot. The crystal has less intensive orange coloration in comparison to the sample $Zn^{100}MoO_4$-bottom which corresponds to the bottom part of the boule. This feature can be explained by lower contamination and defects concentration in the upper part of the boule. The enriched zinc molybdate scintillation elements are presented in Fig. 2.19. The shape of the samples is irregular and can be roughly approximated by rectangular prism. The irregular shape was chosen to keep the crystals mass as high as possible in order to make the bolometric tests more significant. The only condition was to get two flat parallel bases to have a possibility of holding the crystals in the bolometer construction by PTFE clamps.

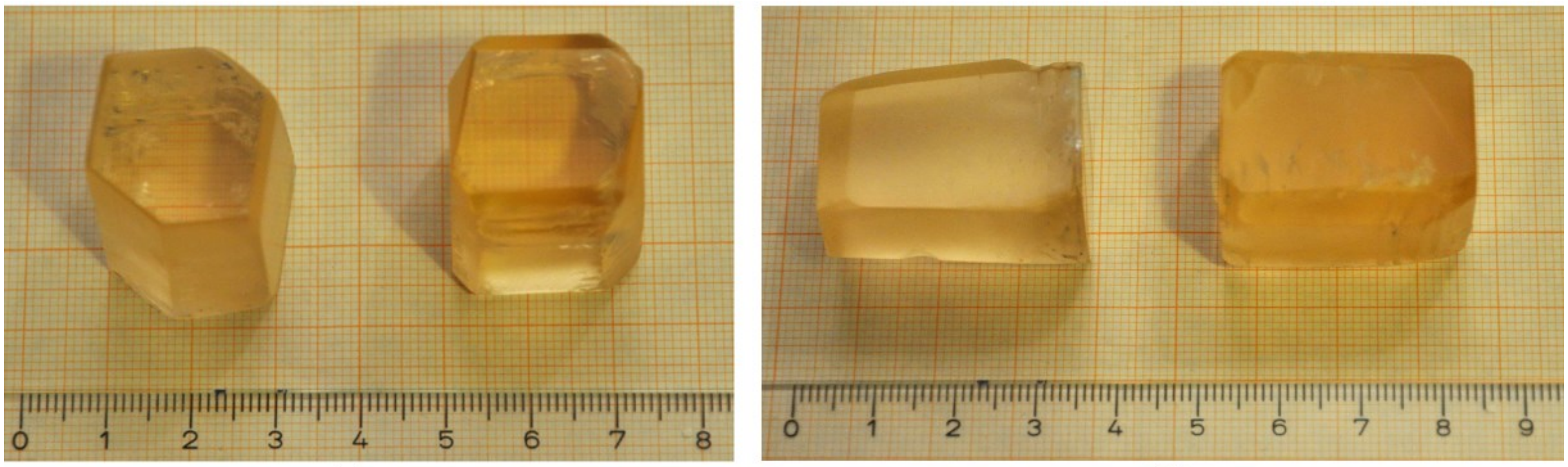

**(a)** **(b)**

Fig. 2.19. Top (a) and side (b) views of the $Zn^{100}MoO_4$ scintillation elements enriched in $^{100}Mo$ to 99.5%. The $Zn^{100}MoO_4$-top sample with a mass of 59.2 g (at the left) was produced from the top part of the boule, while the 62.9 g $Zn^{100}MoO_4$-bottom crystal (at the right) from the bottom part. The scale is in centimeters.

The two $Zn^{100}MoO_4$ crystals were assembled as described in subsection 2.3.1. Both samples were placed inside a single copper holder by using PTFE elements, and only one light detector was used to collect the emitted scintillation light. The high-purity Ge photodetector with a size of Ø50 × 0.25 mm was instrumented with NTD Ge thermistor and mounted at the distance ≈ 2 mm from the top face plane of the $Zn^{100}MoO_4$ crystals. The photograph of the $Zn^{100}MoO_4$ scintillating bolometer array is shown in Fig. 2.20.

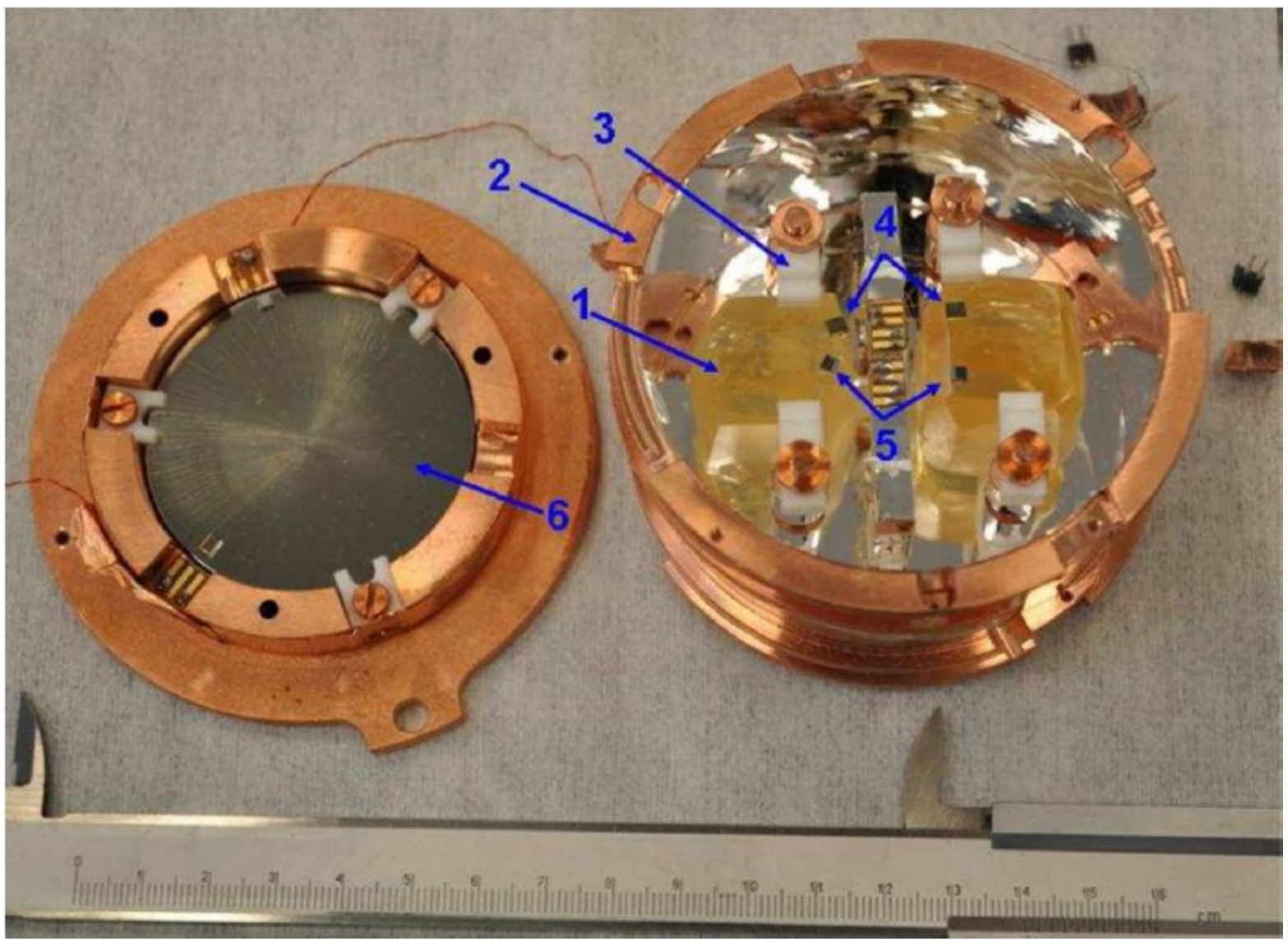

Fig. 2.20. Photograph of the assembled $Zn^{100}MoO_4$ bolometers together with the photodetector: (1) enriched $Zn^{100}MoO_4$ crystals with mass of 59.2 g (left) and 62.9 g (right); (2) copper holder of the detectors with the light reflecting foil (3M VM2000/2002) fixed inside; (3) PTFE elements supporting the crystals; (4) NTD Ge thermistors; (5) heating elements; (6) high-purity Ge photodetector. The scale is in centimeters.

The $Zn^{100}MoO_4$ scintillating bolometers were tested at milli-Kelvin temperatures by using a pulse-tube cryostat housed in the high-power dilution refrigerator operated at the CSNSM [180]. The cryostat was surrounded by passive shield made of low activity lead to suppress environmental gamma background, which provides substantial number of pile-ups due to the slow response of the bolometers based on NTD Ge thermistors.

The NTD Ge thermistors were read-out by room-temperature low-noise electronics, which consists of the DC-coupled voltage-sensitive amplifiers [181] placed inside a Faraday cage. The data streams were recorded by a 16 bit ADC with a sampling frequency of 10 kHz. The bolometers were operated at three base temperatures: 13.7 mK for over 18.3 hours, 15 mK (4.8 h), and 19 mK (24.2 h). The measurements at the 19 mK were performed in order to simulate the typical temperature conditions expected in the EDELWEISS set-up which was then used for the deep underground test at the LSM laboratory (see below).

The bolometric performance of the $Zn^{100}MoO_4$ array was evaluated by measuring thermistor resistance ($R_{bol}$) at the working temperature, amplitude of the signals ($A_{signal}$) for an unitary deposited energy, the full width at half maximum baseline width ($FWHM_{bsl}$), rise ($\tau_R$) and decay ($\tau_D$) times of the pulse. The time properties of the pulses were calculated from 10% to 90% of the signal maximum amplitude in case of rise time, and from 90% to 30% for decay time. The mentioned parameters of the detector performance at the working temperatures 13.7 mK and 19 mK are presented in Table 2.8. Similar performance was obtained with the $Zn^{100}MoO_4$ scintillating bolometers (top and bottom) cooled down to 15 mK.

Table 2.8
The experimental parameters (see text) of the scintillating bolometers based on two enriched $Zn^{100}MoO_4$ crystals with masses of 59.2 g and 62.9 g obtained in aboveground measurements at the working temperatures of 13.7 mK (first row) and 19 mK (second row). The parameters $\tau_R$ and $\tau_D$ were evaluated from the event distribution within the energy ranges of 500–3000 keV and 10–30 keV for the $Zn^{100}MoO_4$ bolometers and for the light detector, respectively.

| Detector | $R_{bol}$ (MΩ) | $A_{signal}$ (μV/MeV) | $FWHM_{bsl}$ (keV) | $\tau_R$ (ms) | $\tau_D$ (ms) |
|---|---|---|---|---|---|
| $Zn^{100}MoO_4$-top | 1.54 | 86.8 | 1.4(1) | 9.0 | 46.3 |
| | 1.17 | 65.0 | 1.8(1) | 8.9 | 48.4 |
| $Zn^{100}MoO_4$-bottom | 1.82 | 95.8 | 1.8(1) | 5.5 | 26.2 |
| | 1.35 | 84.2 | 2.4(1) | 5.8 | 30.7 |
| Light Detector | 0.97 | 409 | 0.28(1) | 2.5 | 14.8 |
| | 0.81 | 336 | 0.37(2) | 2.5 | 15.5 |

The light detector was calibrated by using a weak $^{55}Fe$ source. The energy resolution of the photodetector at 5.9 keV of $^{55}Fe$ was 0.42(2) keV and 0.57(4) keV at the working temperature 13.7 mK and 19 mK, respectively. The $Zn^{100}MoO_4$ detectors were irradiated by gamma quanta from a low activity $^{232}Th$ source in order to determine the bolometers energy scale. The energy spectra obtained with the $Zn^{100}MoO_4$ detectors operated at 13.7 mK are shown in Fig. 2.21. The energy resolution of the bolometers at 2615 keV of $^{208}Tl$ was FWHM = 11(3) keV for the $Zn^{100}MoO_4$-top detector, and FWHM = 15(3) keV for the $Zn^{100}MoO_4$-bottom, for the measurements at the base temperature 13.7 mK (see Table 2.7). However, the statistics of the $^{208}Tl$ peak at 2615 keV is quite low due to the small mass of the crystals and the short duration of the measurements. Therefore, the energy resolution of the detectors was also estimated for the more intensive γ lines presented in the spectra below 1 MeV. For instance, the FWHM at 609 keV peak of $^{214}Bi$ was 5.0(5) for the $Zn^{100}MoO_4$-top detector, and 10(1) keV for the $Zn^{100}MoO_4$-bottom. Previous experience with slow bolometric detectors of large mass demonstrated that the energy resolution of the bolometers is significantly deteriorated by the pile-up effect. Thus, better energy resolution (i.e. definitely less than 10 keV) is expected in the underground set-up, as it was already observed in the underground measurements with natural $ZnMoO_4$ bolometers [173, 141].

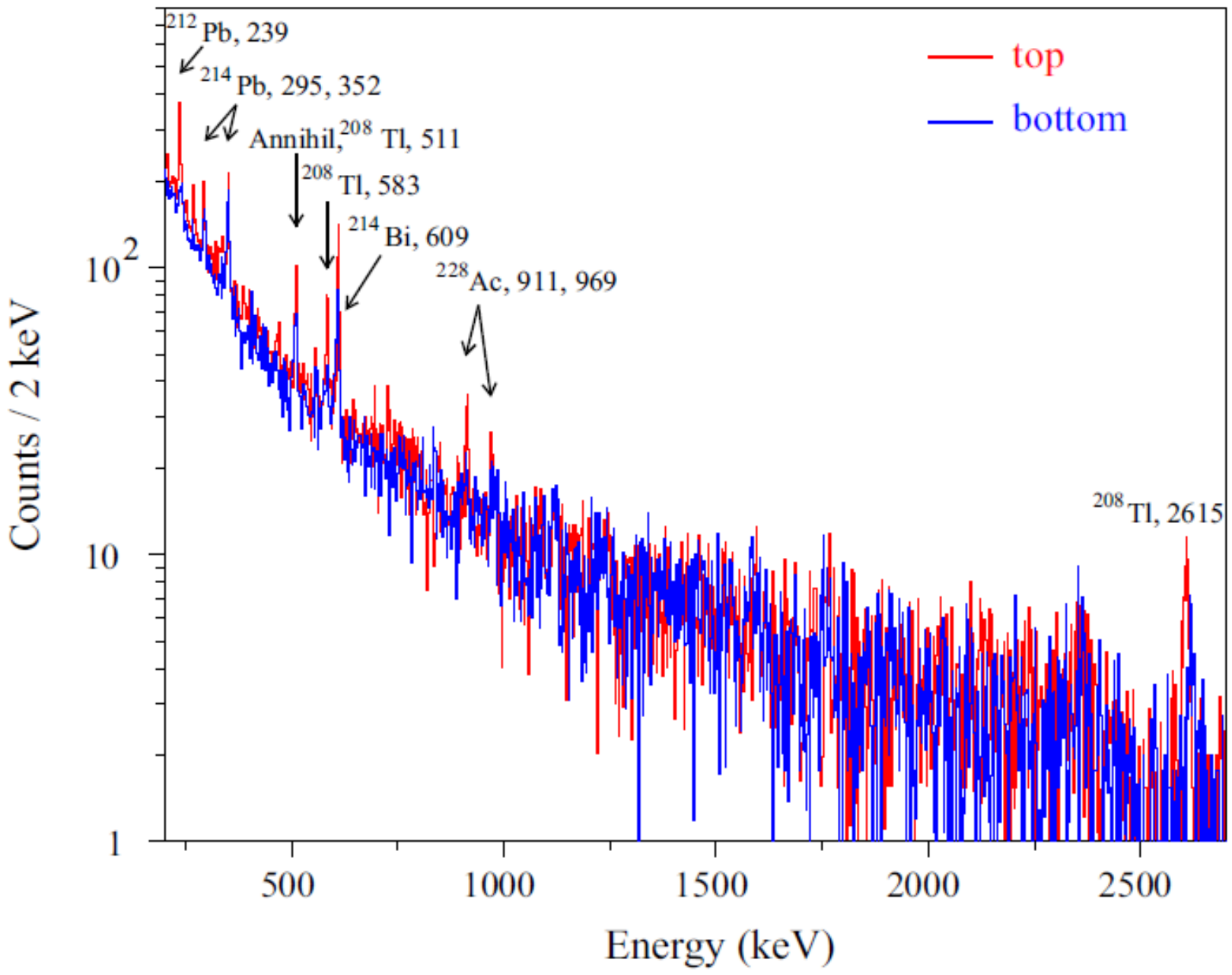

Fig. 2.21. Energy spectra accumulated in aboveground measurements over 18.3 h by using the scintillating bolometers based on two enriched $Zn^{100}MoO_4$ crystals (produced from the top and bottom of the boule) mounted in one copper holder. The detector was operated at 13.7 mK and irradiated by gamma quanta from low activity $^{232}$Th source. The energies of γ peaks are in keV.

Scatter plot of the light versus heat signals for the data measured over 18.3 h in the aboveground set-up with the enriched $Zn^{100}MoO_4$-top detector is presented in Fig. 2.22.

The accumulated data allow to evaluate the light yield for γ quanta (β particles) ($LY_{\gamma(\beta)}$). It was estimated by a fit of the data in the 600 keV–2700 keV energy interval. The similar values were obtained for all the working temperatures: 1.01(11) keV/MeV (at 13.7 mK) and 1.02(11) keV/MeV (at 19 mK) for the $Zn^{100}MoO_4$-top crystal, and 0.93(11) keV/MeV (at 13.7 mK) and 0.99(12) keV/MeV (at 19 mK) for the $Zn^{100}MoO_4$-bottom scintillator. The distributions of the light yield versus the detected heat measured by the $Zn^{100}MoO_4$ scintillating bolometers are shown in Fig. 2.23.

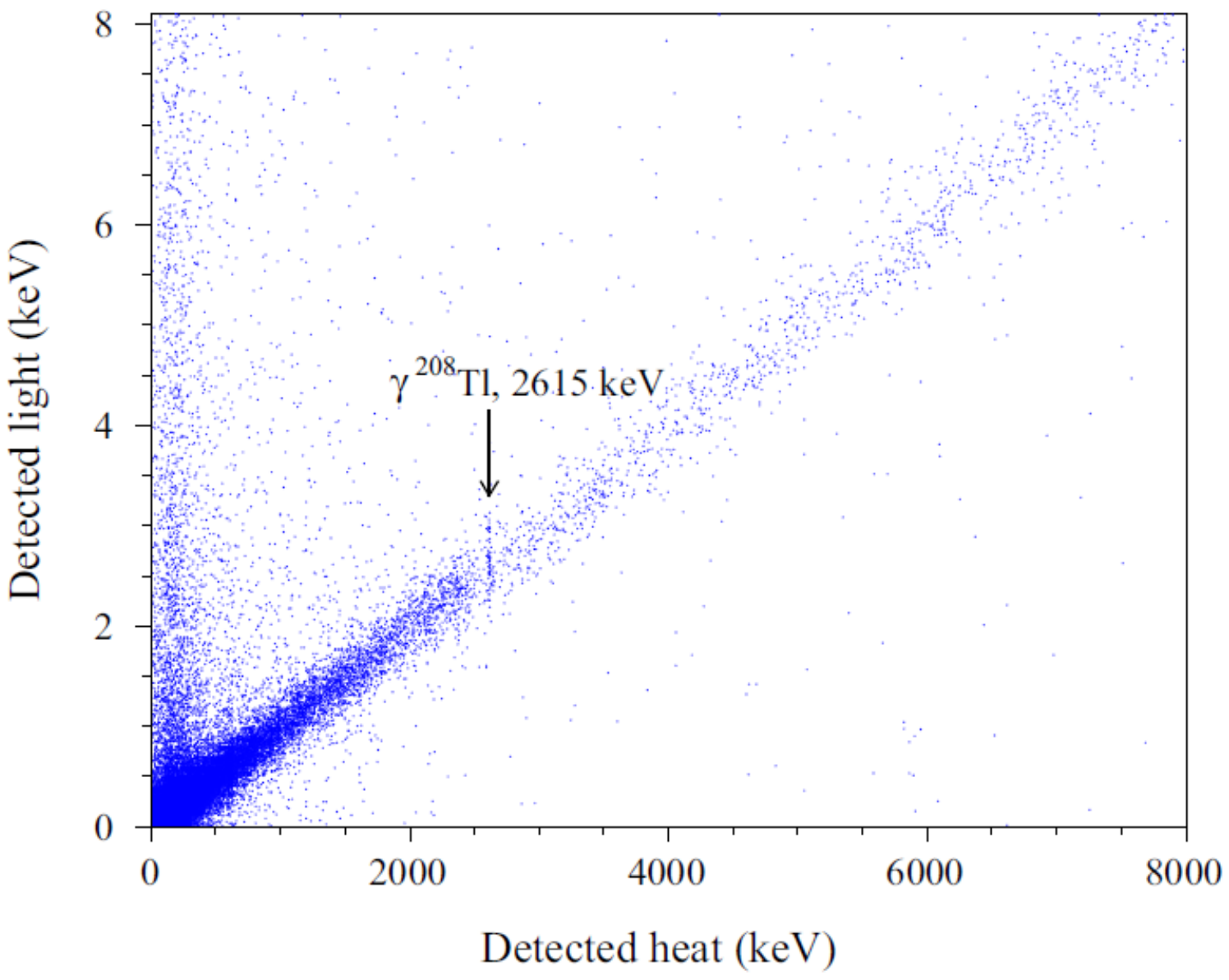

Fig. 2.22. Scatter plot of light versus heat signals measured by the scintillating bolometer based on $Zn^{100}MoO_4$-top crystal at 13.7 mK during 18.3 h of the calibration run with $^{232}$Th source. The band populated by γ(β) events (below 2.6 MeV) and cosmic muons is clearly visible.

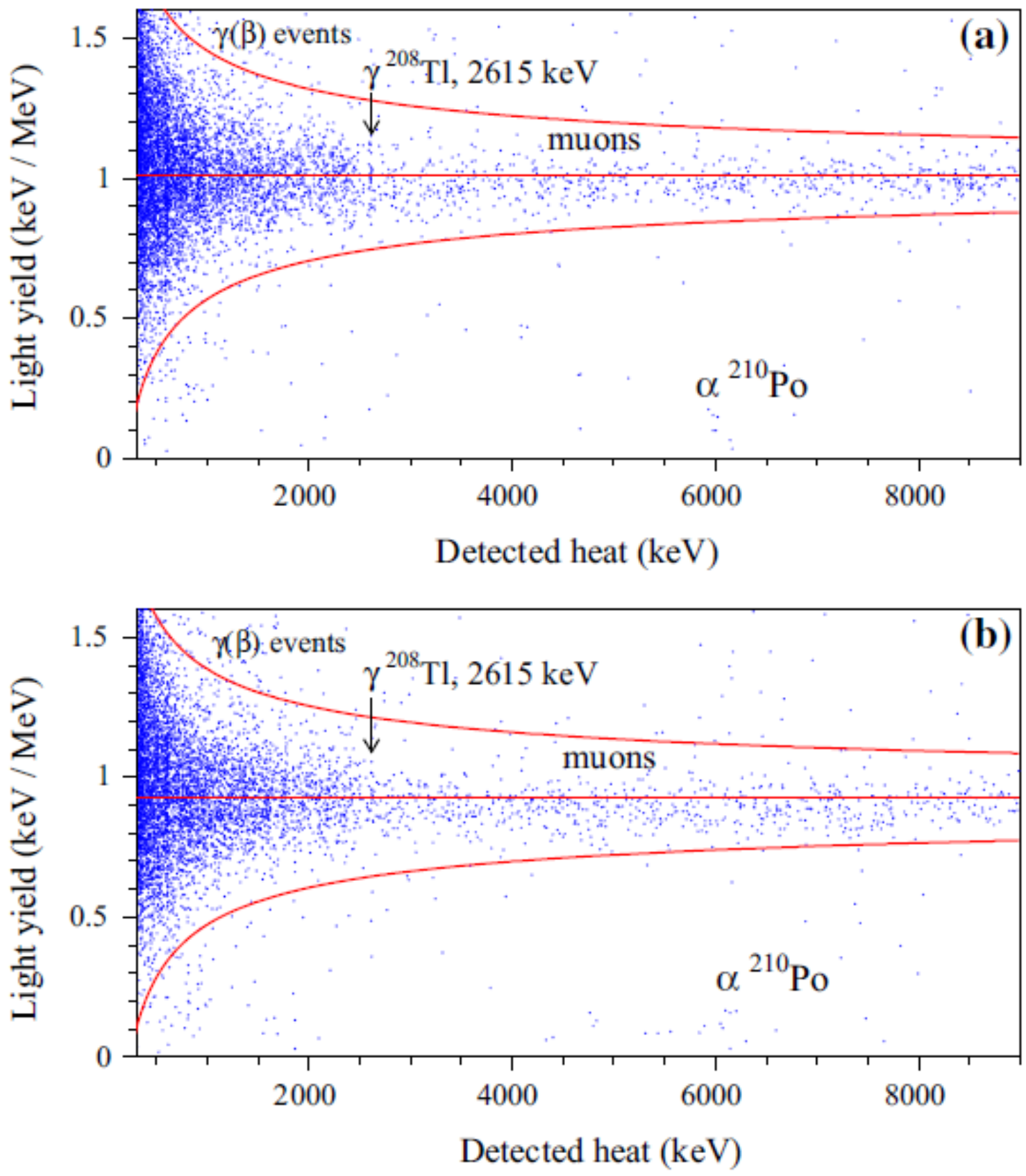

Fig. 2.23. Light yield as a function of the detected heat measured aboveground by using the scintillating bolometers based on $Zn^{100}MoO_4$-top (a) and $Zn^{100}MoO_4$-bottom (b) crystals. The data were accumulated at 13.7 mK during 18.3 h of calibration measurements. The events above 2.6 MeV in the γ(β) region are caused by cosmic muons. The positions of α events related to $^{210}$Po are shifted from the nominal value ($Q_\alpha \approx 5.4$ MeV) due to a different heat response for α particles. The three sigma interval for $LY_{\gamma(\beta)}$ and its mean value are shown by solid red lines.

It should be stressed, that there are few events appear below the γ/β/muon band, at the energies both higher and lower than ~ 3 MeV (see Fig. 2.22 and Fig. 2.23). It is complicated to fully understand the origin of these events due to the low statistics. The possible explanations include the energy-degraded α particles coming from surface contamination and the pile-up effects. In fact, the piled-up events with a fake heat-to-light ratio (i.e. higher than that expected for a β-like interaction) can be produced due to the different time structures of heat and light pulses (which have shorter rise and decay times, as reported in Table 2.8). This effect is negligible underground, where the random coincidences caused by external gamma background are extremely improbable. The problem of the randomly coinciding events will be discussed in detail in Chapter 4.

The two $Zn^{100}MoO_4$ scintillating bolometers have a good reproducibility of their properties. The minor differences in the operational parameters of the two detectors are well within the typical spread observed in this type of devices. Both $Zn^{100}MoO_4$ detectors, despite some difference in the optical quality, have shown practically identical and very promising bolometric and scintillation characteristics.

Finally, the $Zn^{100}MoO_4$ scintillating bolometers were moved deep underground into the EDELWEISS set-up at the LSM. However, operation of the bolometers was not possible due to detector instability and significant noise problem. Therefore, the optimization of $Zn^{100}MoO_4$ detectors in the EDELWEISS set-up is required and foreseen in near future.

### 2.4. Radioactive contamination of zinc molybdate crystals

The radioactive contamination of the natural $ZnMoO_4$ crystals was estimated by analysis of α events selected from the underground measurements, while the aboveground data were used for the enriched $Zn^{100}MoO_4$ samples. In order to stabilize the thermal response of the detectors we used the position of α peak (5.4 MeV) of the internal $^{210}$Po, which was detected in the data for the natural crystals. The α spectra accumulated with the scintillating bolometers based on 313 g and 334 g $ZnMoO_4$ crystals over 851 h and 527 h, respectively, are shown in Fig. 2.24.

The contamination of the crystals by $^{210}$Po was detected through 5.4 MeV α peak, which confirms a broken equilibrium in the radioactive chain. The 313 g $ZnMoO_4$ sample was polluted by $^{226}$Ra (and its daughters $^{222}$Rn, $^{218}$Po, and $^{214}$Bi-$^{214}$Po events), and $^{228}$Th (with its daughter $^{224}$Ra), while the advanced quality $ZnMoO_4$ crystals (336 g and 334 g) produced by recrystallization demonstrated much better level of radiopurity, especially in $^{226}$Ra. Moreover, the 313 g precursor scintillator has a higher surface contamination by $^{210}$Po or/and of the bolometer construction elements which are close to the detector (a peak at 5.3 MeV corresponds to $E_\alpha$ of $^{210}$Po). In addition, unidentified peak around 5.8 MeV was also indicated for 313 g $ZnMoO_4$ crystal and its origin can be related to surface contamination by alpha active radionuclides of unidentified nature.

The activity of internal $^{210}$Po was estimated from the fit of the 5.4 MeV α peak. Three sigma interval (see Table 2.7) centered at the $Q_\alpha$ value was used to calculate the area of the peaks of other radionuclides from U/Th chains. The contribution to the background was estimated in two energy intervals (3.3–4 MeV and 4.35–4.7 MeV) with a flat α continuum in which no peaks are expected. The number of counts excluded with 90% C.L. was calculated by using the Feldman-Cousin procedure [182].

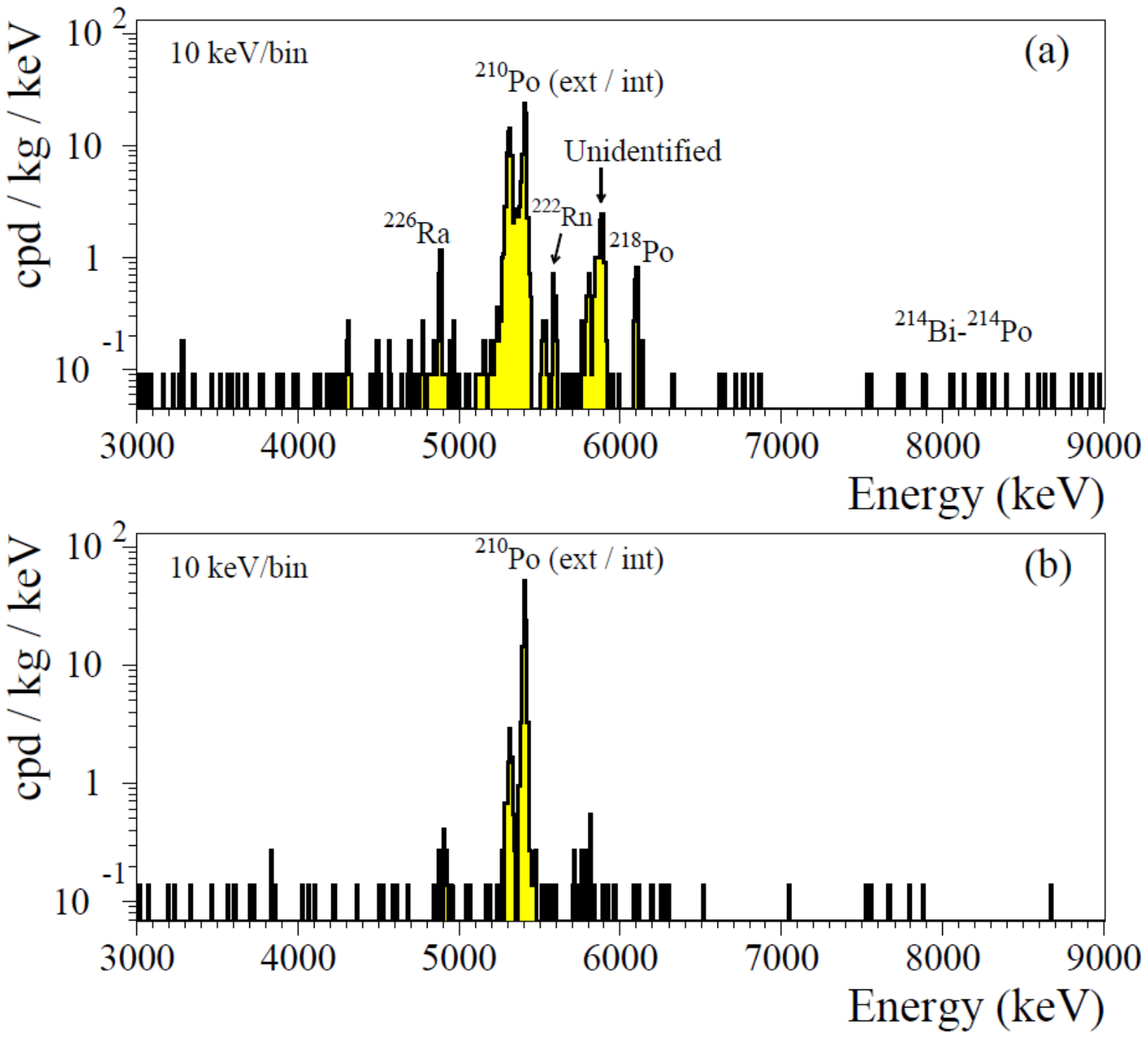

Fig. 2.24. Energy spectra of α particles registered in the underground measurements by using the $ZnMoO_4$ scintillating bolometers based on the (a) 313 g precursor and the (b) 334 g advanced sample operated in the EDELWEISS set-up over 851 h and 527 h, respectively. α peaks with largest area are specified.

Table 2.9
Radioactive contamination of zinc molybdate crystals produced from molybdenum of natural isotopic composition and enriched in $^{100}$Mo. Radioactive contamination of a 329 g $ZnMoO_4$ crystal operated as scintillating bolometer at the LNGS [173] is presented for comparison. The uncertainties are given with 68% C.L., while all the limits are at 90% C.L.

| Nuclide | Activity (mBq/kg) | | | | | |
|---|---|---|---|---|---|---|
| | $Zn^{100}MoO_4$ | | $ZnMoO_4$ | | | |
| | 59.2 g | 62.9 g | 336 g | 334 g | 313 g | 329 g [173] |
| | 42 h | 42 h | 291 h | 527 h | 851 h | 524 h |
| $^{228}$Th | ≤ 0.25 | ≤ 0.21 | ≤ 0.024 | ≤ 0.007 | 0.010(3) | ≤ 0.006 |
| $^{238}$U | ≤ 0.26 | ≤ 0.21 | ≤ 0.008 | ≤ 0.002 | ≤ 0.008 | ≤ 0.006 |
| $^{226}$Ra | ≤ 0.26 | ≤ 0.31 | ≤ 0.021 | ≤ 0.009 | 0.26(5) | 0.27(6) |
| $^{210}$Po | 0.9(3) | 1.1(3) | 0.94(5) | 1.02(7) | 0.62(3) | 0.70(3) |

The obtained results on radioactive contamination of the $ZnMoO_4$ and $Zn^{100}MoO_4$ crystals are presented in Table 2.9. Radioactive contamination of an another $ZnMoO_4$ crystal sample of 329 g, produced from the same boule as the 313 g precursor crystal, is given for comparison. The purification and crystallization techniques used to produce the advanced $ZnMoO_4$ crystals (336 g and 334 g) led to a significant improvement of material radiopurity. The purification and crystallization techniques were particularly efficient to reduce radium contamination: $^{226}$Ra was not detected in the advanced quality samples, while its activity was on the level of 0.3 mBq/kg in both precursor crystals. Finally, the achieved radiopurity level (≤ 0.01 mBq/kg) for $^{228}$Th and $^{226}$Ra gives excellent prospects for the next-generation 0ν2β decay experiments which will explore the inverted hierarchy region of the neutrino mass pattern.

### 2.5. Conclusions and perspectives

Cryogenic scintillating bolometers based on zinc molybdate crystals are one of the most promising detectors to search for neutrinoless double beta decay of $^{100}$Mo thanks to absence of long-lived radioactive isotopes in the crystal compound and comparatively high percentage of molybdenum (43% in weight).

The two-stage molybdenum purification technique: double sublimation with addition of zinc molybdate and recrystallization from aqueous solution of ammonium para-molybdate, was developed in the Nikolaev Institute of Inorganic Chemistry. This purification procedure substantially improves the purity level of molybdenum allowing production of high quality $ZnMoO_4$ crystal scintillators.

$ZnMoO_4$ crystal scintillators with mass up to 1.5 kg were grown from molybdenum with natural isotopic composition by the low-thermal-gradient Czochralski technique. The optical and luminescence properties of the crystals have confirmed an improved quality of the produced crystals.

Enriched in isotope $^{100}$Mo to 99.5% crystal boule ($Zn^{100}MoO_4$) with mass 171 g was developed for the first time. The yield of the crystal boules is in the range of 80%–84%. Such high yield demonstrated the advantage of the low-thermal-gradient Czochralski crystal growing technique. The irrecoverable losses of enriched molybdenum were estimated to be on the level of 4%.

The low-temperature tests of the zinc molybdate crystals (produced from molybdenum with natural isotopic composition and enriched in $^{100}$Mo) demonstrated high performance of the detectors. The radioactive contamination of the advanced $ZnMoO_4$ crystals by $^{228}$Th and $^{226}$Ra is on the level of ≤ 0.01 mBq/kg thanks to the developed methods of molybdenum deep purification and improved crystallization technique.

The obtained results are fully compatible with the next generation neutrinoless double beta decay experiments aiming to approach the inverted hierarchy region of the neutrino mass pattern by using the scintillating bolometers based on the $Zn^{100}MoO_4$ crystals.

# CHAPTER 3

# MONTE CARLO SIMULATION OF ZINC MOLYBDATE CRYOGENIC SCINTILLATING BOLOMETERS

## 3.1. Simulation of light collection from $ZnMoO_4$ crystal scintillators

The scintillation photon collection in $ZnMoO_4$ cryogenic scintillating bolometers was simulated by a Monte Carlo method with the help of the GEANT4 simulation package. The technique of Monte Carlo simulation has been tested with $ZnWO_4$ and $CaWO_4$ crystal scintillators of different shapes, in various conditions of optical contact, reflector material, configuration and surface treatment [183, 184]. In our simulation we used a simplified model of the bolometer which consists of the $ZnMoO_4$ crystal, two germanium light detectors, and cylindrical specular reflector. Schematic view of the detector module is presented in Fig. 3.1.

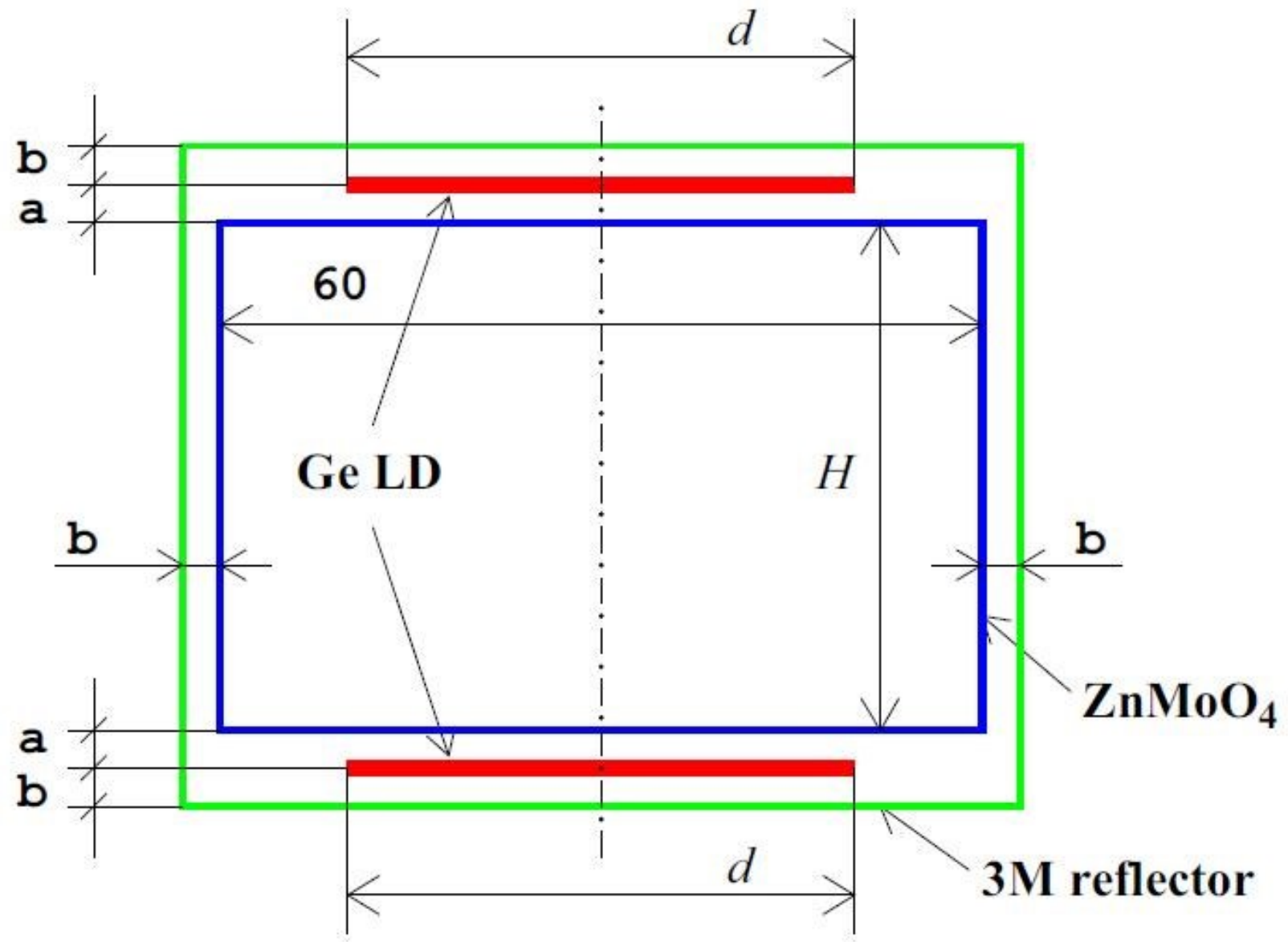

Fig. 3.1. Schematic view of detector module with $ZnMoO_4$ crystal scintillator 60 mm in diameter (or diagonal of hexagonal, octahedron) and height *H* (20 or 40 mm). “Ge LD” denotes germanium light detectors with diameter *d* (20, 40 or 60 mm) fixed at the distance *a* from the $ZnMoO_4$ crystal. Light reflector surrounds the module at the distance *b* from the $ZnMoO_4$ crystal surface and the light detectors, $a = b = 3$ mm.

Light collection efficiency from $ZnMoO_4$ crystals of three different shapes (cylindrical, hexagonal and octagonal), with various sizes (60 mm in diameter or in diagonal, by 20 and 40 mm height) and surface treatment (polished and diffused) was simulated. The $ZnMoO_4$ crystal was viewed from the top and bottom by the two germanium photo-detectors with thickness 0.1 mm and diameters 20, 40 and 60 mm. The reflection coefficient of the specular reflector (3M) was set to 98%. The refraction index and emission spectrum measured in Refs. [164, 171] were taken into account. In our simulation we assumed 300 mm of attenuation length in $ZnMoO_4$ crystals.

Ten thousand vertices uniformly distributed in $ZnMoO_4$ crystal with $10^4$ scintillation photons isotropically distributed in each vertex were generated for each detector geometry.

An example of the GEANT4 simulation output is shown in Fig. 3.2. Results of the simulation are presented in Table 3.1.

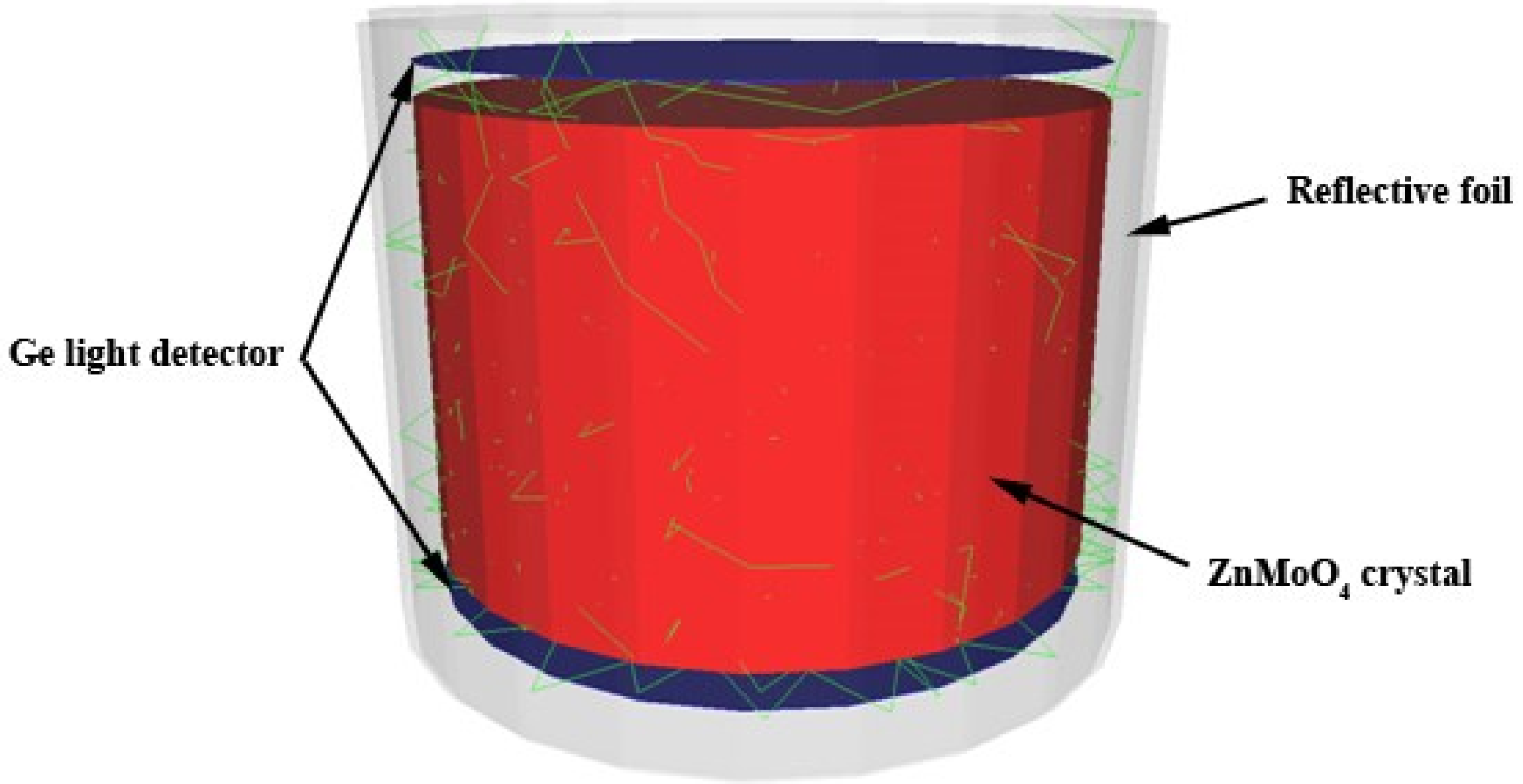

Fig. 3.2. Example of the GEANT4 output of the detector geometry with trajectories of scintillation photons marked by green color.

Table 3.1
Light collection from $ZnMoO_4$ crystal scintillator simulated for different shapes and sizes of the crystal, surface treatments and diameters of photodetectors with the help of the GEANT4 package.

| Shape of $ZnMoO_4$ crystal | Size of crystal | Surface condition | Part of photons (%) reaching photodetectors with diameter (mm) | | |
|---|---|---|---|---|---|
| | | | 20 | 40 | 60 |
| Cylinder | Ø60 × 40 | Polished | 2.6 | 6.7 | 11.8 |
| | | Diffused | 3.6 | 11.0 | 20.5 |
| | Ø60 × 20 | Polished | 4.3 | 10.6 | 16.8 |
| | | Diffused | 7.7 | 21.9 | 36.5 |
| Octahedron | 60 × 40 | Polished | 3.5 | 8.4 | 14.3 |
| | | Diffused | 3.9 | 11.9 | 22.6 |
| | 60 × 20 | Polished | 6.0 | 14.0 | 21.4 |
| | | Diffused | 8.3 | 23.5 | 38.8 |
| Hexagonal | 60 × 40 | Polished | 3.8 | 9.1 | 15.3 |
| | | Diffused | 4.2 | 12.7 | 24.0 |
| | 60 × 20 | Polished | 6.6 | 14.7 | 22.1 |
| | | Diffused | 8.8 | 24.7 | 40.3 |

The best light collection was achieved for the crystal scintillators of hexagonal shape with diffused surface. Polished $ZnMoO_4$ crystals of cylindrical shape show the lowest light collection. Octahedron scintillation elements demonstrate intermediate light collection efficiency and could be a compromise between the strong requirements of minimal losses of the expensive enriched material and a reasonable light collection to be reached.

It was shown that the light collection increased for the $ZnMoO_4$ scintillating crystals with a smaller height thanks to the lower losses of scintillating photons through the absorption in the crystal material. A higher light collection achieved with photodetectors of larger diameter. Diffused surface of the scintillators provided better uniformity of light collection in comparison to polished surface. It should be noted that Lambertian reflection was used to simulate diffused crystal scintillator surfaces. However, real surfaces of inorganic crystals exhibit a combination of Lambertian and specular reflection. Therefore, the results of our calculations can be considered as an upper limit on the light collection efficiency.

We have studied also dependence of light collection on the distance *b* between the specular reflector and side surface of the $ZnMoO_4$ crystal (the same distance was chosen between the specular reflector and germanium photodetector, see Fig. 3.1). Simulations were performed for hexagonal crystal with dimensions 60 × 40 mm and diffused surface. The diameter of the light detectors was chosen to be 60 mm. Other settings were the same as for the simulation before.

As we can see from the Fig. 3.3 the light collection improves with a bigger distance between the light reflector and side surface of the $ZnMoO_4$ crystal and germanium photodetectors with a highest light collection obtained for the distance 11–15 mm.

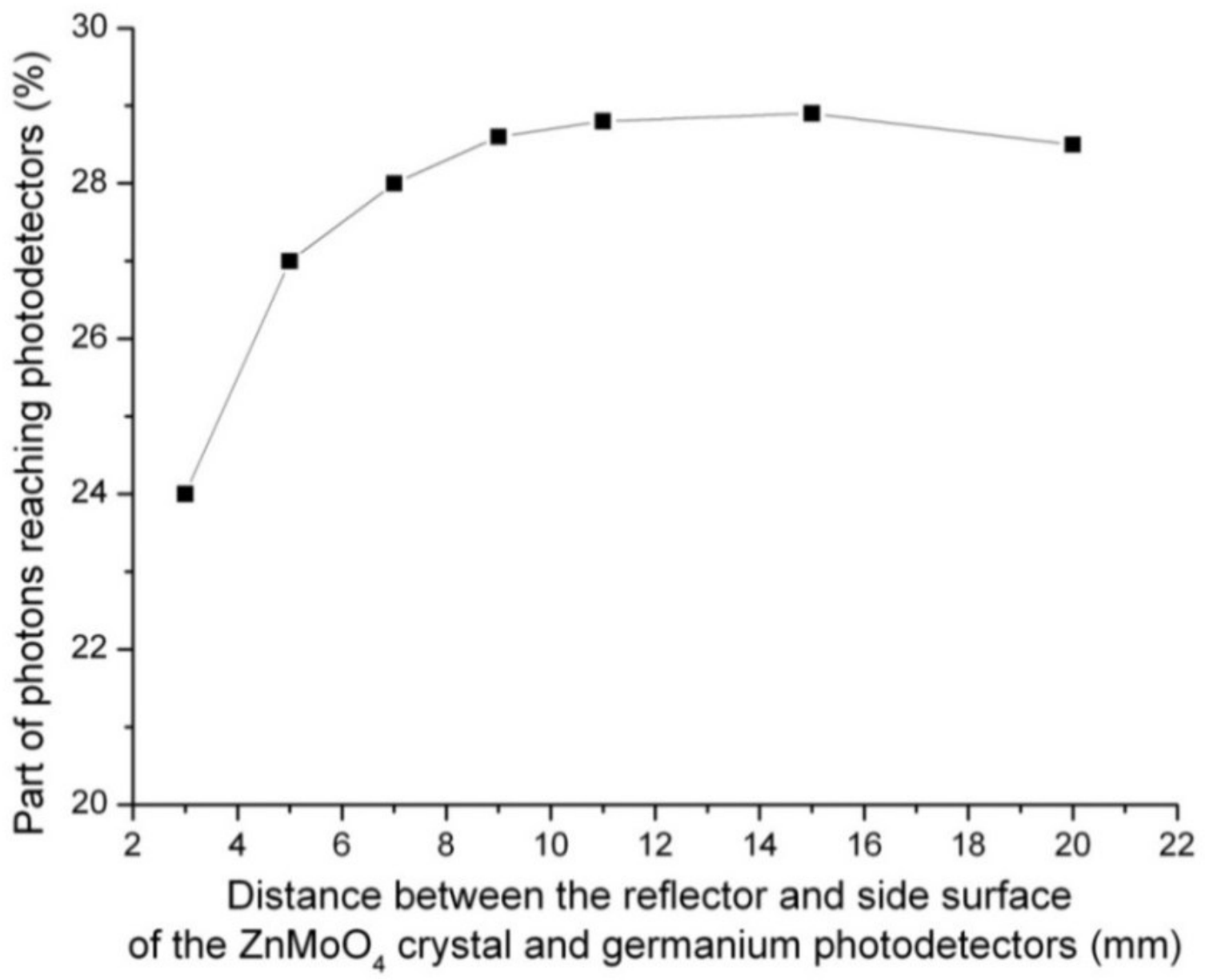

Fig. 3.3. Dependence of the light collection efficiency on the distance *b* (see Fig. 3.1) between the specular reflector and side surface of the $ZnMoO_4$ crystal (for hexagonal crystal 60 × 40 mm with diffused surface) and germanium photodetectors (with 60 mm diameter) simulated with the help of GEANT4 package.

Finally we analyzed the dependence of the light collection on the attenuation length of $ZnMoO_4$ crystal. We selected three values of the attenuation length of 200, 300 and 400 mm. Simulation was performed for the hexagonal crystal 60 × 40 mm with diffused surface with photodetector of 60 mm in diameter. Light reflector surrounded the detector module at the distance $b$ = 3 mm. The results of the simulation are presented in Fig. 3.4. The light collection efficiency is higher for the crystals with larger attenuation length.

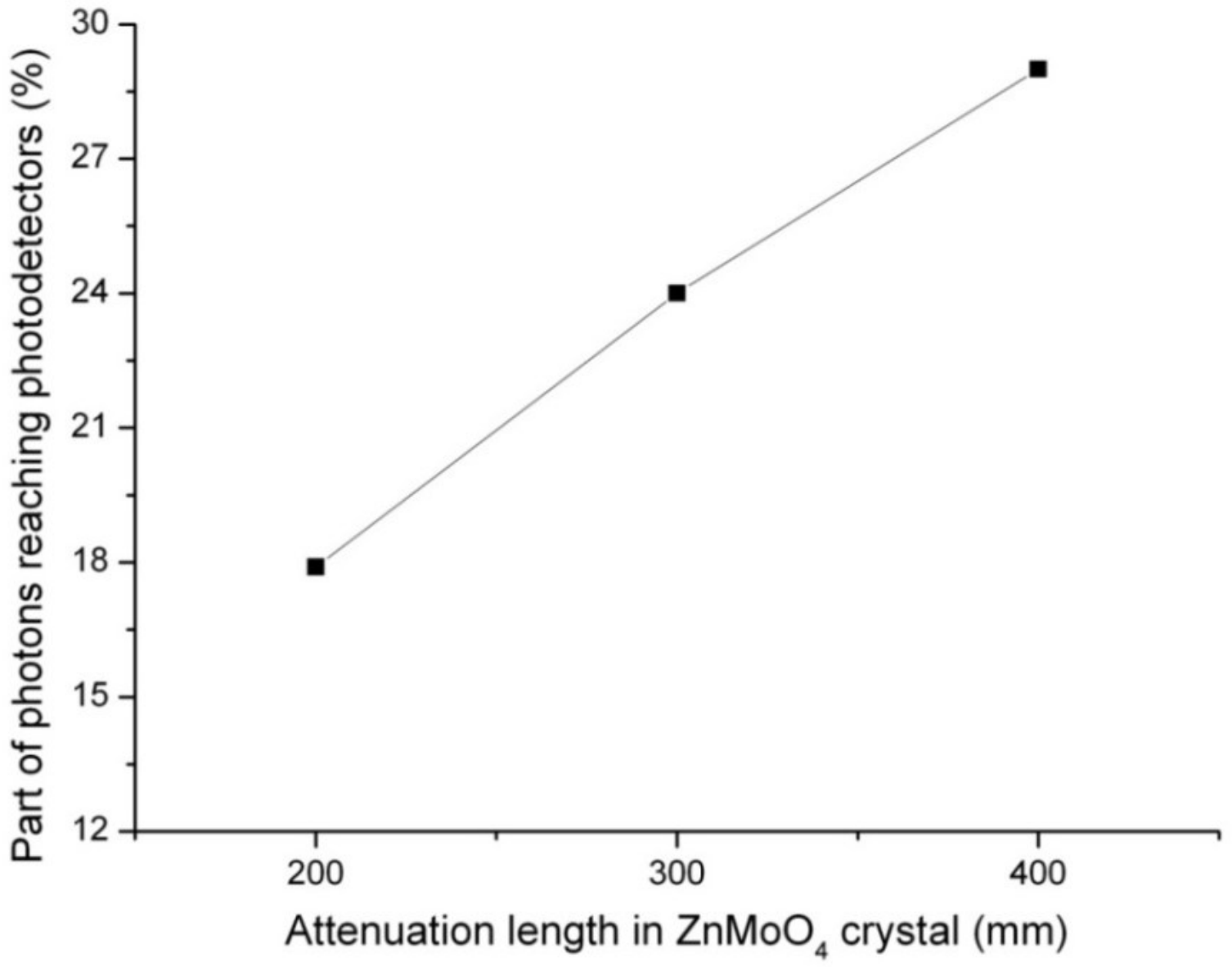

Fig. 3.4. Dependence of the light collection efficiency on the attenuation length in $ZnMoO_4$ crystal (hexagonal 60 × 40 mm, diffused surface) simulated with the help of GEANT4 package.

### 3.2. Simulation of 2ν2β decay in $Zn^{100}MoO_4$ crystals

As it was shown in Chapter 2 the produced enriched zinc molybdate scintillating crystals are of irregular shape to increase mass of enriched isotope. Dependence of the 2ν2β decay detection efficiency on the crystal shape was investigated with the aim to estimate applicability of irregular shape crystals to measure the half-life of $^{100}$Mo relatively to 2ν2β decay.

The best way would be to reproduce the crystal shape as precise as possible. However the shape of the developed $Zn^{100}MoO_4$ scintillators (see Fig. 2.19) is too complicated to be reconstructed by using the standard GEANT4 geometry tools [185]. One could try to build a 3D model and integrate the obtained geometry into the GEANT4 code. We have tried to apply 3D scanning of the crystals to build their geometrical model.

#### 3.2.1. 3D scanning of $Zn^{100}MoO_4$ crystal scintillators

There are several techniques to measure the three-dimensional shape of an object which can be generally divided in a two categories: active and passive methods [186].

In the active techniques the light source is projected on the studied object while a camera capture the footage. The main idea is to establish the correspondence between the points of the object and their mappings on the obtained image. The color of pixels of the captured images represents the distance between the points of the object and the camera plane, from which the partial 3D model can be built (see Fig. 3.5). A complete 3D model can be obtained by taking such depth maps from different angles of the object and combining them together.

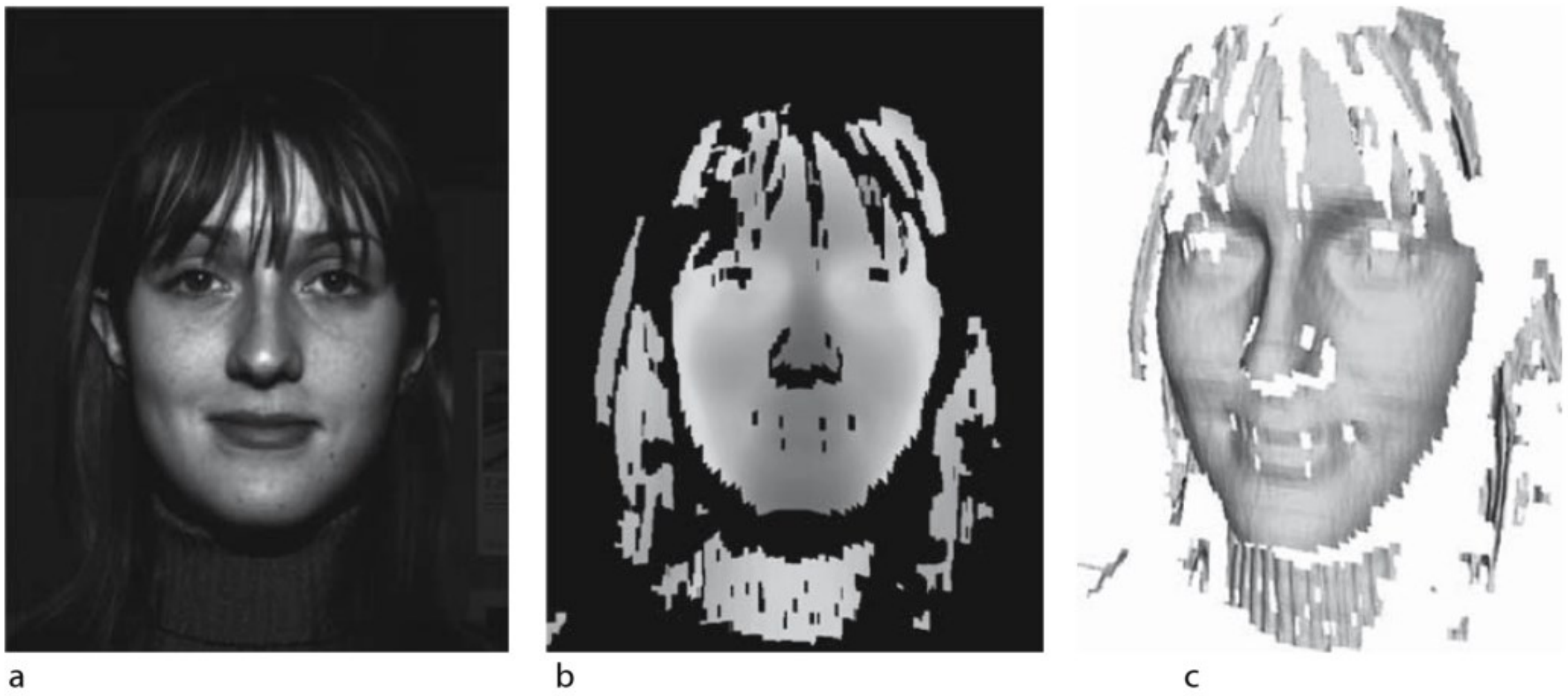

Fig. 3.5. Example of the 3D model acquired with the active technique [186]. The structured light 3D scanner was used to build a model. (a) Photo of the object. (b) Depth map of the object. Darker pixels denote points of the object nearer to the camera. Information lost due to the imperfection of the method is marked with a black color. (c) 3D model of the facial surface obtained from the depth image.

Passive techniques use two or more cameras at different positions to estimate the depth of the object. However, the general principle is again to find the pixels in each image that correspond to the same point of the object. In this case the epipolar geometry constraints are used to reduce the matching candidates [187]. However in our case this technique cannot be applied because correspondence between the points of the object and their mappings on the obtained image is impossible to establish for the textureless objects like scintillation crystals.

To obtain 3D models of the $Zn^{100}MoO_4$ crystal scintillators we used a structured light 3D scanning which is one of the active techniques. The structured light method uses a light projector to project a number of stripe patterns on the studied object. At the same time the camera located at other viewpoint (above, below, left or right of the projector) takes one or several images of each pattern. Special software computes a 3D mesh of the object surface from the images of geometrically distorted stripe patterns due to the surface shape of the object.

We have built a self-made structured light 3D scanner taking into account that professional equipment for 3D scanning costs from at least a few thousand euros [188, 189]. The set-up consists of the Optoma EP719R DLP projector, webcam Logitech HD Webcam C270 and DAVID 3D scanner software [190]. The scanner was placed in the wooden box to have a fixed rectangular angle for the calibration pattern and to prevent undesirable set-up elements movements (see Fig. 3.6). The scanned scintillation crystal was located on the rotating platform near the corner of the box which was covered with a calibration panel. To perform quality scans we put the projector at the minimum possible distance to the crystal. The distance is strongly limited by the projector focal length. The camera was placed on the top of the projector and viewed both the scintillating crystal and calibration panel.

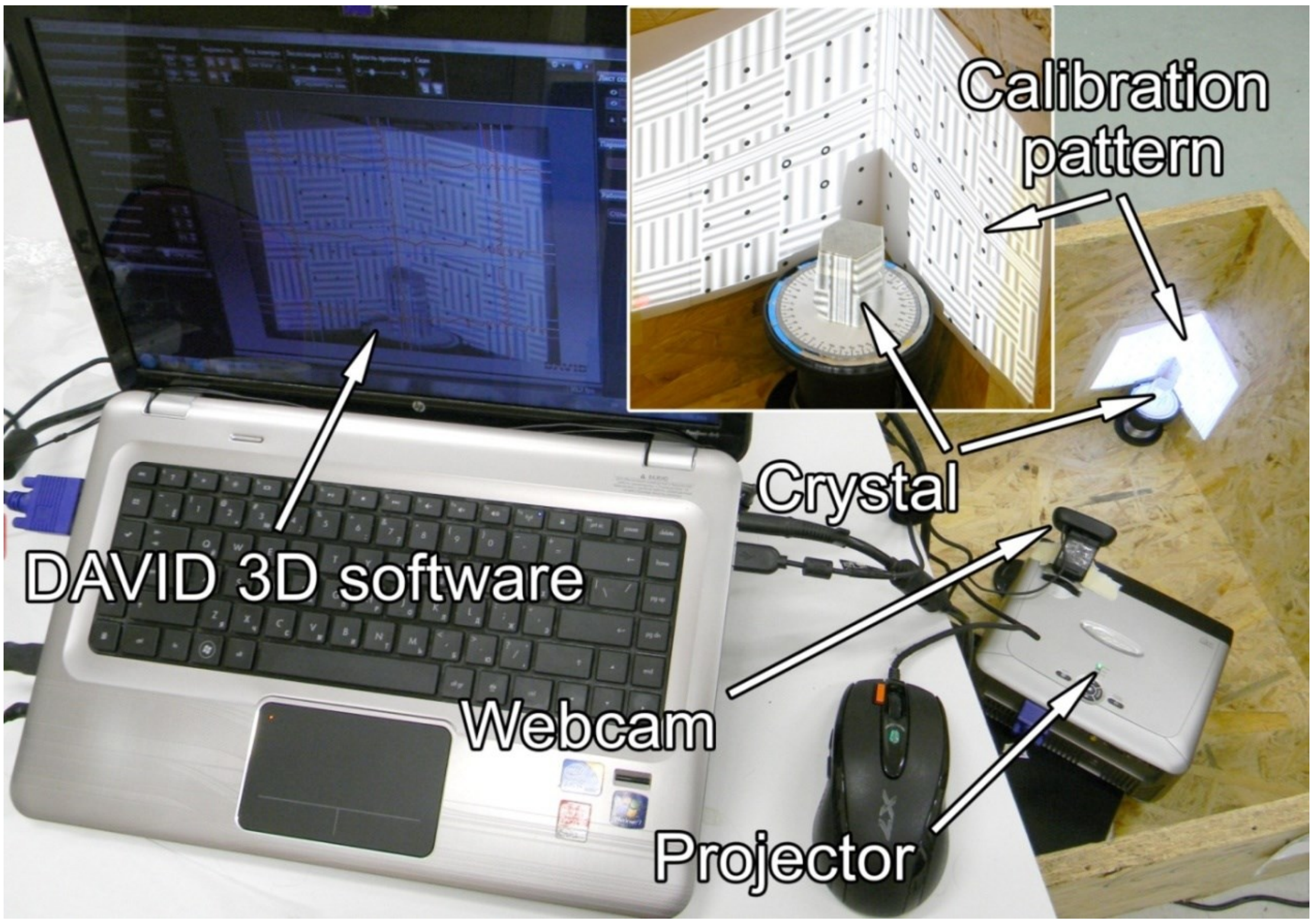

Fig. 3.6. The structured light 3D scanner used to measure the three-dimensional shape of the $Zn^{100}MoO_4$ scintillating crystals. The set-up consists of the projector, webcam and PC with the DAVID 3D scanner software.

The first step was calibration of the 3D scanner without a scanned object. We used the calibration pattern to establish the dependence between the position and orientation of the projector and camera, and to determine the equipment performance.

Afterwards, we put the scintillating crystal onto the rotating platform and performed a 3D scan. However, the reflective and transparent surface of the crystals substantially limits the structured light 3D scanning ability. The light from the projector is reflected away from the camera or right into its optics by the crystal surface. In both cases, the dynamic range of the camera can be exceeded, and depth image will be distorted. There are a few methods to solve this problem [191, 192], while the most common one is coating of the surface. We covered the crystal surface with thin paper masking tape and repeated the scanning procedure.

The problem of reflective and transparent crystal surface was solved in this way; however the scan result was pulsing (see Fig. 3.7). In most cases to avoid waves in the 3D scan the mechanic lens aperture size is used to adjust the image brightness. Since the webcam does not allow to do that, we decreased the projector brightness. Unfortunately this tuning was limited by technical characteristics of the projector and camera, and was not enough to remove the scan waves completely by using the current equipment. So we just proceeded to make a three-dimensional scans of the crystal surface from different angles, by rotating the scintillator for every ten degrees.

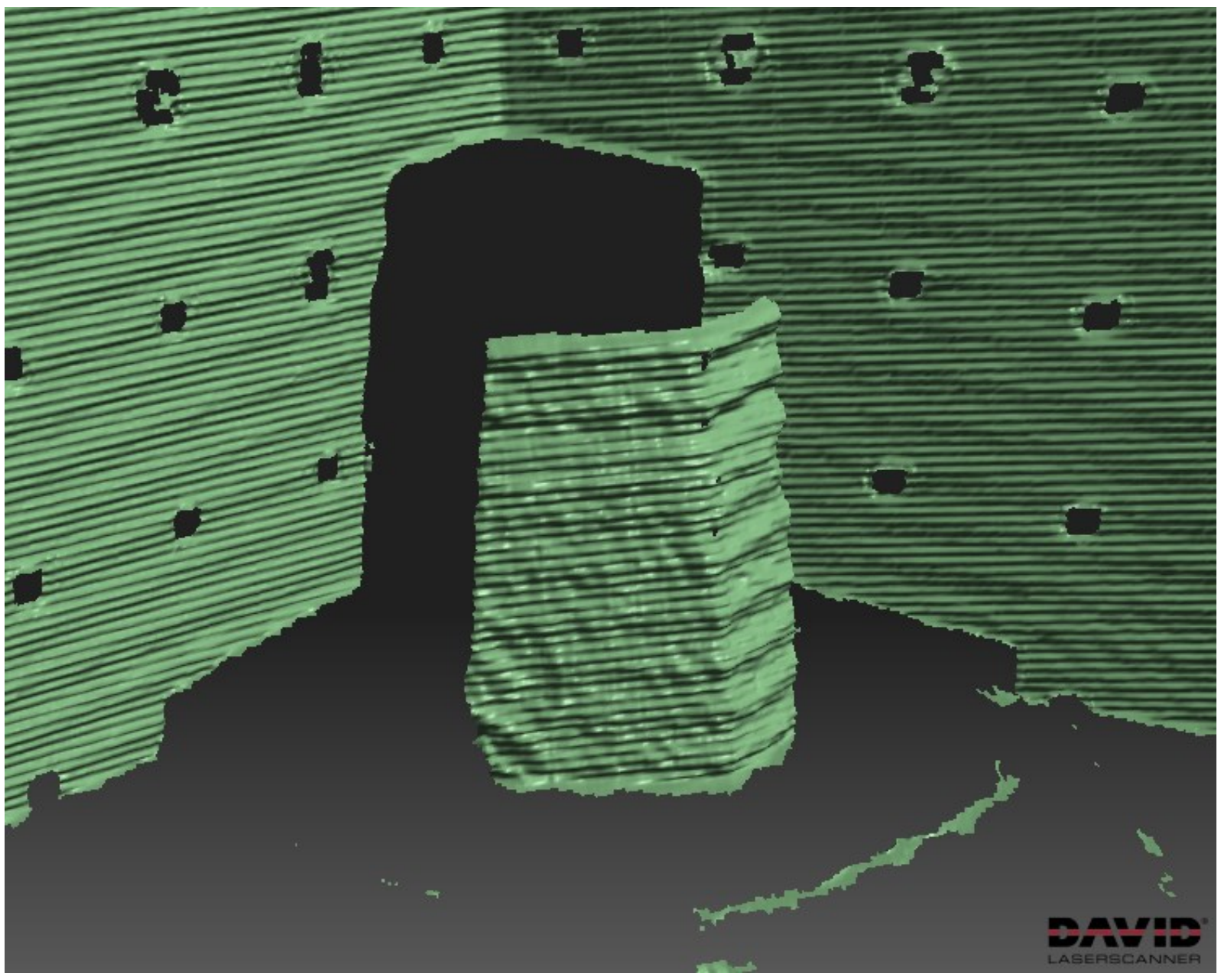

Fig. 3.7. Result of structured light 3D scanning of the crystal surface obtained with a self-made 3D scanner. Waves on the scan result are due to the limited technical characteristics of the camera and projector used in the set-up.

As a result we obtained 36 three-dimensional models of the crystal side surface. The models were cleaned from the scanning artifacts and aligned together one-by-one to reproduce the crystal rotation during the scanning procedure (see Fig. 3.8 (a,b)). Afterwards we merged the aligned 3D models into one complete model. The result of the merging is presented in Fig. 3.8(c).

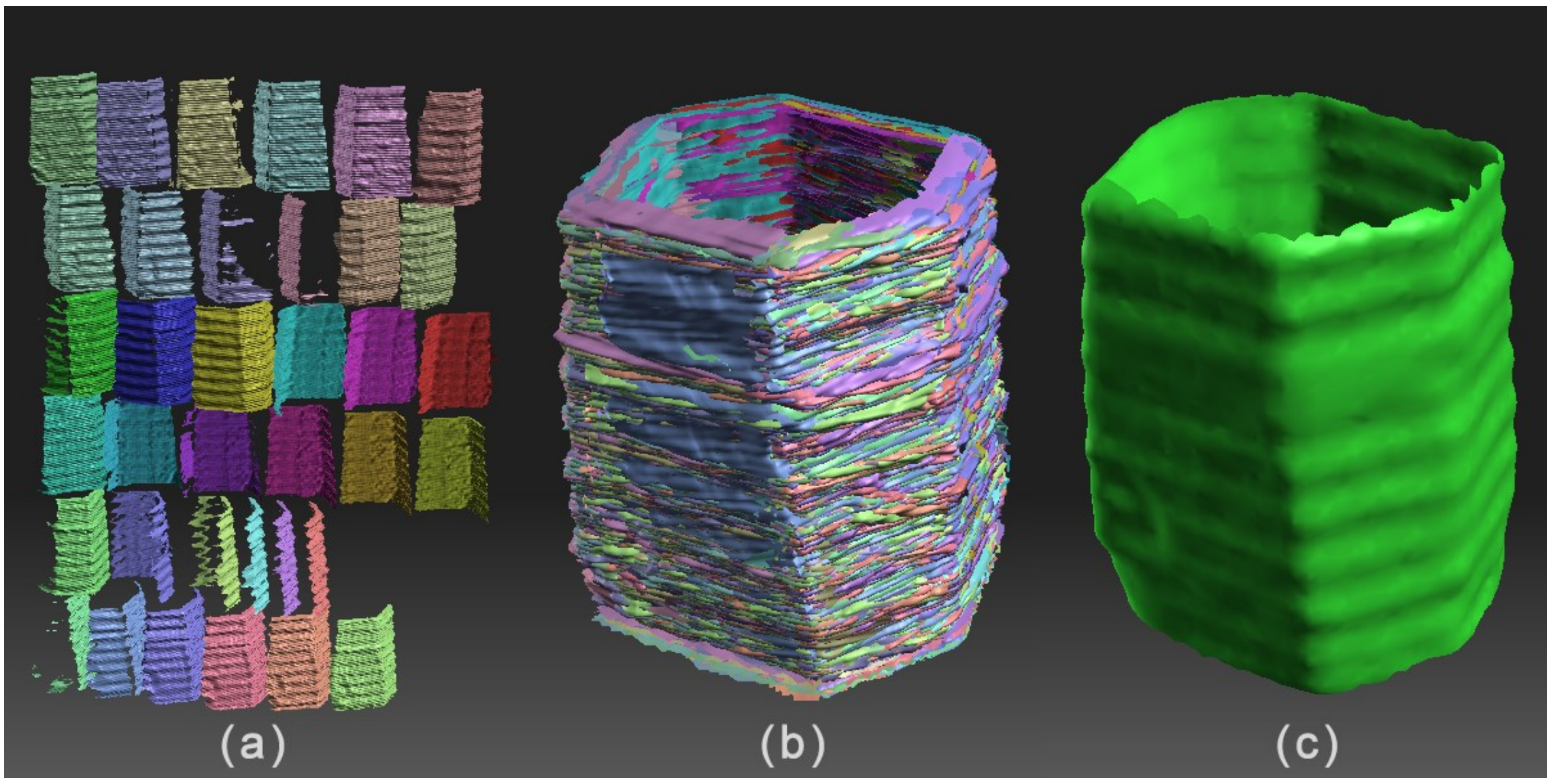

Fig. 3.8. Processing of the obtained 3D scans of the $Zn^{100}MoO_4$ crystal into the complete 3D model: (a) the 36 scans of the crystal side surface cleaned from the scanning artifacts; (b) the aligned 3D models reproducing the crystal side surface; (c) the complete 3D model of the crystal after the shape merging.

Since the 3D model was made only for the side surface of the crystal, the upper and lower planes of the crystal were built manually in the 3D editor software Maya 2008 [193].

Then the obtained 3D model was simplified by reducing the amount of triangles to a reasonable value. The final 3D model of the crystal and photo of the $Zn^{100}MoO_4$ scintillating crystal are presented in Fig. 3.9. The final result of 3D scanning and $Zn^{100}MoO_4$ scintillator has visible shape differences, thus the dependence of double beta decay spectra from the different crystal shapes will be additionally studied in the next subsection.

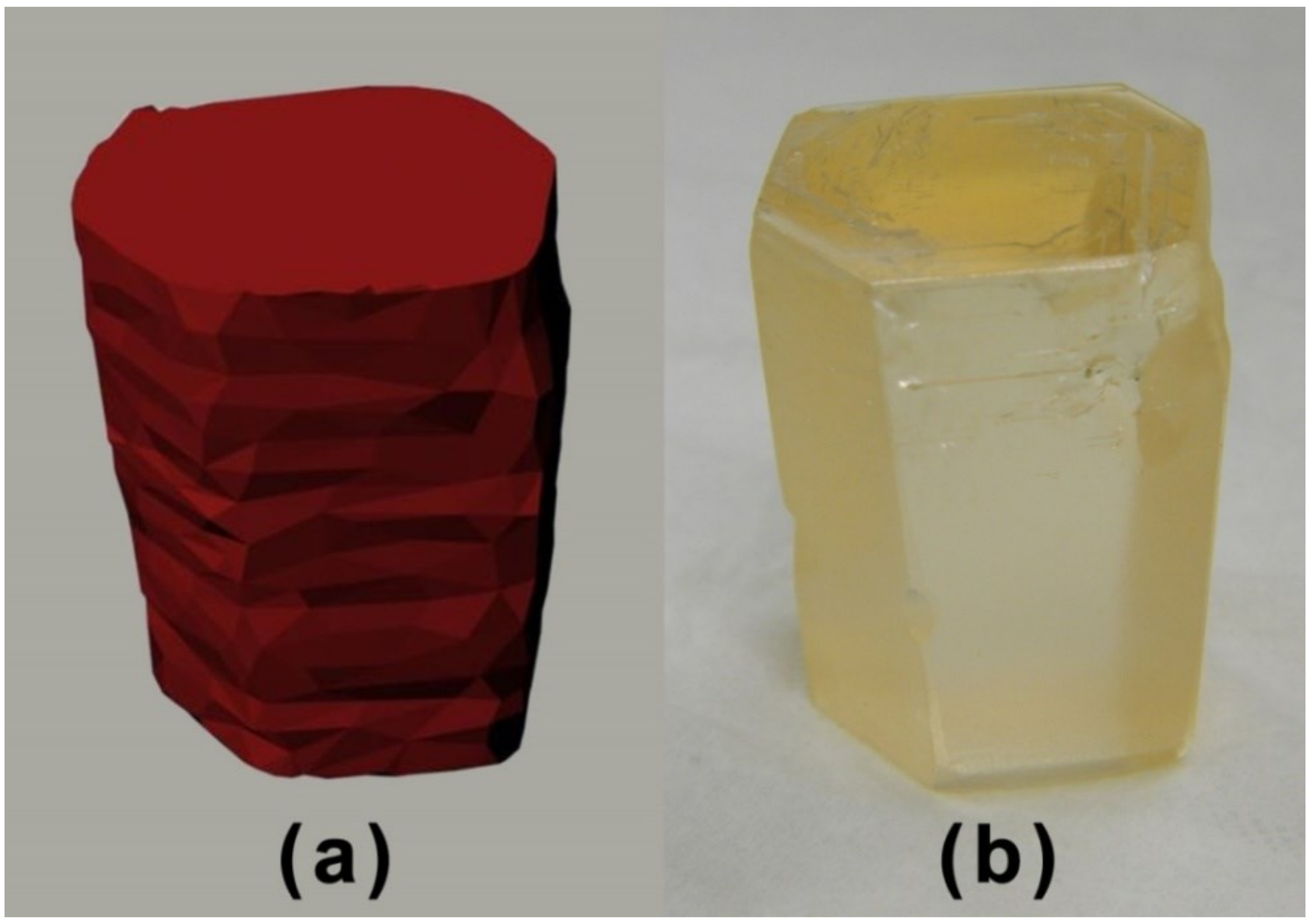

Fig. 3.9. Comparison of the model shape and of 59.2 g $Zn^{100}MoO_4$ crystal scintillator: (a) the final model obtained using the structured light 3D scanner and 3D editor; (b) photo of the $Zn^{100}MoO_4$ crystal scintillator.

### 3.2.1. Simulation of 2ν2β decay in crystal detectors of different shapes

Two-neutrino double beta decay processes were simulated for the $Zn^{100}MoO_4$ scintillator enriched in the isotope $^{100}Mo$ to 100% with three crystal shapes: cuboid, cylinder and hexagonal. The mass and height of the scintillators were fixed with two sets of values for all of the crystal shapes: 59.2 g, 32.2 mm; and 495 g, 40 mm. The first set of values was chosen to simulate the 2ν2β processes in the $Zn^{100}MoO_4$ sample close to the enriched crystals discussed in the previous subsection, while the crystal with a mass of 495 g (Ø60 × 40 mm) expected to be used for the next-generation experiment to search for 0ν2β decay of $^{100}Mo$.

The dependence of 2ν2β decay spectral distribution from the shape of $Zn^{100}MoO_4$ crystal was studied by using the GEANT4 package. For each set of the detectors geometry we generated $10^6$ 2ν2β events uniformly distributed in $Zn^{100}MoO_4$ crystal. The 2ν2β decay of $^{100}Mo$ was simulated both to the ground state and to the first excited $0_1^+$ level of $^{100}Ru$ (see Fig. 1.1 for the decay scheme of $^{100}Mo$). Results of the simulation are presented in Fig. 3.10. The obtained energy spectra are almost the same for the different crystal shapes.

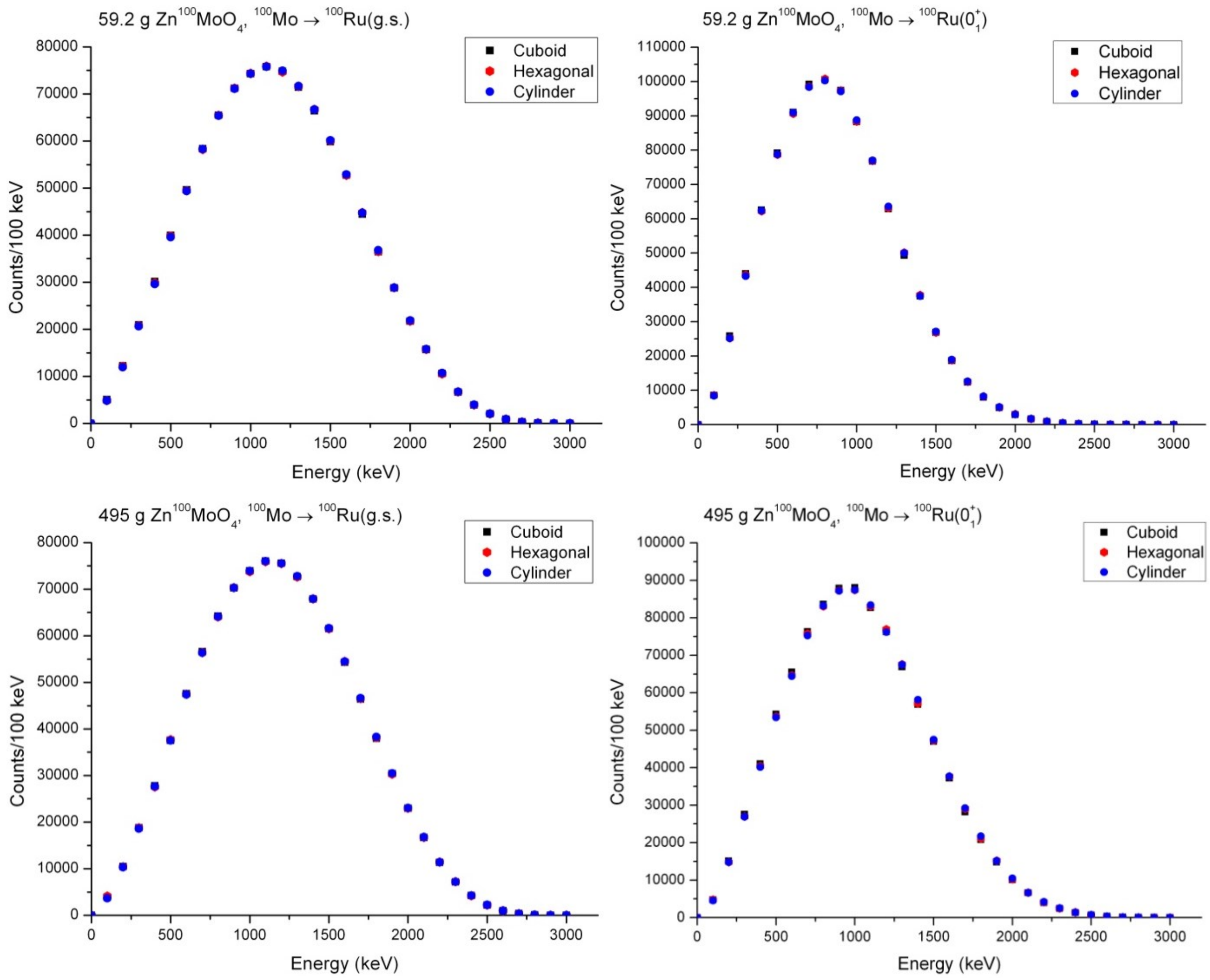

Fig. 3.10. The dependence of 2ν2β decay distribution from the shape of 59.2 g (top panels) and 495 g (bottom panels) Zn$^{100}$MoO$_4$ crystal simulated by using the GEANT4 package. The 2ν2β decay of $^{100}$Mo was simulated to the ground state (left panels) and to the first excited $0_1^+$ level of $^{100}$Ru (right panels).

To analyze the difference between the obtained 2ν2β decay energy distributions for the different crystal shapes we introduced a parameter logically similar to $\chi^2$ and defined as

$$d = \left|F_i - F_j\right| / \sqrt{F_i}, \tag{3.1}$$

where $F_i$ and $F_j$ are the number of counts in the same histogram bin for the cylindrical and cuboid/hexagonal shapes of Zn$^{100}$MoO$_4$ scintillators, respectively. Thus, the cylinder was chosen as a reference shape, and the difference between the simulated 2ν2β decay distributions was studied as cylindrical versus cuboid/hexagonal crystal shapes. The results of the analysis are presented in Fig. 3.11. As one can see, for 2ν2β decay of $^{100}$Mo to the ground state almost all of the points (except the first few bins where the statistics is very low) are in the $d \leq 1$ interval meaning almost no dependence of 2ν2β decay distributions from the crystal shape. Meanwhile, for the two-neutrino double beta decay of $^{100}$Mo to the first excited $0_1^+$ level of $^{100}$Ru the difference increased to (3–4)$d$ interval, which is comparatively higher, but is still rather low. Therefore, we can conclude that the simulation have demonstrated no significant dependence of 2ν2β decay processes in Zn$^{100}$MoO$_4$ scintillators on the crystal detector shape.

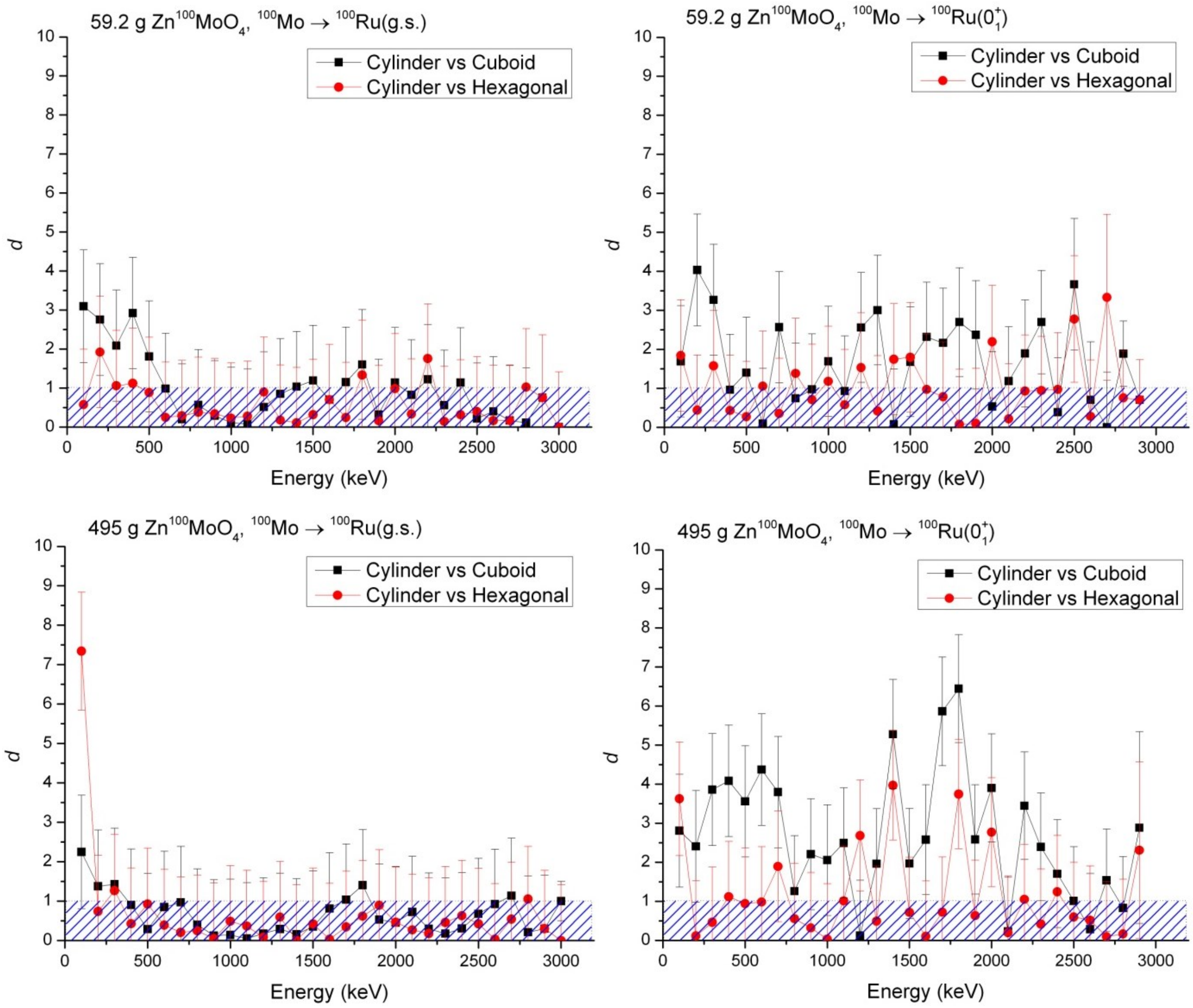

Fig. 3.11. The difference between 2ν2β decay distributions simulated for cylindrical, hexagonal and cuboid shapes of 59.2 g (top panels) and 495 g (bottom panels) Zn$^{100}$MoO$_4$ crystals expressed through the parameter *d* (see text). The calculations were performed for 2ν2β decay of $^{100}$Mo to the ground state (left panels) and to the first excited $0_1^+$ level of $^{100}$Ru (right panels).

### 3.3. Simulation of 48 Zn$^{100}$MoO$_4$ cryogenic scintillating bolometers in the EDELWEISS set-up

The sensitivity of double beta decay experiment depends on the background counting rate as the square root (see Eq. (1.17)), thus decrease of background improves the experimental sensitivity. In the ROI the dominant background source in bolometric experiments is the energy-degraded α continuum with the counting rate of about 0.1–0.05 counts/(keV·kg·yr) [194]. The scintillating bolometers (see subsection 1.3.5.1) allow to reduce this component of background to the level of $10^{-4}$ counts/(keV·kg·yr), thanks to the excellent α/γ(β) particle discrimination with a factor of 99.9% and higher. This corresponds to zero background for the ton × year exposure. However, to demonstrate the perspectives of the future next-generation experiment with Zn$^{100}$MoO$_4$ scintillating bolometers we have performed detailed investigation of background in the ROI.

We have simulated the detector array of 48 $Zn^{100}MoO_4$ scintillating bolometers installed in the EDELWEISS set-up. The brief description of the EDELWEISS set-up can be found in subsection 2.3.1, while the detailed one is in Ref. [177]. The GEANT4 Monte Carlo simulation of the different background sources was performed in order to estimate the counting rate in the region of interest. A typical energy resolution of $ZnMoO_4$ based cryogenic bolometers is about 6 keV at the energy of $^{100}$Mo 0ν2β decay ($Q_{2\beta}$ = 3034.40(17) keV). However, we will consider a more wide ROI interval 2934 keV–3134 keV to increase statistic of the results.

### 3.3.1. Geometry of the EDELWEISS set-up with $Zn^{100}MoO_4$ detectors

A single detector module consists of Ø60 × 40 mm $Zn^{100}MoO_4$ crystal enriched in $^{100}$Mo to 100%. The mass of detector is 495 g. The cylindrical shape of the crystal was chosen to use with a maximum efficiency the available volume of the EDELWEISS set-up. Each $Zn^{100}MoO_4$ crystal is placed on three PTFE pieces at the cavity of cylindrical cup-like copper holder, and fixed by another three PTFE details. The inner surface of the Cu holder is covered with a biaxially-oriented polyethylene terephthalate (BoPET) light reflective foil of 0.08 mm thickness. The top face of $Zn^{100}MoO_4$ crystal is viewed by a Ge light detector Ø44 × 0.15 mm placed inside a copper ring. The cavity of the cylindrical Cu holder is closed from above by a copper cap. The GEANT4 model of the $Zn^{100}MoO_4$ single detector is presented in Fig. 3.12.

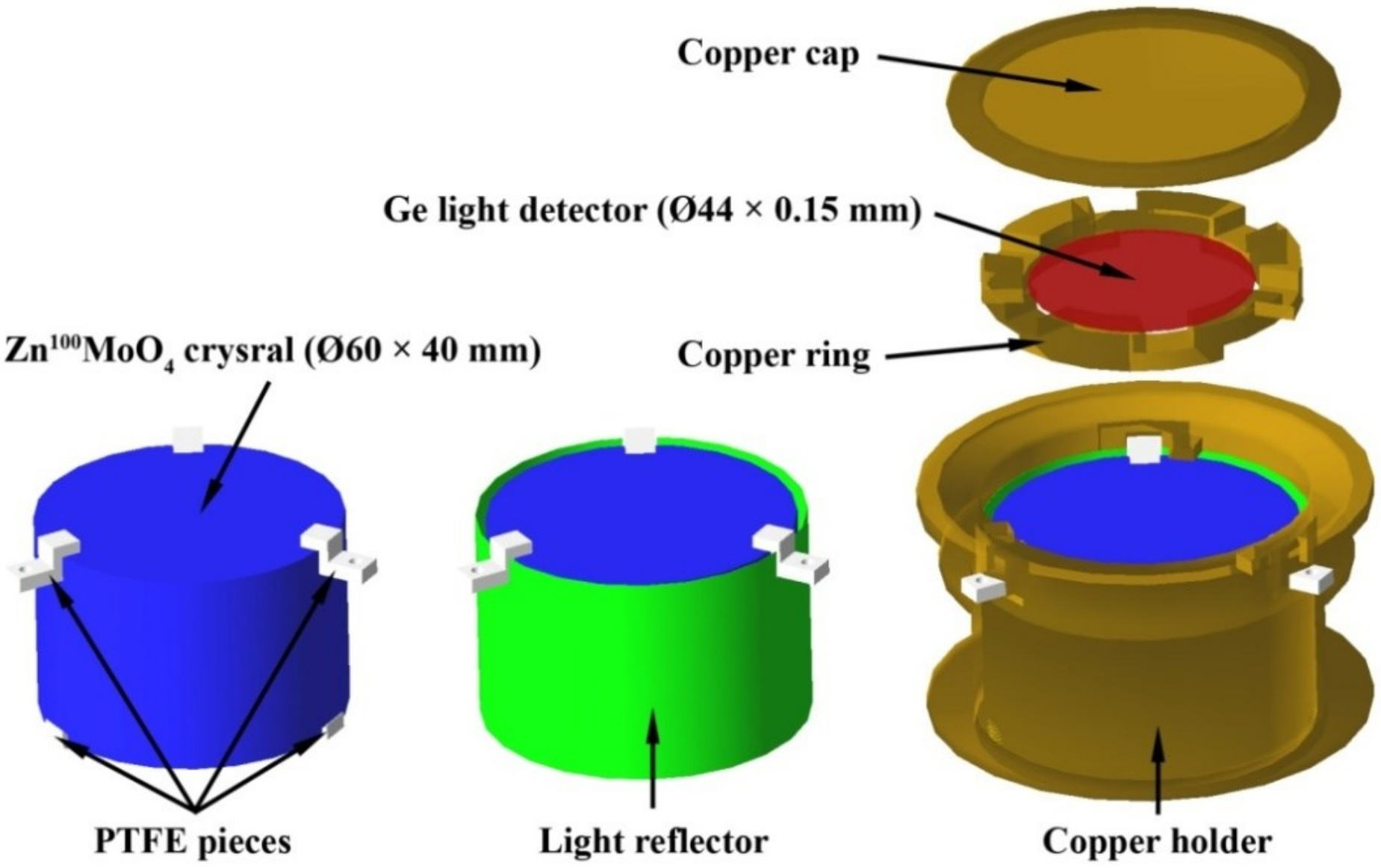

Fig. 3.12. The GEANT4 model of the scintillating bolometer based on $Zn^{100}MoO_4$ crystal Ø60 × 40 mm.

The single modules were arranged into 12 towers, each one consisting of 4 scintillating bolometers. In total the set-up may house an array of 48 $Zn^{100}MoO_4$ detectors with the total mass 23.8 kg (10.2 kg of the isotope $^{100}$Mo). The model of the $Zn^{100}MoO_4$ towers array is shown in Fig. 3.13. The GEANT4 code of the set-up geometry used in our simulations (except the $Zn^{100}MoO_4$ bolometers) was provided by the EDELWEISS collaboration. The GEANT4 model of the EDELWEISS cryostat filled with a detector array of 48 $Zn^{100}MoO_4$ scintillating bolometers is presented in Fig. 3.14.

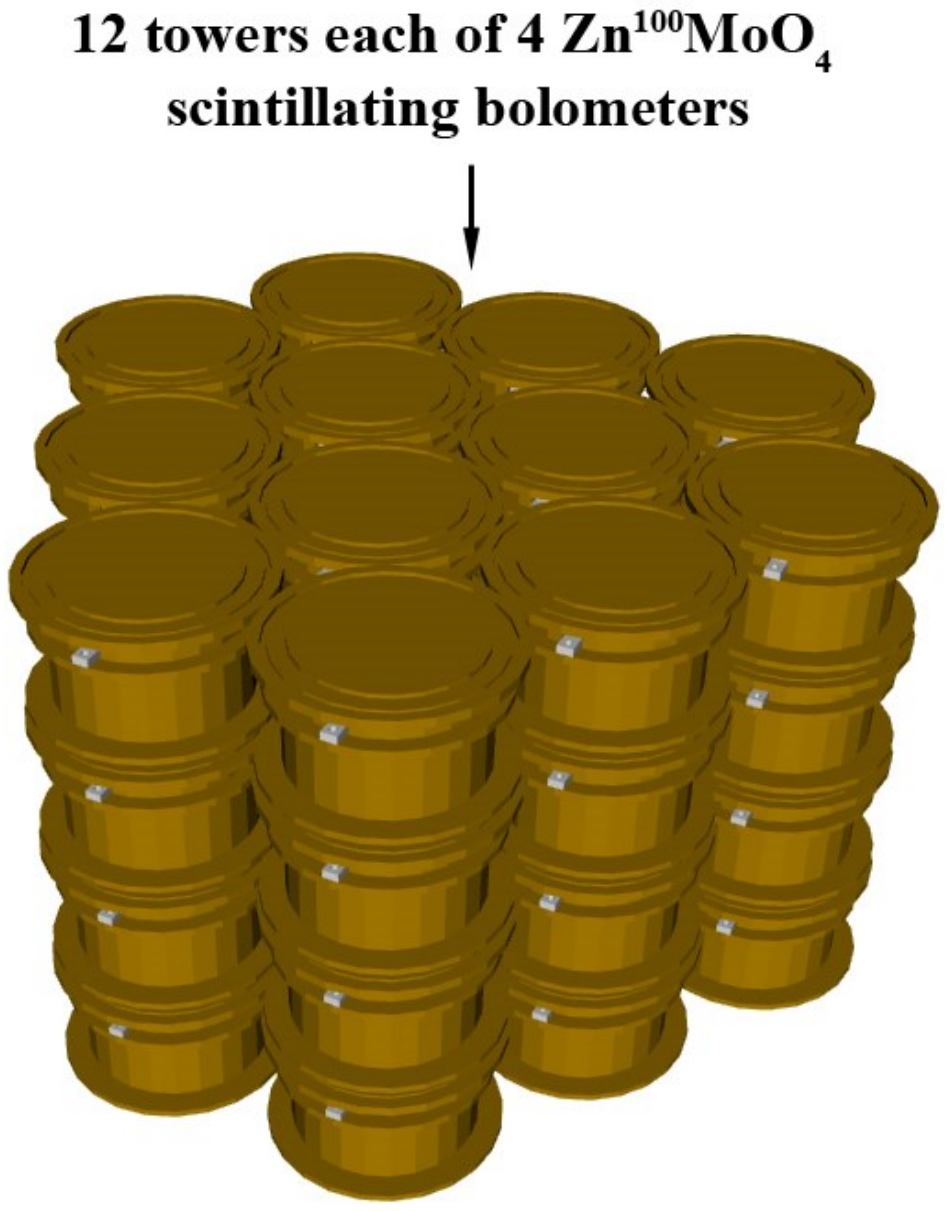

Fig. 3.13. The GEANT4 model of the detector array of 48 scintillating bolometers based on $Zn^{100}MoO_4$ crystals.

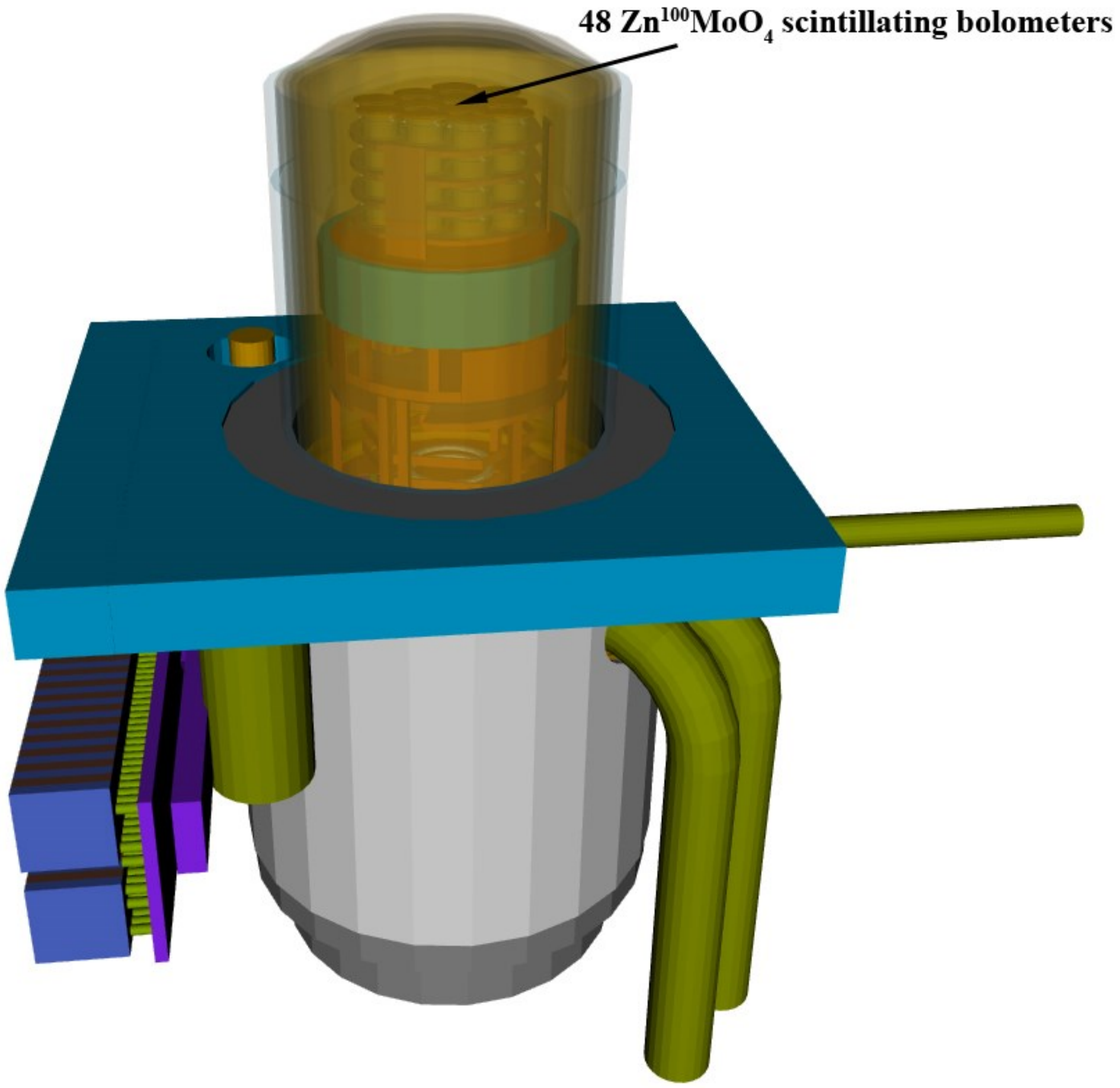

Fig. 3.14. The GEANT4 model of the detector array of 48 $Zn^{100}MoO_4$ scintillating bolometers assembled in the EDELWEISS cryostat.

### 3.3.2. Simulation of $ZnMoO_4$ scintillating bolometer calibration with $^{232}Th$ gamma source in EDELWEISS set-up

We have simulated response of the $ZnMoO_4$ scintillating bolometers installed in the EDELWEISS set-up to gamma quanta of $^{232}Th$ source aiming to compare simulated model with experimental spectrum. The experimental data were obtained with the $ZnMoO_4$ scintillating bolometer Ø50 × 40 mm (334 g). The construction of the scintillating bolometer and detector performances were described in subsection 2.3.2. We built a simplified model of the scintillating bolometers based on 334 g and 336 g $ZnMoO_4$ crystals, as well as 59.2 g and 62.9 g enriched $Zn^{100}MoO_4$ bolometers to reconstruct the set-up geometry during the calibration run. The photograph of the assembled detector and the GEANT4 model are presented in Fig. 3.15.

The isotropic $^{232}Th$ source was made of thoriated tungsten wires with a total mass of 15.2 g which contain 1% of thorium in weight. The activity of the source (≈ 600 Bq) was reduced twice by removing one half of wires before the calibration of 334 g $ZnMoO_4$ detector. The $^{232}Th$ source was placed on the external surface of the 300 K copper screen (see Fig. 3.16) at the distance of about 19 cm from 334 g $ZnMoO_4$ detector. Altogether six copper screens (at 300 K, 100 K, 40 K, 4 K, 1 K and 10 mK) with a thickness of about 3 mm each and 4 cm thick PTFE screen were located between the $ZnMoO_4$ bolometer and $^{232}Th$ source.

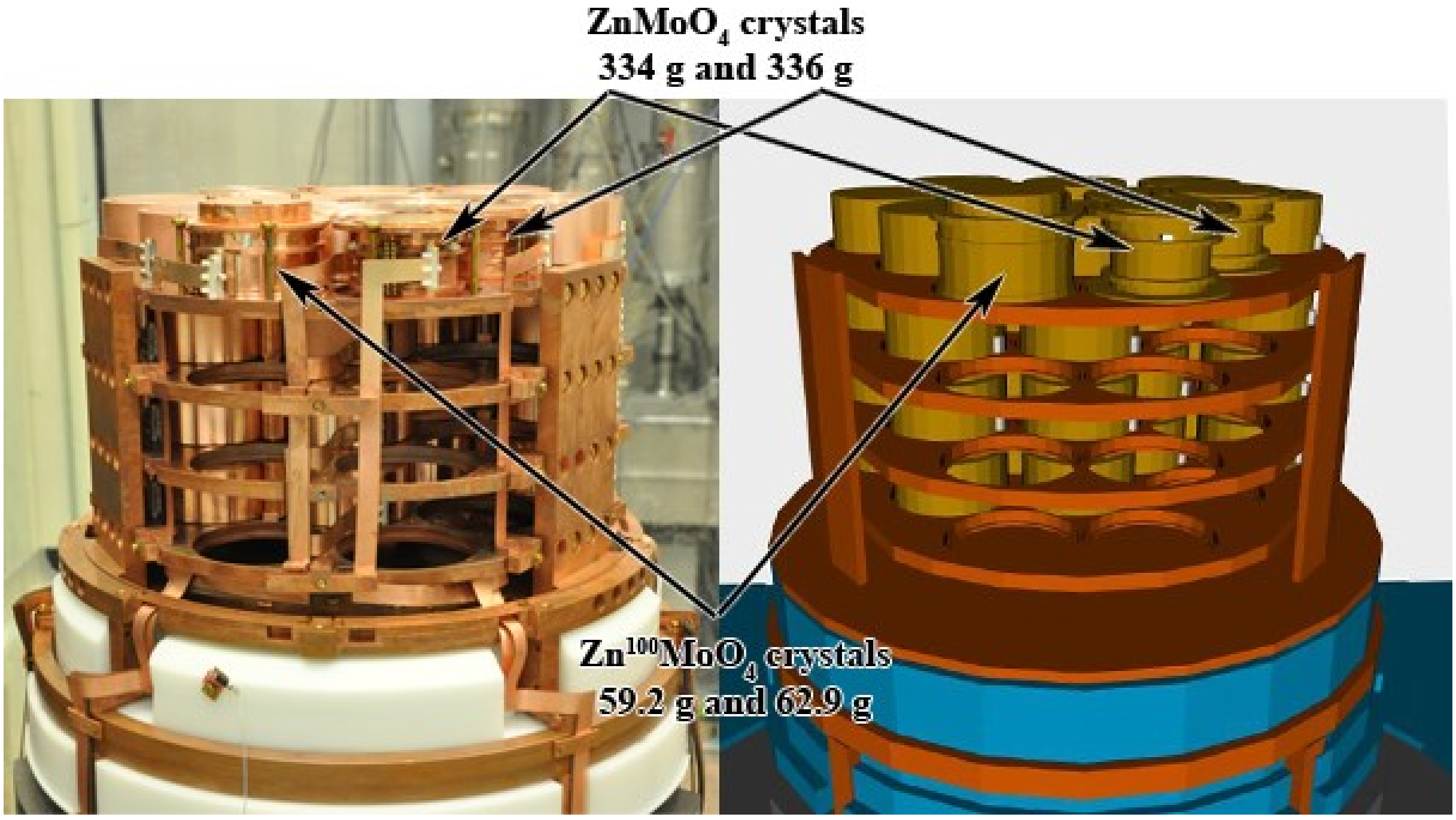

Fig. 3.15. The photograph (left) and the GEANT4 model (right) of the scintillating bolometer based on 334 g $ZnMoO_4$ crystal with a size of Ø50 × 40 mm assembled in the EDELWEISS set-up.

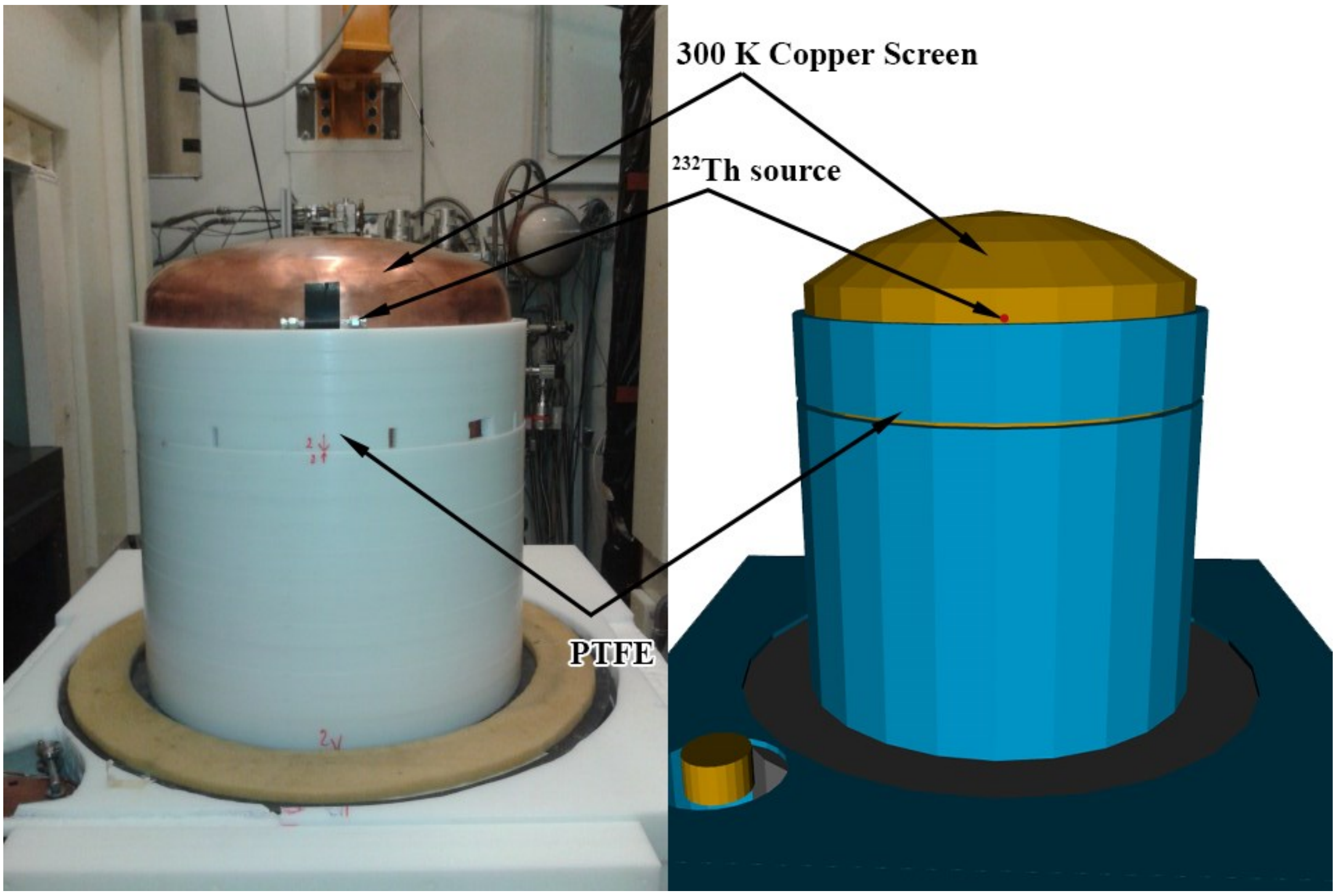

Fig. 3.16. The photograph (left) and the GEANT4 model (right) of the $^{232}$Th source position in the EDELWEISS set-up used to calibrate 334 g $ZnMoO_4$ scintillating bolometer. The source position in the GEANT4 model is marked with a red circle.

The experimental data used for the comparison were accumulated with 334 g $ZnMoO_4$ bolometer during a total time of 13.6 hours of the calibration run. The dead time was estimated as the number of all registered events in all channels multiplied by 1 second. Taking into account 27303 registered events the live time is 6.02 h.

For the GEANT4 simulation we supposed that $^{232}$Th source was in the secular equilibrium with daughter nuclei. We have generated $10^7$ events of $^{232}$Th decay chain using the event generator DECAY0 [195]. The obtained simulation spectrum was blurred by the energy resolution as a function of energy:

$$\mathrm{FWHM(E)} = a + b\mathrm{E}^{c}, \qquad (3.2)$$

where E is the energy; $a$, $b$ and $c$ are coefficients, which were obtained from the experimental $^{232}$Th calibration measurements with 334 g $ZnMoO_4$ detector by fitting the achieved energy resolution values: 9(1) keV at 2614.5 keV and 4.2(6) keV at 911.2 keV (see Fig. 3.17). The simulation results were also normalized on the $^{232}$Th source activity and time of measurements. The comparison of the GEANT4 simulation spectrum and results of measurements is presented in Fig. 3.18.

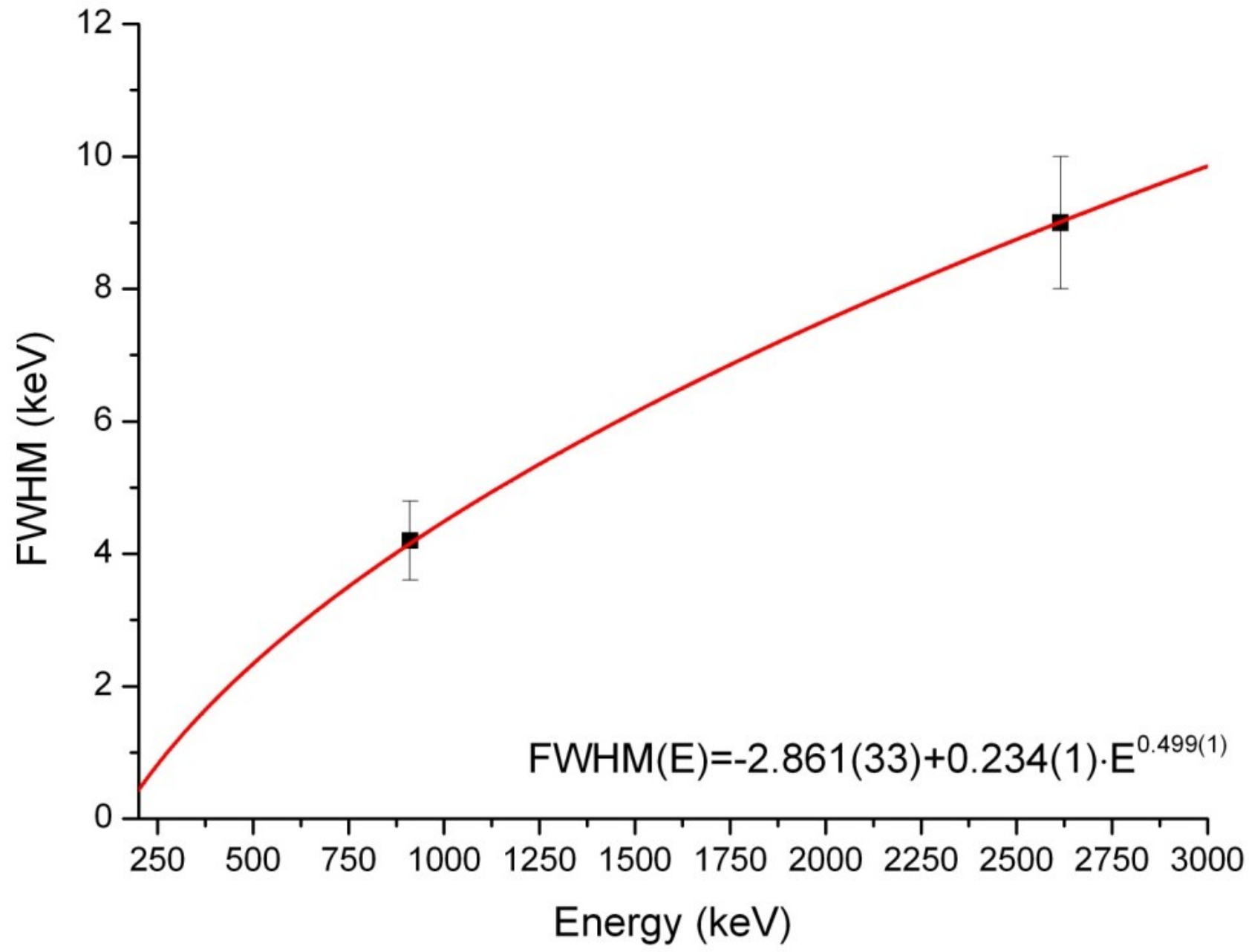

Fig. 3.17. Dependence of the energy resolution of 334 g $ZnMoO_4$ detector from the energy obtained by fitting the experimental energy resolution with a function $\mathrm{FWHM(E)} = a + b\mathrm{E}^{c}$.

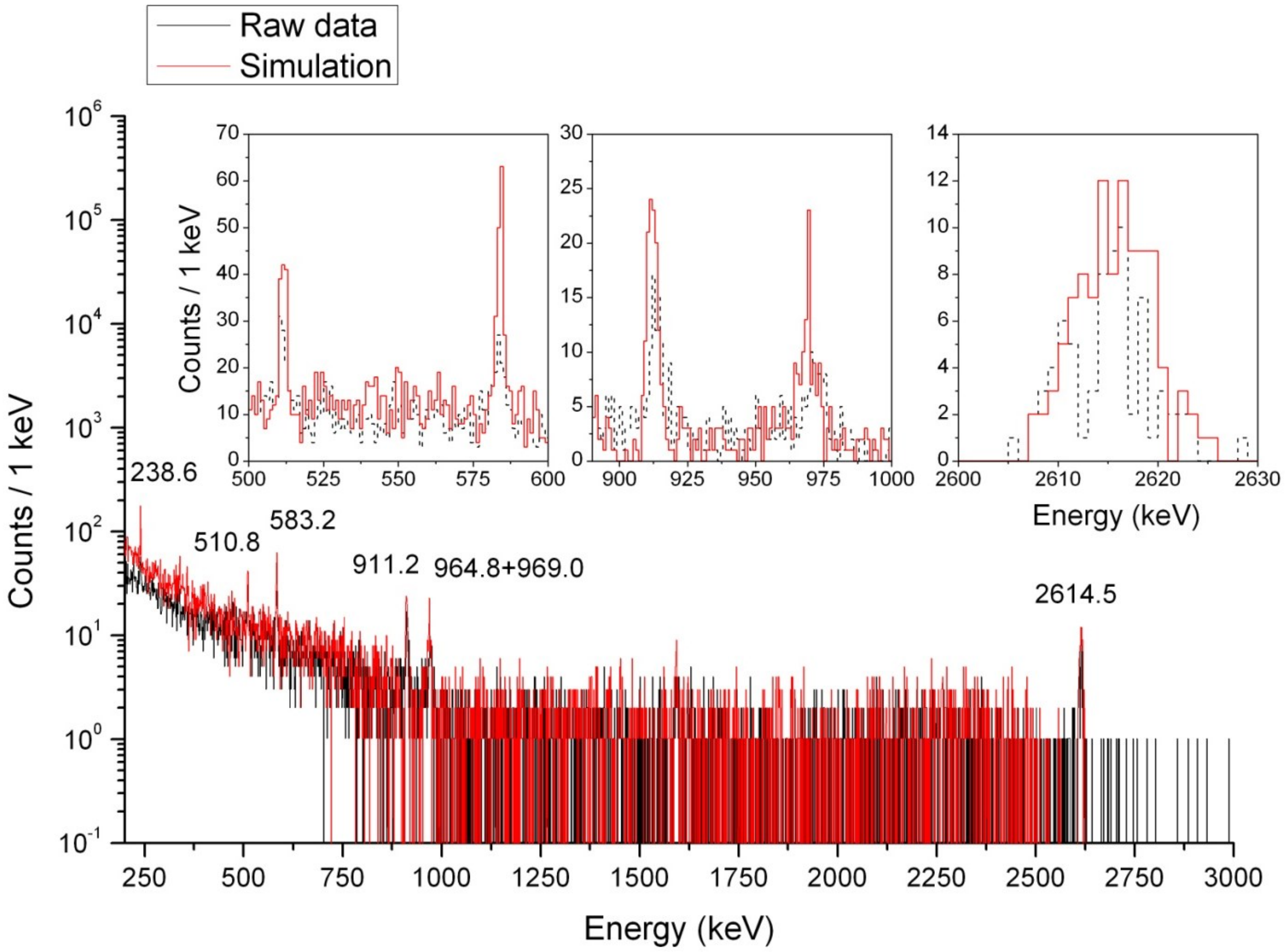

Fig. 3.18. The comparison of the energy spectra simulated using GEANT4 package (red) and accumulated in the underground calibration measurements (black) using the 334 g $ZnMoO_4$ scintillating bolometer in the EDELWEISS set-up. The detector was irradiated by isotropic $^{232}$Th source with the activity of 0.3 kBq. The data were accumulated over 6.02 h of measurements. The energy of γ peaks are in keV.

Monte Carlo simulation reproduces the detector calibration with an overall good quality; however, for the simulated data the counting rate (especially for the most intensive peaks) is higher than that in the experimental data. To estimate the conformity of the simulated and measured data we have compared the area under the most intensive peaks, and the counting rate of each 200 keV excluding peaks as a ratio between the measured and simulated spectra. The result of such comparison is presented in Fig. 3.19.

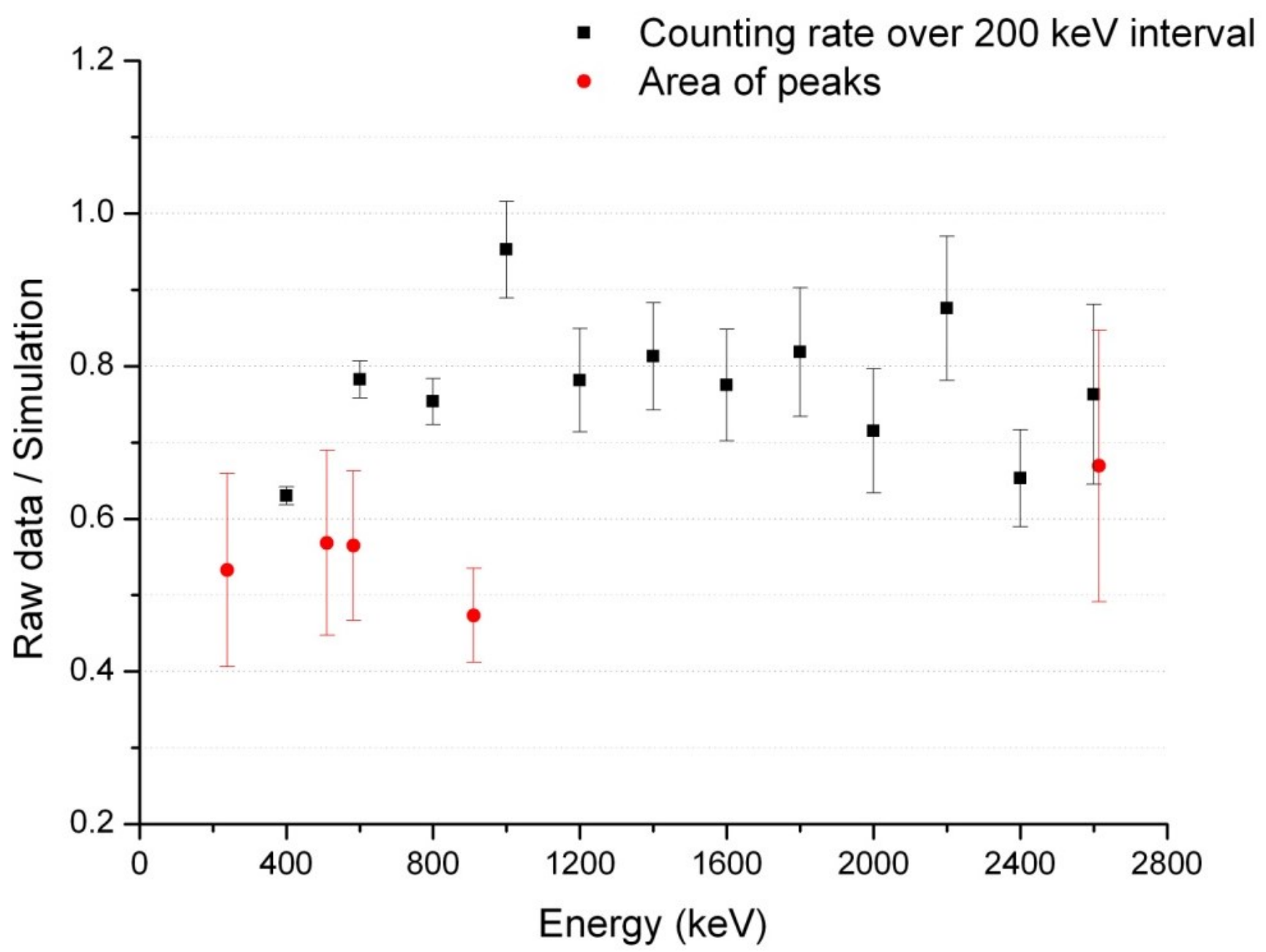

Fig. 3.19. The comparison of the area under the most intensive peaks (red circles) and the number of counts each 200 keV excluding peaks (black boxes) as a ratio between the energy spectrum obtained in the underground calibration measurements using the 334 g $ZnMoO_4$ scintillating bolometer and the simulated one using the GEANT4 package.

As one can see the area of peaks in the experiment is lower than that of the broad energy distribution. This feature can be explained by the pile-up effect in the acquired data which results in leaving of events from peaks to the broad energy distribution. Moreover, the difference between the simulated and measured data can be related to the imprecise knowledge of the $^{232}$Th source activity and position, as well as overall simplified geometry of the GEANT4 model which doesn't take into account some elements of the set-up or describes them very roughly.

### 3.3.3. Internal contamination by Th and Ra

Bulk contamination of $Zn^{100}MoO_4$ crystals by $^{238}$U and $^{232}$Th daughters is expected to be one of the main background components in the ROI. The $Q_{2\beta}$ value of $^{100}$Mo exceeds the natural gamma line of 2615 keV, thus the most dangerous sources of background are β active nuclides $^{208}$Tl ($Q_\beta$ = 4.999 MeV), $^{214}$Bi ($Q_\beta$ = 3.270 MeV) with $^{214}$Po ($Q_\alpha$ = 7.833 MeV), and $^{212}$Bi ($Q_\beta$ = 2.252 MeV) with $^{212}$Po ($Q_\alpha$ = 8.955 MeV). We will consider these sources separately.

$^{208}$Tl is a daughter of $^{228}$Th which decays through the $\beta^-$ decay channel with a half-life of 3.053 minutes. The β particles emitted in the combination with γ quanta of the daughter

nucleus $^{208}Pb$ give contribution to the continuous spectrum which can produce events falling in the ROI.

We have generated $3.594 \cdot 10^5$ events of $^{208}Tl$ decay using the event generator DECAY0, which corresponds to $10^6$ decays of $^{232}Th$ in secular equilibrium taking into account the branching ratio (BR) 35.94%. Here and thereafter we assumed detectors operation in the anticoincidence mode, which selects only events where the energy is deposited in a single crystal. An energy threshold of each detector was set as 50 keV. We have generated events uniformly distributed in one of the $Zn^{100}MoO_4$ crystals located at a bottom edge of detector array as the most conservative case (from the point of view of the anticoincidence efficiency of the multi-detector structure). The energy spectrum of $^{208}Tl$ calculated using the GEANT4 package is presented in Fig. 3.20.

It should be stressed, that background from $^{208}Tl$ can be further effectively suppressed by the detection of α particle ($Q_\alpha$ = 6.207 MeV) emitted by its precursor $^{212}Bi$ with a half-life of $T_{1/2}$ = 3.053 minutes. For instance, 30 minutes (10 half-lives) of vetoing of the crystal in which α particle with the energy of 6.2 MeV was detected, will reduce the background from $^{208}Tl$ by $2^{10} \approx 1000$ times. The dead time due to this method is negligible (less than 1%) for the $^{228}Th$ activity on the level of ~ 10 μBq/kg.

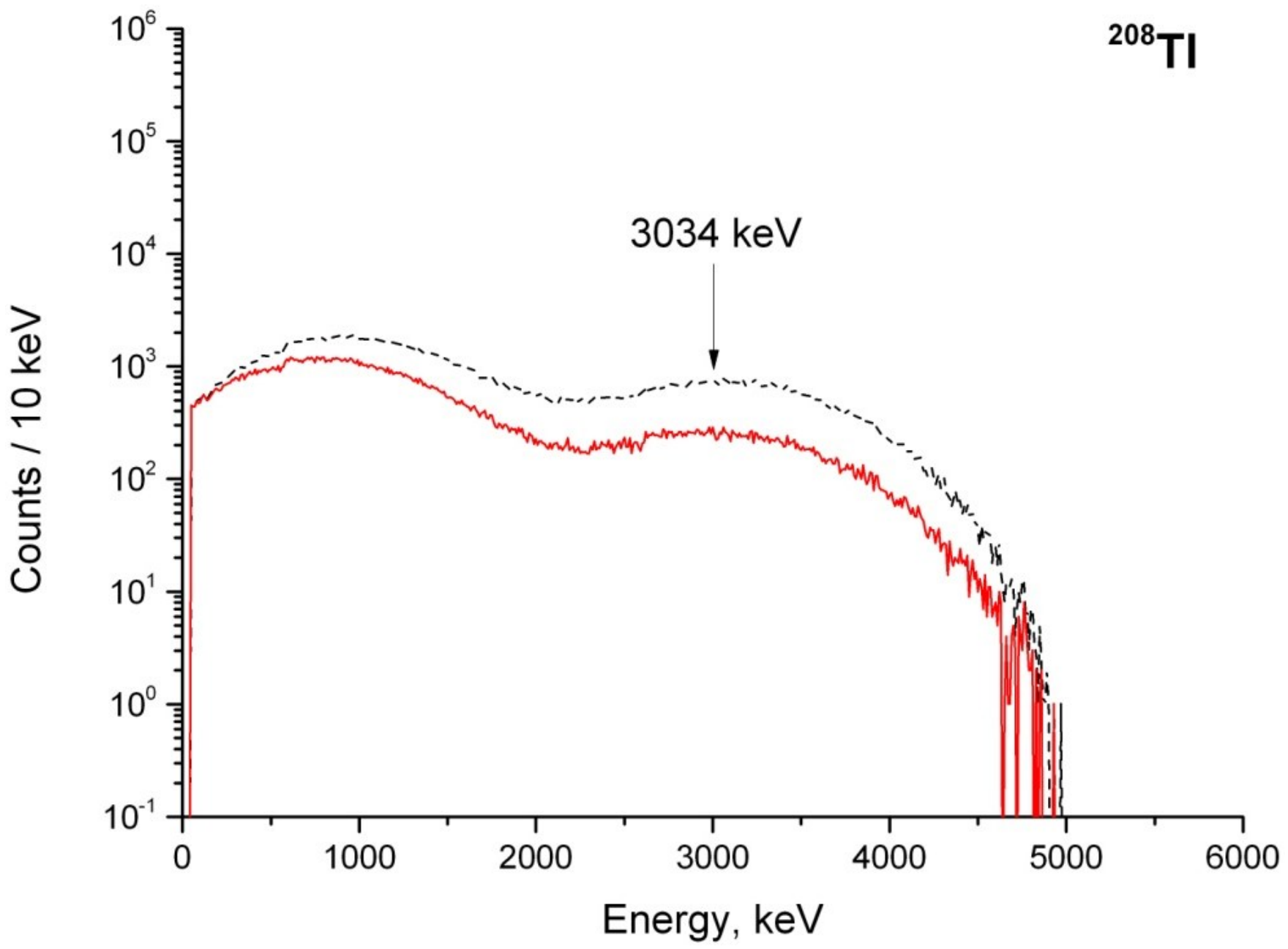

Fig. 3.20. The simulated initial (dashed line) and anticoincidence (solid line) background spectrum induced by $^{208}Tl$ contamination in $Zn^{100}MoO_4$ crystal.

In the $^{238}U$ decay chain the possible sources of background in the ROI are β emitters $^{214}Bi$ and $^{210}Tl$. The nucleus of $^{214}Bi$ decays either through the $\beta^-$ channel to $^{214}Po$ ($Q_\beta$ = 3.270 MeV) with a BR of 99.98%, either through the α decay to $^{210}Tl$ ($Q_\alpha$ = 5.621 MeV) with a BR 0.02%. In the last case the $^{210}Tl$ decays through the $\beta^-$ channel ($Q_\beta$ = 5.482 MeV) with the half-life of 1.3 min. We will not consider this source as a significant contributor to background due to the very low branching ratio. In addition, such background can be easily rejected with a similar delayed coincidence technique used for $^{208}Tl$ source as described above.

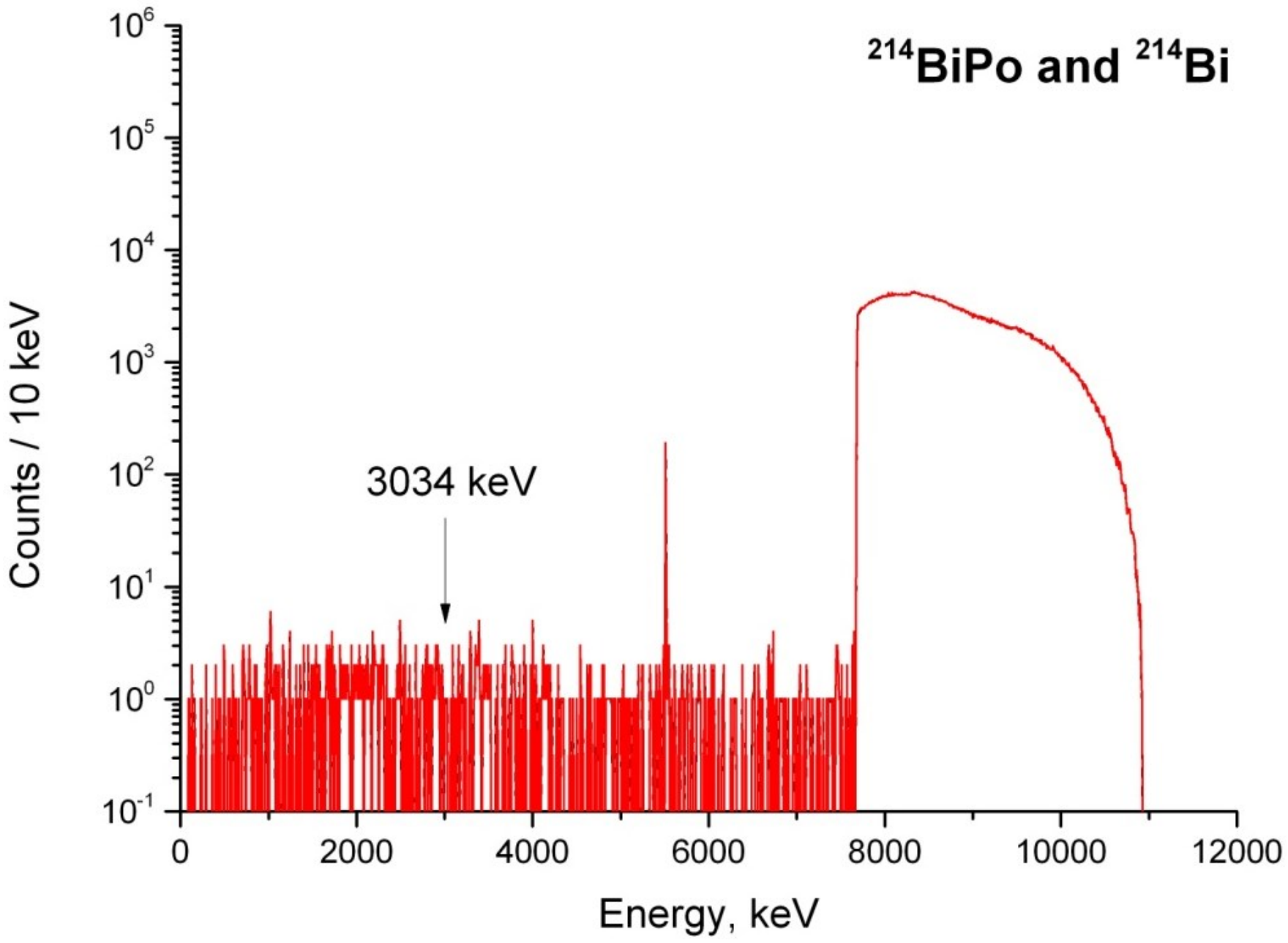

Fig. 3.21. The simulated initial (dashed line) and anticoincidence (solid line) background spectrum caused by $^{214}$Bi α decay and $^{214}$BiPo events in Zn$^{100}$MoO$_4$ crystal.

The $\beta^-$ decay of $^{214}$Bi to $^{214}$Po is followed by α decay of $^{214}$Po with a short half-life ($T_{1/2}$ = 164.3 μs) and energy release of 7.833 MeV. Taking into account the poor time resolution of bolometers (typical rise time in bolometers with NTD thermistors is of the order of ms), these two decays will produce so called $^{214}$BiPo events with the energies mostly higher than 7.8 MeV, except the situation when one of the emitted particle escape the crystal volume. We have simulated $10^6$ events of $^{214}$Bi α decay and $^{214}$BiPo events in Zn$^{100}$MoO$_4$ crystal. The background spectrum is presented in Fig. 3.21.

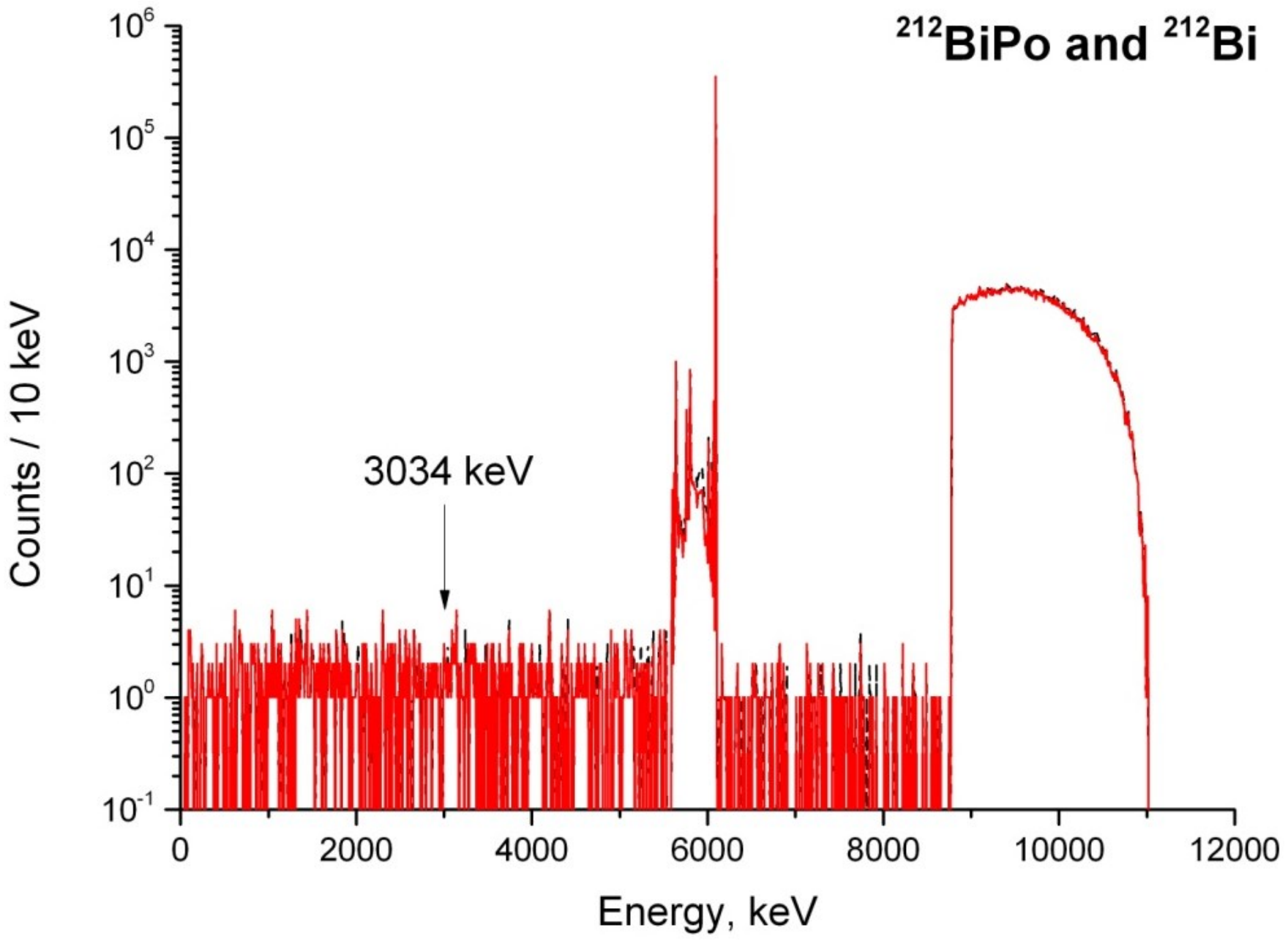

Fig. 3.22. The simulated initial (dashed line) and anticoincidence (solid line) background spectrum induced by $^{212}$BiPo events and α decay of $^{212}$Bi in Zn$^{100}$MoO$_4$ crystal.

Another source of background considered in our simulation is the isotope of $^{212}$Bi from $^{232}$Th decay chain. There are two decay modes for this nucleus: $\beta^-$ decay to $^{212}$Po ($Q_\beta$ = 2.252 MeV, BR = 64.06%) and α decay to $^{208}$Tl ($Q_\alpha$ = 6.208 MeV, BR = 35.94%). The daughter isotope of $^{212}$Po decays through the alpha channel with the half-life $T_{1/2}$ = 299 ns and energy $Q_\alpha$ = 8.955 MeV. These $^{212}$BiPo events are recorded as a single signal in cryogenic bolometers due to their very short half-life. The total energy 8955 keV–11207 keV is much higher than the region of interest 2934 keV–3134 keV, however in case of $^{212}$Bi decay on the surface of $Zn^{100}MoO_4$ scintillator, α particle of $^{212}$Po can partially escape the crystal volume leaving the energy small enough (~ 800 keV) to produce a signal in the region of 0ν2β peak. $10^6$ events of $^{212}$Bi and $^{212}$BiPo decays were generated in $Zn^{100}MoO_4$ crystal. The spectrum of the $^{212}$BiPo events is shown in Fig. 3.22.

### 3.3.4. Surface contamination of $Zn^{100}MoO_4$ crystal

Surface contamination of crystal scintillators is a dangerous source of background because the contamination level can even exceed the one of crystal bulk. A crystal surface can be contaminated by $^{238}$U and $^{232}$Th daughters during a mechanical treatment of material or by contact with contaminated air. It is difficult to measure the depth of the contaminated layer [194, 196], therefore we assume a crystal surface contaminated by $^{238}$U and $^{232}$Th daughters with an exponential depth profile and a mean depth of 5 µm.

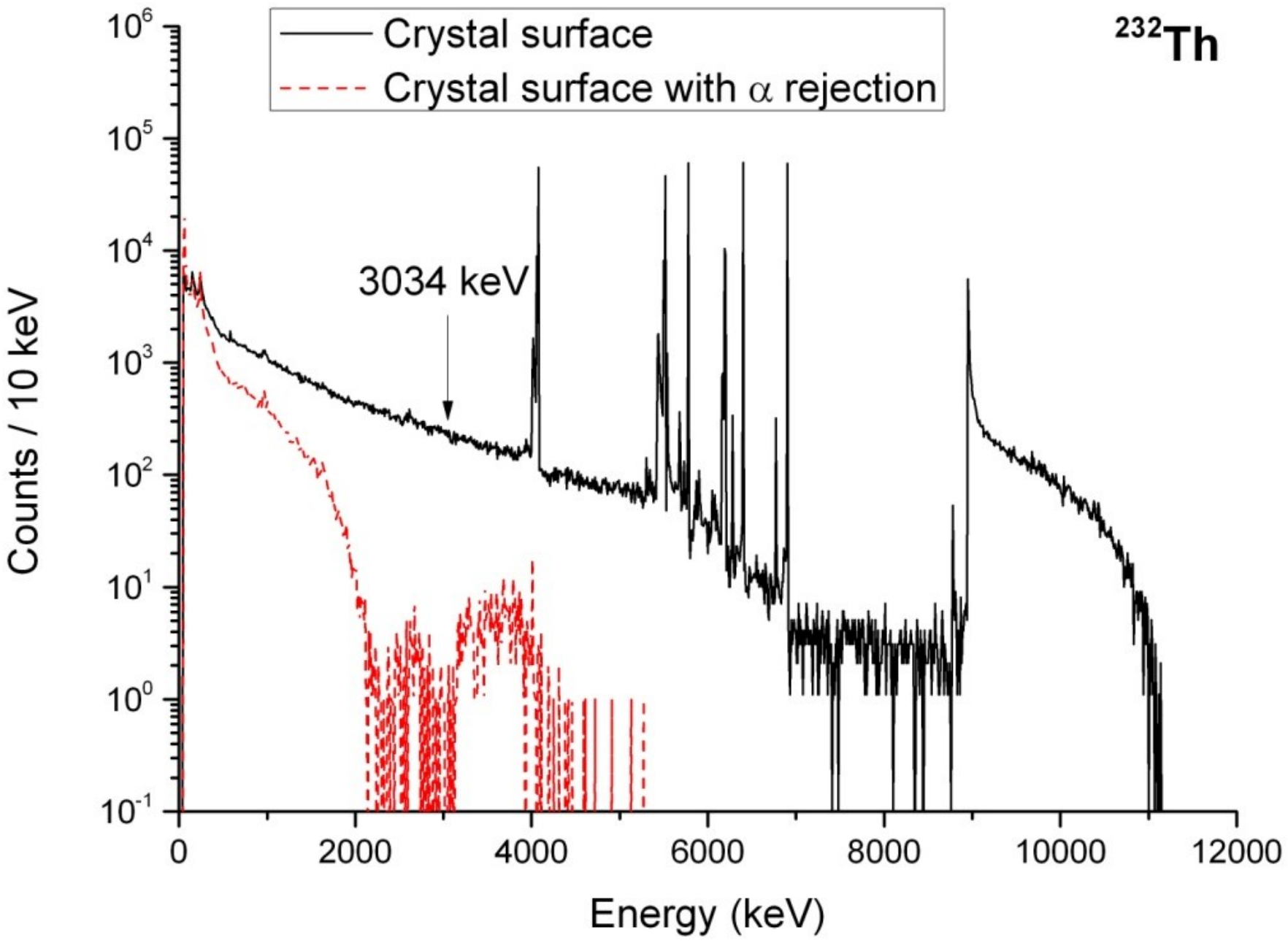

Fig. 3.23. The simulated anticoincidence background spectrum (solid line) and the one with α rejection and suppression with delayed coincidences (dashed line) induced by $^{232}$Th and its daughters contamination in $Zn^{100}MoO_4$ crystal surface with an exponential depth profile and a mean depth of 5 µm.

We have fully simulated $^{232}$Th and $^{238}$U decay chains in $Zn^{100}MoO_4$ crystal surface. We have generated $10^5$ events for each isotope from U/Th decay chains taking into account the branching ratio, but excluding isotopes with the branch chance lower than 0.001%.

Moreover, we have applied a 99.9% rejection efficiency of α events registered in the detector, and $^{208}$Tl suppression by factor $10^3$ using delayed coincidences (see subsection 3.3.3). The anticoincidence background spectrum and the one with α rejection for $^{232}$Th and $^{238}$U chains are presented in Fig. 3.23 and Fig. 3.24, respectively.

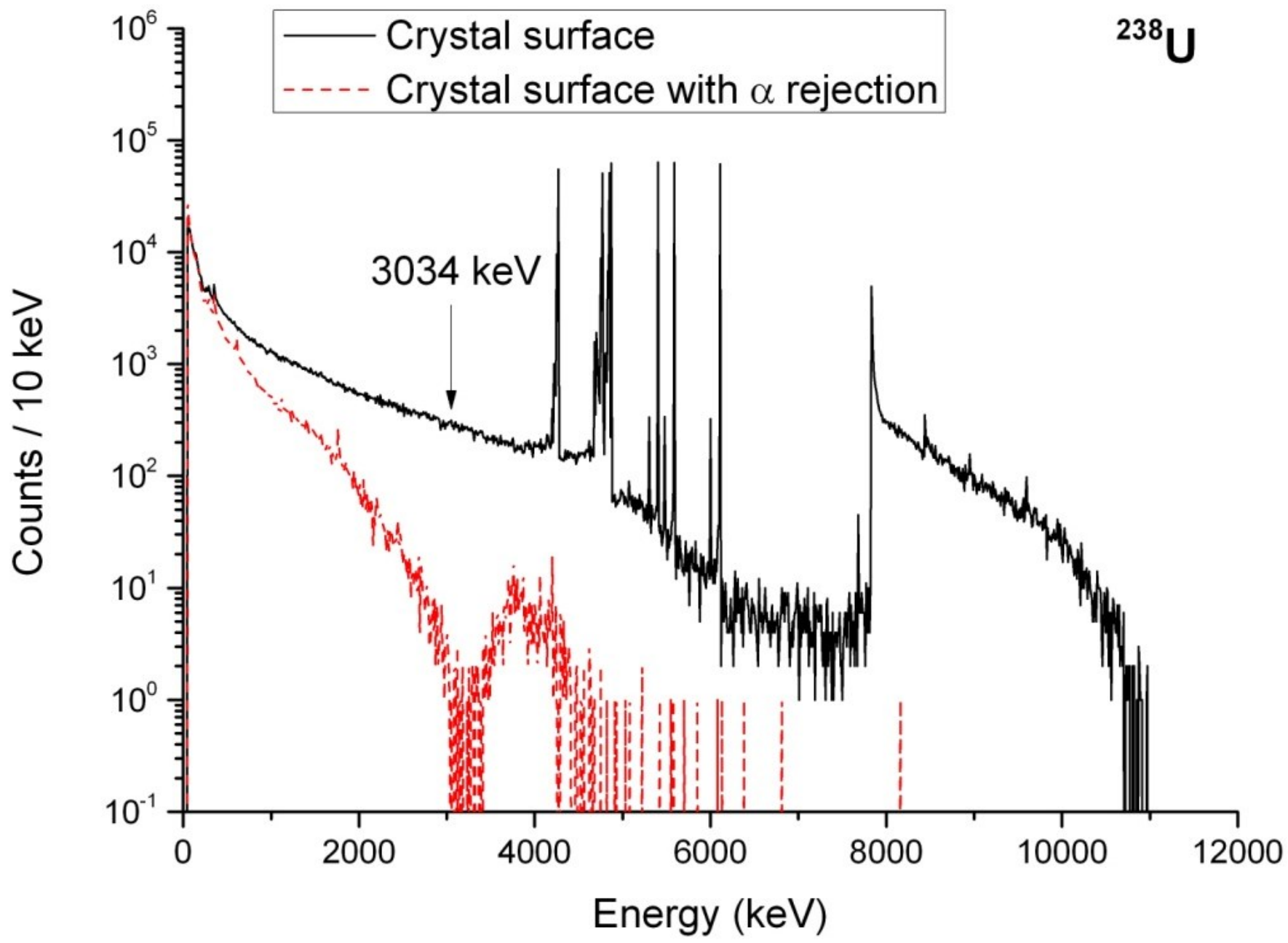

Fig. 3.24. The simulated anticoincidence background spectrum (solid line) and the one with α rejection (dashed line) induced by $^{238}$U and its daughters contamination in Zn$^{100}$MoO$_4$ crystal surface with an exponential depth profile and a mean depth of 5 μm.

### 3.3.5. Contamination of set-up by Th and Ra

We have simulated the radioactive contamination by $^{238}$U and $^{232}$Th daughters of the set-up materials assembled nearest to Zn$^{100}$MoO$_4$ detector: copper (detector module), PTFE (crystal holder) and BoPET (reflective foil). We haven't considered contamination of the Ge wafer because any radioactive decay in the light detector will produce thermal pulses with a different shape than ordinary signal induced by scintillation light.

For each of the set-up materials we generated $10^6$ decays of $^{232}$Th chain in secular equilibrium with daughter nuclei, and $10^6$ events of $^{214}$Bi α decay and $^{214}$BiPo events randomly distributed in material volume. The background spectra for each source and material are presented in Figs. 3.25–3.30.

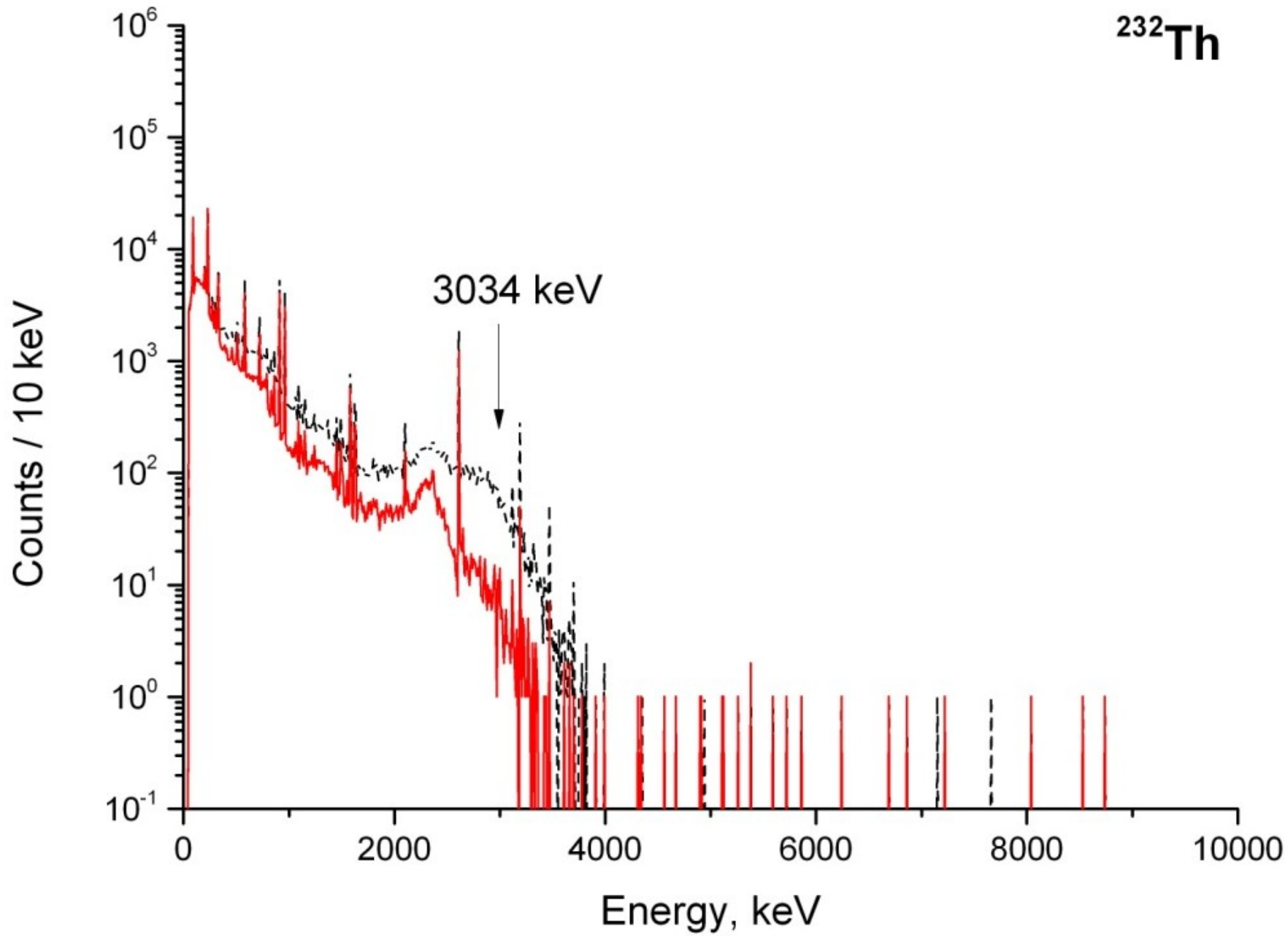

Fig. 3.25. The simulated initial (dashed line) and anticoincidence (solid line) background spectrum caused by $^{232}$Th (in secular equilibrium) contamination in copper holder.

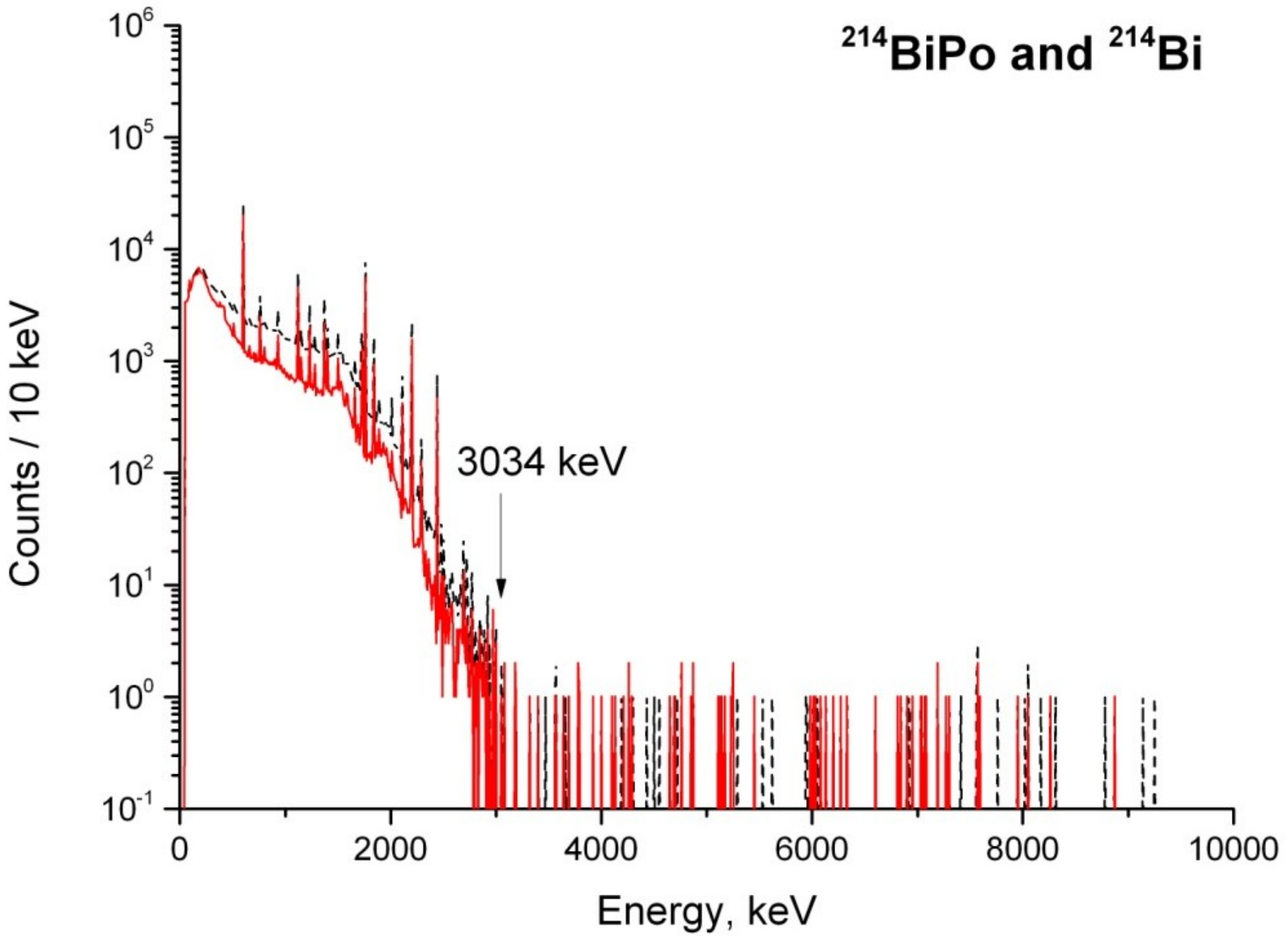

Fig. 3.26. The simulated initial (dashed line) and anticoincidence (solid line) background spectrum caused by $^{214}$Bi α decay and $^{214}$BiPo events in copper holder.

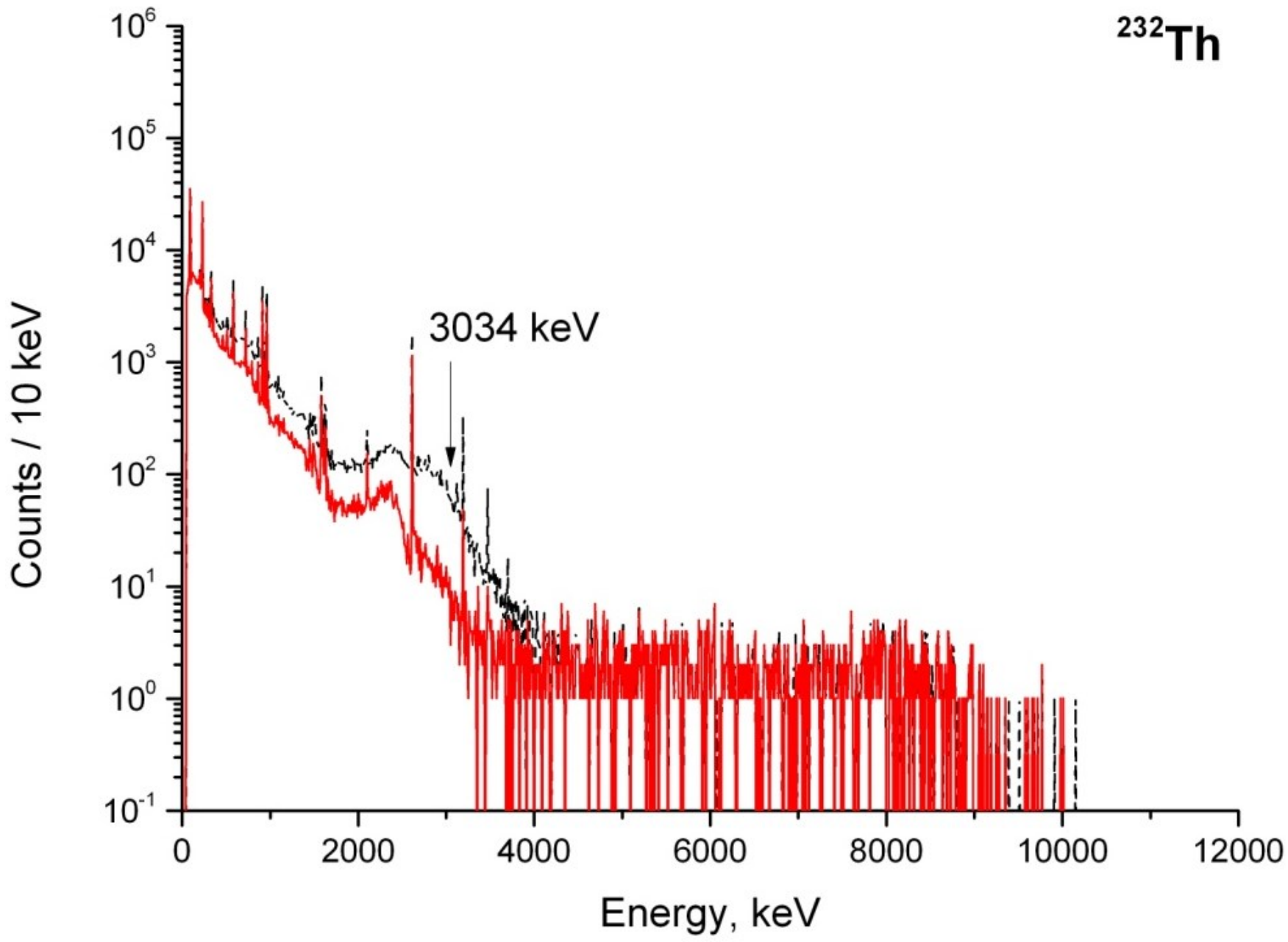

Fig. 3.27. The simulated initial (dashed line) and anticoincidence (solid line) background spectrum caused by $^{232}$Th (in secular equilibrium) contamination in PTFE crystal support details.

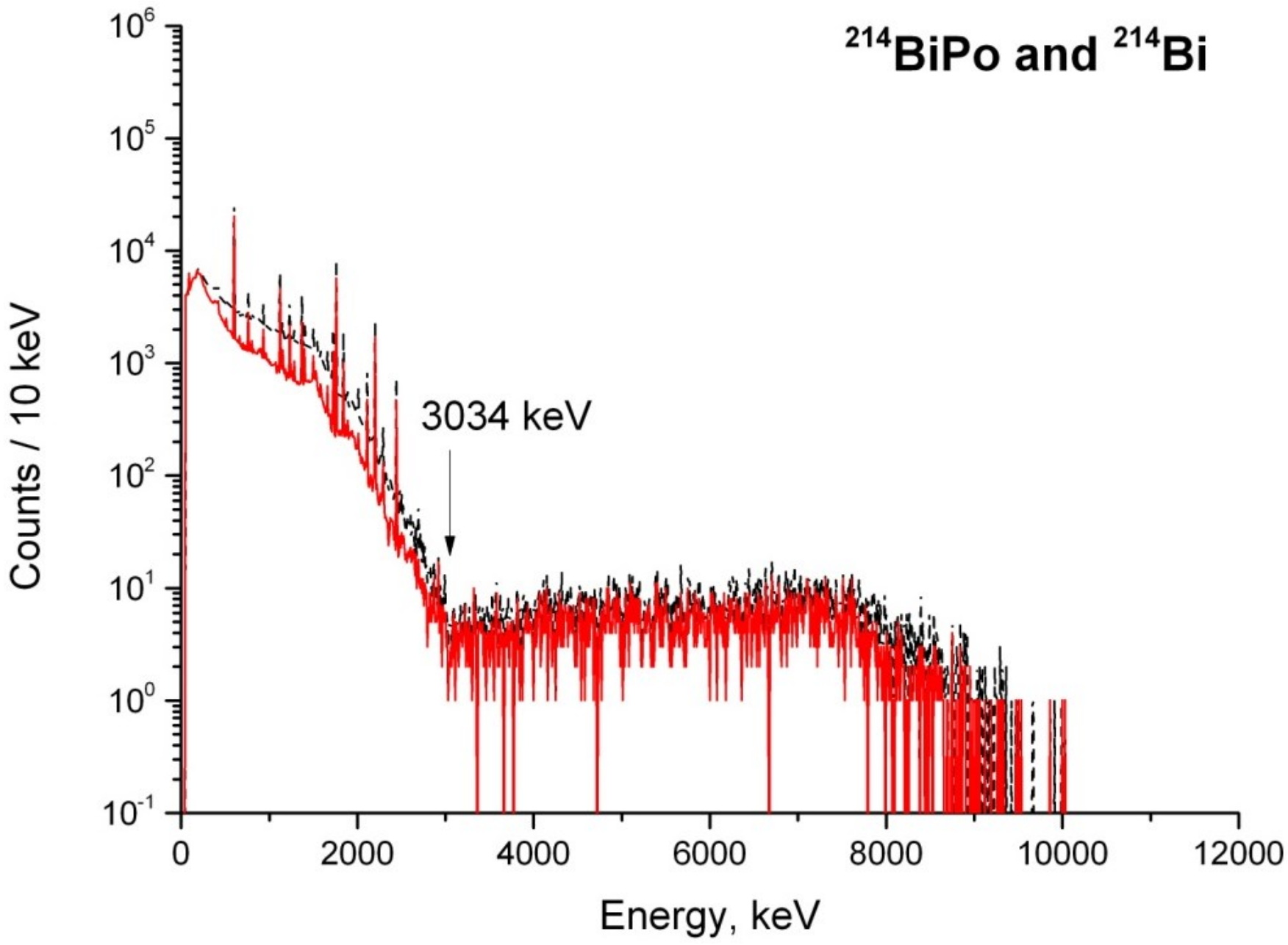

Fig. 3.28. The simulated initial (dashed line) and anticoincidence (solid line) background spectrum caused by $^{214}$Bi α decay and $^{214}$BiPo events in PTFE crystal support details.

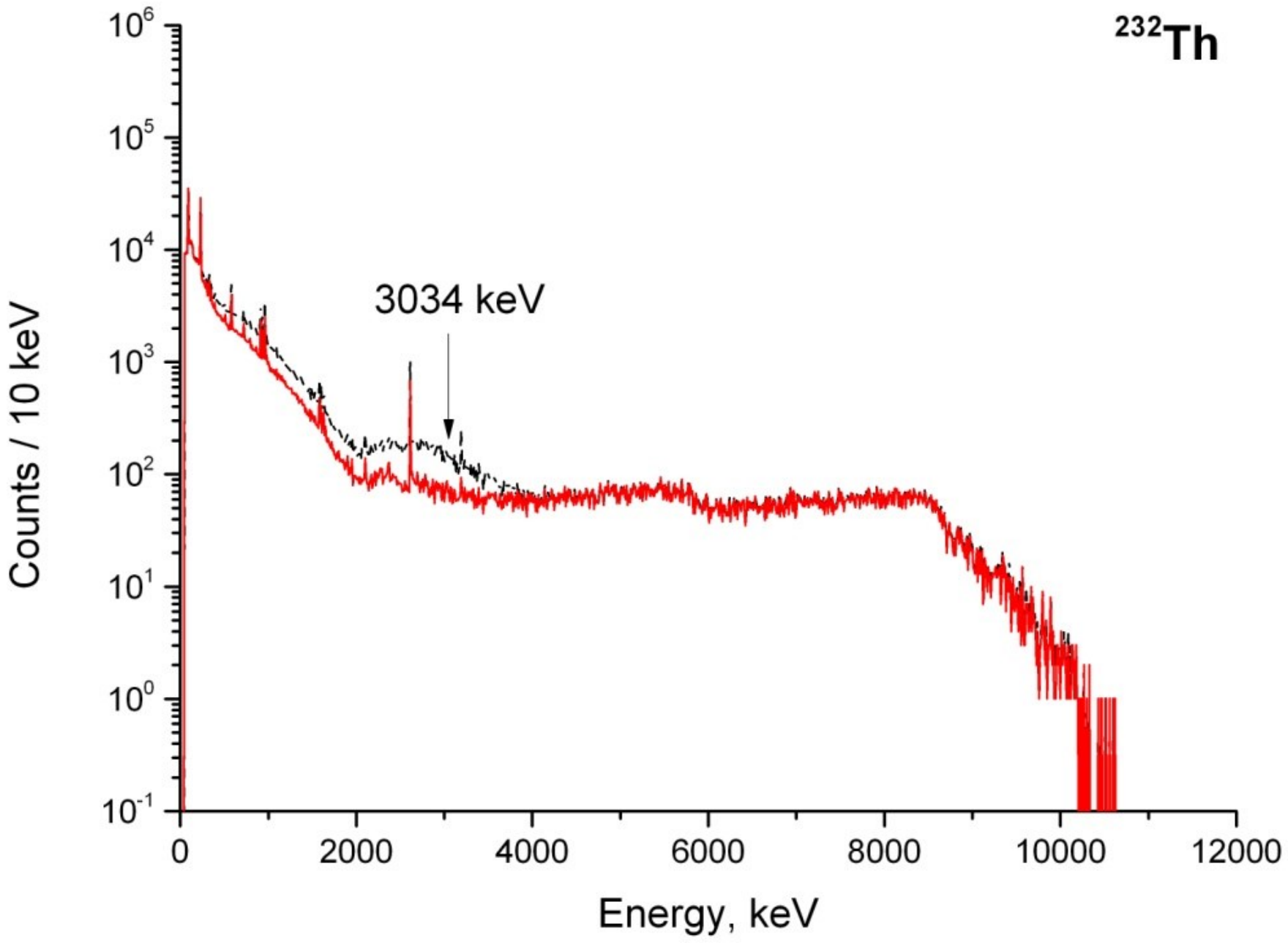

Fig. 3.29. The simulated initial (dashed line) and anticoincidence (solid line) background spectrum caused by $^{232}$Th (in secular equilibrium) contamination in BoPET reflective foil surrounding $Zn^{100}MoO_4$ crystal.

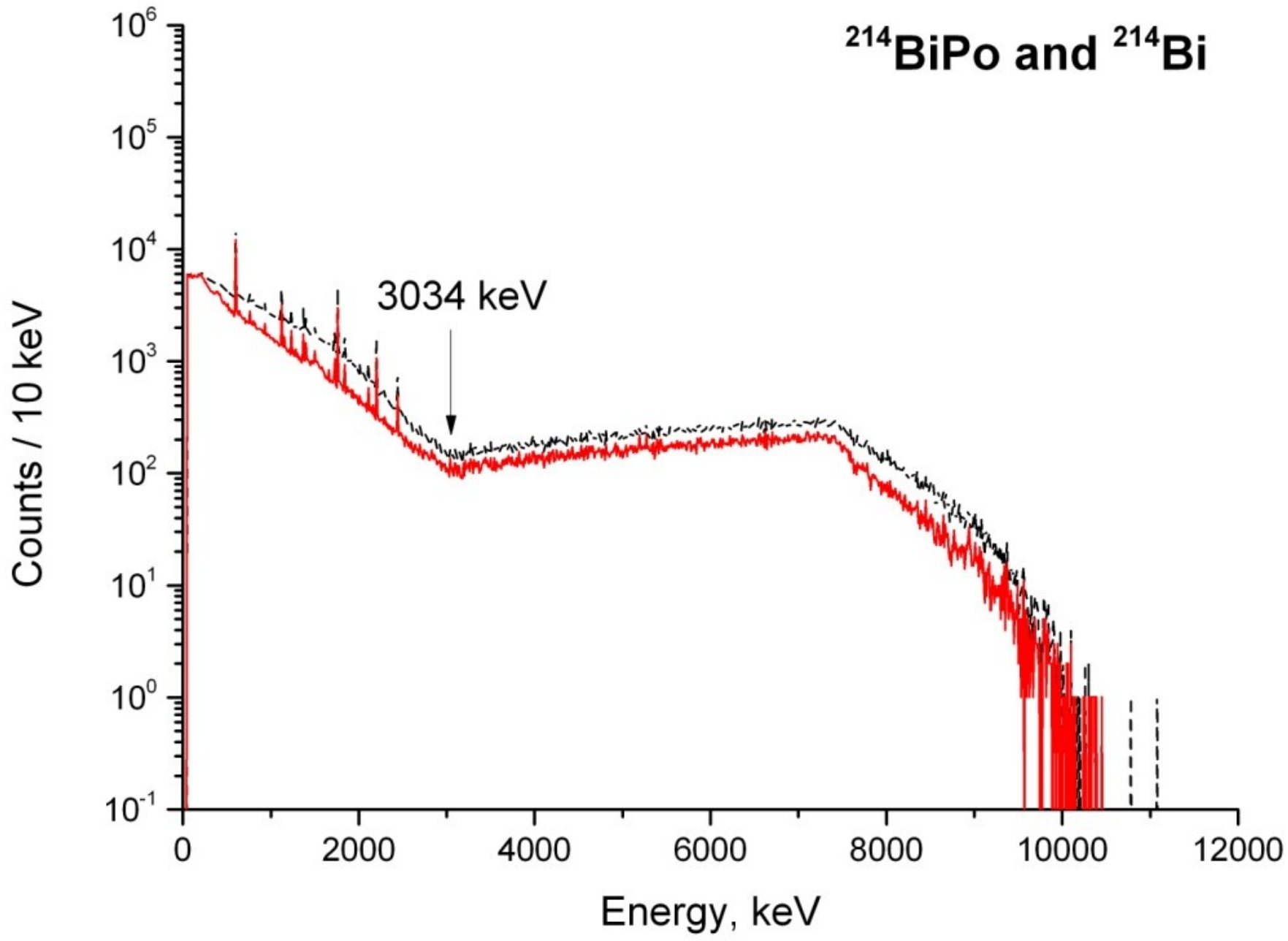

Fig. 3.30. The simulated initial (dashed line) and anticoincidence (solid line) background spectrum caused by $^{214}$Bi α decay and $^{214}$BiPo events in BoPET reflective foil surrounding $Zn^{100}MoO_4$ crystal.

### 3.3.6. Cosmogenic activation of $Zn^{100}MoO_4$ and surrounding materials

The cosmogenic activity induced by interaction of cosmic rays with $Zn^{100}MoO_4$ crystals and copper elements of the set-up was estimated with the help of the COSMO code [197]. In our calculations we assumed 3 months of activation on the Earth surface and 1 year of cooling down in the underground environment. There are two dangerous nuclides which can produce events with the energy $Q_\beta$ > 3035 keV: $^{56}$Co ($Q_\beta$ = 4566 keV, $T_{1/2}$ = 77.27 days) and $^{88}$Y ($Q_\beta$ = 3623 keV, $T_{1/2}$ = 106.65 days). It should be noted that nuclide $^{88}$Y can also appear as daughter of cosmogenic $^{88}$Zr ($T_{1/2}$ = 83.40 days).

To simulate cosmogenic activation in $Zn^{100}MoO_4$ crystal we generated $10^6$ events of $^{88}$Y source using the event generator DECAY0, and $10^6$ decays of $^{56}$Co source with the help of GEANT4 built-in source function. The energy spectra of $^{88}$Y and $^{56}$Co are presented in Fig. 3.31 and Fig. 3.32 respectively.

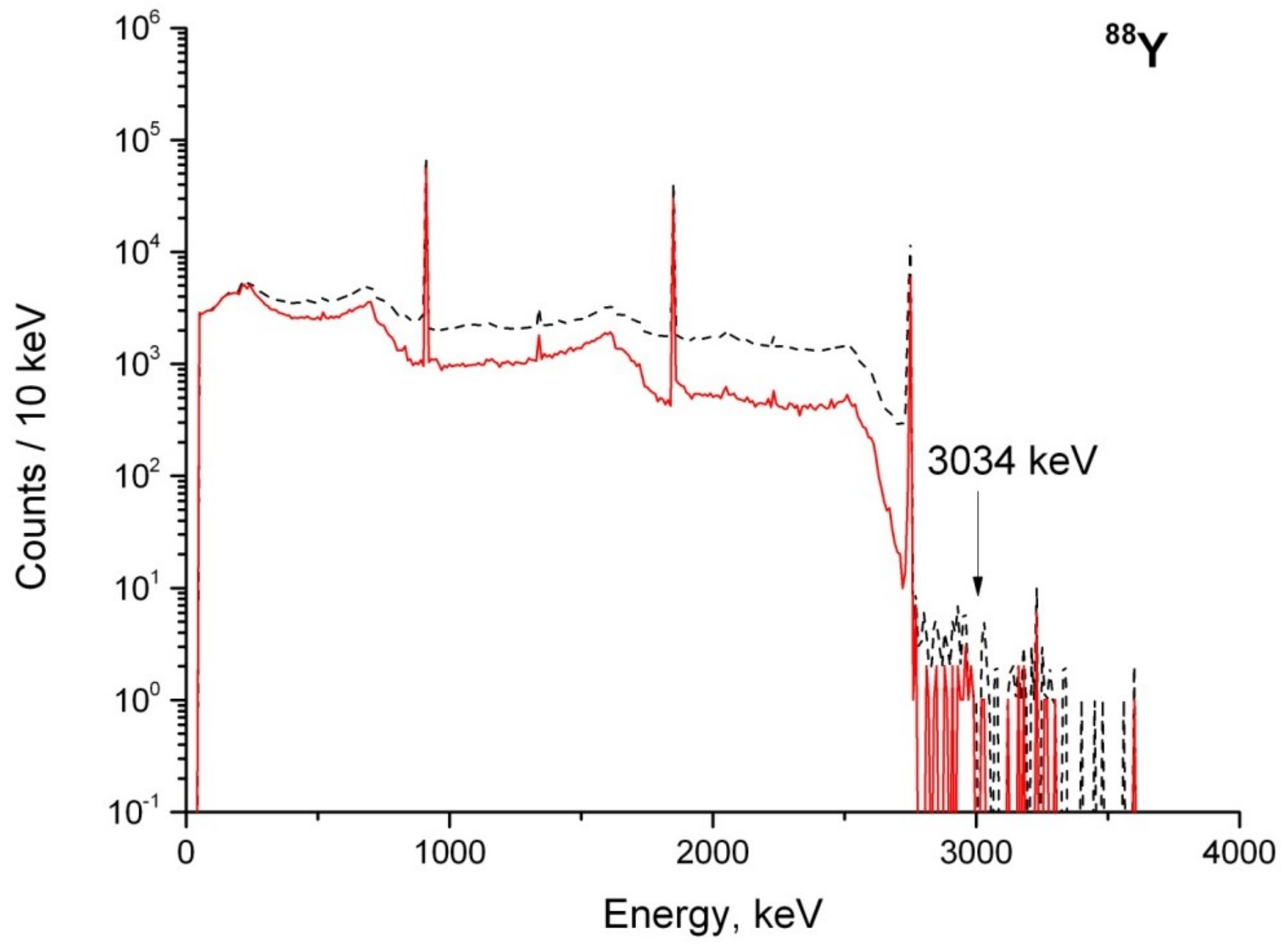

Fig. 3.31. The simulated initial (dashed line) and anticoincidence (solid line) background spectrum of $^{88}$Y produced by cosmogenic activation in $Zn^{100}MoO_4$ crystal.

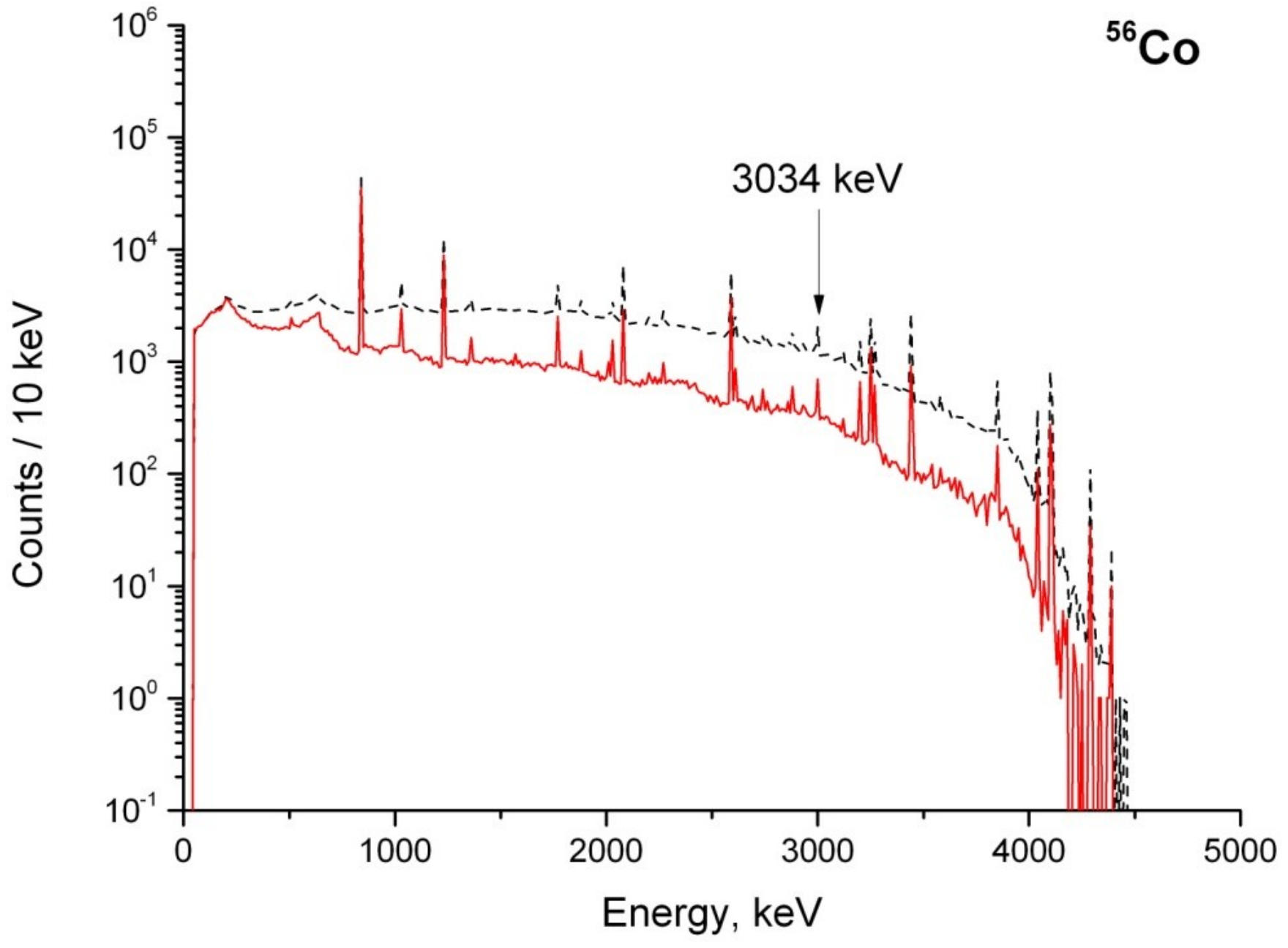

Fig. 3.32. The simulated initial (dashed line) and anticoincidence (solid line) background spectrum of $^{56}$Co produced by cosmogenic activation in $Zn^{100}MoO_4$ crystal.

In case of copper holder the only dangerous product of cosmogenic activation is $^{56}$Co. The energy spectrum obtained by generating $10^6$ decays of $^{56}$Co is shown in Fig. 3.33.

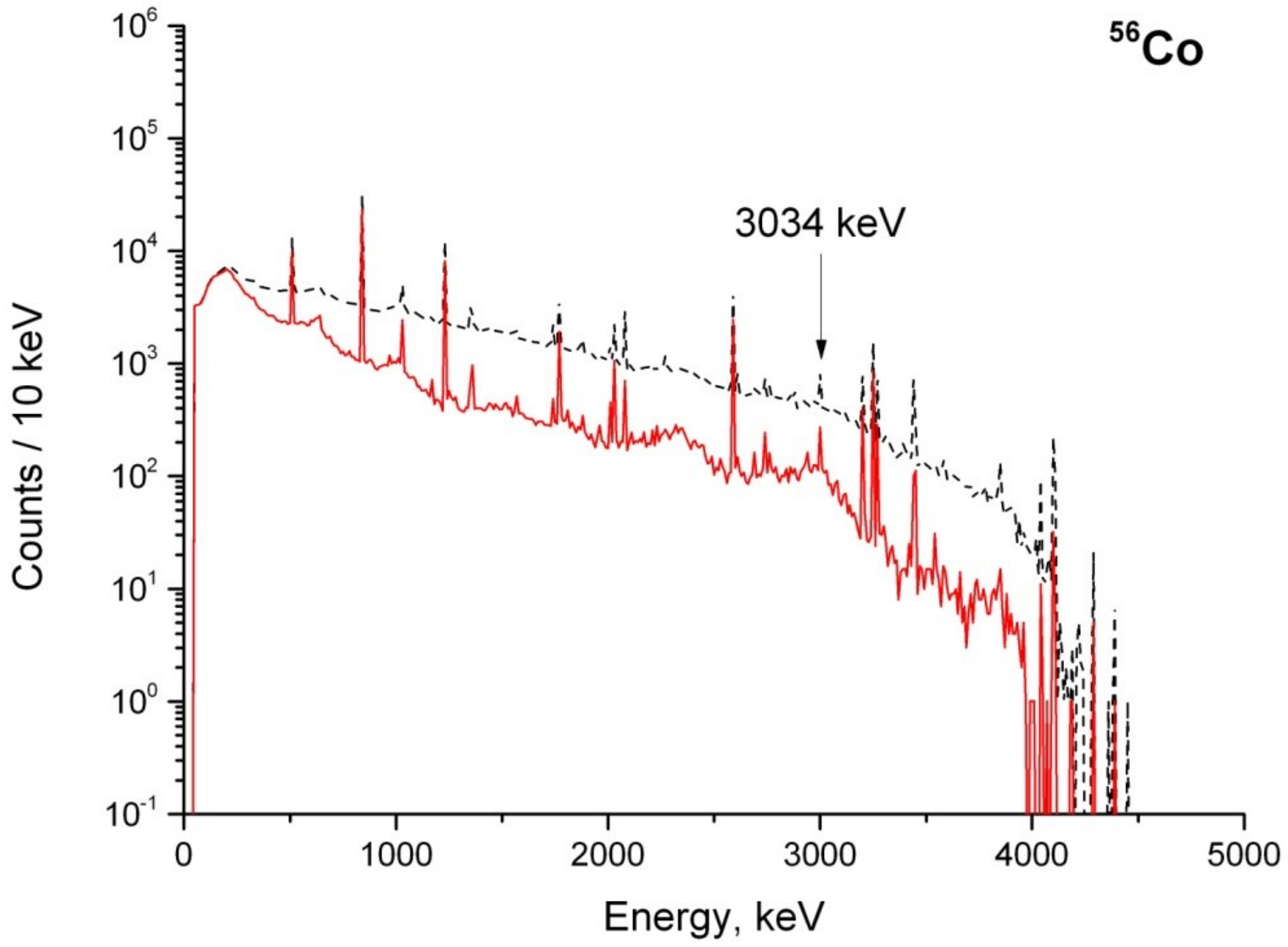

Fig. 3.33. The simulated initial (dashed line) and anticoincidence (solid line) background spectrum of $^{56}$Co produced by cosmogenic activation in the copper holder.

### 3.3.7. Simulation of neutrinoless double beta decay of $^{100}$Mo

We have generated $10^6$ events of 0ν2β decay of $^{100}$Mo in Zn$^{100}$MoO$_4$ crystal using the event generator DECAY0. The energy spectrum of 0ν2β decay of $^{100}$Mo simulated using the GEANT4 package is presented in Fig. 3.34.

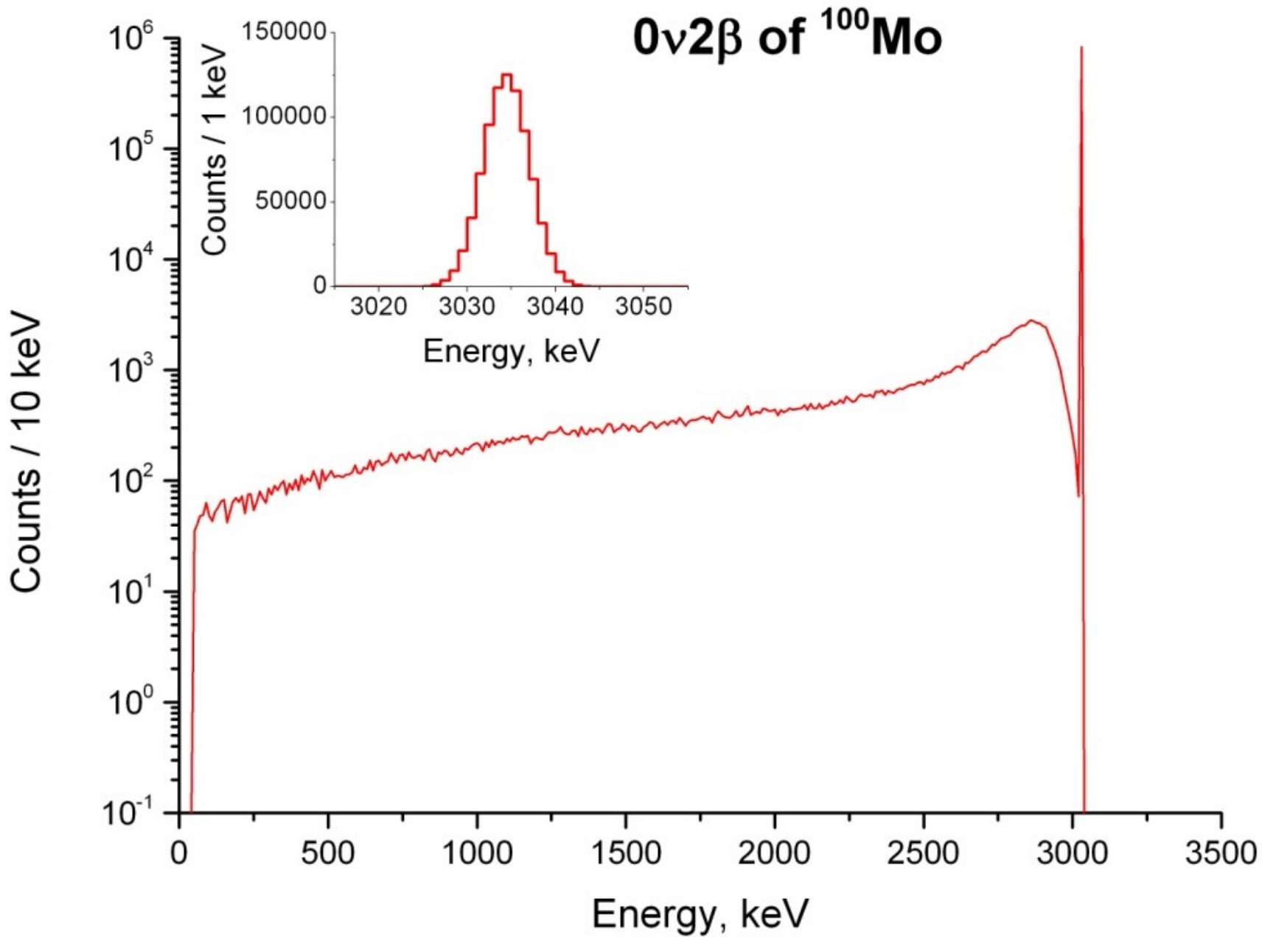

Fig. 3.34. The simulated spectrum of 0ν2β decay of $^{100}$Mo in Zn$^{100}$MoO$_4$ crystal. (Inset) The 0ν2β peak with FWHM = 6 keV.

### 3.3.8. Total background evaluation

We have estimated the total background level that could be reached with the sources mentioned in the previous subsections. The radioactive contamination of copper and PTFE was taken from Ref. [153]. The activities for surface contamination of $Zn^{100}MoO_4$ crystal were used the same as in Ref. [198]. The cosmogenic activity was evaluated as an average activity during 5 years of underground measurements, assuming 3 months of activation on the Earth surface and one year of cooling down in the underground environment before the experiment. The background counting rates at the ROI are presented in Table 3.2. A total background spectrum is shown in Fig. 3.35.

Table 3.2
Monte Carlo simulated background contributions in $Zn^{100}MoO_4$ scintillating bolometer in the EDELWEISS set-up. Background was calculated after anticoincidence method application in the ROI interval 2934 keV–3134 keV.

| Position | Source of background | Activity (μBq/kg) | Background (counts/(keV·kg·yr)) |
|---|---|---|---|
| $Zn^{100}MoO_4$ crystal bulk | $^{208}$Tl | 10 ($^{232}$Th) | $8.0 \times 10^{-6}$ |
| | $^{214}$Bi | 10 | $3.1 \times 10^{-8}$ |
| | $^{212}$Bi | 10 ($^{232}$Th) | $5.1 \times 10^{-8}$ |
| | $^{88}$Y | 0.3 | $6.3 \times 10^{-7}$ |
| | $^{56}$Co | 0.06 | $6.2 \times 10^{-5}$ |
| $Zn^{100}MoO_4$ crystal surface | $^{232}$Th | 0.5 | $1.2 \times 10^{-5}$ |
| | $^{238}$U | 2.4 | $1.5 \times 10^{-4}$ |
| Cu holder | $^{232}$Th | 20 | $1.3 \times 10^{-6}$ |
| | $^{214}$Bi | 70 | $1.5 \times 10^{-7}$ |
| | $^{56}$Co | 0.2 | $6.6 \times 10^{-5}$ |
| PTFE clamps | $^{232}$Th | 100 | $9.6 \times 10^{-6}$ |
| | $^{214}$Bi | 60 | $7.5 \times 10^{-7}$ |
| BoPET reflective foil | $^{232}$Th | 100 | $7.5 \times 10^{-5}$ |
| | $^{214}$Bi | 60 | $2.1 \times 10^{-5}$ |
| **Total** | | | $4.1 \times 10^{-4}$ |

It should be stressed, that background from bulk and surface contamination of $Zn^{100}MoO_4$ crystal by $^{208}$Tl is suppressed by factor $10^3$ obtained with delayed coincidences (see subsection 3.3.3). Moreover, the reported background rates from α particles takes into account the suppression by pulse-shape discrimination with efficiency 99.9%.

The total background rate in the region of interest is $4.1 \times 10^{-4}$ counts/(keV·kg·yr) for $Zn^{100}MoO_4$ crystal scintillators, which corresponds to 2.5 counts/(ton·yr) in a 6 keV window centered at the 0ν2β peak position.

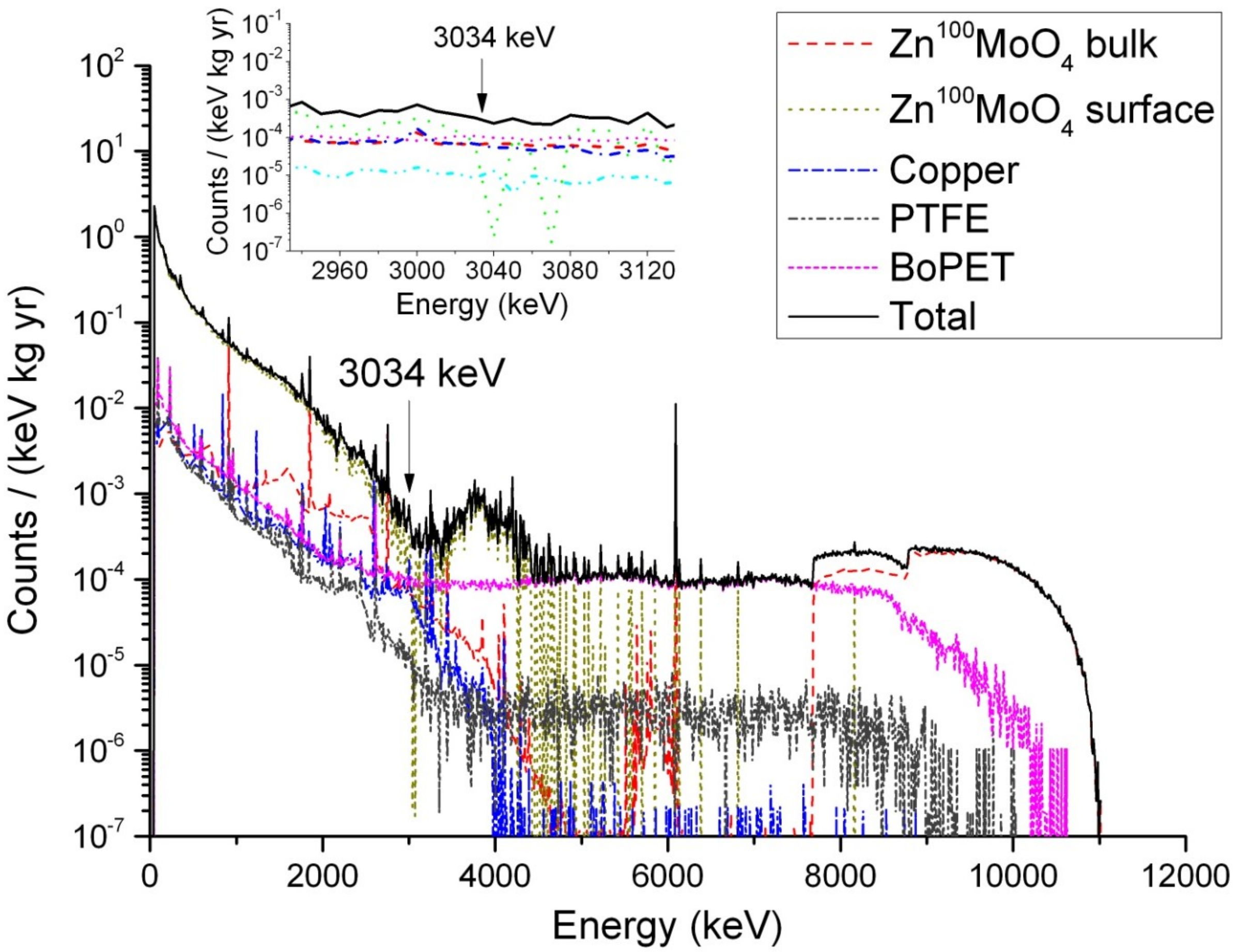

Fig. 3.35. Background spectra induced by $^{232}$Th (and daughters), $^{238}$U (and daughters), $^{88}$Y and $^{56}$Co bulk and surface contamination in the Zn$^{100}$MoO$_4$ crystal, internal contamination of Cu holder, PTFE clamps and BoPET reflective foil.

### 3.4. Conclusions and perspectives

The performed GEANT4 simulation of the light collection from ZnMoO$_4$ crystal scintillators demonstrated the advantage of hexagonal crystal shape over the cylindrical one, as well as the diffused crystal surface over the polished one. Therefore, the best light collection in the scintillating bolometers can be achieved by using a hexagonal crystal shape (or at least the octahedral as an intermediate variant) with a diffused surface. Moreover, the photodetectors should have a reasonable size of the order of the crystal diameter for higher collection of scintillation photons. Finally, an optimal distance between the light reflector and ZnMoO$_4$ crystal surface is about 11 mm–15 mm.

The dependence of energy spectra on shape of ZnMoO$_4$ crystals was investigated, taking into account the irregular shape of the first samples of Zn$^{100}$MoO$_4$ crystals. 3D scanning was applied to build geometrical model of the Zn$^{100}$MoO$_4$ crystals. The final 3D model has visible shape differences in comparison to the Zn$^{100}$MoO$_4$ crystal scintillator, which can be explained by imperfect equipment used for the scanning. However, the Monte Carlo simulation of 2ν2β decay processes in Zn$^{100}$MoO$_4$ crystals of different shape (hexagonal, octahedral, and cylindrical) demonstrated no significant dependence of the obtained spectra from the crystal shape, which allows to use a simplified model of the crystal for the GEANT4 simulations in case of irregular scintillator shape.

Monte Carlo simulation of 48 $Zn^{100}MoO_4$ crystal scintillators working as scintillating bolometers in the EDELWEISS set-up was performed. Main sources of background in the ROI were simulated as the contamination of $Zn^{100}MoO_4$ crystal and nearest materials (copper holder, PTFE clamps, and BoPET reflective foil) by $^{238}U$ and $^{232}Th$ daughters, and cosmogenic activation. A total background rate in the region of interest is $4.1 \times 10^{-4}$ counts/(keV·kg·yr) for $Zn^{100}MoO_4$ crystal scintillators, which makes scintillating bolometers based on $Zn^{100}MoO_4$ crystals an extremely perspective double beta decay detectors able to explore the inverted hierarchy region of neutrino mass pattern.

# CHAPTER 4

# REJECTION OF THE RANDOMLY COINCIDENT EVENTS IN 0ν2β DECAY EXPERIMENTS WITH $ZnMoO_4$ CRYOGENIC BOLOMETERS

## 4.1. Randomly coincident events in cryogenic bolometers as a source of background

A disadvantage of the low temperature bolometers is their poor time resolution, which can lead to a considerable background at the energy $Q_{2\beta}$ due to a random coincidences of the events with a lower energy, especially caused by the unavoidable two-neutrino 2β decay [ 199 , 200 ]. This problem is extremely important for the low-background cryogenic experiments aiming to search for 0ν2β decay of $^{100}$Mo, because of the short two-neutrino double beta decay half-life $T_{1/2}(^{100}\text{Mo}) = 7.1 \times 10^{18}$ yr [201].

As it will be shown below, randomly coincident events in $Zn^{100}MoO_4$ cryogenic bolometers can be even the main source of background in future large scale high radiopurity bolometric experiments.

### 4.1.1. Background from random coincidences of 2ν2β decay events

To estimate background level from randomly coincident (rc) two-neutrino 2β decay events in cryogenic bolometers we should built energy distribution of these randomly coinciding 2ν2β events.

The energy spectrum of β particles emitted in 2ν2β decay can be described by the two-dimensional distribution $\rho_{12}(t_1, t_2)$ [5]:

$$\rho_{12}(t_1,t_2) = e_1 p_1 F(t_1,Z) \cdot e_2 p_2 F(t_2,Z) \cdot (t_0 - t_1 - t_2)^5, \quad (4.1)$$

where $t_i$ is the kinetic energy of the $i$-th electron (all energies here are in units of the electron mass $m_0c^2$), $t_0$ is the energy available in the 2β process, $p_i$ is the momentum of the $i$-th electron $p_i = \sqrt{t_i(t_i+2)}$ (in units of $m_0c$), and $e_i = t_i + 1$. The Fermi function $F(t,Z)$, which takes into account the influence of the electric field of the nucleus on the emitted electrons, is defined as

$$F(t,Z) = const \cdot p^{2s-2} \exp(\pi\eta) \, |\Gamma(s+i\eta)|^2, \quad (4.2)$$

where $s = \sqrt{1-(\alpha Z)^2}$, $\eta = \alpha Z e/p$, $\alpha = 1/137.036$, $Z$ is the atomic number of the daughter nucleus ($Z > 0$ for $2\beta^-$ and $Z < 0$ for $2\beta^+$ decay), and $\Gamma$ is the gamma function.

The distribution $\rho(t)$ for the sum of electron energies $t = t_1 + t_2$ is obtained by integration:

$$\rho(t) = \int_0^t \rho_{12}(t - t_2, t_2) dt_2. \quad (4.3)$$

The distribution $\rho(t)$ for $^{100}$Mo is shown in Fig. 4.1.

The Primakoff–Rosen (PR) approximation for the Fermi function $F(t,Z) \sim e/p$ [202], which is adequate for $Z > 0$, allows us to simplify Eq. (4.1) to the expression

$$\rho_{12}^{\text{PR}}(t_1,t_2) = (t_1+1)^2 (t_2+1)^2 (t_0 - t_1 - t_2)^5 \quad (4.4)$$

and to obtain the formula for $\rho(t)$ analytically:

$$\rho^{\text{PR}}(t) = t(t_0 - t)^5 (t^4 + 10t^3 + 40t^2 + 60t + 30). \quad (4.5)$$

The energy distribution for two randomly coincident 2ν2β decays $\rho_{\mathrm{rc}}(t)$ can be obtained by numerical convolution

$$\rho_{\mathrm{rc}}(t) = \int_0^t \rho(t-x)\rho(x)dx, \qquad (4.6)$$

or with Monte Carlo method by sampling energy releases in two independent 2ν2β events in accordance with the distribution (4.3) and adding them. Assuming an ideal energy resolution of the detector, the energy spectrum obtained by sampling $10^8$ coincident 2ν2β events for $^{100}$Mo is shown in Fig. 4.1. This distribution can be approximated by the following compact expression, similar to that reported in Eq. (4.5):

$$\rho_{\mathrm{rc}}(t) = t^3(2t_0 - t)^{10} \sum_{i=0}^{8} a_i t^i. \qquad (4.7)$$

The coefficients $a_i$ depend on the isotopes and its values for $^{82}$Se, $^{100}$Mo, $^{116}$Cd, and $^{130}$Te (as the most promising 2β isotopes for bolometric experiments) are given in Table 4.1. The approximation for $^{100}$Mo is shown in Fig. 4.1.

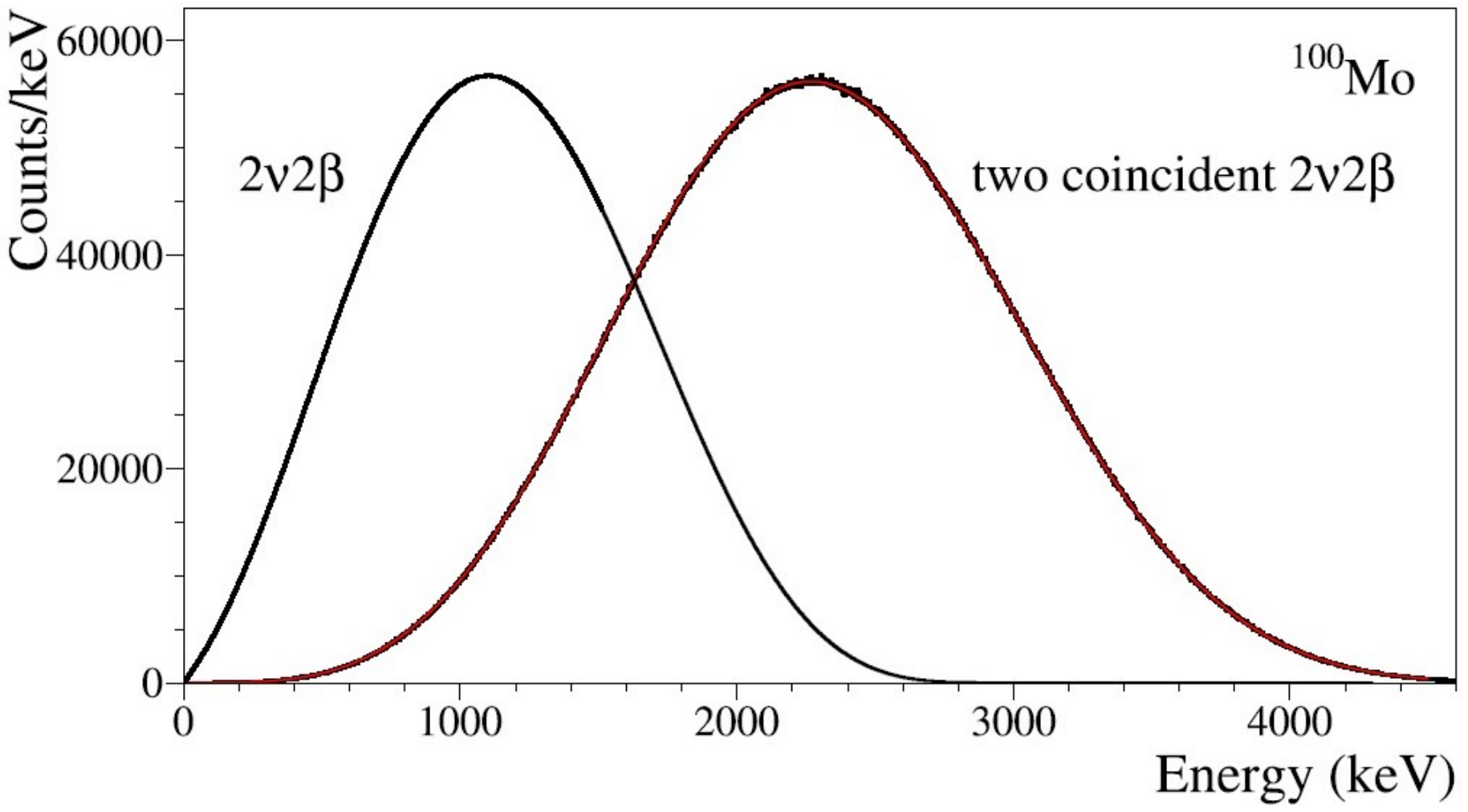

Fig. 4.1. The energy distribution for the sum of energies of two electrons emitted in 2ν2β decay of $^{100}$Mo, and energy spectrum of $10^8$ two randomly coinciding 2ν2β events for $^{100}$Mo obtained by Monte Carlo sampling. The approximation of the random coincidence spectrum by the expression (4.7) is shown by the solid red line.

Table 4.1
The values of the coefficients $a_i$ in Eq. (4.7) for the energy distribution of two randomly coinciding 2ν2β events for $^{82}$Se, $^{100}$Mo, $^{116}$Cd, and $^{130}$Te.

| $a_i$ | Isotope | | | |
|---|---|---|---|---|
| | $^{82}$Se | $^{100}$Mo | $^{116}$Cd | $^{130}$Te |
| $a_0$ | 3446.59 | 5827.48 | 20093.8 | 15145.6 |
| $a_1$ | −7746.37 | −14399.7 | −1318.65 | 5554.78 |
| $a_2$ | 22574.7 | 34128.1 | 69134.6 | 39930.0 |
| $a_3$ | −16189.1 | −23815.5 | −31971.3 | −13338.5 |
| $a_4$ | 8467.01 | 11271.7 | 17976.8 | 7689.45 |
| $a_5$ | −2156.83 | −2711.31 | −3486.40 | −813.887 |
| $a_6$ | 337.172 | 390.396 | 406.762 | −80.3126 |
| $a_7$ | −28.9146 | −30.8774 | −30.5846 | 19.1035 |
| $a_8$ | 1 | 1 | 1 | −1 |

The random coincidence counting rate $I_{rc}$ in a chosen energy interval $\Delta E$ is determined by the time resolution of the detector $\tau$ and the counting rate for single 2ν2β decay events $I_0$:

$$I_{rc} = \tau \cdot I_0^2 \cdot \varepsilon = \tau \cdot \left( \frac{\ln 2N}{T_{1/2}^{2\nu 2\beta}} \right)^2 \cdot \varepsilon, \tag{4.8}$$

where $N$ is the number of 2β decaying nuclei under investigation, and $\varepsilon$ is the probability of registration of events in the energy interval $\Delta E$. In Eq. (4.8) and in the following, we assume that if two events occur in the detector within a temporal interval lower than the time resolution $\tau$, they give rise to a single signal with amplitude equal to the sum of the amplitudes expected for the two separated signals. The calculated probabilities at the energy $Q_{2\beta}$ of the 0ν2β decay for the $\Delta E$ = 1 keV interval are equal to $\varepsilon = 3.5 \times 10^{-4}$ for $^{82}$Se and $\varepsilon = 3.3 \times 10^{-4}$ for $^{100}$Mo, $^{116}$Cd, and $^{130}$Te.

Counting rates of two randomly coincident 2ν2β events for the detectors with typical large mass bolometer volume of 100 cm$^3$ at $Q_{2\beta}$ for different 2β candidates and compounds are presented in Table 4.2. In our calculations we suppose that isotopic enrichment for $^{82}$Se, $^{100}$Mo, and $^{116}$Cd is on the level of 100 %, while for $^{130}$Te natural abundance (34.08 %) is taken. Time resolution of one detector is set to $\tau$ = 1 ms. The reported rates scale linearly with the time resolution.

Table 4.2
Counting rate of two randomly coinciding 2ν2β events in cryogenic Zn$^{82}$Se, $^{40}$Ca$^{100}$MoO$_4$, Zn$^{100}$MoO$_4$, $^{116}$CdWO$_4$, and TeO$_2$ detectors of 100 cm$^3$ volume. Isotopic enrichment for $^{82}$Se, $^{100}$Mo, and $^{116}$Cd is assumed to be 100 %, for $^{130}$Te the natural abundance (34.08 %) is taken. $C$ is the mass concentration of the isotope of interest, $\rho$ is the density of the material (g/cm$^3$), $N$ is the number of 2β candidate nuclei in one detector, and $B_{rc}$ is the counting rate at $Q_{2\beta}$ (counts/(keV·kg·yr)) under the assumption of 1 ms time resolution of the detector.

| Isotope | $T_{1/2}^{2\nu 2\beta}$ (yr) [62] | Detector ($\rho$) | $C$ | $N$ | $B_{rc}$ |
|---|---|---|---|---|---|
| $^{82}$Se | $9.2 \times 10^{19}$ | Zn$^{82}$Se (5.65) | 55.6 % | $2.31 \times 10^{24}$ | $5.9 \times 10^{-6}$ |
| $^{100}$Mo | $7.1 \times 10^{18}$ | $^{40}$Ca$^{100}$MoO$_4$ (4.35) | 49.0 % | $1.28 \times 10^{24}$ | $3.8 \times 10^{-4}$ |
| | | Zn$^{100}$MoO$_4$ (4.3) | 43.6 % | $1.13 \times 10^{24}$ | $2.9 \times 10^{-4}$ |
| $^{116}$Cd | $2.8 \times 10^{19}$ | $^{116}$CdWO$_4$ (8.0) | 31.9 % | $1.32 \times 10^{24}$ | $1.4 \times 10^{-5}$ |
| $^{130}$Te | $6.8 \times 10^{20}$ | TeO$_2$ (5.9) | 27.2 % | $0.76 \times 10^{24}$ | $1.1 \times 10^{-8}$ |

The counting rate $B_{rc}$ of background events at $Q_{2\beta}$ due to the random coincidences of 2ν2β decay signals depends on the time resolution of the detector $\tau$, the energy resolution $R$, the volume of the detector $V$, and the abundance or enrichment $\delta$ of the candidate nuclei contained in the detector with the following relation:

$$B_{rc} \sim \tau \cdot R \cdot \left( T_{1/2}^{2\nu 2\beta} \right)^{-2} \cdot V^2 \cdot \delta^2. \tag{4.9}$$

Overall, from the Table 4.2 we can conclude that the most crucial random coincidence background of 2ν2β decay events is for $^{100}$Mo due to its relatively short 2ν2β half-life. Obviously, any other source of background with high enough energy can contribute due to the random coincidences. Such background level will be estimated in the next subsection.

### 4.1.2. Background from randomly coinciding events in $ZnMoO_4$ cryogenic bolometers

To estimate the background level from randomly coinciding events in $ZnMoO_4$ bolometers we should take into account other sources of background (in addition to the 2ν2β decay) which can contribute to the region of interest due to random coincidences. One of such possible sources is external gamma background. We have estimated its contribution taking into account the level of background already achieved in the CUORICINO cryogenic detector [93]. A simplified model of the CUORICINO background was taken from [93] in the energy interval 300 keV–2620 keV. The background below 300 keV was extrapolated by an exponential function, while above 2620 keV it was taken to be zero (see Fig. 4.2). Monte Carlo simulated energy spectrum of two randomly coincident events from external gamma background is presented in Fig. 4.2.

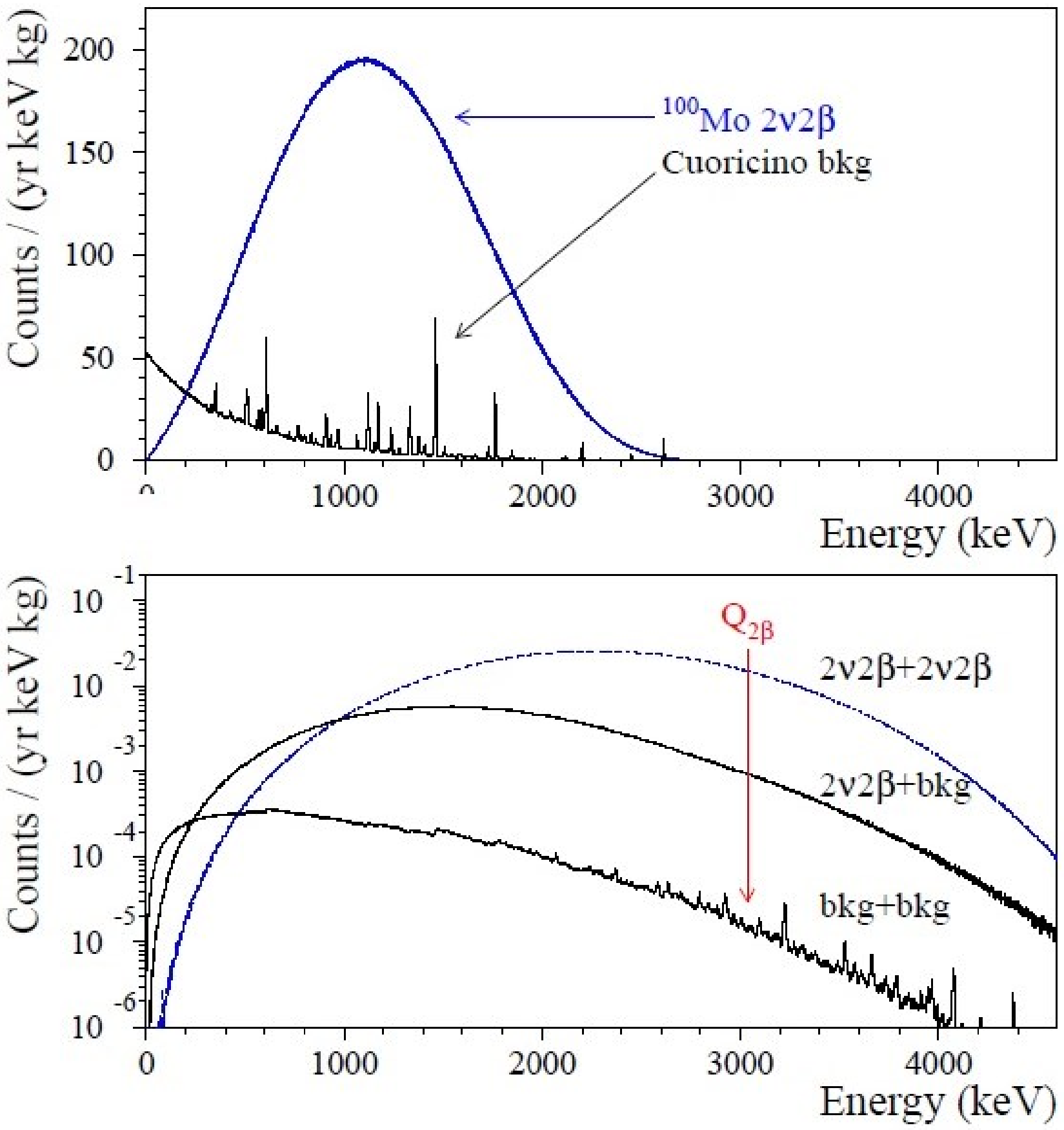

Fig. 4.2. Distribution of the sum of energies of two electrons emitted in 2ν2β decay of $^{100}Mo$ and model of the background energy spectrum from external gamma quanta [93] (upper panel). Monte Carlo simulated energy spectra of two randomly coincident 2ν2β events, coincidence of 2ν2β events with external gamma events, and randomly coinciding external gamma events (lower panel).

We also built the energy distribution of the randomly coinciding 2ν2β events by using the approach already described in subsection 4.1.1, but now under the assumption that two events are not resolved in the time interval 45 ms. This time interval was chosen taking into account typical rise-times observed in large mass bolometers, like those operated in the CUORICINO experiment. The energy spectra of $10^9$ Monte Carlo generated 2ν2β decay events of $^{100}$Mo and of two randomly coinciding 2ν2β events are presented in Fig. 4.2. The rate of the randomly coinciding 2ν2β events was calculated for the $Zn^{100}MoO_4$ crystal with a 100% enrichment by $^{100}$Mo and typical size of Ø60 × 40 mm by using Eq. (4.8), where time resolution of the detector is $\tau = 45$ ms, the number of 2β decaying nuclei $N = 1.3 \times 10^{24}$, half-life of the two-neutrino double beta decay of $^{100}$Mo $T_{1/2}^{2\nu2\beta} = 7.1 \times 10^{18}$ yr.

Finally we simulated coincidences of the CUORICINO background with the 2ν2β decay of $^{100}$Mo as another possible source of background (see Fig. 4.2). Supposing radiopure $Zn^{100}MoO_4$ crystal scintillators and a cryostat with the level of radioactive contamination already achieved in the CUORICINO set-up, we can conclude from the Fig. 4.2, that the main contribution to the background is expected to be from the 2ν2β decay of $^{100}$Mo. The total counting rate due to the randomly coinciding 2ν2β decay events and external gamma events in the region of interest is estimated as ≈ 0.016 counts/(keV·kg·yr) in the 10 keV ROI around 3034 keV for Ø60 × 40 mm $Zn^{100}MoO_4$ cryogenic bolometer with time resolution 45 ms.

### 4.2. Generation of randomly coinciding signals

Single and randomly coincident signals were generated by using pulse profiles and noise baselines accumulated with 313 g $ZnMoO_4$ crystal scintillator operated as a cryogenic scintillating bolometer with Ge light detector [203] at the Centre de Sciences Nucléaires et de Sciences de la Matière (CSNSM, Orsay, France) and at the Laboratoire Souterrain de Modane (LSM, Modane, France). Data of the two measurements have been used: the first one performed at the CSNSM with a sampling rate of 5 kSPS (kilosamples per second) both for the light and heat channels, and the second one at the LSM with a sampling rate 1.9841 kSPS for both channels. A few hundred pulse profiles of heat and light signals with the energy of a few MeV were selected from the CSNSM data. Ten thousand of baseline samples were selected from the LSM data.

Sums of the few hundred pulse profiles both from heat and light channels were fitted with the following phenomenological function:

$$f_S(t) = A \cdot \left(e^{-t/\tau_1} + e^{-t/\tau_2} - e^{-t/\tau_3} - e^{-t/\tau_4}\right), \tag{4.10}$$

where $A$ is the amplitude, $\tau_1$, $\tau_2$, $\tau_3$ and $\tau_4$ are the time constants.

Amplitude and time constants used for the fit were set manually as initial parameters. Then the pulse profiles were adjusted by the least squares method using Levenberg–Marquardt algorithm. Amplitude of the $f_S(t)$ function was normalized on the maximum amplitude of the pulse.

To generate randomly coinciding signals in the region of the $Q_{2\beta}$ value of $^{100}$Mo, the amplitude of the first pulse $A_1$ was obtained by sampling the 2ν2β distribution for $^{100}$Mo, while the amplitude of the second pulse was chosen as $A_2 = Q_{2\beta} - A_1 + \Delta E$, where $\Delta E$ is a random component in the energy interval [−5, +5] keV (which is a typical energy resolution of cryogenic bolometers).

Ten thousand coinciding signals (built by using the data accumulated with the sampling rate 1.9841 kSPS) were randomly generated in the time interval from 0 to $3.3 \cdot \tau_R$ ($\Delta t = [0, 3.3 \cdot \tau_R]$), where $\tau_R$ is the rise-time of the signals (defined as the time needed to increase the pulse amplitude from 10% to 90% of its maximum). As it will be demonstrated in the section 4.3.2, the rejection efficiency of randomly coinciding signals (RE, defined as the part of the pile-up events rejected by pulse-shape discrimination) reaches almost its maximal value when the time interval of consideration exceeds (3–4)$\tau_R$. Ten thousand single signals were also generated.

We define signal-to-noise ratio as the ratio of the maximum signal amplitude to the standard deviation of the noise baseline, and set its value to 30 for light signals and 900 for heat signals as typical values achieved with $ZnMoO_4$ scintillating bolometers. Examples of the generated single light and heat pulses are presented in Fig. 4.3.

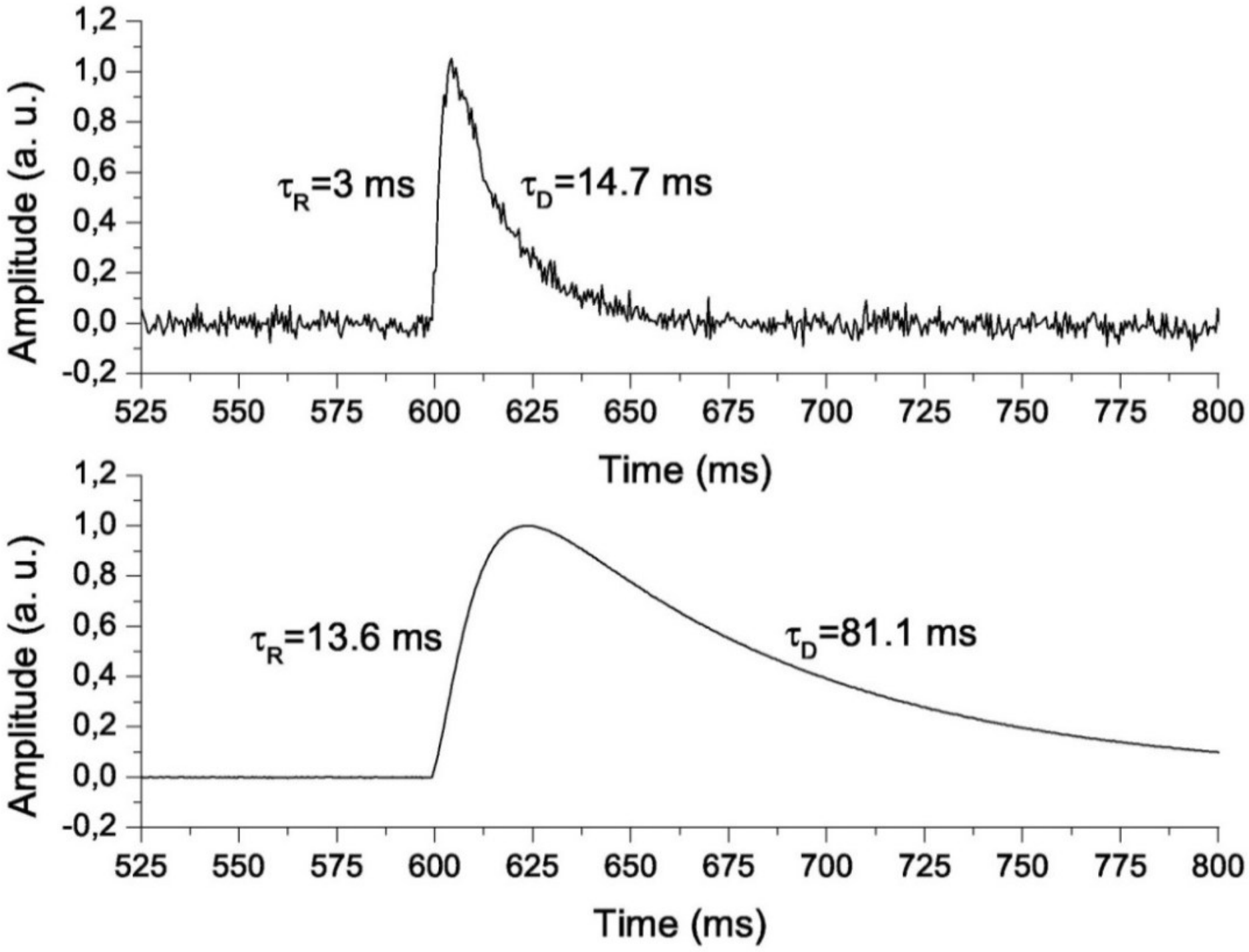

Fig. 4.3. Examples of the generated light (upper panel) and heat (lower panel) single pulses. $\tau_R$ and $\tau_D$ denote rise- and decay-times, respectively.

### 4.3. Methods of pulse-shape discrimination

To discriminate randomly coincident events in cryogenic bolometers we have applied three techniques: mean-time method, $\chi^2$ approach, and front edge analysis. We demanded 95% efficiency in accepting single pulses as potentially good 0ν2β signals. As a first step of pulse-shape analysis, one should develop a method to determine the time of signal origin..

#### 4.3.1. Reconstruction of the time origin of the events

To reconstruct the time origin of each event from our data the following procedure was used:

1. Preliminary search for the presence of a signal by a very simple algorithm, which searches for a channel where the signal amplitude exceeds a certain level (typically about one third of the signal maximum value);

2. Summation of the data over a certain number of channels (typically over 2-6 channels for the light signals (see Fig. 4.4), depending on the time structure of signal and noise data; this procedure was not necessary to use for the heat signals thanks to the high signal-to-noise ratio);

3. Calculation of the standard deviation of the integrated signal baseline fluctuations;

4. Search for the pulse start under the request that the signal exceeds the threshold set on a certain number of standard deviations of the baseline and – in a case of heat signals – the amplitude in the following two channels increases channel by channel, with the pulse start taken one channel earlier from the calculated one.

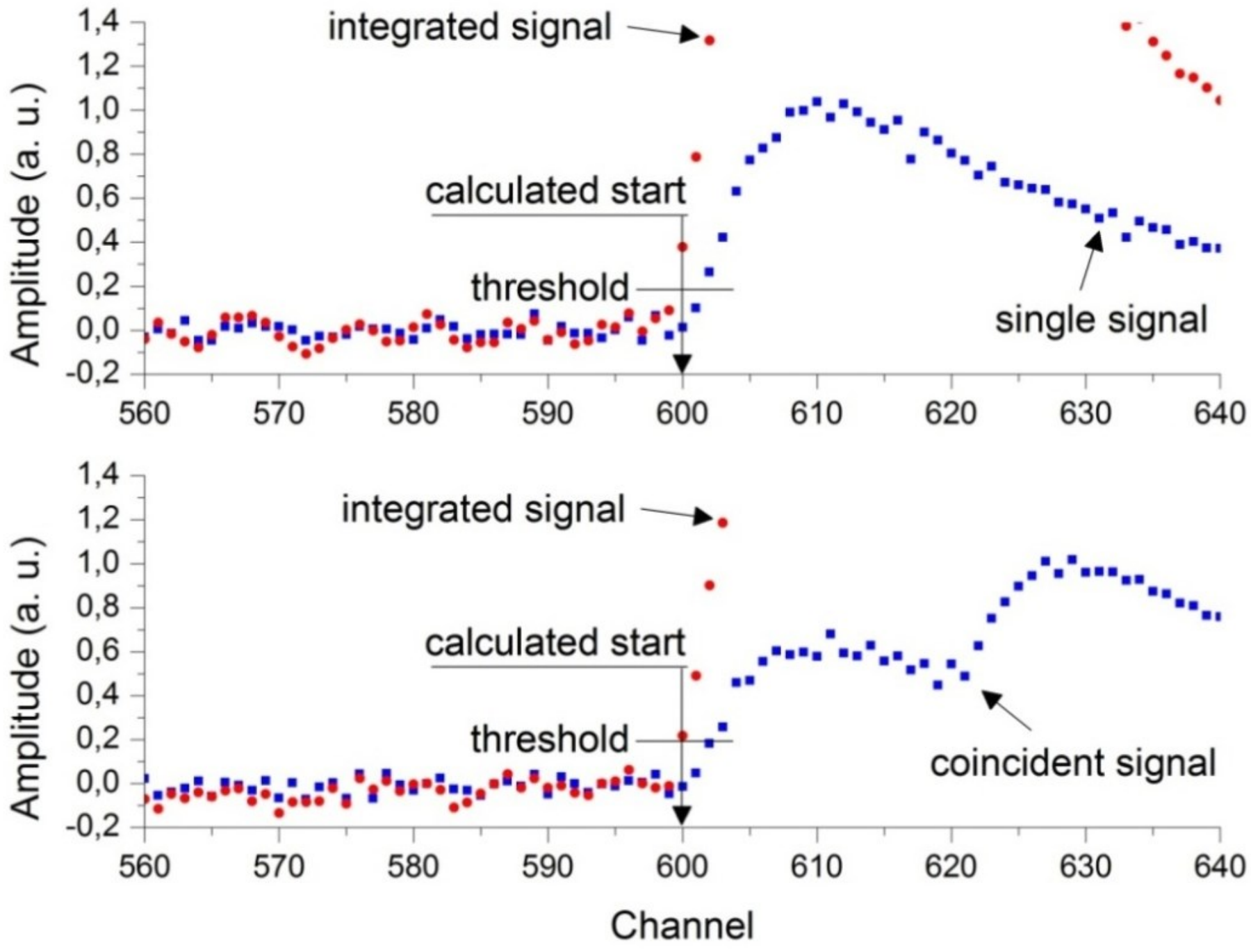

Fig. 4.4. Integrated signals (red circles) of the single (upper panel) and coincident (lower panel) light pulses (blue squares) used for the determination of the pulse origin. Pulse start was calculated under the request that integrated signal exceeds the threshold set as four standard deviations of its baseline.

We have performed optimization of the algorithm taking into account time properties of the signals and noise baselines, sampling rate and signal-to-noise ratio. For example, optimization of the algorithm to reconstruct the time origin of the light signals over the number of channels for data summation is presented in Fig. 4.5.

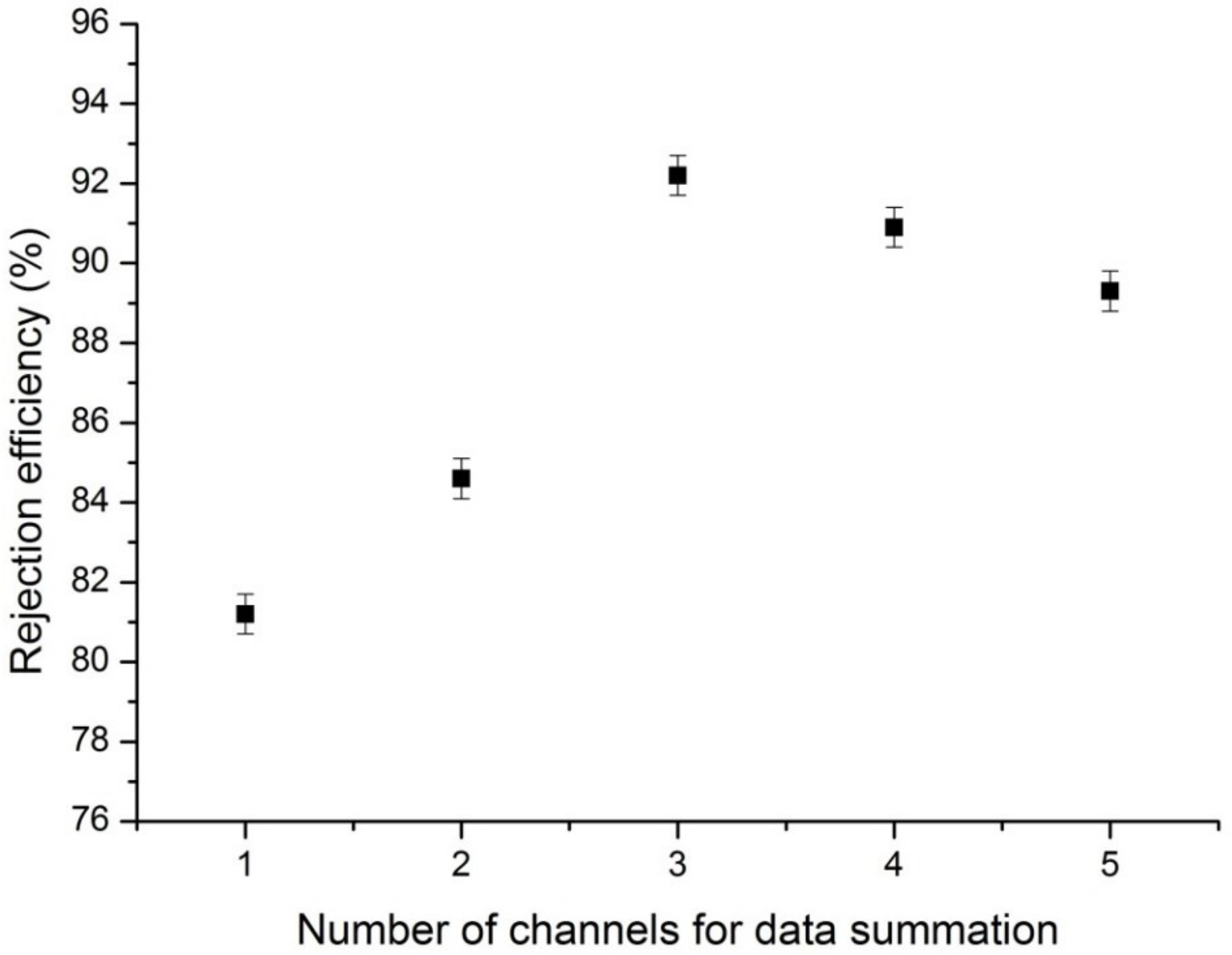

Fig. 4.5. Dependence of the rejection efficiency (by using the mean-time method) on the number of channels to sum the data. Integrated signals were used in the algorithm to reconstruct the time origin of events. The analysis was performed for the ten thousand single and coincident light pulses with 3 ms rise-time.

### 4.3.2. Mean-time method

For each pulse $f(t_k)$ we calculate the parameter $\langle t \rangle$ (mean-time) with the following formula:

$$\langle t \rangle = \sum f(t_k) t_k / \sum f(t_k), \quad (4.11)$$

where the sum is over time channels $k$, starting from the origin of a pulse and up to a certain time.

To analyze efficiency of the pulse-shape discrimination we have chosen the time interval $\Delta t$ where the randomly coincident signals were generated. Six sets of ten thousand single and randomly coinciding light (with $\tau_R = 3$ ms) and heat events (with $\tau_R = 13.6$ ms) were generated in the time intervals between two coincident pulses ranging from 0 to a maximum value, varying from 1 to about 6 pulse rise-times. The results of the analysis are presented in Fig. 4.6. The uncertainties of the rejection efficiency were estimated by applying mean-time method to three sets of light and heat data generated by using three sets of different noise baseline profiles (about 3500 profiles in the each set). As we can see from Fig. 4.6 the rejection efficiency of randomly coinciding signals reaches its maximal value when the time interval $\Delta t$ is larger than (3–4)$\tau_R$ and remains unchanged with an increasing of $\Delta t$. As a conclusion all further analysis was done by using data generated in the time interval $\Delta t = [0, 3.3 \cdot \tau_R]$.

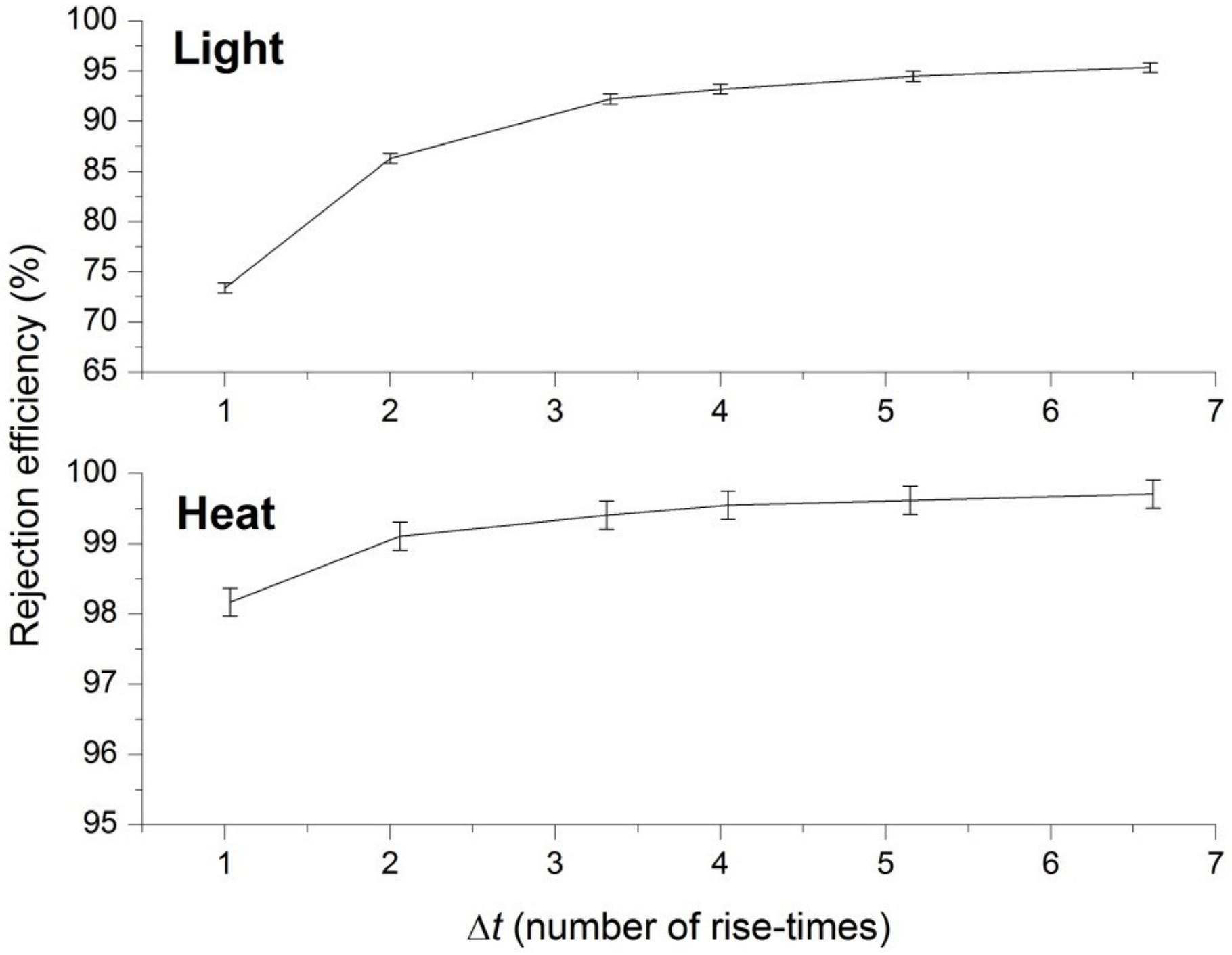

Fig. 4.6. Dependence of the rejection efficiency (by using the mean-time method) for heat and light channels on the time interval $\Delta t$ where the randomly coinciding signals were generated.

Typical distributions of the mean-time parameters for single and coincident events are presented in Fig. 4.7. Under the requirement to detect 95% of single events as potentially good signals, the rejection efficiency of randomly coinciding pulses is 92.2%.

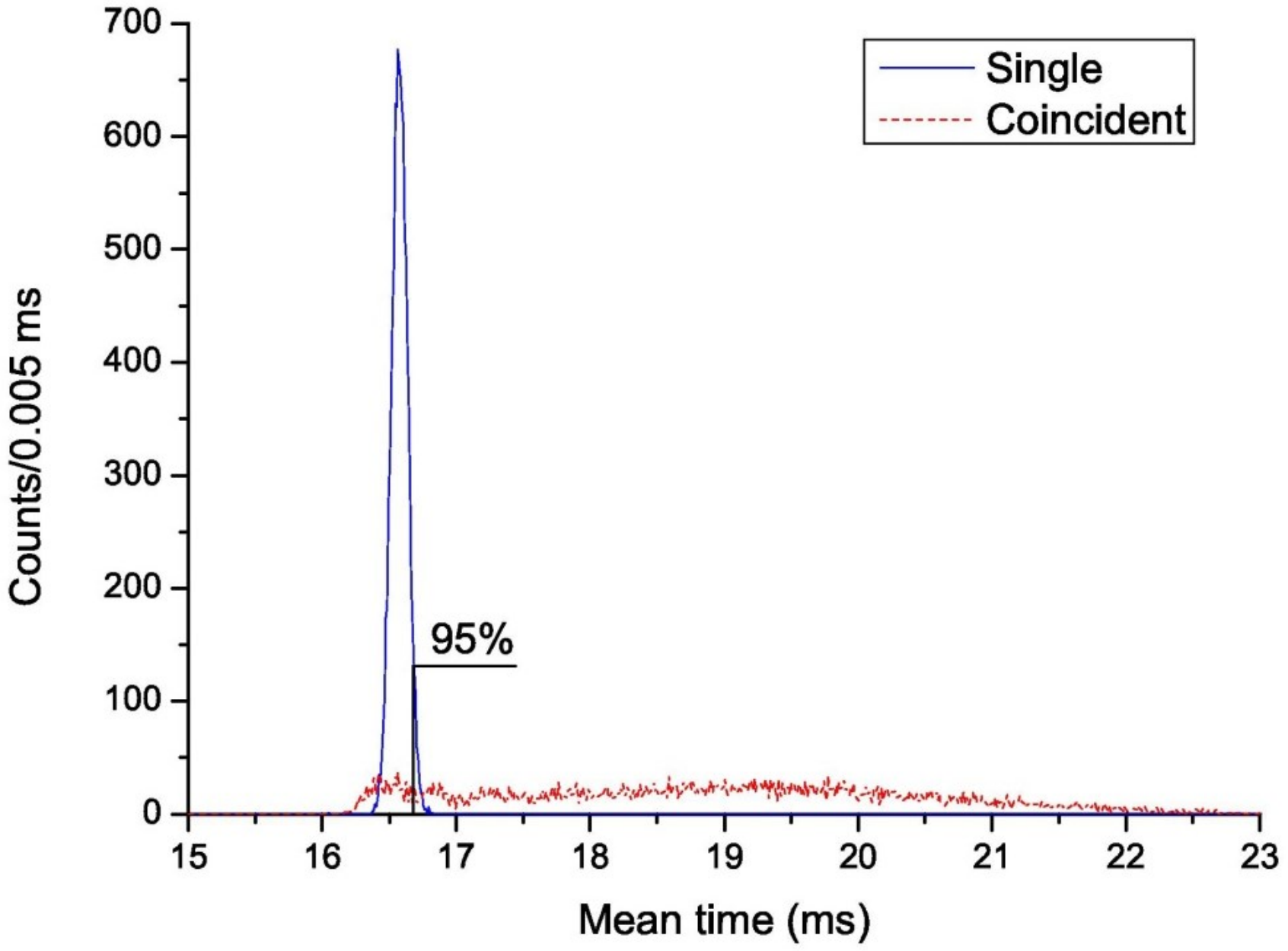

Fig. 4.7. Distribution of the mean-time parameter for single and coincident light pulses with a rise-time 3 ms. Rejection efficiency of coinciding pulses is 92.2%. The events left from the line are accepted as single events (95% of single events). 7.8% of pile-up events are in the "single" event region due to incorrect start finding and / or too small time difference between two coinciding signals.

From Eq. 4.11 we could expect that for the mean-time method the rejection efficiency of pulse-shape discrimination depends on the choice of the time interval used to calculate a discrimination parameter. The results of the mean-time parameter optimization are presented in Fig. 4.8. The rejection efficiency has its maximum value when the mean-time parameter is calculated from the signal origin to the 30th channel which approximately corresponds to the pulse decay time. Other pulse-shape discrimination methods were optimized in a similar way.

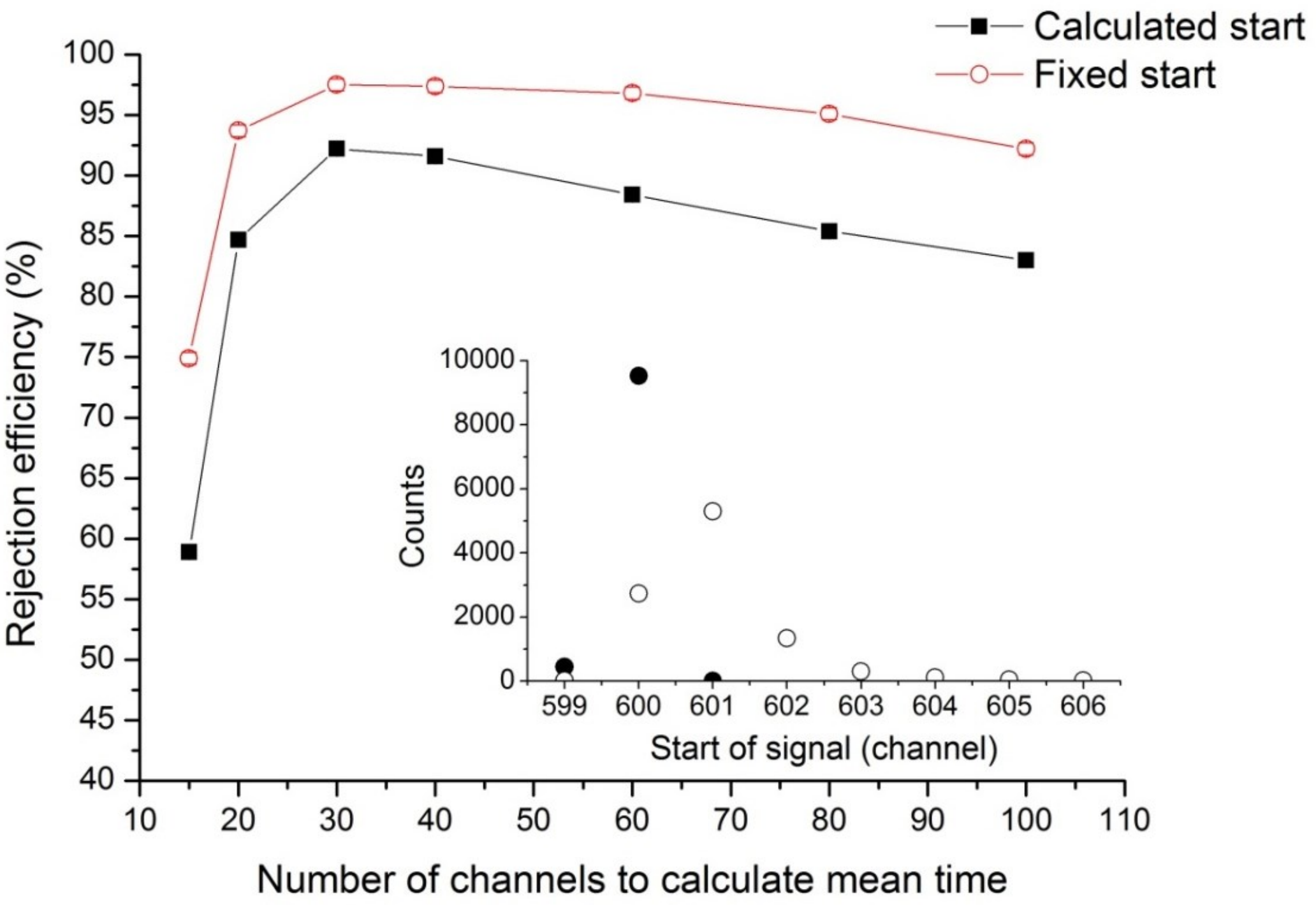

Fig. 4.8. Dependence of the rejection efficiency of the mean-time method on the number of channels used to calculate the parameter $\langle t \rangle$. The analysis was performed for the ten thousand single and coincident light signals with 3 ms rise-time. The rejection efficiency of randomly coinciding pulses is 92.2% for the cases when the start of the signals was found by algorithm described in subsection 4.3.1 (squares), and 97.5% using the known start position (circles). One channel is 0.504 ms. (Inset) Distribution of the calculated by our algorithm start positions for the sets of single (filled circles) and randomly coinciding (open circles) events.

Optimization of the rejection efficiency presented in Fig. 4.8 also demonstrates the importance to develop and optimize the algorithm to find correctly the start of the signals (see subsection 4.3.1). This problem is especially significant for the light pulses for which the signal-to-noise ratio is comparatively low. Obviously rejection efficiency is substantially higher when the start position of each pulse is known from the generation algorithm (97.5% for known start versus 92.2% for the calculated one). The distributions of the calculated start positions for the sets of single and randomly coinciding events are shown in Inset of Fig. 4.8. The distribution demonstrates that correct finding of the start position for the randomly coinciding signals (especially if the first pulse is small) is more complicated than that for the single events.

### 4.3.3. $\chi^2$ method

The following formula was used to calculate the $\chi^2$ parameter for each pulse $f(t_k)$:

$$\chi^2 = \sum(f(t_k) - f_S(t_k))^2, \quad (4.12)$$

where the sum is over time channels $k$, starting from the origin of pulse and up to a certain time, $f_S(t)$ is the shape of the signal defined by Eq. (4.10). The number of channels to calculate the $\chi^2$ parameter has been optimized to reach maximal rejection efficiency (see Fig. 4.9). The results of the $\chi^2$ method with comparison between different methods will be presented in subsection 4.4.

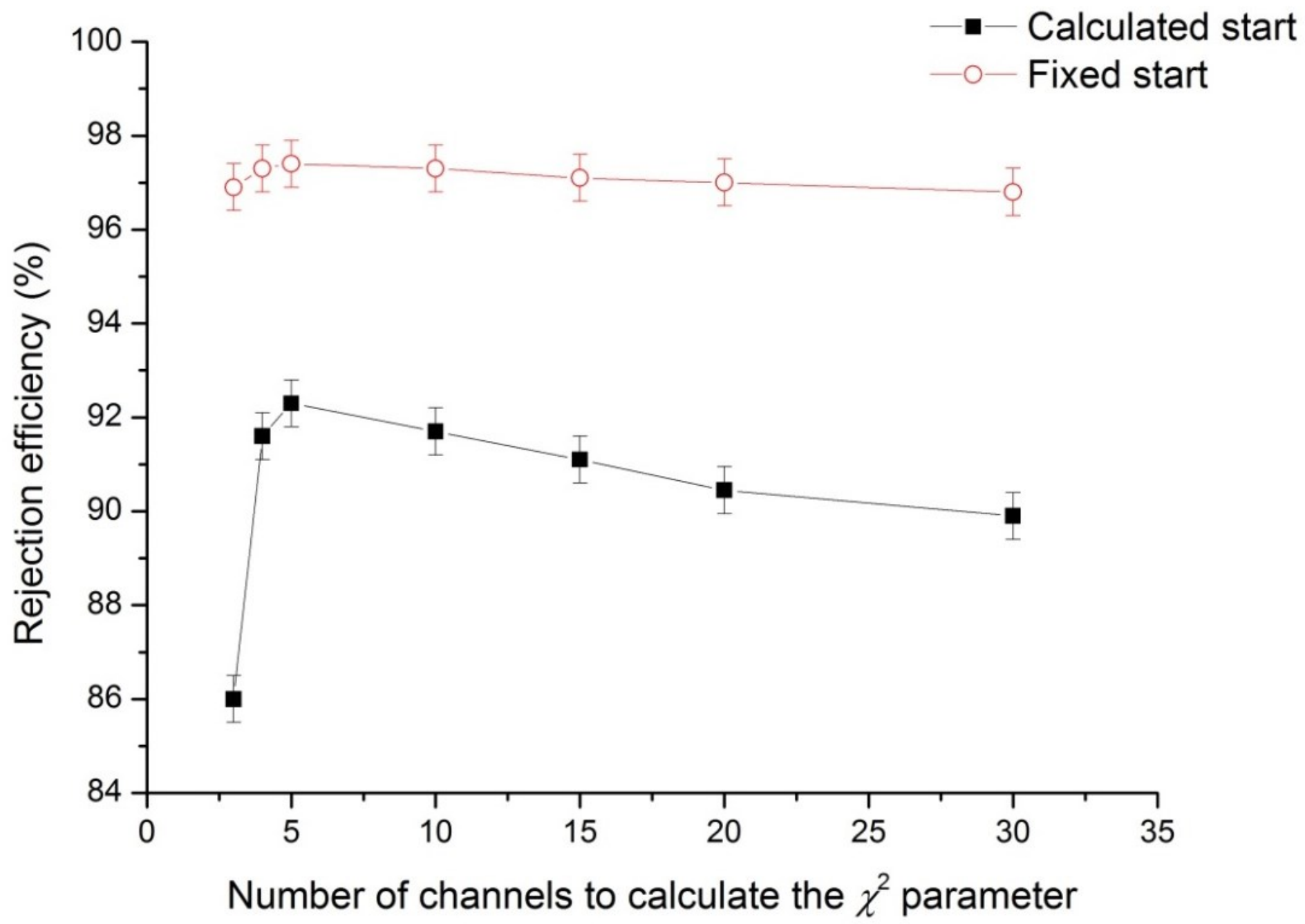

Fig. 4.9. Dependence of the rejection efficiency of the $\chi^2$ method on the number of channels used to calculate the $\chi^2$ parameter. The analysis was performed for the ten thousand single and coincident light signals with 3 ms rise-time. The rejection efficiency of randomly coinciding pulses is 92.3% for the cases when the start of the signals was found by algorithm described in subsection 4.3.1 (squares), and 97.4% using the known start position (circles). One channel is 0.504 ms.

### 4.3.4. Front edge analysis

The front edge parameter was defined as the time between two points on the pulse front edge with amplitudes $Y_1$% and $Y_2$% of the pulse amplitude. To provide the maximal rejection efficiency of coincident events we optimized the front edge parameter $Y_1$. For instance, the highest rejection efficiency for heat pulses with $\tau_R = 13.6$ ms was reached with the front edge parameter determined as the time between the signal origin and the time where the signal amplitude reaches $Y_2 = 90$% of its maximum (RE = 99.3%). For the light pulses due to the low signal-to-noise ratio we also applied the band-pass filter in addition to the optimization of the $Y_1$ parameter. The results of the front edge parameter optimization for light pulses with the rise-time $\tau_R = 3$ ms are presented in Fig. 4.10. Comparison between the different methods will be presented in subsection 4.4.

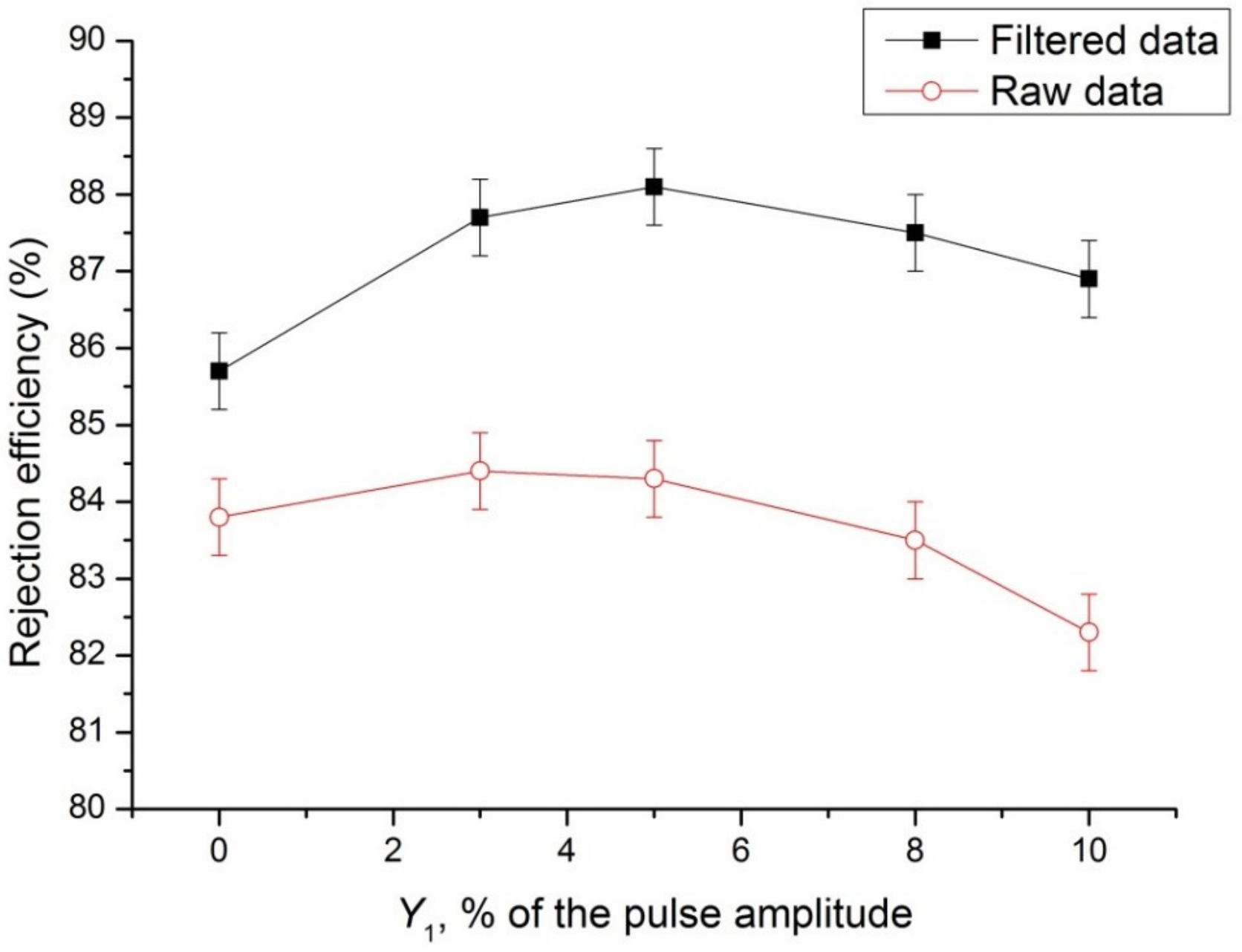

Fig. 4.10. Dependence of the rejection efficiency of the front edge method on the parameter $Y_1$ used to calculate signal front edge. The analysis was performed for the ten thousand filtered (by using band-pass filter) and raw, single and coinciding light pulses with 3 ms rise-time. The highest rejection efficiency of pile-up events was achieved for filtered pulses (RE = 88.1%).

However, the rejection efficiency of the front edge method is limited due to the fraction of randomly coinciding events with a small first (with the amplitude $A_1$ below $Y_1$) or second pulse (with the low amplitude, and appearing well after the first signal maximum).

### 4.4. Comparison of rejection efficiency of the pulse-shape discrimination methods

We have applied the three described above methods to ten thousand generated single and coincidence events both for heat and light channels. Comparison of the obtained results is presented in Table 4.3. The rejection efficiency of the methods was calculated with pulses start positions found by our algorithms, and using known start positions from the generation procedure. All the methods give rejection efficiency 88%–92% by using the light signals with a rise-time of 3 ms and around 99% for the much slower heat signals with a rise-time of 13.6 ms. One can conclude that the signal-to-noise ratio has a crucial role in the pulse-shape discrimination of randomly coinciding events in cryogenic bolometers. Analysis of signals with lower level of noise allows to reach a much higher rejection efficiency even with slower heat signals. Dependence of the rejection efficiency (by using the mean-time method) on the signal-to-noise ratio for heat signals confirms this assumption (see Fig. 4.11).

We also analyzed the dependence of the rejection efficiency on the rise-time of light pulses under an assumption that the pulse amplitude is not changed by shortening the rise-time (see Fig. 4.12). One could expect that for faster signal any of the methods should give a higher efficiency of pulse-shape discrimination. However, the tendency of the rejection efficiency improvement for faster signals is rather weak. For example, the rejection efficiency for pulses with the 2 ms rise-time is even worse in comparison to slower signals with the rise-times of 3 ms and 4.5 ms. Such result can be explained by a not high enough sampling rate (1.9841 kSPS) used for the data acquisition. Indeed, the pulse profiles acquired with the sampling rate are too discrete: for instance, the front edge of the signals with the rise-time 2 ms is represented by only 4 points.

Table 4.3
Rejection efficiency of randomly coinciding 2ν2β events by pulse-shape discrimination of light and heat signals for the two conditions of the signal start determination, i.e. (1) start of the signals known from the generation procedure, (2) start position found by the pulse profile analysis.

| Channel, rise-time | Start position | Mean-time method, % | Front edge analysis, % | $\chi^2$ method, % |
|---|---|---|---|---|
| Light, 3 ms | Known | 97.5 ± 0.5 | 96.4 ± 0.5 | 97.4 ± 0.5 |
| | Found | 92.2 ± 0.5 | 88.1 ± 0.5 | 92.3 ± 0.5 |
| Heat, 13.6 ms | Known | 99.4 ± 0.2 | 99.4 ± 0.2 | 99.4 ± 0.2 |
| | Found | 99.3 ± 0.2 | 99.3 ± 0.2 | 99.3 ± 0.2 |

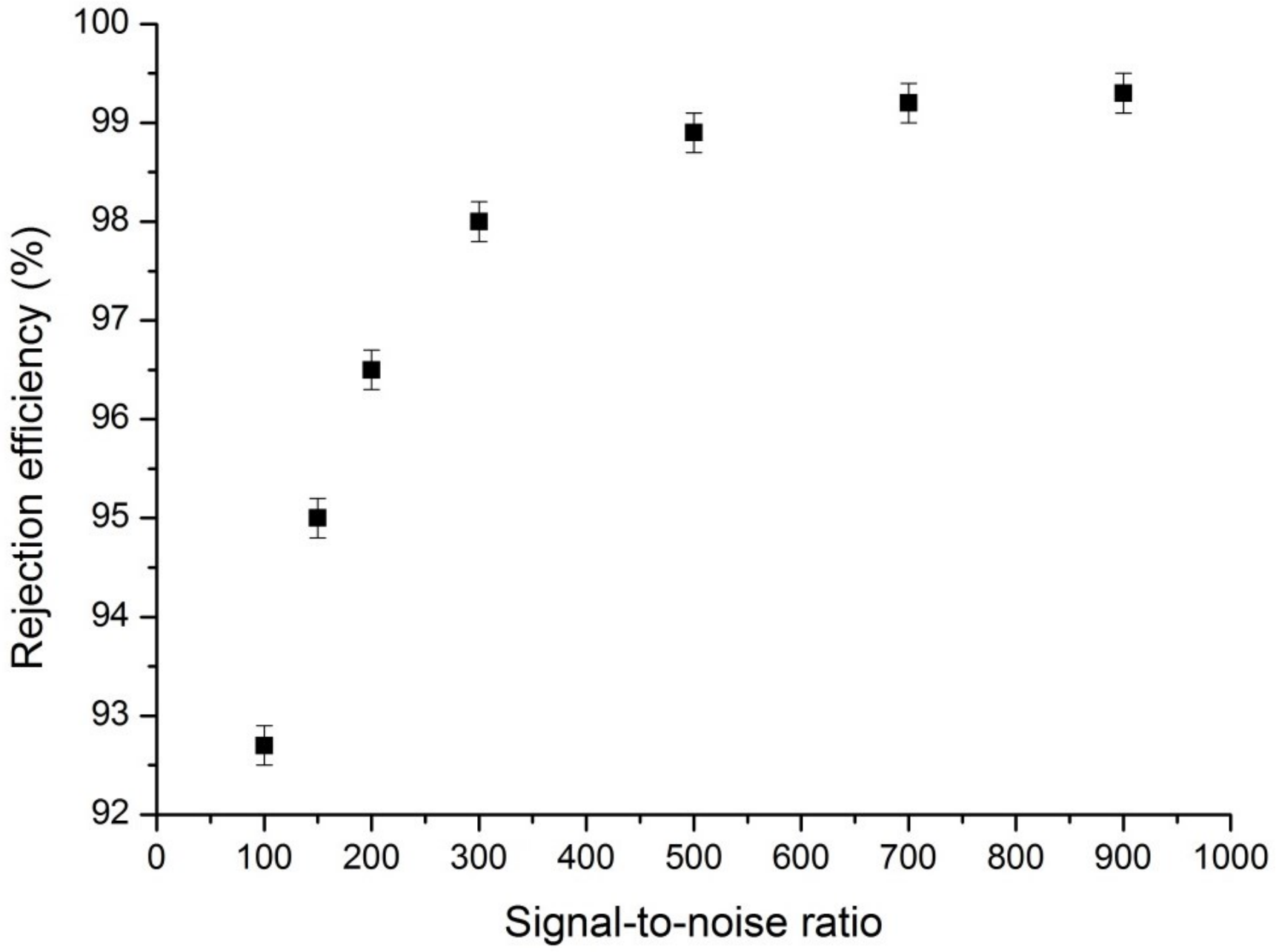

Fig. 4.11. Dependence of the rejection efficiency (using the mean-time method) on the signal-to-noise ratio for the heat channel.

The importance of the data acquisition sampling rate for the pulse-shape analysis was proved by the generation of single and coincident events with a two times lower sampling rate. The sampling rate of the noise baselines was decreased by factor 2 by averaging two neighbor channels to one. As one can see from Fig. 4.12 the rejection efficiency decreased.

Following the previous discussion at the end of subsection 4.1.2, we calculated how the developed pile-up rejection methods would improve the estimated background level of ≈ 0.016 counts/(keV·kg·yr) in an enriched $Zn^{100}MoO_4$ detector based on a crystal size of Ø60 × 40 mm. We remind that this value was obtained assuming that pulses with a time delay between them larger than 45 ms would be recognized as two signals, while two events within this time interval would produce one signal with an amplitude approximately equal to the sum of the individual signals. The resulting background level can be decreased by 99.3% in the heat channel and by 92.3% in the light channel, as these are our best pile-up rejection efficiencies with unknown pulse start position (see Table 4.3) [3] . The final estimated background level after the pile-up rejection using pulse-shape discrimination in the heat channel is ≈ $1.1 \times 10^{-4}$ counts/(keV·kg·yr). A higher value is obtained using the light channel because of the worse rejection efficiency. Improvement of the signal-to-noise ratio and application of digitizers with a high enough sampling frequency (both in the heat and light channel) are needed for the future high sensitivity cryogenic experiments to provide rejection of the background from randomly coinciding events.

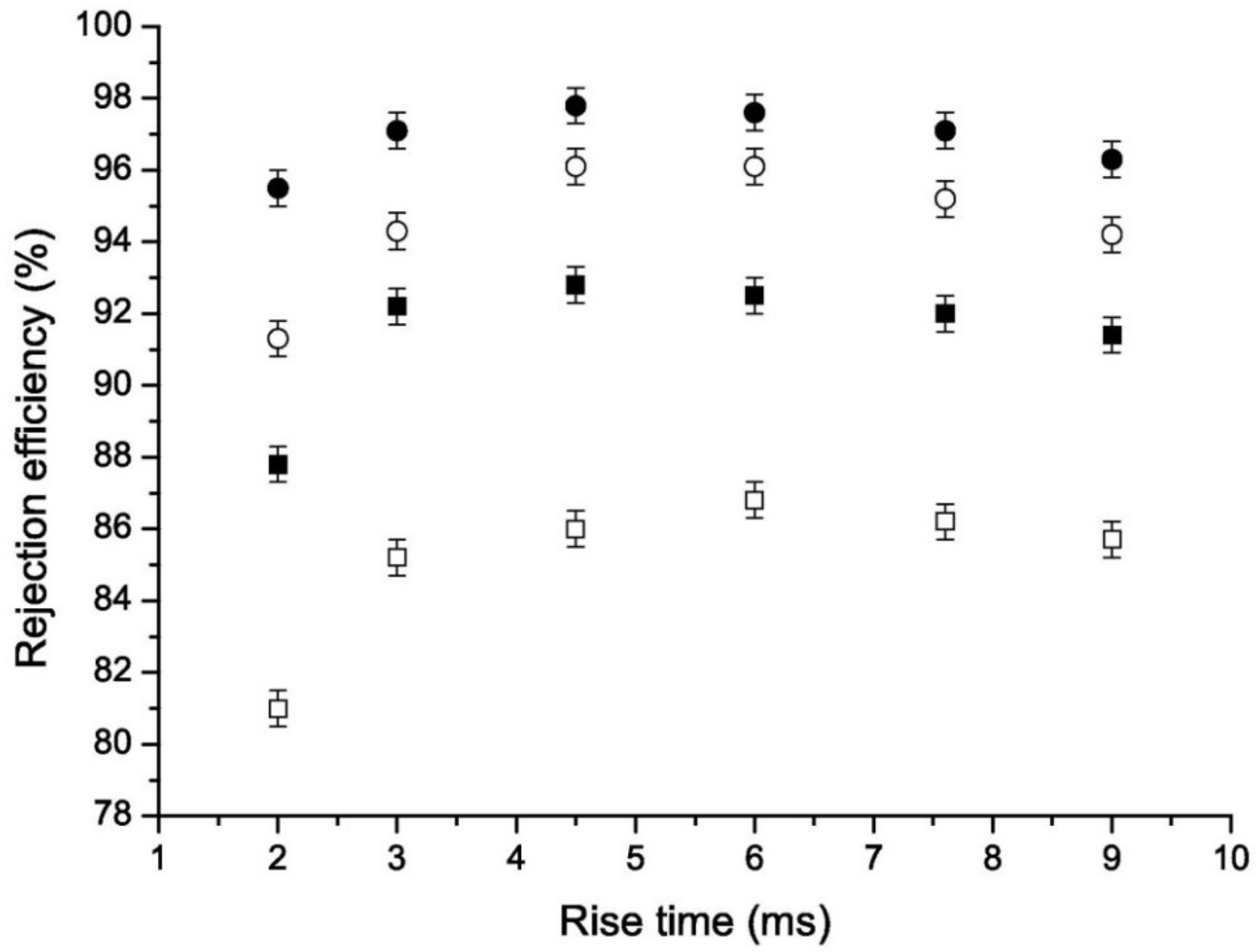

Fig. 4.12. Dependence of the rejection efficiency for the light pulses (by using the mean-time method) on the rise-time, signal-to-noise ratio and data acquisition sampling rate. The filled squares (circles) represent the data with a signal-to-noise ratio of 30 (100) acquired with a sampling rate of 1.9841 kSPS, while the open markers show results for the same signals acquired with a sampling rate of 0.9921 kSPS.

### 4.5. Application of the pulse-shape discrimination method

The developed pulse-shape discrimination technique was applied to the coinciding events. These events were accumulated with the 313 g $ZnMoO_4$ crystal scintillator operated as cryogenic scintillating bolometer at the LSM. Sampling rate of the data acquisition was 1.9841 kSPS. During the 19 hours of calibration measurement with a $^{232}$Th source we found six coincident gamma events from a heat channel which gave a signal in the energy range 2.8 MeV–3.2 MeV (see Fig. 4.13). We also selected 59 single heat pulses with the energy 2.6 MeV, and 1000 noise baselines from the same run.

[3] We safely assume that the rejection efficiency in the 45 ms time interval is the same as in the shorter $3.3 \cdot \tau_R$ time interval since the rejection efficiency starts to saturate, as it can be appreciated in Fig. 4.6.

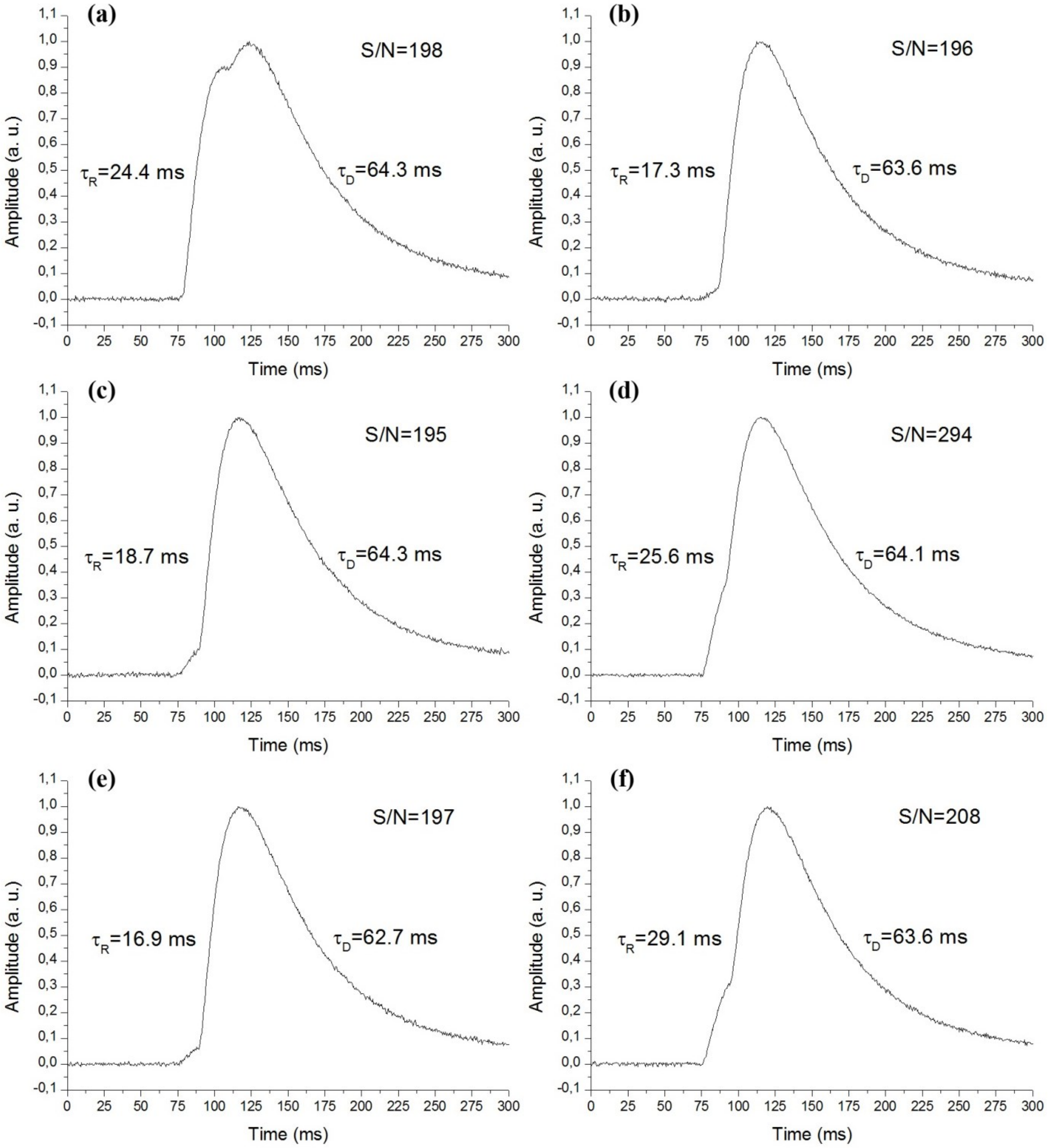

Fig. 4.13. Coinciding events accumulated in the heat channel of the 313 g $ZnMoO_4$ cryogenic scintillating bolometer at the LSM. S/N means signal-to-noise ratio, $\tau_R$ and $\tau_D$ denote rise- and decay-times, respectively. Amplitude of the pulses was normalized at its maximum.

Sum of 59 single pulse profiles was fitted by using function $f_S(t)$ defined in Eq. (4.10). Taking into account the average signal-to-noise ratio of the pulses, we fix the value of the signal-to-noise ratio at 200. Thousand single heat pulses and thousand coincident events were generated by using the procedure described in section 4.2.

Mean-time method was applied to reject the coinciding events. In order to achieve the highest possible rejection efficiency we performed an optimization of the time interval used to calculate the mean-time parameter. Under the requirement to accept 95% of single pulses as potentially good signals, we achieved the rejection efficiency of the generated coincident events on the level of (97.5 ± 0.6)%.

Accuracy of the mean-time method used for generated pulses was also checked on the 59 single events. With a same threshold on the mean-time parameter as for the generated signals we accepted exactly 95% of the single events (56 of the 59 events).

As a final goal we applied the pulse-shape discrimination technique to the coinciding events visible by eye. Rejection efficiency was 100% (see Fig. 4.14), i.e., all of the coinciding events were rejected with the help of the developed algorithm.

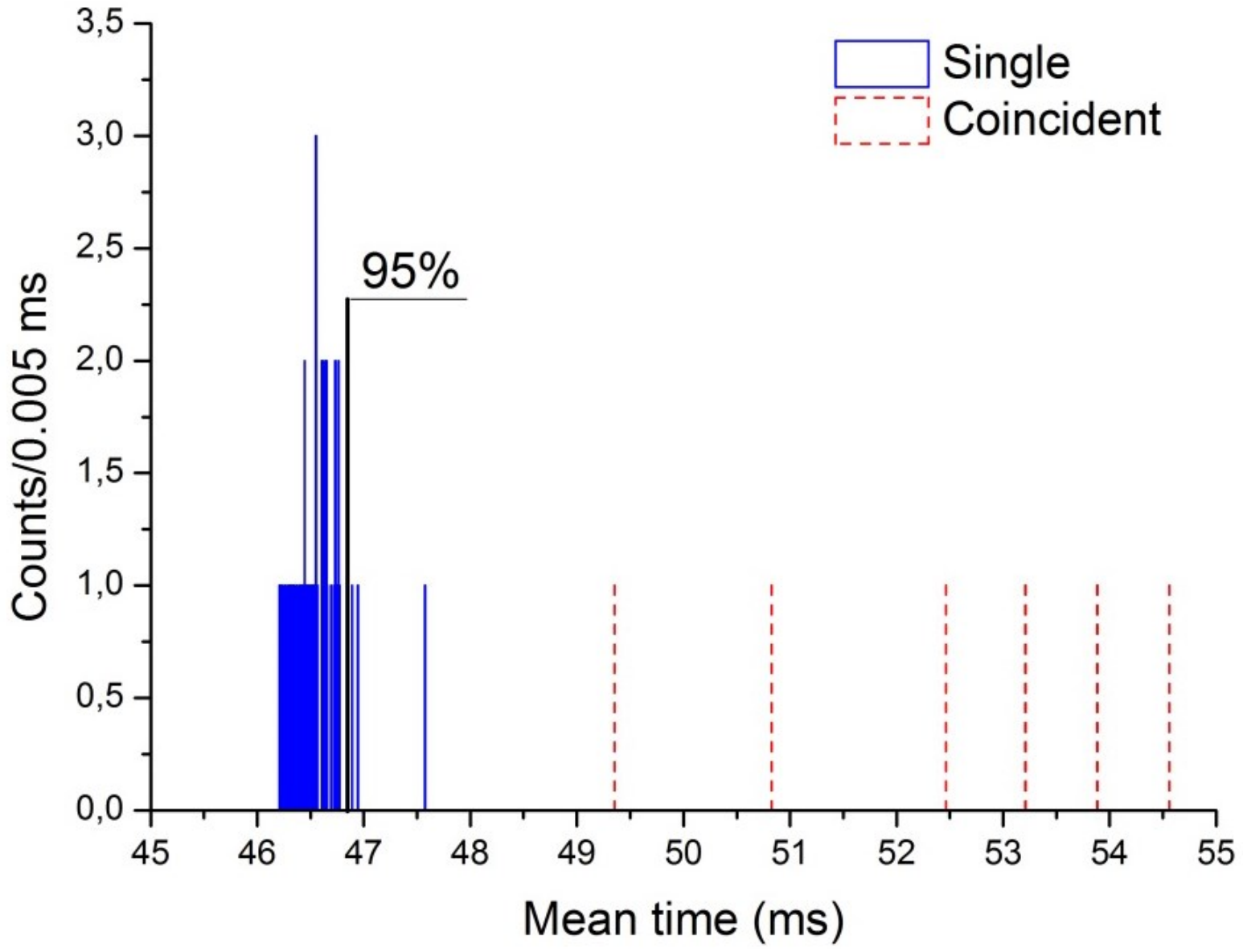

Fig. 4.14. Distribution of the mean-time parameter for the 59 single and 6 coinciding heat pulses. The events left from line are accepted as single events (95% of single events). All of the six coinciding pulses (dashed lines at the right) were rejected by using the mean-time method.

### 4.6. Conclusions and perspectives

Random coincidence of events (particularly from 2ν2β decay) could be one of the main sources of background in cryogenic bolometers to search for 0ν2β decay because of their poor time resolution. The background is expected to be especially significant in experiments with $^{100}$Mo due to comparatively short half-life of 2ν2β decay. Moreover, while other sources of background can be reduced by shielding of the set-up, underground placement of the experimental set-up, purification of detector materials, decreasing of the noise level, and application of anti-coincidence method, the random coincidence of 2ν2β events is an irremovable source of background, as it is related to the presence of the isotope under investigation in the detector compound. However, the background can be effectively suppressed with the help of pulse-shape discrimination.

The simulated randomly coinciding 2ν2β decay events were discriminated with an efficiency at the level of 99% by applying the mean-time method and $\chi^2$ approach to the heat signals from $ZnMoO_4$ cryogenic bolometer with a rise-time of 13.6 ms and a signal-to-noise ratio of 900, and at the level of 92% for the light signals with 3 ms rise-time and signal-to-noise ratio of 30. The front edge analysis provides similar results for the heat pulses, but slightly worse for the light signals due to much lower signal-to-noise ratio and low sampling frequency.

The signal-to-noise ratio is an important feature to reject randomly coinciding events, particularly in $ZnMoO_4$ due to the comparatively low light yield, which leads to a rather low signal-to-noise ratio in the light channel.

Development of algorithms to reconstruct the origin of a signal with as the highest possible accuracy is requested to improve the rejection capability of any pulse-shape discrimination technique. The sampling rate of the data acquisition should be high enough to provide effective pulse-shape discrimination of randomly coinciding events. Finally, any pulse-shape discrimination methods should be optimized taking into account certain detector performance to reduce the background effectively.

The developed pulse-shape discrimination technique was applied to the coinciding heat signals accumulated with the 313 g $ZnMoO_4$ cryogenic scintillating bolometer. All of the six pile-up pulses selected "by eye" from the data accumulated in a calibration run with a $^{232}$Th gamma source by the 313 g $ZnMoO_4$ cryogenic scintillating bolometer were rejected. Higher statistics of the coinciding events is necessary to improve the accuracy of the rejection efficiency estimation.

The counting rate of $Zn^{100}MoO_4$ cryogenic scintillating bolometers due to the random coincidences of events (particularly of the irrecoverable $2\nu2\beta$ decay) can be reduced from ≈ 0.02 counts/(keV·kg·yr) to the level ≈ $10^{-4}$ counts/(keV·kg·yr), which makes $Zn^{100}MoO_4$ cryogenic scintillating bolometers one of the most promising detectors to search for neutrinoless double beta decay at a level of sensitivity high enough to probe the inverted hierarchy region of the neutrino mass pattern [153].

# CONCLUSIONS

The recent progress in $ZnMoO_4$ crystal production was achieved by developing a two-stage molybdenum purification technique using double sublimation with addition of zinc molybdate and recrystallization from aqueous solutions of ammonium para-molybdate. This purification method combined with the low-thermal-gradient Czochralski technique gave a significant improvement of the $ZnMoO_4$ crystal quality. Natural $ZnMoO_4$ crystals with a mass of ~ 0.3 kg and two $Zn^{100}MoO_4$ scintillators enriched in $^{100}Mo$ up to 99.5% with a mass of 59 g and 63 g were produced. The output of the crystal boules was at the level of 80%–84%, the irrecoverable losses of enriched molybdenum were in the ranges of a few percent thanks to the advantage of the low-thermal-gradient Czochralski crystal growing technique. Recently a zinc molybdate crystal boule enriched in $^{100}Mo$ with a mass of ~1.4 kg was grown by the low-thermal-gradient Czochralski technique.

The optical and luminescence properties of the produced crystals confirmed improved quality of the detectors. The cryogenic bolometric measurements with the natural and enriched zinc molybdate crystals performed aboveground and underground demonstrated their excellent characteristics. The developed purification techniques significantly improved the radioactive contamination of $ZnMoO_4$ crystals by $^{228}Th$ and $^{226}Ra$ to the level of ≤ 0.005 mBq/kg.

Simulation of the light collection from $ZnMoO_4$ bolometers has shown the advantages of hexagonal (octahedral) crystal shape in comparison to cylindrical. To improve the light collection efficiency and uniformity the scintillator surface should be diffused. The Monte Carlo simulation of 2ν2β decay processes in $Zn^{100}MoO_4$ crystals of different shapes (hexagonal, octahedral, and cylindrical) demonstrated no significant dependence of the obtained spectra from the crystal shape.

Monte Carlo simulation of 48 $Zn^{100}MoO_4$ scintillating bolometers (with a mass 495 g, size Ø60 × 40 mm) in the EDELWEISS set-up was performed. The contamination of $Zn^{100}MoO_4$ crystal and nearest materials (copper holder, PTFE clamps, BoPET reflective foil) by $^{238}U$ and $^{232}Th$ daughters, and cosmogenic activation was simulated. A total background rate in the region of interest is $4.1 \times 10^{-4}$ counts/(keV·kg·yr) for $Zn^{100}MoO_4$ crystal scintillators.

Random coincidence of events (particularly of 2ν2β decay) in cryogenic bolometers could be one of the main sources of background due to the poor time resolution. The background due to the random coincidence of 2ν2β events is an irremovable background, as it is related to the presence of the isotope under investigation in the detector compound. However, this background can be reduced from ≈ 0.02 counts/(keV·kg·yr) to the level of $\approx 10^{-4}$ counts/(keV·kg·yr) at $Q_{2\beta}$ by applying developed pulse-shape discrimination. It should be stressed that the signal-to-noise ratio is one of the most important features to reject randomly coinciding events. Accurate algorithms to reconstruct the origin of signals are also of high importance. Moreover, the sampling rate of the data acquisition should be high enough to provide an effective rejection of randomly coinciding events. Any pulse-shape discrimination methods should be optimized taking into account certain detector performance.

Total background rate of $Zn^{100}MoO_4$ scintillating bolometers can be estimated as ≈ 5 × $10^{-4}$ counts/(keV·kg·yr) at $Q_{2\beta}$, including ≈ 4 × $10^{-4}$ counts/(keV·kg·yr) from radioactive sources and ≈ $10^{-4}$ counts/(keV·kg·yr) from random coincidence of events.

The current progress on the zinc molybdate crystal production, the performed low-temperature bolometric measurements and Monte Carlo simulation of the cryogenic scintillating bolometers allows us to conclude that cryogenic scintillating bolometers based on $Zn^{100}MoO_4$ crystals are excellent candidates for the next-generation large-scale experiment to search for neutrinoless double beta decay of $^{100}Mo$ with the aim to explore the inverted hierarchy region of neutrino mass pattern.

# ANNEX

Schemes of $^{238}U$ and $^{232}Th$ decay chains are courtesy provided by V.I. Tretyak.

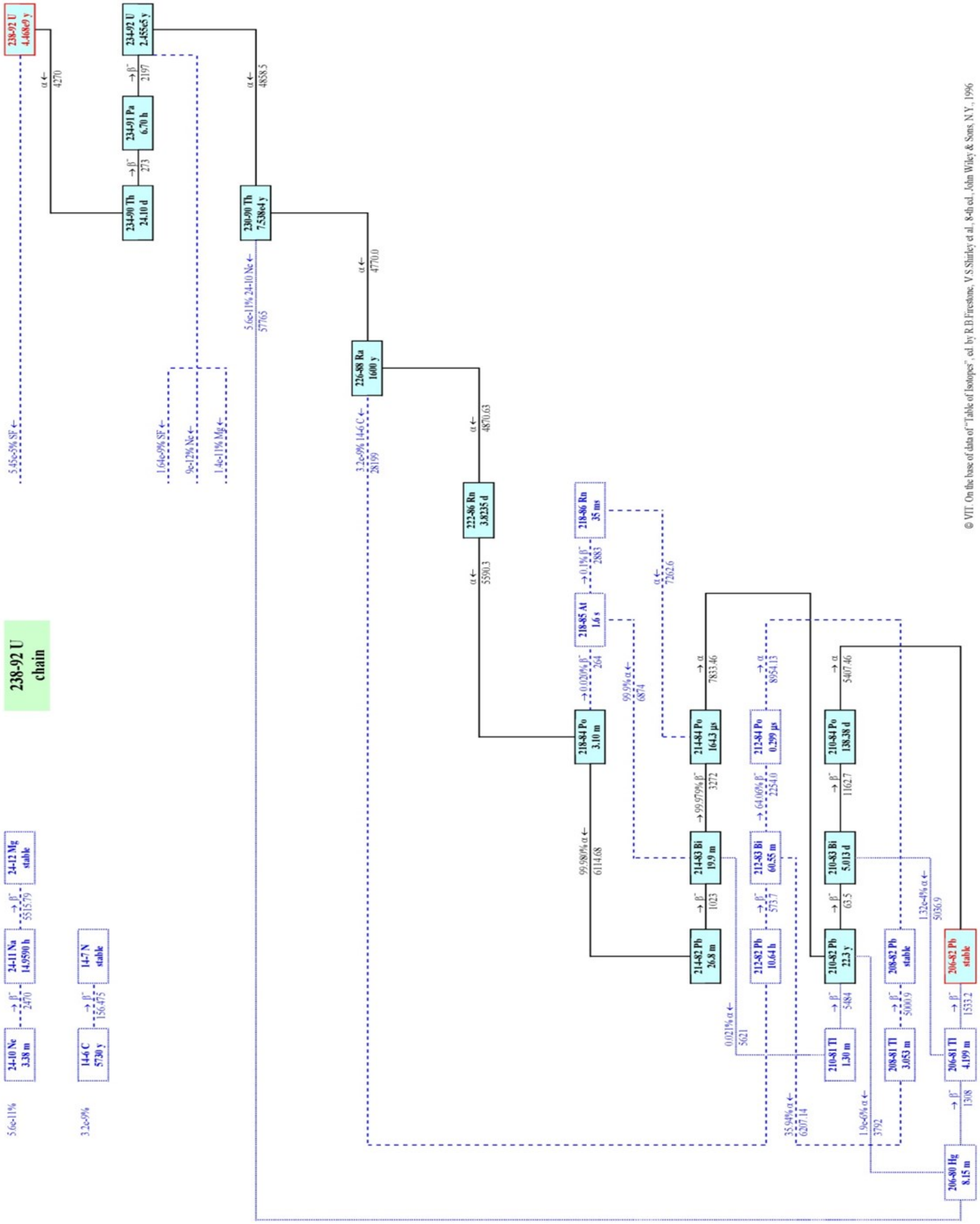

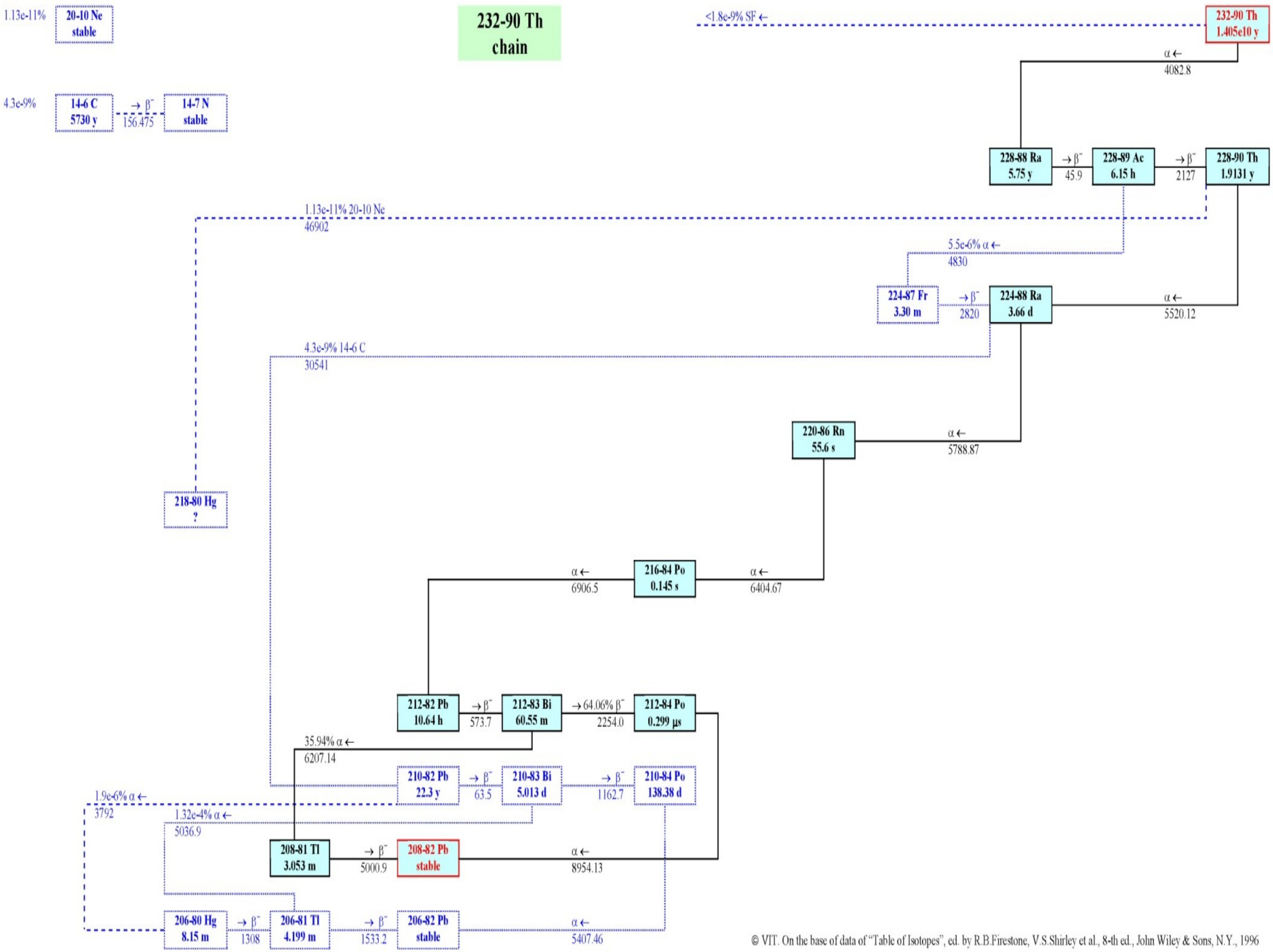
232-90 Th chain
232-90 Th 1.405e10 y
<1.8e-9% SF ←
α ← 4082.8
228-88 Ra 5.75 y
→ β⁻ 45.9
228-89 Ac 6.15 h
→ β⁻ 2127
228-90 Th 1.9131 y
1.13e-11% 20-10 Ne 46902
5.5e-6% α ← 4830
224-87 Fr 3.30 m
→ β⁻ 2820
224-88 Ra 3.66 d
α ← 5520.12
4.3e-9% 14-6 C 30541
220-86 Rn 55.6 s
α ← 5788.87
218-80 Hg ?
216-84 Po 0.145 s
α ← 6906.5
α ← 6404.67
212-82 Pb 10.64 h
→ β⁻ 573.7
212-83 Bi 60.55 m
→ 64.06% β⁻ 2254.0
212-84 Po 0.299 μs
35.94% α ← 6207.14
210-82 Pb 22.3 y
→ β⁻ 63.5
210-83 Bi 5.013 d
→ β⁻ 1162.7
210-84 Po 138.38 d
1.9e-6% α ← 3792
1.32e-4% α ← 5036.9
208-81 Tl 3.053 m
→ β⁻ 5000.9
208-82 Pb stable
α ← 8954.13
206-80 Hg 8.15 m
→ β⁻ 1308
206-81 Tl 4.199 m
→ β⁻ 1533.2
206-82 Pb stable
α ← 5407.46
1.13e-11%
20-10 Ne stable
4.3e-9%
14-6 C 5730 y
→ β⁻ 156.475
14-7 N stable